\documentclass[11pt]{article}

\usepackage[numbers]{natbib}
\usepackage{amsmath}
\usepackage{amssymb}
\usepackage{mathrsfs}
\usepackage{bbm}
\usepackage{amsthm}
\usepackage{xcolor}
\usepackage{tikz}
\usetikzlibrary{arrows.meta,positioning,calc}
\usepackage[section]{placeins}
\usepackage{array}
\usepackage{booktabs}
\usepackage{tabularx}
\usepackage{hyperref}
\usepackage{algorithm}
\usepackage{algpseudocode}
\algdef{SE}[DATASTRUCTURE]{DataStructure}{EndDataStructure}[1]
  {\textbf{data structure} \textsc{#1}}{\textbf{end data structure}}
\algdef{SE}[MEMBERS]{Members}{EndMembers}
  {\textbf{members}}{\textbf{end members}}
\usepackage{enumitem}\usepackage[margin=1in]{geometry}

\hypersetup{colorlinks=true,citecolor=blue,linkcolor=blue,hypertexnames=false}
\allowdisplaybreaks

\newcommand{\paperTitle}{Solving Linear Programs in the Matrix Multiplication Time}
\newcommand{\paperAuthor}{Zhao Song\thanks{\texttt{magic.linuxkde@gmail.com}.}}
 
\newcommand{\R}{\mathbb R}
\newcommand{\E}{\mathbb E}
\renewcommand{\d}{\mathrm{d}}
\DeclareMathOperator{\rank}{rank}

\newcommand{\Tmat}{\mathcal{T}_{\mathrm{mat}}}
\DeclareMathOperator{\new}{new}
\DeclareMathOperator{\old}{old}
\newcommand{\lev}{\mathrm{lev}}
\newcommand{\cnt}{\mathrm{cnt}}
\DeclareMathOperator{\nnz}{nnz}
\DeclareMathOperator{\poly}{poly}
\DeclareMathOperator{\OPT}{OPT}

\definecolor{lpBlue}{HTML}{255E91}
\definecolor{lpTeal}{HTML}{167A74}
\definecolor{lpOrange}{HTML}{AD5C21}
\definecolor{lpPurple}{HTML}{71528D}
\definecolor{lpGray}{HTML}{5C6670}

\tikzset{
  lpfig/base/.style={font=\normalsize,line width=.65pt,>=Stealth},
  lpfig/box/.style={draw=lpBlue,fill=lpBlue!5,rounded corners=2pt,
    align=center,inner sep=7pt},
  lpfig/teal/.style={draw=lpTeal,fill=lpTeal!7},
  lpfig/orange/.style={draw=lpOrange,fill=lpOrange!7},
  lpfig/purple/.style={draw=lpPurple,fill=lpPurple!6},
  lpfig/arrow/.style={->,line width=.85pt,draw=lpGray},
  lpfig/emph/.style={->,line width=1.3pt,draw=lpTeal},
  lpfig/note/.style={align=center,text=lpGray},
  lpfig/dot/.style={circle,fill=lpBlue,inner sep=2pt},
  lpfig/tree/.style={draw=lpBlue,fill=white,rounded corners=2pt,
    minimum width=1.25cm,minimum height=.55cm,align=center},
  lpfig/cell/.style={draw=lpGray,minimum width=.8cm,minimum height=.55cm,
    inner sep=2pt,align=center},
  lpfig/title/.style={font=\bfseries,align=center}
}

\theoremstyle{plain}
\newtheorem{theorem}{Theorem}[section]
\newtheorem{lemma}[theorem]{Lemma}
\newtheorem{definition}[theorem]{Definition}

\newtheorem{corollary}[theorem]{Corollary}
\theoremstyle{definition}
\newtheorem{openproblem}[theorem]{Open Problem}
\newtheorem{remark}[theorem]{Remark}

\begin{document}

\date{October 5, 2026}
\title{\paperTitle}
\author{\paperAuthor}
\maketitle
\begin{abstract}
We give a randomized algorithm for feasible bounded-box linear programs with $n$ variables and $d\le n=O(d)$ linearly independent equality constraints. For additive accuracy $\epsilon$, its expected running time is $O((n^\omega+n^2\log n)\log(1/\epsilon))$, where $\omega$ is the matrix multiplication exponent. This answers the matrix-multiplication-time open question posted in Song's Ph.D. thesis~\cite{son19} in this setting.
\end{abstract}
\newpage
 
\section{Introduction}\label{sec:square_problem}

Linear programming asks us to optimize a linear function subject to linear
constraints. For a program with $n$ variables and $d$ independent equality
constraints, we study arithmetic running-time bounds that reflect both the
cost of matrix multiplication and the sparsity of the constraint matrix.
The central question is how closely the cost of solving a linear program
can approach the cost of the linear algebra associated with its input.

The search for efficient LP algorithms began with Dantzig's simplex
method~\cite{d51}. Klee and Minty~\cite{km72} demonstrated its exponential
worst-case behavior under standard pivot rules. Khachiyan
established polynomial-time solvability using the ellipsoid method~\cite{k79}.
Karmarkar introduced a polynomial-time interior-point algorithm~\cite{k84}.
Renegar developed a path-following method~\cite{r88}, and Nesterov and
Nemirovskii established the theory of self-concordant barriers and path
following~\cite{nn89,nn91,nn94}.
Further developments include homogeneous self-dual formulations~\cite{ytm94}
and alternative potential functions and search directions~\cite{prt02}.

Progress toward faster LP algorithms has relied on reducing both the
number of iterations and the cost of the linear algebra in each iteration. Vaidya accelerated LP algorithms through inverse maintenance and fast
matrix multiplication~\cite{v87,v89b}, while volumetric and cutting-plane
methods provided another approach~\cite{v89a,lsw15,jlsw20}. Lee and Sidford
combined improved path following with efficient inverse maintenance to
exploit the rank and sparsity of the constraint matrix~\cite{ls13,ls14,ls15,ls19}.

For dense LPs, Cohen, Lee, and Song~\cite{cls19} developed a stochastic
central path method with projection maintenance. Their running-time
exponent matched the matrix-multiplication exponent known at the time. Subsequent work
developed deterministic~\cite{b20} and oblivious-sketching~\cite{sy21}
variants, extended the approach to empirical risk minimization~\cite{lsz19},
and improved dynamic inverse maintenance for general LPs~\cite{jswz21}.
For sufficiently tall dense LPs, Brand, Lee, Sidford, and Song~\cite{blss20}
obtained nearly linear time. Brand, Lee, Liu, Saranurak, Sidford, Song, and Wang~\cite{bllsssw21} subsequently
improved the bounds for LPs with two-sided constraints through robust
stochastic path following and dynamic Lewis-weight maintenance.
More recently, Song~\cite{s26b} reduced the logarithmic overhead in the
Lee--Sidford linear-system-solve bound while retaining its square-root
dependence on rank, assuming an interior point and a certified magnitude bound.

\paragraph{Fast matrix multiplication.}
The running time of dense LP algorithms is commonly expressed in terms
of the cost of matrix multiplication. Let $\omega$ denote the exponent of square matrix multiplication
in the arithmetic-operation model. Strassen~\cite{s69} showed that two
$2\times2$ matrices can be multiplied with seven scalar multiplications,
giving $\omega\leq\log_2 7<2.808$. Further reductions relied on algebraic
constructions: Pan's trilinear aggregation~\cite{p78}, approximate bilinear
algorithms~\cite{bcrl79}, Bini's conversion from approximate to exact
algorithms~\cite{b80}, and Sch\"onhage's asymptotic sum inequality~\cite{s81}.
Strassen's laser method~\cite{s87} and the Coppersmith--Winograd
construction~\cite{cw90} led to $\omega<2.376$.

A series of works sharpened the analysis of these constructions. Analyses
of higher tensor powers by Stothers~\cite{s10}, Vassilevska
Williams~\cite{w12}, and Le Gall~\cite{g14} improved the bound on $\omega$.
The refined laser method of Alman and Vassilevska Williams~\cite{aw21},
asymmetric hashing of Duan, Wu, and Zhou~\cite{dwz23}, and further asymmetric
analyses~\cite{wxxz24,adwxxz25} reduced it to $\omega<2.371339$.
Dupont et al.~\cite{dekmrszawb26} used modern optimization and AlphaEvolve
to obtain $\omega<2.371177$.

The square exponent alone does not describe every matrix product used by
LP algorithms. For rectangular products, write $\omega(1,k,1)$ for the
exponent of multiplying an $n\times n^k$ matrix by an $n^k\times n$ matrix,
and let $\alpha:=\sup\{k\geq0:\omega(1,k,1)=2\}$ be the dual exponent.
Early rectangular bounds were developed by Lotti and Romani~\cite{lr83}
and Coppersmith~\cite{c97}. Le Gall~\cite{g12}, Le Gall and Urrutia~\cite{gu18},
and Vassilevska Williams, Xu, Xu, and Zhou~\cite{wxxz24} successively improved
the known lower bound on $\alpha$ to $0.30298$, $0.31389$, and $0.321334$.

Recent advances improve both square and rectangular products, with
different assumptions on the underlying field. OpenAI's staggered
extraction preprint~\cite{o26c} gives $\omega<2.371054887$ over every fixed
field. A separate construction~\cite{o26b} gives $\omega<2.258$ and
$\alpha>0.465$ in characteristic zero, as well as
$\omega(1,0.709,1)<2.092$. A later argument using polynomial multiplication
and Strassen's asymptotic spectrum~\cite{s88,o26a} gives $\omega\leq9/4$
over $\mathbb C$. Uzna\'nski~\cite{uzn26} extended this approach to
rectangular products in characteristic zero, obtaining $\alpha\geq1/2$
and $\omega(1,k,1)\leq1+k+1/(4k)$ for $k\geq1/2$.

The algorithmic form and field dependence of these bounds have also been
clarified. Pass~\cite{p26} gave a constructive proof using polynomial
packings and explicit recursive matrix-multiplication identities.
Amanov, van den Berg, and Zuiddam~\cite{abz26} gave an exposition extending
the square and rectangular bounds to every field, and proved that their
specified degeneration ingredients cannot yield better bounds.
Shpilka~\cite{s26} gave an elementary, self-contained proof with explicit
algorithms. These bounds allow an arbitrarily small additive loss in the
arithmetic exponent.

\subsection{Key research questions}\label{sec:key_research_questions}

A central open question is whether linear programming can be solved in
matrix-multiplication time~\cite{son19}, even if matrix multiplication
could be performed in $O(n^2)$ time. Several dense LP bounds in
Table~\ref{tab:lp_running_time} contain an $n^{5/2-\alpha/2}$ term and a fixed
power above $n^2$, in addition to $n^\omega$. These additional terms can
prevent the running time from following further improvements in $\omega$.
For general $d\leq n$, the $nd+d^{5/2}$ bound also leaves a question about
sparsity: its $nd$ term does not decrease when the matrix has fewer nonzero
entries. We therefore ask:
\begin{enumerate}[leftmargin=*]
\item Can dense linear programs be solved in matrix-multiplication time,
up to logarithmic and accuracy factors, even if $\omega$ approaches $2$?
\item For general $d\leq n$, can the $nd$ term in the running time be
reduced when the constraint matrix is sparse, while retaining efficient
dependence on $d$?
\end{enumerate}
Our two results address these questions in different parameter regimes.

\subsection{Our results}\label{sec:our_results}

\paragraph{Dense linear programming.}
Our first result, developed in Part~\ref{part:dense_lp}, answers the
question from Song's thesis~\cite{son19} affirmatively, up to logarithmic and accuracy factors, for feasible
bounded-box linear programs with $d\leq n=O(d)$. The running time has no
polynomial term above $n^2$ other than $n^\omega$, so it continues to follow
improvements in the square matrix-multiplication exponent. The formal
statement appears in Section~\ref{sec:lp_reduction}.

\begin{theorem}[Informal version of Theorem~\ref{thm:lp_reduction}]\label{thm:main_informal}
Let $C_0$ be a fixed constant, and let $A\in\R^{d\times n}$,
$b\in\R^d$, and $c,\ell,u\in\R^n$ satisfy $\rank(A)=d$ and
$d\leq n\leq C_0d$. This is the $n=O(d)$ setting, including
instances with $n-d=\Theta(n)$. Suppose $\ell<u$ and the bounded-box
linear program
\[
 \OPT:=\min\{c^\top x:Ax=b,\ \ell\leq x\leq u\}
\]
is feasible. For any accuracy parameter $0<\epsilon\leq1$,
a randomized algorithm terminates almost surely and returns a vector
$x_{\rm out}$ satisfying
\[
 \ell\leq x_{\rm out}\leq u,\qquad
 \|Ax_{\rm out}-b\|_\infty\leq\epsilon,\qquad
 c^\top x_{\rm out}\leq\OPT+\epsilon.
\]
Its expected exact real-arithmetic time is
\[
 O((n^\omega+n^2\log n)\log(nW/\epsilon)),
\]
where
\[
 W:=\max\{2,\|A\|_1,\|c\|_\infty,\|\ell\|_\infty,\|u\|_\infty\},
\]
and $\|A\|_1:=\sum_{i=1}^d\sum_{j=1}^n|A_{ij}|$ is the entrywise $\ell_1$ norm.
\end{theorem}

\noindent\textbf{Remark (Comparison with Cohen, Lee, and Song).}
With the currently known bounds $\omega\leq9/4$ and $\alpha\geq1/2$
of~\cite{uzn26}, both our bound and that of Cohen, Lee, and
Song~\cite{cls19} have dimension exponent $9/4$, under the exponent
convention of Table~\ref{tab:lp_running_time} and suppressing logarithmic
and accuracy factors. Any improvement in $\omega$ below $9/4$ improves
our exponent. If we continue to use $1/2$ as the lower bound on $\alpha$, the
term $n^{5/2-\alpha/2}$ in~\cite{cls19} still yields an $n^{9/4}$
contribution to the resulting upper bound.

The additive $n^2\log n$ term is lower order than $n^\omega$ for every fixed
$\omega>2$; at $\omega=2$, it adds only a logarithmic factor under the
exponent convention of Table~\ref{tab:lp_running_time}. In particular, the
bound contains neither a separate dependence on $\alpha$ nor an additional
fixed power above $n^2$. Section~\ref{sec:lp_proofs} gives the dense LP
algorithm and its analysis. The supporting data structures are developed
in Sections~\ref{sec:lp_exact_ip}, \ref{sec:lp_proposal_tree},
\ref{sec:lp_exact_l2_sampler}, and~\ref{sec:lp_entry_state}.

\paragraph{Sparse-input linear programming.}
Part~\ref{part:sparse_lp} presents our second result. It applies to
bounded-box linear programs with general $d\leq n$ and makes the running time depend on the number of nonzero entries
of the constraint matrix. Its assumptions and full statement appear in
Section~\ref{sec:tall_sparse_lp}.

\begin{theorem}[Informal version of Theorem~\ref{thm:tall_sparse_bounded_box_lp}]\label{thm:tall_main_informal}
Let $A\in\R^{d\times n}$ have full row rank and no zero columns, with
$n\geq d\geq2$, and suppose $\ell<u$ and the program
$\OPT:=\min\{c^\top x:Ax=b,\ \ell\leq x\leq u\}$ is feasible.
For $0<\epsilon\leq1$, there is an algorithm returns a vector $x_{\rm out}$ satisfying
\[
 \ell\leq x_{\rm out}\leq u,\qquad
 \|Ax_{\rm out}-b\|_\infty\leq\epsilon,\qquad
 c^\top x_{\rm out}\leq\OPT+\epsilon.
\]
Its expected running time is
\[
 \widetilde O((\sqrt{\nnz(A)nd}+d^{5/2})
 \poly \log(nW/\epsilon)).
\]
Here $\nnz(A)$ is the number of nonzero entries of $A$, and
\[
 W:=\max\{2,\|A\|_{\max},\|b\|_\infty,\|c\|_\infty,
 \|\ell\|_\infty,\|u\|_\infty,
 \frac{\max_{i\in[n]}\{u_i-\ell_i\}}{\min_{i\in[n]}\{u_i-\ell_i\}}\},
\]
where $\|A\|_1:=\sum_{i=1}^d\sum_{j=1}^n|A_{ij}|$ is the entrywise $\ell_1$ norm.
\end{theorem}

Suppressing logarithmic and accuracy factors, the bound is
$\sqrt{\nnz(A)nd}+d^{5/2}$. For dense matrices with $\nnz(A)=\Theta(nd)$,
this is $nd+d^{5/2}$. When $\nnz(A)=O(n)$, the bound becomes
$O(n\sqrt d+d^{5/2})$. More generally, the first term improves on $nd$ by
a factor of $\sqrt{nd/\nnz(A)}$; the $d^{5/2}$ term remains. Thus this
result complements the dense bound by exploiting sparsity for general
$d\leq n$. Table~\ref{tab:lp_running_time} places both results alongside
selected previous bounds.

\begin{table}[!ht]
\centering
\begin{tabular}{|l|l|l|l|}
\hline
Year & Authors & Reference & Running time \\
\hline
1979 & Khachiyan & \cite{k79} & $n^4$ \\
1984 & Karmarkar & \cite{k84} & $n^{7/2}$ \\
1987 & Vaidya & \cite{v87} & $n^3$ \\
1989 & Vaidya & \cite{v89b} & $n^{5/2}$ \\
2015 & Lee, Sidford & \cite{ls15} & $n^{5/2}$ \\
2019 & Cohen, Lee, Song & \cite{cls19}
  & $n^\omega+n^{5/2-\alpha/2}+n^{2+1/6}$ \\
2019 & Lee, Song, Zhang & \cite{lsz19}
  & $n^\omega+n^{5/2-\alpha/2}+n^{2+1/6}$ \\
2019 & Lee, Sidford & \cite{ls19} & $\sqrt d T_{\rm solve}\log^{13}n$ \\
2020 & Brand & \cite{b20}
  & $n^\omega+n^{5/2-\alpha/2}+n^{2+1/6}$ \\
2020 & Brand, Lee, Sidford, Song
  & \cite{blss20} & $nd+d^3$ \\
2021 & Song, Yu & \cite{sy21}
  & $n^\omega+n^{5/2-\alpha/2}+n^{2+1/6}$ \\
2021 & Jiang, Song, Weinstein, Zhang & \cite{jswz21}
  & $n^\omega+n^{5/2-\alpha/2}+n^{2+1/18}$ \\
2021 & \begin{tabular}[t]{@{}l@{}}Brand, Lee, Liu, Saranurak, \\ Sidford, Song, Wang\end{tabular}
  & \cite{bllsssw21} & $nd+d^{5/2}$ \\
2026 & Song & \cite{s26b} & $\sqrt d T_{\rm solve}\log^4n$ \\
2026 & This work & Theorem~\ref{thm:main_informal}
  & $n^\omega+n^2\log n$ \\
2026 & This work & Theorem~\ref{thm:tall_main_informal}
  & $\sqrt{\nnz(A)nd}+d^{5/2}$ \\
\hline
\end{tabular}
\caption{Selected arithmetic running times for linear programming.
Here $n$ is the number of variables and $d$ is the rank of the
equality-constraint matrix. Rows involving $d$ retain the rank dependence
for $d\leq n$; the other rows give the dense $d=\Theta(n)$ bounds.
The quantity $T_{\rm solve}\geq\nnz(A)+n$ is the arithmetic cost of computing
$(ADA^\top)^{-1}q$ for $A\in\R^{d\times n}$, an input positive diagonal
matrix $D\in\R^{n\times n}$, and $q\in\R^d$, including the required matrix
and vector operations.
The bounds in~\cite{ls19,s26b} assume a supplied interior point;
the bound in~\cite{s26b} also assumes a certified input-magnitude bound.
We omit asymptotic constants, precision factors, and logarithmic factors
except those displayed. Matrix-multiplication terms include an $n^{o(1)}$
allowance. The rank-dependent bounds apply to standard-form LPs
in~\cite{blss20} and to LPs with two-sided coordinate bounds
in~\cite{ls19,bllsssw21,s26b}.}
\label{tab:lp_running_time}
\end{table}

\subsection{Open problems}
\label{sec:open_problems}

Beyond the goal of solving LPs in input-sparsity time, a major open
question in fast linear programming is whether a strongly polynomial
algorithm exists~\cite{son19}. We first formulate the input-sparsity
target for bounded-box LPs: time linear in the number of nonzero entries,
plus a matrix-multiplication term depending only on the rank.

\begin{openproblem}[Input sparsity and matrix multiplication]\label{op:input_sparsity_lp}
Under the assumptions of Theorem~\ref{thm:tall_main_informal}, can we obtain
the same feasibility and objective guarantees, with probability at least
$2/3$, in expected
\[
 \widetilde O((\nnz(A)+d^\omega)\poly \log(nW/\epsilon))
\]
arithmetic operations, including initialization and construction of
$x_{\rm out}$? Here $W$ is defined in Theorem~\ref{thm:tall_main_informal},
$\widetilde O$ suppresses logarithmic factors in the dimensions, and
$d^\omega$ uses the same exponent convention as above.
\end{openproblem}

This asks whether the dependence on $\sqrt{\nnz(A)nd}$ and $d^{5/2}$
can be reduced to $\nnz(A)$ and $d^\omega$, respectively. Such a bound
would combine linear dependence on the number of nonzero entries with
matrix-multiplication time in the number of equality constraints.

The second question asks for a more fundamental improvement: can the
arithmetic-operation count depend only on the problem dimensions?
Polynomial-time solvability does not settle this question, since a
polynomial running-time bound may still depend on the bit lengths of the
coefficients. Our bounds retain a dependence on numerical scale and
accuracy. In his 1998 list of eighteen mathematical problems for the
twenty-first century, Smale included linear programming as
Problem~9~\cite{s98}. His stated formulation asks for a polynomial-time
algorithm over the real numbers that decides feasibility of linear
inequalities. The strongly polynomial formulation below asks for an exact
solution to a rational LP, with an arithmetic-operation count independent
of coefficient bit lengths and a polynomial bound on intermediate encoding
lengths.

\begin{openproblem}[Strongly polynomial linear programming]\label{op:strongly_polynomial_lp}
Is there an algorithm that solves every rational linear program with
$n$ variables and $d$ constraints using $\poly(n,d)$
elementary arithmetic operations and comparisons, independently of the
bit lengths of its coefficients? The algorithm must return an exact
optimal solution or correctly report infeasibility or unboundedness,
and the binary encoding length of every intermediate rational number
must remain polynomial in the input encoding length.
\end{openproblem}

For this question, the number of operations must depend only on the
problem dimensions. In particular, obtaining such a bound requires
removing the dependence on numerical scale and accuracy from the
operation count while still recovering an exact solution.

The third question concerns the number of linear-system solves needed
for LP. Song~\cite{s26b} conjectures that the remaining $\log^4 n$
overhead in his linear-system-solve bound can be removed.

\begin{openproblem}[Removing logarithmic overhead in linear-system solves]
\label{op:lp_linear_system_solves}
Under the assumptions of Theorem~1.1 of~\cite{s26b}, including a supplied
interior point and a finite certified magnitude bound $U$, can an LP
with $n$ variables and equality constraints $Ax=b$, where
$A\in\R^{d\times n}$ has rank $d$, be solved using
\[
 O(\sqrt d\log(nU/\epsilon))
\]
linear-system solves in $ADA^\top$ for positive diagonal matrices $D$?
The algorithm should return a feasible $x_{\rm out}$ with
$c^\top x_{\rm out}\leq\OPT+\epsilon$ with constant probability.
Here $U$ has the meaning specified in the cited theorem.
\end{openproblem}

\section{Preliminaries}\label{sec:preliminaries}

In Section~\ref{subsec:prelim_notation}, we fix notation and specify the arithmetic model.
In Section~\ref{sec:basic_probability_tools}, we apply the Chernoff bound to prove concentration of squared Gaussian norms.
In Section~\ref{subsec:prelim_bregman}, we introduce the Bregman divergence used to analyze coordinate updates.
In Section~\ref{subsec:prelim_matrix_bounds}, we give the arithmetic costs of matrix multiplication and inversion.

\subsection{Notation}\label{subsec:prelim_notation}
For a positive integer $n$, write $[n]:=\{1,\ldots,n\}$. Vectors are column vectors. We write ${\bf 0}_n$ and ${\bf 1}_n$ for the length-$n$ all-zeros and all-ones vectors, respectively. For an event or condition $E$, its indicator $\mathbbm{1}[E]$ equals $1$ if $E$ holds and $0$ otherwise. The superscript $\top$ denotes transpose.

For $x,y\in\R^n$, their Euclidean inner product is $\langle x,y\rangle:=x^\top y=\sum_{i=1}^n x_i y_i$. We use the vector norms $
 \|x\|_1:=\sum_{i=1}^n |x_i|$, $
 \|x\|_2:=(\sum_{i=1}^n x_i^2)^{1/2}$, $
 \|x\|_\infty:=\max_{i\in[n]}\{|x_i|\}$,
called the $\ell_1$, Euclidean (or $\ell_2$), and $\ell_\infty$ norms, respectively.

For $A\in\R^{m\times n}$, define $
 \|A\|_{\rm op}:=\sup_{\|x\|_2=1}\|Ax\|_2$, $
 \|A\|_F:=(\sum_{i=1}^m\sum_{j=1}^n A_{ij}^2)^{1/2}$, $
 \|A\|_1:=\sum_{i=1}^m\sum_{j=1}^n|A_{ij}|$.
These are the spectral (Euclidean operator), Frobenius, and entrywise $\ell_1$ norms, respectively.

For $A\in\R^{m\times n}$, its range (column space) is $\operatorname{range}(A):=\{Ax:x\in\R^n\}\subseteq\R^m$, and its kernel (null space) is $\ker(A):=\{x\in\R^n:Ax={\bf 0}_m\}$. We write $\operatorname{rank}(A):=\dim(\operatorname{range}(A))$ for its rank.

We write $I$ for the identity matrix and $e_i$ for the $i$th standard basis vector, with dimensions determined by context. For index sets $J$ and $K$, $x_J$ denotes a subvector and $A_{J,K}$ denotes a submatrix; a colon in an index position means all indices in that position. The notation $\operatorname{diag}(x)$ denotes the diagonal matrix with diagonal $x$.

For $A\in\R^{m\times n}$ and $B\in\R^{p\times q}$, define
\[
\mathrm{blockDiag}(A,B):=
\begin{pmatrix}
A & 0_{m\times q}\\
0_{p\times n} & B
\end{pmatrix}
\in\R^{(m+p)\times(n+q)}.
\]
When padding $A\in\R^{n\times n}$ to dimension $n'\ge n$, the notation $\mathrm{blockDiag}(A,0)$ uses a zero block of size $(n'-n)\times(n'-n)$; for $n'=n$, it equals $A$.

For vectors $x,y\in\R^n$, we define their coordinatewise product $xy\in\R^n$ by $(xy)_i:=x_i y_i$ for every $i\in[n]$. Vector inequalities such as $x>{\bf 0}_n$ are interpreted coordinatewise.

We write $\Pr[\mathcal{A}]$ for the probability of an event $\mathcal{A}$ and $\E[X]$ for the expectation of a random variable $X$. The notation $\log$ denotes the natural logarithm; other bases are indicated explicitly, as in $\log_2$.

\paragraph{Arithmetic model}
We count exact real arithmetic, positive square roots, and independent
standard Gaussian and uniform draws.
Dense multiplication and inversion of order $n$ cost $O(n^\omega)$,
with the exponent allowance stated in Section~\ref{sec:square_problem}.
The inversion bound follows from the standard block reduction to multiplication.

\subsection{Probability bounds}
\label{sec:basic_probability_tools}
We use the exponential-moment form of the Chernoff
bound~\cite{c52}. The next corollary applies this bound to
squared Gaussian norms.

\begin{lemma}[Chernoff bound]\label{lem:chernoff_bound}
Let $X$ be a real-valued random variable. For every $a\in\R$ and $\lambda>0$,
\begin{equation}\label{eq:chernoff_bound}
\Pr[X\ge a]\le e^{-\lambda a}\E[e^{\lambda X}],\qquad
\Pr[X\le a]\le e^{\lambda a}\E[e^{-\lambda X}],
\end{equation}
whenever the respective expectation is finite. If $X=\sum_{i=1}^k X_i$ for independent random variables $X_1,\ldots,X_k$, then
$\E[e^{\theta X}]=\prod_{i=1}^k\E[e^{\theta X_i}]$
whenever these expectations are finite.
\end{lemma}
\begin{proof}
For $\lambda>0$, monotonicity of the exponential gives
$\Pr[X\ge a]=\Pr[e^{\lambda X}\ge e^{\lambda a}]$.
Markov's inequality gives the first bound in Eq.~\eqref{eq:chernoff_bound}.
Applying that bound to $-X$ at threshold $-a$ gives the second bound.
The product formula follows from independence.
\end{proof}

\begin{corollary}[Gaussian squared-norm concentration]\label{cor:gaussian_squared_norm}
Let $k\ge1$ be an integer, and let $g\in\R^k$ have independent standard Gaussian entries.
For every $0<\delta<1$,
\[
\Pr[|\|g\|_2^2/k-1|\ge\delta]\le2e^{-k\delta^2/8}.
\]
\end{corollary}
\begin{proof}
Define $Z:=\|g\|_2^2$. Direct integration of the standard Gaussian density gives
$\E[e^{\theta g_i^2}]=(1-2\theta)^{-1/2}$ for $\theta<1/2$.
Independence therefore gives
\begin{equation}\label{eq:gaussian_squared_norm_mgf}
\E[e^{\theta Z}]=(1-2\theta)^{-k/2}\qquad(\theta<1/2).
\end{equation}
For $0<\delta<1$, we have
\begin{align}
\delta-\log(1+\delta)&=\int_0^\delta\frac{t}{1+t}\d t\ge\frac{\delta^2}{4},\notag\\*
-\delta-\log(1-\delta)&=\int_0^\delta\frac{t}{1-t}\d t\ge\frac{\delta^2}{2}.\label{eq:gaussian_logarithm_bounds}
\end{align}
The identities follow by integration. The first inequality uses $1+t\le2$,
and the second uses $1-t\le1$.
Set $\lambda_+:=\delta/(2(1+\delta))$ and
$\lambda_-:=\delta/(2(1-\delta))$. For the upper tail,
\begin{align*}
\Pr[Z\ge(1+\delta)k]
&\le e^{-\lambda_+(1+\delta)k}\E[e^{\lambda_+Z}]
=e^{-\lambda_+(1+\delta)k}(1-2\lambda_+)^{-k/2}\\*
&=e^{-k(\delta-\log(1+\delta))/2}
\le e^{-k\delta^2/8}.
\end{align*}
The first step uses Lemma~\ref{lem:chernoff_bound}.
The second uses Eq.~\eqref{eq:gaussian_squared_norm_mgf}.
The third substitutes $\lambda_+$, and the fourth uses the first inequality in Eq.~\eqref{eq:gaussian_logarithm_bounds}.
For the lower tail,
\begin{align*}
\Pr[Z\le(1-\delta)k]
&\le e^{\lambda_-(1-\delta)k}\E[e^{-\lambda_-Z}]
=e^{\lambda_-(1-\delta)k}(1+2\lambda_-)^{-k/2}\\*
&=e^{-k(-\delta-\log(1-\delta))/2}
\le e^{-k\delta^2/4}.
\end{align*}
The first step uses Lemma~\ref{lem:chernoff_bound}.
The second uses Eq.~\eqref{eq:gaussian_squared_norm_mgf}.
The third substitutes $\lambda_-$, and the fourth uses the second inequality in Eq.~\eqref{eq:gaussian_logarithm_bounds}.
The union bound gives
\begin{align*}
\Pr[|Z/k-1|\ge\delta]
&\le\Pr[Z\ge(1+\delta)k]+\Pr[Z\le(1-\delta)k]\\*
&\le e^{-k\delta^2/8}+e^{-k\delta^2/4}
\le2e^{-k\delta^2/8}.
\end{align*}
The second step uses the two tail estimates just proved.
The third uses $e^{-k\delta^2/4}\le e^{-k\delta^2/8}$.
This proves Corollary~\ref{cor:gaussian_squared_norm}.
\end{proof}

\subsection{Bregman divergence}\label{subsec:prelim_bregman}
Bregman divergences originate in Bregman's work on convex feasibility and
convex programming~\cite{b67}.

\begin{definition}[Bregman divergence]\label{def:bregman_divergence}
Let $\Omega\subseteq\R^n$ be an open convex set, and let
$h:\Omega\to\R$ be a differentiable strictly convex function.
The Bregman divergence generated by $h$ is
\[
 \mathscr{J}_h(x,y):=h(x)-h(y)-\langle\nabla h(y),x-y\rangle,\qquad x,y\in\Omega.
\]
\end{definition}
It measures the gap between $h(x)$ and the affine approximation to $h$ at
$y$. Convexity gives $\mathscr{J}_h(x,y)\ge0$, with equality only when $x=y$ by
strict convexity, although it is generally asymmetric and need not satisfy
the triangle inequality. The choice $h(x)=\|x\|_2^2/2$ gives
$\mathscr{J}_h(x,y)=\|x-y\|_2^2/2$. Mirror descent uses this divergence in place of
a quadratic proximity term to adapt updates to the geometry specified by
$h$~\cite{bt03}.
Bregman geometry also appears in high-precision computation of
$\ell_p$-Lewis weights: Song~\cite{s26c} combines fourth-moment control
with a restarted accelerated Bregman-gradient method for $p\ge4$.
In that setting, a separable reference function gives local relative
smoothness and strong convexity, enabling acceleration after a global
initialization phase.
In Section~\ref{sec:lp_potential_and_entry_updates}, we use
$h(v):=\sum_{i=1}^n\beta_2(v_i)$, where $\beta_2$ is defined in
Definition~\ref{def:lp_beta_2}; the potential in Eq.~\eqref{eq:lp_potential}
is $\mathscr{J}(v)=\mathscr{J}_h(v_*,v)$, with the root $v_*$ in the first argument.
Here $v_*$ is fixed, while $v$ is the point at which $h$ is
linearized. Since $h$ is separable, $\mathscr{J}(v)$ is a sum
of coordinate contributions, the form used to analyze
coordinate updates in Section~\ref{sec:lp_potential_and_entry_updates}.

\subsection{Matrix multiplication bounds}\label{subsec:prelim_matrix_bounds}

The running-time analysis uses fast matrix multiplication
for dense products and matrix inversion. We state the
corresponding costs in the exact real-arithmetic model
specified above.

We write $\Tmat(a,b,c)$ for the arithmetic cost of multiplying an $a\times b$ matrix by a $b\times c$ matrix.
We write $\omega=\omega(1,1,1)$ for the square matrix-multiplication exponent. The characteristic-zero arithmetic bound in \cite{uzn26,o26a} gives
$
 2\leq\omega\leq9/4.
$
Multiplication and nonsingular inversion of order $q$ cost $q^{\omega+o(1)}$ arithmetic operations; the inversion bound follows from the standard block reduction to matrix multiplication~\cite{bh74}.

\clearpage
\part{Linear programming in matrix multiplication time}\label{part:dense_lp}

This part establishes the dense LP guarantee of
Theorem~\ref{thm:main_informal}. For feasible bounded-box linear programs
with $d\leq n=O(d)$, the expected running time is
\[
 O((n^\omega+n^2\log n)\log(nW/\epsilon))
\]
in the exact real-arithmetic model, with the input scale and
approximation guarantees specified in that theorem.

Section~\ref{sec:lp_proofs} presents the dense LP algorithm and its analysis.
Sections~\ref{sec:lp_exact_ip}, \ref{sec:lp_proposal_tree},
\ref{sec:lp_exact_l2_sampler}, and~\ref{sec:lp_entry_state} develop the
supporting data structures for exact inner products, dyadic proposals,
exact $\ell_2$ sampling, and entry states, respectively.
Section~\ref{sec:lp_more_potentials} studies an alternative arctangent
potential and its properties.

 \section{Dense LP algorithm and its analysis}\label{sec:lp_proofs}

In Section~\ref{sec:lp_definitions}, we define the functions $\beta_1$ and $\beta_2$.
In Sections~\ref{sec:lp_root} and~\ref{sec:lp_one_step}, we characterize the central root and show that a small residual suffices for \textsc{OneStep} to preserve feasibility and reduce the complementarity target.
In Section~\ref{sec:lp_potential_and_entry_updates}, we bound the change in potential under a coordinate update.
In Sections~\ref{sec:lp_normalization} and~\ref{sec:lp_convergence}, we bound the size of the iterates and prove convergence of \textsc{IterativeSolve}.
In Sections~\ref{sec:lp_implementation_and_base_sampling} and~\ref{sec:lp_sampling_validity}, we implement base sampling and justify the coordinate sampling law for adaptive iterates.
In Section~\ref{sec:lp_iterative_solve_work}, we bound the expected work of \textsc{IterativeSolve}.
In Section~\ref{sec:lp_reduction}, we construct a strictly feasible starting point and prove the dense LP guarantee.

\subsection{Definitions of \texorpdfstring{$\beta_1$ and $\beta_2$}{beta1 and beta2}}\label{sec:lp_definitions}
The function $\beta_1$ encodes the complementarity condition in the fixed-point equation used by \textsc{OneStep}. Functions of vectors act coordinatewise.

\begin{definition}[The function $\beta_1$]\label{def:lp_beta_1}
Let $\gamma_1\ge2$ be a fixed constant.
We use a function $\beta_1:\R\to\R$ satisfying the following properties.
\begin{itemize}
\item[(a)] \textbf{Regularity and normalization.}
The function $\beta_1$ is twice continuously differentiable and strictly convex,
with $\beta_1(0)=\beta_1'(0)=0$.
\item[(b)] \textbf{Complementarity identity.} For every $\zeta\in\R$,
\[
 (\beta_1(\zeta)+\gamma_1)^2-\zeta^2=\gamma_1^2.
\]
\item[(c)] \textbf{Efficient evaluation.}
The value $\beta_1(\zeta)$ for $\zeta\in\R$ can be evaluated exactly in $O(1)$
operations in the arithmetic model of Section~\ref{sec:preliminaries}.
\end{itemize}
\end{definition}

The coordinate solver measures progress with a second function, $\beta_2$. Its curvature controls the potential, and its derivative and inverse determine each coordinate update.

\begin{definition}[The function $\beta_2$]\label{def:lp_beta_2}
Let $\gamma_1\ge2$ be the fixed constant in Definition~\ref{def:lp_beta_1},
and fix constants $3/4\le\gamma_2\le1\le\gamma_3\le2$.
We use a function $\beta_2:\R\to\R$ satisfying the following properties.
\begin{itemize}
\item[(a)] \textbf{Regularity and normalization.}
The function $\beta_2$ is twice continuously differentiable and strictly convex,
with $\beta_2(0)=\beta_2'(0)=0$.
\item[(b)] \textbf{Curvature bounds.} For every $\zeta\in\R$,
\[
 \frac{\gamma_2}{(\gamma_1+|\zeta|)^2}\le\beta_2''(\zeta)
 \le\frac{\gamma_3}{(\gamma_1+|\zeta|)^2}.
\]
\item[(c)] \textbf{Efficient mirror map.}
Writing $I_{\beta_2}:=\beta_2'(\R)$, the values $\beta_2'(\zeta)$ for $\zeta\in\R$ and
$(\beta_2')^{-1}(\theta)$ for $\theta\in I_{\beta_2}$ can each be evaluated exactly in $O(1)$
operations in the arithmetic model of Section~\ref{sec:preliminaries}.
\end{itemize}
\end{definition}

The constants $\gamma_1,\gamma_2,\gamma_3$ remain fixed throughout the
algorithm and its analysis. The bounds below hold uniformly over the stated
ranges of $\gamma_2$ and $\gamma_3$.
Only the two maps in (c), rather than the value of $\beta_2$, are evaluated
by the algorithms.

The Kadison--Singer problem \cite{ks59} has many formulations. In 2013, Marcus, Spielman, and Srivastava~\cite{mss15} proved the existence of a signing satisfying the discrepancy bound in Weaver's formulation~\cite{w04}. A naive algorithm enumerates all possible signings. Whether such a signing could be found in polynomial time was a long-standing open question. In recent independent and concurrent work, Ezeunala and Jiang~\cite{ej26}, Kathuria~\cite{k26}, and Song and Yue~\cite{sy26} gave polynomial-time algorithms for this problem. 
In particular, our dense LP algorithm is inspired by the work of Song and Yue \cite{sy26} on the Kadison--Singer problem. Their square-root budget appears in the convex conjugate of our function $\beta_1$: $\beta_1^*(\zeta)=\gamma_1(1-\sqrt{1-\zeta^2})$ for $|\zeta|\le 1$. Evaluating the budget at $\beta_1'(\zeta)$ gives $\sqrt{1-(\beta_1'(\zeta))^2}=\gamma_1/\sqrt{\zeta^2+\gamma_1^2}$ for $\zeta\in\mathbb{R}$.
The square of this quantity controls the curvature of $\beta_2$ in Eq.~\eqref{eq:lp_derivative_squared_contraction}.

The complementarity identity determines $\beta_1$ uniquely and gives an explicit formula for its derivatives.

\begin{lemma}[A choice of $\beta_1$]\label{lem:lp_beta_1_example}
Let $\beta_1(\zeta):=\sqrt{\zeta^2+\gamma_1^2}-\gamma_1$. The following properties hold.
\begin{itemize}
\item[(a)] The function $\beta_1$ is the unique function satisfying
properties (a), (b), and (c) of Definition~\ref{def:lp_beta_1}.
\item[(b)] For every $\zeta\in\R$, its derivatives satisfy
\[
 \beta_1'(\zeta)=\frac{\zeta}{\sqrt{\zeta^2+\gamma_1^2}},\qquad
 \beta_1''(\zeta)=\frac{\gamma_1^2}{(\zeta^2+\gamma_1^2)^{3/2}}>0.
\]
\item[(c)] For every $\zeta\in\R$, $0\le\beta_1(\zeta)\le|\zeta|$.
\end{itemize}
\end{lemma}
\begin{proof}
{\bf Proof of Part (b).}
Direct differentiation gives the displayed derivative formulas.

{\bf Proof of Part (a).}
For Definition~\ref{def:lp_beta_1}-(a), the formulas in Part~(b) apply.
Their denominators are positive because $\gamma_1\ge2$, so $\beta_1$ is twice
continuously differentiable. The positive second derivative gives strict
convexity, and substitution at $\zeta=0$ gives $\beta_1(0)=\beta_1'(0)=0$.
For Definition~\ref{def:lp_beta_1}-(b), we have
\[
 (\beta_1(\zeta)+\gamma_1)^2-\zeta^2
 =(\sqrt{\zeta^2+\gamma_1^2})^2-\zeta^2
 =\gamma_1^2,
\]
where the first step substitutes the formula for $\beta_1$, and the second
step simplifies the square.
For Definition~\ref{def:lp_beta_1}-(c), the formula uses a constant number of arithmetic operations and
one positive square root, all permitted in the stated model.

For uniqueness, let $\widetilde\beta_1$ satisfy Definition~\ref{def:lp_beta_1}.
By convexity and normalization,
$\widetilde\beta_1(\zeta)\ge\widetilde\beta_1(0)+\widetilde\beta_1'(0)\zeta=0$.
Thus $\widetilde\beta_1(\zeta)+\gamma_1>0$, and property (b) gives
$\widetilde\beta_1(\zeta)+\gamma_1=\sqrt{\zeta^2+\gamma_1^2}$.
Subtracting $\gamma_1$ proves uniqueness.

{\bf Proof of Part (c).}
The bounds $\gamma_1\le\sqrt{\zeta^2+\gamma_1^2}\le\gamma_1+|\zeta|$ give
$0\le\beta_1(\zeta)\le|\zeta|$.
\end{proof}

For $\beta_2$, the logarithmic choice below gives a mirror map and inverse that can both be evaluated by rational arithmetic.

\begin{lemma}[A choice of $\beta_2$]\label{lem:lp_beta_2_example}
Let $\beta_2(\zeta):=|\zeta|/\gamma_1-\log(1+|\zeta|/\gamma_1)$. The following properties hold.
\begin{itemize}
\item[(a)] The function $\beta_2$ satisfies properties (a), (b), and (c)
of Definition~\ref{def:lp_beta_2}.
\item[(b)] For every $\zeta\in\R$, $\beta_2'(\zeta)=\zeta/(\gamma_1(\gamma_1+|\zeta|))$.
\item[(c)] Its derivative has image $I_{\beta_2}=(-1/\gamma_1,1/\gamma_1)$,
and $(\beta_2')^{-1}(\theta)=\gamma_1^2\theta/(1-\gamma_1|\theta|)$ for $|\theta|<1/\gamma_1$.
\end{itemize}
\end{lemma}
\begin{proof}
{\bf Proof of Parts (a) and (b).}
For Definition~\ref{def:lp_beta_2}-(a), we use the integral representation
\[
 \beta_2(\zeta)=\int_0^{|\zeta|}\frac{\zeta_3}{\gamma_1(\gamma_1+\zeta_3)}\d \zeta_3.
\]
It gives $\beta_2(0)=0$ and $\lim_{\zeta\to0}\beta_2(\zeta)/\zeta=0$, hence $\beta_2'(0)=0$.
Differentiating separately on $\zeta>0$ and $\zeta<0$ gives the formula in
Part~(b). Moreover,
$\lim_{\zeta\to0}\beta_2'(\zeta)/\zeta=1/\gamma_1^2$, so
$\beta_2''(0)=1/\gamma_1^2$.
Differentiating away from zero therefore yields the continuous formula
\[
 \beta_2''(\zeta)=\frac{1}{(\gamma_1+|\zeta|)^2}>0\qquad(\zeta\in\R).
\]
This proves twice continuous differentiability and strict convexity,
and also proves Definition~\ref{def:lp_beta_2}-(b), since $\gamma_2\le1\le\gamma_3$.

{\bf Proof of Part (c).}
The derivative $\beta_2'$ is strictly increasing, with limits $\pm1/\gamma_1$
at $\pm\infty$. It is therefore a bijection onto
$I_{\beta_2}=(-1/\gamma_1,1/\gamma_1)$.
Solving $\theta=\beta_2'(\zeta)$, using that $\theta$ and $\zeta$ have the same sign,
gives the inverse formula in Part~(c).
Its denominator is positive on $I_{\beta_2}$.
Each formula uses $O(1)$ arithmetic operations and absolute values,
proving Definition~\ref{def:lp_beta_2}-(c) and completing Part~(a).
\end{proof}

\subsection{Projection geometry and the central root}\label{sec:lp_root}

The primal and dual constraints restrict the update to complementary subspaces. Their orthogonal projections determine the matrix and right-hand side of the fixed-point equation.

\begin{lemma}[Projection geometry and scaled neighborhood]\label{lem:lp_projection}
Let $\gamma_1\ge2$ and $\beta_1$ be as in Definition~\ref{def:lp_beta_1}.
Set $\epsilon_{1,t}:=0.1$ and $\epsilon_{2,t}:=0.05$. Let $\mathscr{B}\in\R^{d\times n}$ have full row rank, where $n\ge3$, and let
$\mathscr{D},y\in\R^d$, $c,x,s\in\R^n$, and $t>0$ satisfy
$\mathscr{B}x=\mathscr{D}$, $\mathscr{B}^\top y+s=c$, $x,s>{\bf 0}_n$, and
$t\le x_is_i\le(1+\epsilon_{1,t})t$ for every $i\in[n]$.
Set $X:=\operatorname{diag}(x)$ and define
$\mathscr{M}:=(\mathscr{B}X^2\mathscr{B}^\top)^{-1}$, $H:=X\mathscr{B}^\top \mathscr{M}\mathscr{B}X$, $\overline{H}:=I-H$, and $\mathscr{A}:=\overline{H}-H$. We define $\mu:=xs$.
We define $\mu_t:=xs/t$ and $b:=-\gamma_1 \overline{H}(\mu_t-{\bf 1}_n)$.
Then the following guarantees hold.
\begin{itemize}
\item[(a)] $H,\overline{H}$ are complementary orthogonal projections, $\mathscr{A}^\top \mathscr{A}=I$,
$\operatorname{range}(H)=\operatorname{range}(X\mathscr{B}^\top)$, and $\ker(H)=\ker(\mathscr{B}X)$.
\item[(b)] For every $i\in[n]$, $1\le\mu_{t,i}\le1+\epsilon_{1,t}$.
\item[(c)] The vector $b$ satisfies $\|b\|_2\le\gamma_1 n/2$.
\end{itemize}
\end{lemma}
\begin{proof}
{\bf Proof of Part (a).}
Full row rank of $\mathscr{B}$ and positivity of $X$ make $\mathscr{B}X^2\mathscr{B}^\top$ positive
definite, so $\mathscr{M}$ is well defined and symmetric. Direct multiplication gives
$H=H^\top$ and $H^2=H$. The factorization of $H$ shows that its range is
contained in $\operatorname{range}(X\mathscr{B}^\top)$, and
$HX\mathscr{B}^\top=X\mathscr{B}^\top$ gives the reverse inclusion. Its kernel is the
orthogonal complement of this range, namely $\ker(\mathscr{B}X)$. Thus $\overline{H}=I-H$
is the complementary orthogonal projection and $\mathscr{A}^\top \mathscr{A}=(\overline{H}-H)^2=I$.

{\bf Proof of Part (b).}
The bounds on $\mu_{t,i}$ in Part~(b) follow by dividing the
input neighborhood by $t>0$.

{\bf Proof of Part (c).}
We have
\[
 \|b\|_2\le\gamma_1\|\mu_t-{\bf 1}_n\|_2\le\gamma_1\epsilon_{1,t}\sqrt n\le\gamma_1 n/2,
\]
where the first step follows from $b=-\gamma_1 \overline{H}(\mu_t-{\bf 1}_n)$ and $\|\overline{H}\|_{\rm op}\le1$;
the second step follows from $|\mu_{t,i}-1|\le\epsilon_{1,t}$;
and the third step uses $\epsilon_{1,t}=0.1$ and $n\ge1$.
\end{proof}

We show that the projected equations have a unique fixed point. It determines two positive scaling factors in each coordinate: their product is one, and their projections enforce the scaled constraints.

\begin{lemma}\label{lem:lp_root}
Use the assumptions and notation of Lemma~\ref{lem:lp_projection}.
The following guarantees hold.
\begin{itemize}
\item[(a)] There is a unique solution $v_*$ of $v_*=\mathscr{A}\beta_1(v_*)+b$.
\item[(b)] Let $\Delta \in \{-1,+1\}$. Define
$
 \mu_i^\Delta:=( \sqrt{v_{*,i}^2+\gamma_1^2}+\Delta v_{*,i}) /{\gamma_1}.
$ 
Then $\mu^+>{\bf 0}_n$, $\mu^->{\bf 0}_n$, $\mu^+\mu^-={\bf 1}_n$,
$H\mu^+=H{\bf 1}_n$, and $\overline{H}\mu^-=\overline{H}\mu_t$.
\item[(c)] We have $\sum_{i=1}^n\beta_1(v_{*,i})\le\gamma_1\epsilon_{1,t}^2n/(2(1+\epsilon_{1,t}))$.
\item[(d)] We have $\|v_*\|_1<3\gamma_1 n/8$.
\item[(e)] We have 
$\|v_*\|_\infty<\gamma_1\epsilon_{1,t}n/2$.
\end{itemize}
\end{lemma}
\begin{proof}
{\bf Proof of Part (a).}
Minimize
\[
 \mu_t^\top\nu-\sum_{i=1}^n\log\nu_i
 \quad\text{subject to}\quad H\nu=H{\bf 1}_n,\quad\nu>{\bf 0}_n.
\]
The point ${\bf 1}_n$ is feasible. Each scalar function $\mu_{t,i}z-\log z$
is bounded below and tends to infinity as $z\downarrow0$ or $z\to\infty$.
Consequently a sublevel set through ${\bf 1}_n$ is compact inside the positive
orthant. The strictly convex objective has a unique minimizer. Its first
order condition is $\overline{H}(\mu_t-\nu^{-1})={\bf 0}_n$. Here $\nu^{-1}$ denotes the coordinatewise reciprocal of $\nu$. Define
$v_*:=(\gamma_1/2)(\nu-\nu^{-1})$. Then
$\sqrt{v_{*,i}^2+\gamma_1^2}=(\gamma_1/2)(\nu_i+\nu_i^{-1})$, so
$\beta_1(v_*)=(\gamma_1/2)(\nu+\nu^{-1}-2{\bf 1}_n)$. The two projected constraints imply
$Hv_*=-H\beta_1(v_*)$ and $\overline{H}v_*=\overline{H}\beta_1(v_*)+b$, which together give $v_*=\mathscr{A}\beta_1(v_*)+b$.
For uniqueness, suppose that $\widehat{v}_*\ne v_*$ is another solution. We have
\[
 \|\widehat{v}_*-v_*\|_2
 =\|\mathscr{A}(\beta_1(\widehat{v}_*)-\beta_1(v_*))\|_2
 \le\|\beta_1(\widehat{v}_*)-\beta_1(v_*)\|_2<\|\widehat{v}_*-v_*\|_2,
\]
where the first step subtracts $v_*=\mathscr{A}\beta_1(v_*)+b$ from $\widehat{v}_*=\mathscr{A}\beta_1(\widehat{v}_*)+b$;
the second step uses $\|\mathscr{A}\|_{\rm op}=1$ from Lemma~\ref{lem:lp_projection}-(a);
and the third step applies the mean value theorem coordinatewise, using
$|\beta_1'(\zeta)|=|\zeta|/\sqrt{\zeta^2+\gamma_1^2}<1$ from
Lemma~\ref{lem:lp_beta_1_example}-(b) and $\widehat{v}_*\ne v_*$.
This contradiction proves uniqueness and completes Part~(a).

{\bf Proof of Part (b).}
For the root constructed in Part~(a), the identities
$v_*=(\gamma_1/2)(\nu-\nu^{-1})$ and
$\sqrt{v_{*,i}^2+\gamma_1^2}=(\gamma_1/2)(\nu_i+\nu_i^{-1})$
give $\mu^+=\nu$ and $\mu^-=\nu^{-1}$.
Since $\nu>{\bf 0}_n$, both vectors are positive and their coordinatewise
product is ${\bf 1}_n$. The constraint $H\nu=H{\bf 1}_n$ gives
$H\mu^+=H{\bf 1}_n$, and the first order condition
$\overline{H}(\mu_t-\nu^{-1})={\bf 0}_n$ gives
$\overline{H}\mu^-=\overline{H}\mu_t$. This proves Part~(b).

{\bf Proof of Part (c).}
We have
\begin{equation}\label{eq:lp_root_sum}
 \mu_t^\top\nu+{\bf 1}_n^\top\nu^{-1}=n+{\bf 1}_n^\top\mu_t\le(2+\epsilon_{1,t})n.
\end{equation}
The equality follows by expanding
$(\nu-{\bf 1}_n)^\top(\nu^{-1}-\mu_t)=0$ and using $\nu^\top\nu^{-1}=n$.
Here $\nu-{\bf 1}_n\in\operatorname{range}(\overline H)$ and
$\nu^{-1}-\mu_t\in\operatorname{range}(H)$ are orthogonal.
The inequality uses $\mu_{t,i}\le1+\epsilon_{1,t}$ from
Lemma~\ref{lem:lp_projection}-(b).

Rearranging the equality in Eq.~\eqref{eq:lp_root_sum} gives
\begin{equation}\label{eq:lp_root_deviation_identity}
 \sum_{i=1}^n(\nu_i+\nu_i^{-1}-2)
 =\sum_{i=1}^n(\mu_{t,i}-1)(1-\nu_i).
\end{equation}
Since $\nu_i+\nu_i^{-1}-2=(\nu_i-1)^2/\nu_i$, Cauchy--Schwarz gives
\[
 \sum_{i=1}^n(\nu_i+\nu_i^{-1}-2)
 \le (\sum_{i=1}^n(\mu_{t,i}-1)^2\nu_i)^{1/2}
 (\sum_{i=1}^n(\nu_i+\nu_i^{-1}-2))^{1/2}.
\]
If the sum on the left is zero, the claimed bound is immediate.
Otherwise, squaring and dividing by this positive sum yields
\[
 \sum_{i=1}^n(\nu_i+\nu_i^{-1}-2)
 \le\sum_{i=1}^n(\mu_{t,i}-1)^2\nu_i
 \le\epsilon_{1,t}\sum_{i=1}^n(\mu_{t,i}-1)\nu_i.
\]
The second step uses $0\le\mu_{t,i}-1\le\epsilon_{1,t}$ and $\nu_i>0$.
Substituting Eq.~\eqref{eq:lp_root_deviation_identity} into the right side
and collecting the deviation sum gives
\[
 (1+\epsilon_{1,t})\sum_{i=1}^n(\nu_i+\nu_i^{-1}-2)
 \le\epsilon_{1,t}\sum_{i=1}^n(\mu_{t,i}-1)
 \le\epsilon_{1,t}^2n.
\]
The last step again uses $\mu_{t,i}-1\le\epsilon_{1,t}$.
Dividing by $1+\epsilon_{1,t}>0$ and adding $2n$ gives
\begin{equation}\label{eq:lp_root_reciprocal_sum}
 \sum_{i=1}^n(\nu_i+\nu_i^{-1})
 \le(2+\frac{\epsilon_{1,t}^2}{1+\epsilon_{1,t}})n.
\end{equation}
Consequently,
\[
 \sum_{i=1}^n\beta_1(v_{*,i})
 =\frac{\gamma_1}{2}\sum_{i=1}^n(\nu_i+\nu_i^{-1}-2)
 \le\frac{\gamma_1\epsilon_{1,t}^2n}{2(1+\epsilon_{1,t})}.
\]
The equality uses the formula for $\beta_1(v_*)$ in Part~(a).
The inequality uses Eq.~\eqref{eq:lp_root_reciprocal_sum}, proving Part~(c).

{\bf Proof of Part (d).}
The identity $|v_{*,i}|^2=\beta_1(v_{*,i})(\beta_1(v_{*,i})+2\gamma_1)$
and Cauchy--Schwarz give
\[
 \|v_*\|_1
 \le ((\sum_{i=1}^n\beta_1(v_{*,i}))
 (\sum_{i=1}^n\beta_1(v_{*,i})+2\gamma_1n))^{1/2}.
\]
Substituting Part~(c) into both nonnegative factors yields
\[
 \|v_*\|_1
 \le\gamma_1n(\frac{\epsilon_{1,t}^2}{2(1+\epsilon_{1,t})}
 (\frac{\epsilon_{1,t}^2}{2(1+\epsilon_{1,t})}+2))^{1/2}
 =\frac{\gamma_1\epsilon_{1,t}(2+\epsilon_{1,t})n}{2(1+\epsilon_{1,t})}.
\]
The equality uses $\epsilon_{1,t}^2+4(1+\epsilon_{1,t})=(2+\epsilon_{1,t})^2$.
Since $(2+\epsilon_{1,t})/(2(1+\epsilon_{1,t}))<1$, we obtain
$\|v_*\|_1<\gamma_1\epsilon_{1,t}n<3\gamma_1n/8$.
The last inequality uses $\epsilon_{1,t}=0.1$, proving Part~(d).

{\bf Proof of Part (e).}
Let $\nu$ be the minimizer from Part~(a), and define
$g_j:=\mu_{t,j}\nu_j+\nu_j^{-1}-1-\mu_{t,j}$ for every $j\in[n]$.
Eq.~\eqref{eq:lp_root_sum} gives $\sum_{j=1}^n g_j=0$.
For every $j\in[n]$, we have
\begin{equation}\label{eq:lp_root_energy_lower}
 g_j\ge-(\sqrt{\mu_{t,j}}-1)^2
 \ge-\frac{(\mu_{t,j}-1)^2}{4}\ge-\frac{\epsilon_{1,t}^2}{4}.
\end{equation}
The first step applies AM--GM to $\mu_{t,j}\nu_j+\nu_j^{-1}$.
The second step uses
$\sqrt{\mu_{t,j}}-1=(\mu_{t,j}-1)/(\sqrt{\mu_{t,j}}+1)$ and $\mu_{t,j}\ge1$.
The third step uses $\mu_{t,j}-1\le\epsilon_{1,t}$.

Suppose that $|v_{*,i}|\ge\gamma_1\epsilon_{1,t}n/2$ for some $i\in[n]$, and set
$r:=|v_{*,i}|/\gamma_1$. Then $r\ge\epsilon_{1,t}n/2>\epsilon_{1,t}$, since $n\ge3$.
Part~(a) gives $\nu_i=\sqrt{1+r^2}+r$ or $\nu_i=\sqrt{1+r^2}-r$.
Minimizing the expression for $g_i$ over $1\le\mu_{t,i}\le1+\epsilon_{1,t}$ gives
\[
 g_i\ge(2+\epsilon_{1,t})(\sqrt{1+r^2}-1)-\epsilon_{1,t}r.
\]
Indeed, when $\nu_i\ge1$, the minimum is $2(\sqrt{1+r^2}-1)$,
which is at least the displayed bound because $\sqrt{1+r^2}-1\le r$.
When $\nu_i\le1$, the minimum equals the displayed bound.
Rationalizing the square root and then using $\sqrt{1+r^2}\le1+r$ gives
\[
 g_i\ge\frac{(2+\epsilon_{1,t})r^2}{\sqrt{1+r^2}+1}-\epsilon_{1,t}r
 \ge\frac{(2+\epsilon_{1,t})r^2}{2+r}-\epsilon_{1,t}r
 =\frac{2r(r-\epsilon_{1,t})}{2+r}.
\]
The last step is simple algebra. The expression
$2r(r-\epsilon_{1,t})/(2+r)=2r(1-(2+\epsilon_{1,t})/(2+r))$
is increasing for $r\ge\epsilon_{1,t}$, since both factors on the right
are nonnegative and increasing. Thus
\begin{equation}\label{eq:lp_root_energy_large}
 g_i\ge\frac{\epsilon_{1,t}^2n(n-2)}{4+\epsilon_{1,t}n}\ge\frac{\epsilon_{1,t}^2n(n-1)}{2(4+\epsilon_{1,t}n)}>\frac{\epsilon_{1,t}^2(n-1)}4.
\end{equation}
The first step uses $r\ge\epsilon_{1,t}n/2$ and the preceding monotonicity.
The second step uses $n-2\ge(n-1)/2$ for $n\ge3$.
The third step uses $4+\epsilon_{1,t}n<2n$, since $\epsilon_{1,t}=0.1$ and $n\ge3$.
Combining Eqs.~\eqref{eq:lp_root_energy_lower} and~\eqref{eq:lp_root_energy_large} yields
\[
 \sum_{j=1}^n g_j>
 \frac{\epsilon_{1,t}^2(n-1)}4-\frac{(n-1)\epsilon_{1,t}^2}4=0.
\]
This contradicts $\sum_{j=1}^n g_j=0$ and proves Part~(e).
\end{proof}

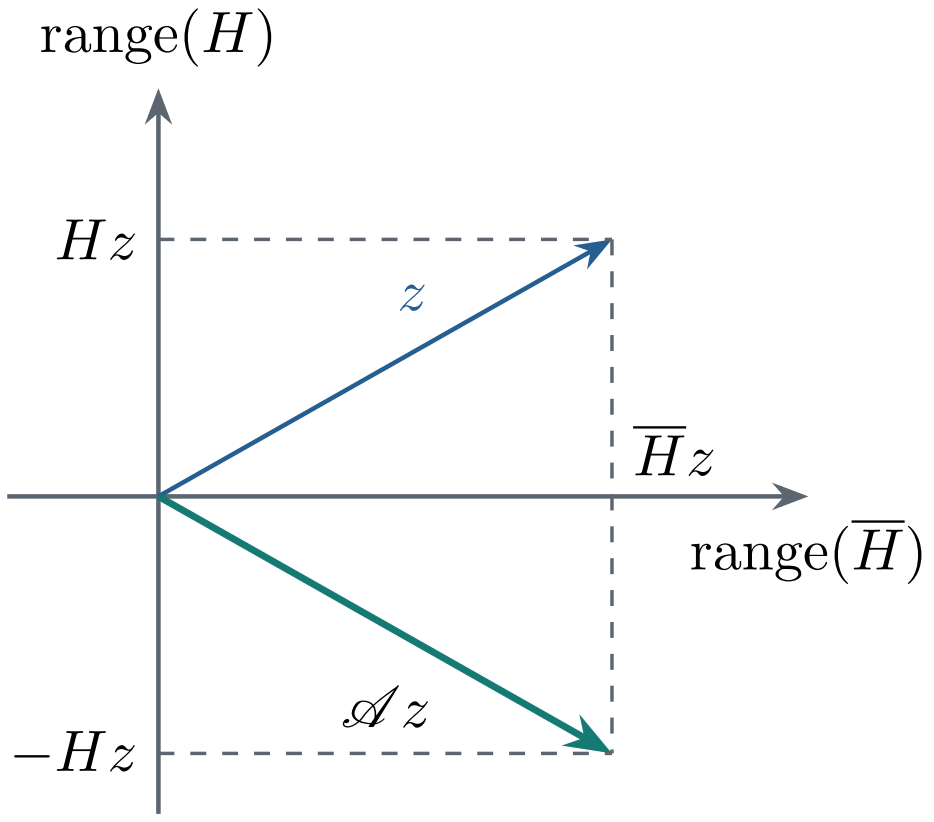
\begin{figure}[!htbp]
\centering
\begin{tikzpicture}[lpfig/base]
\draw[lpfig/arrow] (0,0)--(5.3,0) node[below] {$\operatorname{range}(\overline{H})$};
\draw[lpfig/arrow] (1,-2.1)--(1,2.7) node[above] {$\operatorname{range}(H)$};
\coordinate (o) at (1,0); \coordinate (v) at (4,1.7); \coordinate (av) at (4,-1.7);
\draw[lpfig/arrow,lpBlue] (o)--(v) node[pos=.64,above left] {$z$};
\draw[lpfig/emph] (o)--(av) node[pos=.65,below left] {$\mathscr{A}z$};
\draw[dashed,lpGray] (v)--(4,0)--(av);
\draw[dashed,lpGray] (1,1.7)--(v);
\draw[dashed,lpGray] (1,-1.7)--(av);
\node[left] at (1,1.7) {$Hz$}; \node[left] at (1,-1.7) {$-Hz$};
\node[above right] at (4,0) {$\overline{H}z$};
\end{tikzpicture}
\caption{The projection geometry underlying \textsc{OneStep}. The diagram illustrates the orthogonal decomposition in Lemma~\ref{lem:lp_projection}; $z$ is an arbitrary vector.}
\label{fig:lp_projection}
\end{figure}

\subsection{Residual stability and OneStep}\label{sec:lp_one_step}

The solver can test the residual of a candidate vector without knowing the fixed point. To use this test as a stopping rule, we bound the distance to the fixed point in terms of the residual.

\begin{lemma}[Residual stability]\label{lem:lp_residual_stability}
Let $\|\mathscr{A}\|_{\rm op}\le1$. Let $v_* \in \R^n$ denote the vector that $v_*=\mathscr{A}\beta_1(v_*)+b$. For vector $v \in \R^n$, we define
$v^{\new}:=\mathscr{A}\beta_1(v)+b \in \R^n$, and $\delta_v:=v^{\new}-v \in \R^n$, with $\|\delta_v\|_2\le\eta$. If $2\eta(\|v_*\|_\infty+\gamma_1)<\gamma_1^2$, then
\[
 \|v-v_*\|_2\le\frac{2\eta(\|v_*\|_\infty+\gamma_1)^2}{\gamma_1^2-2\eta(\|v_*\|_\infty+\gamma_1)}.
\]
\end{lemma}
\begin{proof}
We have
\begin{align}
 2\langle v-v_*,\delta_v\rangle+\|\delta_v\|_2^2
 &=\|v-v_*+\delta_v\|_2^2-\|v-v_*\|_2^2\notag\\
 &=\|\mathscr{A}(\beta_1(v)-\beta_1(v_*))\|_2^2-\|v-v_*\|_2^2\notag\\
 &\le\|\beta_1(v)-\beta_1(v_*)\|_2^2-\|v-v_*\|_2^2\notag\\
 &=-\sum_{i=1}^n((v_i-v_{*,i})^2-(\beta_1(v_i)-\beta_1(v_{*,i}))^2),\label{eq:lp_stability_nonexpansive}
\end{align}
where the first step follows by expanding the squared norm;
the second step follows from $\delta_v=\mathscr{A}\beta_1(v)+b-v$ and $v_*=\mathscr{A}\beta_1(v_*)+b$;
the third step follows from $\|\mathscr{A}\|_{\rm op}\le1$;
and the fourth step follows by writing the squared Euclidean norms as sums over coordinates.

For real $\zeta_1,\zeta_2$, we have
\begin{equation}\label{eq:lp_scalar_gap}
 (\zeta_1-\zeta_2)^2-(\beta_1(\zeta_1)-\beta_1(\zeta_2))^2
 =\frac{2\gamma_1^2(\zeta_1-\zeta_2)^2}{\sqrt{\zeta_1^2+\gamma_1^2}\sqrt{\zeta_2^2+\gamma_1^2}+\zeta_1\zeta_2+\gamma_1^2}
 \ge\frac{\gamma_1^2(\zeta_1-\zeta_2)^2}{(|\zeta_1|+\gamma_1)(|\zeta_2|+\gamma_1)},
\end{equation}
where the first step follows by substituting the formula in Lemma~\ref{lem:lp_beta_1_example}
and rationalizing; and the second step follows from
$\sqrt{\zeta_1^2+\gamma_1^2}\le|\zeta_1|+\gamma_1$, $\sqrt{\zeta_2^2+\gamma_1^2}\le|\zeta_2|+\gamma_1$, and
$\zeta_1\zeta_2+\gamma_1^2\le(|\zeta_1|+\gamma_1)(|\zeta_2|+\gamma_1)$.

Assume $\|v-v_*\|_2>0$. For every $i\in[n]$, we have
\begin{equation}
 |v_i|\le |v_i-v_{*,i}|+|v_{*,i}|\le \|v-v_*\|_2+|v_{*,i}|\le \|v-v_*\|_2+\|v_*\|_\infty,\label{eq:lp_stability_coordinate_bound}
\end{equation}
where the first step follows from the triangle inequality;
the second step uses $|v_i-v_{*,i}|\le\|v-v_*\|_2$;
and the third step uses $|v_{*,i}|\le\|v_*\|_\infty$.
We have
\begin{align*}
 2\|v-v_*\|_2\eta
 &\ge 2\|v-v_*\|_2\|\delta_v\|_2\\
 &\ge -2\langle v-v_*,\delta_v\rangle\\
 &\ge -2\langle v-v_*,\delta_v\rangle-\|\delta_v\|_2^2\\
 &\ge \sum_{i=1}^n((v_i-v_{*,i})^2-(\beta_1(v_i)-\beta_1(v_{*,i}))^2)\\
 &\ge \sum_{i=1}^n\frac{\gamma_1^2(v_i-v_{*,i})^2}{(|v_i|+\gamma_1)(|v_{*,i}|+\gamma_1)}\\
 &\ge \sum_{i=1}^n\frac{\gamma_1^2(v_i-v_{*,i})^2}{(|v_i|+\gamma_1)(\|v_*\|_\infty+\gamma_1)}\\
 &\ge \sum_{i=1}^n\frac{\gamma_1^2(v_i-v_{*,i})^2}{(\|v-v_*\|_2+\|v_*\|_\infty+\gamma_1)(\|v_*\|_\infty+\gamma_1)}\\
 &=\frac{\gamma_1^2\|v-v_*\|_2^2}{(\|v-v_*\|_2+\|v_*\|_\infty+\gamma_1)(\|v_*\|_\infty+\gamma_1)}.
\end{align*}
The first step uses $\|\delta_v\|_2\le\eta$.
The second step follows from Cauchy--Schwarz.
The third step uses $\|\delta_v\|_2^2\ge0$.
The fourth step follows from Eq.~\eqref{eq:lp_stability_nonexpansive}.
The fifth step applies Eq.~\eqref{eq:lp_scalar_gap} coordinatewise.
The sixth step uses $|v_{*,i}|\le\|v_*\|_\infty$.
The seventh step uses Eq.~\eqref{eq:lp_stability_coordinate_bound}.
The eighth step uses $\sum_{i=1}^n(v_i-v_{*,i})^2=\|v-v_*\|_2^2$.
Since $\gamma_1^2-2\eta(\|v_*\|_\infty+\gamma_1)>0$ by hypothesis,
rearranging proves the claimed bound. The case $\|v-v_*\|_2=0$ is immediate.
\end{proof}

\begin{algorithm}[!ht]
\caption{\textsc{OneStep}}\label{alg:one_step}
\begin{algorithmic}[1]
\Procedure{OneStep}{$\mathscr{B}\in\R^{d\times n},x\in\R_{>0}^n,y\in\R^d,s\in\R_{>0}^n,t\in\R_{>0}$} \Comment{Theorem~\ref{thm:lp_one_step}}
\State $X\gets\operatorname{diag}(x)$; $\mathscr{M}\gets(\mathscr{B}X^2\mathscr{B}^\top)^{-1}$.
\State $\mu\gets xs$; $\mu_t\gets\mu/t$.
\State $H\gets X\mathscr{B}^\top \mathscr{M}\mathscr{B}X$; $\overline{H}\gets I-H$.
\State $\mathscr{A}\gets\overline{H}-H$; $b\gets-\gamma_1\overline{H}(\mu_t-{\bf 1}_n)$.
\State $v_{\mathrm{out}}\gets\Call{IterativeSolve}{\mathscr{A},b}$. \Comment{Algorithm~\ref{alg:iterative_solve}}
\State $\delta_{v_{\mathrm{out}}}\gets \mathscr{A}\beta_1(v_{\mathrm{out}})+b-v_{\mathrm{out}}$.
\State $\mu_i^\Delta\gets(\sqrt{v_{\mathrm{out},i}^2+\gamma_1^2}+\Delta v_{\mathrm{out},i})/\gamma_1$ for $i\in[n]$ and $\Delta\in\{-1,+1\}$.
\State $\delta_x\gets\mu^++H\delta_{v_{\mathrm{out}}}/\gamma_1$; $\delta_s\gets\mu^--\overline{H}\delta_{v_{\mathrm{out}}}/\gamma_1$.
\State $\delta_y\gets \mathscr{M}\mathscr{B}X(\delta_s-\mu_t)$.
\State $x^{\new}\gets X\delta_x$. \Comment{$x$ move}
\State $y^{\new}\gets y-t\delta_y$. \Comment{$y$ move}
\State $s^{\new}\gets tX^{-1}\delta_s$. \Comment{$s$ move}
\State $t^{\new}\gets(1-\epsilon_{2,t})t$. \Comment{$t$ move}
\State \Return $(x^{\new},y^{\new},s^{\new},t^{\new})$.
\EndProcedure
\end{algorithmic}
\end{algorithm}

The residual tolerance is chosen so that \textsc{OneStep} preserves strict feasibility and places every complementarity product in the neighborhood of the smaller target.

\begin{theorem}[Correctness of OneStep]\label{thm:lp_one_step}
Use the matrices and scaled variables of Lemma~\ref{lem:lp_projection}.
Define $t^{\new}:=(1-\epsilon_{2,t})t$. Let $c_1:=256$. 
In dimension $n\ge3$, any returned vector $v_{\mathrm{out}}$ with residual
$\|\mathscr{A}\beta_1(v_{\mathrm{out}})+b-v_{\mathrm{out}}\|_2\le\gamma_1/(c_1n^2)$
makes Algorithm~\ref{alg:one_step} satisfy the following guarantees.
\begin{itemize}
\item[(a)] The returned primal-dual pair is strictly feasible.
\item[(b)] The output satisfies $t^{\new}\le x_i^{\new}s_i^{\new}\le(1+\epsilon_{1,t})t^{\new}$
for every $i\in[n]$.
\end{itemize}
\end{theorem}
\begin{proof}
{\bf Proof of Part (a).}
Define $\delta_{v_{\mathrm{out}}}:=\mathscr{A}\beta_1(v_{\mathrm{out}})+b-v_{\mathrm{out}}$ and the raw positive pair
\[
 \mu^\Delta:=\frac{\sqrt{v_{\mathrm{out}}^2+\gamma_1^2}+\Delta v_{\mathrm{out}}}{\gamma_1},\qquad \Delta\in\{-1,+1\}.
\]

We have
\[
 H\delta_{v_{\mathrm{out}}}=-H\beta_1(v_{\mathrm{out}})-Hv_{\mathrm{out}},\qquad \overline{H}\delta_{v_{\mathrm{out}}}=\overline{H}\beta_1(v_{\mathrm{out}})-\gamma_1 \overline{H}(\mu_t-{\bf 1}_n)-\overline{H}v_{\mathrm{out}},
\]
where these formulas follow by projecting the residual and using
$Hb={\bf 0}_n$, $H\mathscr{A}=-H$, $\overline{H}\mathscr{A}=\overline{H}$, and $b=-\gamma_1 \overline{H}(\mu_t-{\bf 1}_n)$.

Define the corrected pair by $\delta_x:=\mu^++H\delta_{v_{\mathrm{out}}}/\gamma_1$ and $\delta_s:=\mu^--\overline{H}\delta_{v_{\mathrm{out}}}/\gamma_1$.
We have
\begin{equation}\label{eq:lp_corrected_constraints}
 H\delta_x=H{\bf 1}_n,\qquad \overline{H}\delta_s=\overline{H}\mu_t.
\end{equation}
These identities follow by substituting the definitions of the corrected pair
and the projected residual formulas.

We have
\[
 \mathscr{B}X\delta_x=\mathscr{B}X{\bf 1}_n=\mathscr{D},
\]
where the first step follows from $H\delta_x=H{\bf 1}_n$ and
$\ker(H)=\ker(\mathscr{B}X)$; and the second step follows from $\mathscr{B}x=\mathscr{D}$.

For $\delta_y:=\mathscr{M}\mathscr{B}X(\delta_s-\mu_t)$, we have
\[
 X\mathscr{B}^\top\delta_y=H(\delta_s-\mu_t)=\delta_s-\mu_t,
\]
where the first step follows from the definitions of $\delta_y$ and $H$;
and the second step follows because $\delta_s-\mu_t\in\operatorname{range}(X\mathscr{B}^\top)$
and $H$ projects onto this space.

We have
\[
 \mathscr{B}^\top(y-t\delta_y)+tX^{-1}\delta_s
 =\mathscr{B}^\top y+tX^{-1}\mu_t=c,
\]
where the first step follows by substituting
$\mathscr{B}^\top\delta_y=X^{-1}(\delta_s-\mu_t)$; and the second step follows from
$tX^{-1}\mu_t=s$ and $\mathscr{B}^\top y+s=c$.

It remains to prove positivity and centrality; residual smallness alone
must not be used without a size bound. Define $\eta:=\gamma_1/(c_1n^2)$.

Lemma~\ref{lem:lp_root}-(e) gives
\[
 \|v_*\|_\infty+\gamma_1
 <\gamma_1(1+\epsilon_{1,t}n/2)\le\gamma_1n/2.
\]
The last step uses $\epsilon_{1,t}=0.1$ and $n\ge3$.
Consequently,
$2\eta(\|v_*\|_\infty+\gamma_1)\le\gamma_1^2/(c_1n)<\gamma_1^2/2$,
so Lemma~\ref{lem:lp_residual_stability} applies and its denominator exceeds $\gamma_1^2/2$.
We obtain
\[
 \|v_{\mathrm{out}}-v_*\|_2
 \le\frac{4\eta(\|v_*\|_\infty+\gamma_1)^2}{\gamma_1^2}
 \le\eta n^2=\gamma_1/c_1.
\]
The second step uses $\|v_*\|_\infty+\gamma_1\le\gamma_1n/2$,
and the equality substitutes $\eta=\gamma_1/(c_1n^2)$.
The triangle inequality now gives
\begin{equation}\label{eq:lp_approximate_root_coordinate_bound}
 2|v_{\mathrm{out},i}|/\gamma_1
 \le\epsilon_{1,t}n+2/c_1.
\end{equation}
Here we used $2\|v_*\|_\infty/\gamma_1<\epsilon_{1,t}n$
and the distance estimate.

We have
\[
 \min\{\mu_i^+,\mu_i^-\}\ge1/(2|v_{\mathrm{out},i}|/\gamma_1+1)>\eta/\gamma_1.
\]
The first step uses $\mu_i^+\mu_i^-=1$ and
$\max\{\mu_i^+,\mu_i^-\}\le2|v_{\mathrm{out},i}|/\gamma_1+1$.
The second step uses Eq.~\eqref{eq:lp_approximate_root_coordinate_bound},
$\epsilon_{1,t}n+1+2/c_1<c_1n^2$ for $n\ge3$, and
$\eta/\gamma_1=1/(c_1n^2)$.

Each correction has magnitude at most $\eta/\gamma_1$, so positivity follows.

{\bf Proof of Part (b).}
We can show
\begin{equation}\label{eq:lp_raw_pair_sum_bound}
 \mu_i^++\mu_i^-=\frac{2\sqrt{v_{\mathrm{out},i}^2+\gamma_1^2}}{\gamma_1}\le\frac{2(|v_{\mathrm{out},i}|+\gamma_1)}{\gamma_1}=\frac{2|v_{\mathrm{out},i}|}{\gamma_1}+2.
\end{equation}
The first step uses the definitions of $\mu_i^+$ and $\mu_i^-$.
The second step uses $v_{\mathrm{out},i}^2+\gamma_1^2\le(|v_{\mathrm{out},i}|+\gamma_1)^2$.
The third step expands the fraction.

We have
\begin{align}
 |\delta_{x,i}\delta_{s,i}-1|
 &\le(\mu_i^++\mu_i^-)\eta/\gamma_1+\eta^2/\gamma_1^2
 \le(2|v_{\mathrm{out},i}|/\gamma_1+2)\eta/\gamma_1+\eta^2/\gamma_1^2\notag\\*
 &\le\frac{\epsilon_{1,t}n+2+2/c_1}{c_1n^2}+\frac1{c_1^2n^4}\le\frac1{c_1n}+\frac1{c_1^2n^4}< \epsilon_{0,t}.\label{eq:lp_corrected_product_bound}
\end{align}
where the first step follows by expanding the product, using
$\mu_i^+\mu_i^-=1$, and bounding both corrections by $\eta/\gamma_1$;
the second step follows from Eq.~\eqref{eq:lp_raw_pair_sum_bound};
the third step follows from Eq.~\eqref{eq:lp_approximate_root_coordinate_bound}
and $\eta/\gamma_1=1/(c_1n^2)$;
the fourth step uses $\epsilon_{1,t}n+2+2/c_1\le n$ for $n\ge3$;
and the fifth step uses $n\ge3$, the definition of $c_1$, and $\epsilon_{0,t}:=1/128$.

The returned vectors satisfy $x_i^{\new}s_i^{\new}=t\delta_{x,i}\delta_{s,i}$.
Thus Eq.~\eqref{eq:lp_corrected_product_bound} gives $(1-\epsilon_{0,t})t\le x_i^{\new}s_i^{\new}\le (1+\epsilon_{0,t})t$.
Since $t^{\new}=(1-\epsilon_{2,t})t$, we have
\[
 t^{\new}\le (1- \epsilon_{0,t}) t\le x_i^{\new}s_i^{\new}\le(1+ \epsilon_{0,t})t\le(1+\epsilon_{1,t})t^{\new}.
\]
The first step uses $1-\epsilon_{2,t}\le (1-\epsilon_{0,t})$; the second and third steps use
Eq.~\eqref{eq:lp_corrected_product_bound}; and the fourth step uses $1+\epsilon_{0,t}\le(1+\epsilon_{1,t})(1-\epsilon_{2,t})$.
This proves the asserted neighborhood.
\end{proof}

\subsection{Potential contraction and entry updates}\label{sec:lp_potential_and_entry_updates}

To quantify progress toward the fixed point, we compare the residual direction with the displacement from the root. Their inner product controls both the potential and the squared residual norm.

\begin{lemma}[Nonlinear contraction dominates the potential]\label{lem:lp_potential_contraction}
Let $\mathscr{A}\in\R^{n\times n}$ and $b\in\R^n$, and let $\gamma_1\ge2$ and
$\beta_1$ be as in Definition~\ref{def:lp_beta_1}.
Assume $\|\mathscr{A}\|_{\rm op}\le1$, $\|b\|_2\le\gamma_1 n/2$, and that a root
$v_*=\mathscr{A}\beta_1(v_*)+b$ exists with $\|v_*\|_1\le3\gamma_1 n/8$. Symmetry of $\mathscr{A}$ is not required.
Let $\beta_2$ satisfy Definition~\ref{def:lp_beta_2} with the same $\gamma_1$
and fixed $3/4\le\gamma_2\le1\le\gamma_3\le2$.
For $h(v):=\sum_{i=1}^n\beta_2(v_i)$, define the potential using
Definition~\ref{def:bregman_divergence} with the root in the first argument:
\begin{equation}\label{eq:lp_potential}
 \mathscr{J}(v):=\mathscr{J}_h(v_*,v)=\sum_{i=1}^n(\beta_2(v_{*,i})-\beta_2(v_i)-\beta_2'(v_i)(v_{*,i}-v_i)).
\end{equation}
For $v^{\new}:=\mathscr{A}\beta_1(v)+b$ and $\delta_v:=v^{\new}-v$, the following guarantees hold.
\begin{itemize}
\item[(a)] The potential satisfies $\mathscr{J}(v)\ge0$.
\item[(b)] Its initial value satisfies $\mathscr{J}({\bf 0}_n)\le3\gamma_3 n/8$.
\item[(c)] The residual satisfies $\langle v-v_*,\delta_v\rangle\le -\frac{\gamma_1^2}{2\gamma_3}\mathscr{J}(v)-\tfrac12\|\delta_v\|_2^2$.
\end{itemize}
\end{lemma}
\begin{proof}
{\bf Proof of Part (a).}
Property (a) of Definition~\ref{def:lp_beta_2} gives $\mathscr{J}(v)\ge0$.

{\bf Proof of Part (b).}
Integrating the upper bound in Definition~\ref{def:lp_beta_2}-(b) from zero gives
$|\beta_2'(\zeta)|\le\gamma_3|\zeta|/(\gamma_1(\gamma_1+|\zeta|))\le\gamma_3/\gamma_1$.
Together with Definition~\ref{def:lp_beta_2}-(a), a second integration gives
$0\le\beta_2(\zeta)\le\gamma_3|\zeta|/\gamma_1$.

We have
\[
 \mathscr{J}({\bf 0}_n)=\sum_{i=1}^n\beta_2(v_{*,i})
 \le\gamma_3\|v_*\|_1/\gamma_1\le3\gamma_3 n/8,
\]
where the first step uses $\beta_2(0)=\beta_2'(0)=0$;
the second step uses $\beta_2(\zeta)\le\gamma_3|\zeta|/\gamma_1$;
and the third step uses $\|v_*\|_1\le3\gamma_1 n/8$.

{\bf Proof of Part (c).}
For real $\zeta$, we have
\begin{equation}\label{eq:lp_derivative_squared_contraction}
 1-\beta_1'(\zeta)^2=\frac{\gamma_1^2}{\zeta^2+\gamma_1^2}
 \ge\frac{\gamma_1^2}{(\gamma_1+|\zeta|)^2}
 \ge\frac{\gamma_1^2}{\gamma_3}\beta_2''(\zeta).
\end{equation}
The first step substitutes the derivative in Lemma~\ref{lem:lp_beta_1_example}.
The second step uses $\zeta^2+\gamma_1^2\le(\gamma_1+|\zeta|)^2$.
The third step uses Definition~\ref{def:lp_beta_2}-(b).

For real $\zeta_1,\zeta_2$, the fundamental theorem of calculus gives
$\beta_1(\zeta_1)-\beta_1(\zeta_2)=\int_{\zeta_2}^{\zeta_1}\beta_1'(\zeta_3)\d \zeta_3$.
Cauchy--Schwarz therefore gives
\begin{equation}\label{eq:lp_scalar_cauchy_schwarz}
 (\beta_1(\zeta_1)-\beta_1(\zeta_2))^2
 \le(\zeta_1-\zeta_2)\int_{\zeta_2}^{\zeta_1}\beta_1'(\zeta_3)^2\d \zeta_3.
\end{equation}
This holds also when $\zeta_1<\zeta_2$, by reversing the integration limits.
We have
\begin{align}
 &(\zeta_1-\zeta_2)^2-(\beta_1(\zeta_1)-\beta_1(\zeta_2))^2
 \ge(\zeta_1-\zeta_2)\int_{\zeta_2}^{\zeta_1}(1-\beta_1'(\zeta_3)^2)\d \zeta_3
 \ge\frac{\gamma_1^2}{\gamma_3}(\zeta_1-\zeta_2)\int_{\zeta_2}^{\zeta_1}\beta_2''(\zeta_3)\d \zeta_3\notag\\
 &=\frac{\gamma_1^2}{\gamma_3}(\zeta_1-\zeta_2)(\beta_2'(\zeta_1)-\beta_2'(\zeta_2))
 \ge\frac{\gamma_1^2}{\gamma_3}(\beta_2(\zeta_2)-\beta_2(\zeta_1)-\beta_2'(\zeta_1)(\zeta_2-\zeta_1)).
 \label{eq:lp_scalar_difference_contraction}
\end{align}
The first step uses Eq.~\eqref{eq:lp_scalar_cauchy_schwarz}.
The second step integrates Eq.~\eqref{eq:lp_derivative_squared_contraction};
if $\zeta_1<\zeta_2$, both the integral and its prefactor change sign.
The third step uses the fundamental theorem of calculus.
The fourth step uses
$(\zeta_1-\zeta_2)(\beta_2'(\zeta_1)-\beta_2'(\zeta_2))
=\mathscr{J}_{\beta_2}(\zeta_1,\zeta_2)+\mathscr{J}_{\beta_2}(\zeta_2,\zeta_1)$
and nonnegativity of the Bregman divergence from
Definition~\ref{def:bregman_divergence}.

The identity $v_*=\mathscr{A}\beta_1(v_*)+b$ gives
$v-v_*+\delta_v=\mathscr{A}(\beta_1(v)-\beta_1(v_*))$.
We have
\begin{align*}
 2\langle v-v_*,\delta_v\rangle+\|\delta_v\|_2^2
 &=\|v-v_*+\delta_v\|_2^2-\|v-v_*\|_2^2\\
 &\le\|\beta_1(v)-\beta_1(v_*)\|_2^2-\|v-v_*\|_2^2
 \le-\frac{\gamma_1^2}{\gamma_3}\mathscr{J}(v).
\end{align*}
The first step expands the squared norm.
The second step uses $v-v_*+\delta_v=\mathscr{A}(\beta_1(v)-\beta_1(v_*))$ and $\|\mathscr{A}\|_{\rm op}\le1$.
The third step sums Eq.~\eqref{eq:lp_scalar_difference_contraction}
over $\zeta_1=v_i$ and $\zeta_2=v_{*,i}$.
Rearranging proves Part~(c).

\end{proof}

The algorithm moves one coordinate at a time in the variable $\beta_2'(v_i)$. We check that the move stays in the range of $\beta_2'$ and bound its effect on the potential.

\begin{lemma}[A coordinate update is well defined]\label{lem:lp_coordinate_update}
Use the hypotheses and potential of Lemma~\ref{lem:lp_potential_contraction}, and define $v^{\new}:=\mathscr{A}\beta_1(v)+b \in \R^n$ and $\delta_v:=v^{\new}-v \in \R^n$. Let $c:=2$.
Fix one coordinate $i\in[n]$. Define $v_{{\rm drift},i}:=\gamma_1+|v_i|$, $v_{{\rm eng},i}:=v_{{\rm drift},i}^2+\delta_{v,i}^2$, and
$v_{{\rm inv},i}:=\delta_{v,i}/(c v_{{\rm eng},i})$, where $\delta_{v,i}$ is the $i$th coordinate of $\delta_v$. Define $v_{\rm tra}\in\R^n$ by $v_{{\rm tra},i}:=(\beta_2')^{-1}(\beta_2'(v_i)+v_{{\rm inv},i})$ and $v_{{\rm tra},j}:=v_j$ for every $j\in[n]\setminus\{i\}$. The coordinate update has the following guarantees.
\begin{itemize}
\item[(a)] The value $\beta_2'(v_i)+v_{{\rm inv},i}$ belongs to $I_{\beta_2}=\beta_2'(\R)$,
so $v_{\rm tra}$ is well defined.
\item[(b)] The potential change satisfies
\[
 \mathscr{J}(v_{\rm tra})-\mathscr{J}(v)\le\frac{(v_i-v_{*,i})\delta_{v,i}}{c v_{{\rm eng},i}}
 +\frac{3\delta_{v,i}^2}{8c v_{{\rm eng},i}}.
\]
\item[(c)] The updated coordinate satisfies
\[
 \frac34\le\frac{2c\gamma_2}{2c\gamma_2+1}
 \le\frac{\gamma_1+|v_{{\rm tra},i}|}{\gamma_1+|v_i|}
 \le\frac{2c\gamma_2}{2c\gamma_2-1}\le\frac32.
\]
\end{itemize}
\end{lemma}
\begin{proof}
{\bf Proof of Part (a).}
Since $v_{{\rm eng},i}=v_{{\rm drift},i}^2+\delta_{v,i}^2\ge2v_{{\rm drift},i}|\delta_{v,i}|$ and $c=2$,
we have $|v_{{\rm inv},i}|\le1/(2cv_{{\rm drift},i})=1/(4v_{{\rm drift},i})$.
By Definition~\ref{def:lp_beta_2}-(a)--(b), $\beta_2'$ is strictly increasing
and has finite limits at both infinities.
The distances from $\beta_2'(v_i)$ to the two endpoints of $I_{\beta_2}$ are
$\int_{-\infty}^{v_i}\beta_2''(\zeta_3)\d \zeta_3$ and
$\int_{v_i}^{\infty}\beta_2''(\zeta_3)\d \zeta_3$.
The lower curvature bound gives at least $\gamma_2/v_{{\rm drift},i}$ for each distance:
integration along the ray away from zero gives exactly this lower bound,
and integration along the other ray gives at least as large a lower bound.
Since $\gamma_2\ge3/4$, we have $|v_{{\rm inv},i}|<\gamma_2/v_{{\rm drift},i}$.
Thus the entire segment from $\beta_2'(v_i)$ to $\beta_2'(v_i)+v_{{\rm inv},i}$
lies in $I_{\beta_2}$, proving Part~(a).

{\bf Proof of Part (b).}
For $\tau\in[0,1]$, define
$s(\tau):=(\beta_2')^{-1}(\beta_2'(v_i)+\tau v_{{\rm inv},i})$, and set
$\kappa_i:=v_{{\rm drift},i}|v_{{\rm inv},i}|/\gamma_2$.
Part~(a) makes $s(\tau)$ well defined, and
$0\le\kappa_i\le1/(2c\gamma_2)\le1/3$.
We have
\begin{equation}\label{eq:lp_coordinate_reciprocal_path}
 |\frac1{\gamma_1+|s(\tau)|}-\frac1{v_{{\rm drift},i}}|
 \le\frac{|\beta_2'(s(\tau))-\beta_2'(v_i)|}{\gamma_2}
 =\frac{\tau|v_{{\rm inv},i}|}{\gamma_2}.
\end{equation}
The first step integrates the absolute derivative of
$x\mapsto1/(\gamma_1+|x|)$ between $v_i$ and $s(\tau)$, splitting at zero
if necessary, and uses Definition~\ref{def:lp_beta_2}-(b).
The second step uses the definition of $s(\tau)$.
Since $1-\kappa_i\tau\ge2/3>0$, we obtain
\begin{equation}\label{eq:lp_coordinate_dual_curvature}
 \frac1{\beta_2''(s(\tau))}
 \le\frac{(\gamma_1+|s(\tau)|)^2}{\gamma_2}
 \le\frac{v_{{\rm drift},i}^2}{\gamma_2(1-\kappa_i\tau)^2},
\end{equation}
where the first step uses Definition~\ref{def:lp_beta_2}-(b), and the second
step rearranges Eq.~\eqref{eq:lp_coordinate_reciprocal_path}.

Only coordinate $i$ changes. As a function of its dual variable
$\theta=\beta_2'(v_i)$, its contribution to $\mathscr{J}$ has first derivative
$(\beta_2')^{-1}(\theta)-v_{*,i}$ and second derivative
$1/\beta_2''((\beta_2')^{-1}(\theta))$, by the chain rule and the inverse
function theorem. Taylor's formula with an integral remainder gives
\begin{equation}\label{eq:lp_coordinate_taylor_remainder}
 \mathscr{J}(v_{\rm tra})-\mathscr{J}(v)
 =(v_i-v_{*,i})v_{{\rm inv},i}+v_{{\rm inv},i}^2\int_0^1\frac{1-\tau}{\beta_2''(s(\tau))}\d\tau.
\end{equation}
We have
\[
 \int_0^1\frac{1-\tau}{(1-\kappa_i\tau)^2}\d\tau
 =\int_0^1\frac{\tau}{1-\kappa_i\tau}\d\tau
 \le\frac1{2(1-\kappa_i)},
\]
where the first step integrates by parts using
$(\tau/(1-\kappa_i\tau))'=1/(1-\kappa_i\tau)^2$;
and the second step uses $1-\kappa_i\tau\ge1-\kappa_i>0$ and
$\int_0^1\tau\d\tau=1/2$.
Together with Eq.~\eqref{eq:lp_coordinate_dual_curvature}, this gives
\begin{equation}\label{eq:lp_coordinate_remainder_bound}
 v_{{\rm inv},i}^2\int_0^1\frac{1-\tau}{\beta_2''(s(\tau))}\d\tau
 \le\frac{v_{{\rm drift},i}^2v_{{\rm inv},i}^2}{2\gamma_2(1-\kappa_i)}
 =\frac{v_{{\rm drift},i}^2\delta_{v,i}^2}{2c v_{{\rm eng},i}(c\gamma_2v_{{\rm eng},i}-v_{{\rm drift},i}|\delta_{v,i}|)},
\end{equation}
where the second step substitutes the definitions of $v_{{\rm inv},i}$ and $\kappa_i$.
The denominator is positive because $\kappa_i\le1/3$.
Moreover,
\[
 c\gamma_2v_{{\rm eng},i}-v_{{\rm drift},i}|\delta_{v,i}|
 \ge\frac32(v_{{\rm drift},i}^2+\delta_{v,i}^2)-v_{{\rm drift},i}|\delta_{v,i}|
 =\frac32(|\delta_{v,i}|-v_{{\rm drift},i}/3)^2+\frac43v_{{\rm drift},i}^2
 \ge\frac43v_{{\rm drift},i}^2,
\]
where the first step uses $c=2$ and $\gamma_2\ge3/4$;
the second step completes the square; and the third step drops the
nonnegative square.
Substituting this bound into Eq.~\eqref{eq:lp_coordinate_remainder_bound}
bounds the remainder by $3\delta_{v,i}^2/(8c v_{{\rm eng},i})$.
Eq.~\eqref{eq:lp_coordinate_taylor_remainder} and
$v_{{\rm inv},i}=\delta_{v,i}/(c v_{{\rm eng},i})$ now prove Part~(b).

{\bf Proof of Part (c).}
Since $s(1)=v_{{\rm tra},i}$, Eq.~\eqref{eq:lp_coordinate_reciprocal_path} gives
\[
 \frac1{v_{{\rm drift},i}}(1-\frac1{2c\gamma_2})
 \le\frac1{\gamma_1+|v_{{\rm tra},i}|}
 \le\frac1{v_{{\rm drift},i}}(1+\frac1{2c\gamma_2}),
\]
where both steps use $|v_{{\rm inv},i}|\le1/(2cv_{{\rm drift},i})$ from Part~(a).
All quantities are positive, so taking reciprocals proves the two middle
bounds in Part~(c). Finally, $2c\gamma_2\ge3$ gives the outer bounds
$3/4$ and $3/2$.
\end{proof}

\subsection{Size and normalization bounds}\label{sec:lp_normalization}

The shrink process and its potential decrease are proved in
Lemma~\ref{lem:lp_shrink} of Section~\ref{sec:lp_entry_state}.
The following lemma gives the size bounds at the completed iteration starts of the implemented algorithm.

\begin{lemma}[Normalization and intermediate size]\label{lem:lp_normalization}
Assume the hypotheses of Lemma~\ref{lem:lp_potential_contraction} and use Algorithm~\ref{alg:iterative_solve}.
Let $n=2^{L_1}$, with $L_1\ge1$, denote its padded dimension, and set
$L_6:=2^{10}\gamma_1^2n^2$ and $L_2:=L_1+4$.
At every iteration start reached before a return or restart, with
$\mathcal{N}_1:=\sum_{i=1}^n(\gamma_1+|v_i|)^2$ and $\mathcal{N}_2:=\|\delta_v\|_2^2$,
the following guarantees hold.
\begin{itemize}
\item[(a)] The total base weight satisfies $\mathcal{N}_1<2^{6}\gamma_1^2n^2$.
\item[(b)] The squared residual norm satisfies $\mathcal{N}_2<2^{7}\gamma_1^2n^2$.
\item[(c)] The normalization bound is $\mathcal{N}_1+6\mathcal{N}_2<L_6=2^{10}\gamma_1^2n^2$.
\item[(d)] Every intermediate coordinate, including before shrink, has magnitude
less than $8\gamma_1 n$, so the buckets of normalized magnitude $2|v_i|/\gamma_1$
use only levels in $[L_2]$.
\end{itemize}
\end{lemma}
\begin{proof}
{\bf Proof of Part (a).}
The initial vector is zero and satisfies the norm bounds of
Lemma~\ref{lem:lp_shrink}-(c). At each completed iteration, a
$\mathsf{null}$ or $\mathsf{fail}$ outcome leaves the vector unchanged; a coordinate
update is followed by shrink, which restores those bounds. This
induction holds for every selected coordinate and does not use its
sampling probability. An interrupted iteration does not lead to
another iteration start in the same trial.

We have
\[
 \mathcal{N}_1\le(\gamma_1 n+\|v\|_1)^2\le(57/8)^2\gamma_1^2n^2<2^{6}\gamma_1^2n^2,
\]
where the first step follows from the nonnegativity of $\gamma_1+|v_i|$ for every $i$;
the second step follows from $\|v\|_1\le(49/8)\gamma_1 n$ in Lemma~\ref{lem:lp_shrink};
and the third step follows from $(57/8)^2<2^{6}$.

{\bf Proof of Part (b).}
Let $Z:=\sum_{i=1}^n|v_i|\mathbbm{1}[|v_i|\ge\gamma_1]$.
We have
\[
 \|v\|_2^2\le Z^2+n\gamma_1^2\le\frac{1713}{64}\gamma_1^2n^2<(21/4)^2\gamma_1^2n^2,
\]
where the first step separates the coordinates with $|v_i|\ge\gamma_1$,
whose squared sum is at most $Z^2$, from those with $|v_i|<\gamma_1$,
whose squared sum is at most $n\gamma_1^2$;
the second step uses $Z\le(41/8)\gamma_1 n$ from Lemma~\ref{lem:lp_shrink}-(c)
and $n\ge2$, giving $(41/8)^2+1/n\le1713/64$;
and the third step uses $1713<1764$.
Consequently,
\[
 \|\delta_v\|_2\le2\|v\|_2+\|b\|_2<11\gamma_1 n,
\]
where the first step follows from the triangle inequality applied to
$\delta_v=\mathscr{A}\beta_1(v)+b-v$, together with $\|\mathscr{A}\|_{\rm op}\le1$ and $0\le\beta_1(v_i)\le|v_i|$;
and the second step uses $\|v\|_2<(21/4)\gamma_1 n$ and
$\|b\|_2\le\gamma_1 n/2$, together with $2\cdot(21/4)+1/2=11$.
Squaring and using $11^2=121<2^{7}$ proves Part~(b).

{\bf Proof of Part (c).}
Parts~(a) and~(b), together with $2^{6}+6\cdot2^{7}=832<2^{10}$, prove the claim.

{\bf Proof of Part (d).}
We have
\[
 |v_{{\rm tra},i}|\le\frac32(\gamma_1+(41/8)\gamma_1 n)-\gamma_1
 =\gamma_1(123n/16+1/2)<8\gamma_1 n,
\]
where the first step uses Lemma~\ref{lem:lp_coordinate_update}-(c) and the bound
$\|v\|_\infty\le(41/8)\gamma_1 n$ from Lemma~\ref{lem:lp_shrink}-(c);
the second step simplifies the expression; and the third step uses $n\ge2$.
The other coordinates are unchanged, and shrink only reduces magnitudes.
Since $\gamma_1 2^{L_2-1}=8\gamma_1 n$, all intermediate coordinates lie
in the buckets of Definition~\ref{def:lp_entry_state_invariant}.
\end{proof}

\subsection{Convergence and work budgets}\label{sec:lp_convergence}

Over one trial, the expected decrease in potential pays for both the squared residuals and the shrink updates. This gives a bound on the chance that a completed trial fails its residual test.

\begin{theorem}[Convergence and work budgets of IterativeSolve]\label{thm:lp_convergence}
Assume the hypotheses of Lemma~\ref{lem:lp_potential_contraction}, and consider one trial of
Algorithm~\ref{alg:iterative_solve}, ending at its first return or restart.
Let $n=2^{L_1}$, with $L_1\ge1$, be its padded dimension, and define
$L_5:=2^{12}n^2$, $L_6:=\gamma_1^2L_5/4$, and $L_7:=64L_5L_1$.
Write $\delta_v:=\mathscr{A}\beta_1(v)+b-v$.
In the sketched branch, a state is good in the sense of
Lemma~\ref{lem:lp_ensemble}; in the exact branch, every state is good.
Let $\epsilon_{\rm ds}$ upper bound the probability that the trial
reaches a bad iteration-start state before the operation cutoff.

For $0\le k<L_7$, let $J_k$ indicate that the trial reaches iteration
$k$ with a good current state and without any earlier bad state or
cutoff, and let $v_k$ be its vector at that time. Let $J_{L_7}$ indicate
that all $L_7$ iterations and the final residual test are completed
before the cutoff without a bad iteration-start state; $v_{L_7}$ is
the tested vector. Let $n_k^{\rm shrink}$ count the shrink changes
completed in iteration $k$ before stopping, counting repetitions,
and set it to zero when $J_k=0$. Terms multiplied by $J_k$ are defined
to be zero when $J_k=0$. Let $\mathcal F_k$ contain the sketches and
the history through the corresponding iteration-start inspection,
with stopped histories kept fixed. These indicators are used only in
the analysis. The following guarantees hold.
\begin{itemize}
\item[(a)] For $0\le k<L_7$,
\[
\begin{aligned}
 \E[J_{k+1}\mathscr{J}(v_{k+1})\mid\mathcal F_k]
 &\le(1-1/(2L_5))J_k\mathscr{J}(v_k)-\frac{J_k\|\delta_{v_k}\|_2^2}{16L_6}
 -\frac1{16}\E[n_k^{\rm shrink}\mid\mathcal F_k].
\end{aligned}
\]
\item[(b)] The probability that the trial completes without a cutoff
and fails its exact residual test is at most $1/16+\epsilon_{\rm ds}$.
\item[(c)] The expected squared residual budget before stopping satisfies
$\sum_{k=0}^{L_7-1}\E[J_k\|\delta_{v_k}\|_2^2]\le12L_6 n$.
\item[(d)] The expected number of shrink changes before stopping satisfies
$\E[\sum_{k=0}^{L_7-1}n_k^{\rm shrink}]\le16n$.
\end{itemize}
\end{theorem}
\begin{proof}
{\bf Proof of Part (a).}
If $J_k=0$, all terms are zero. Suppose $J_k=1$ and fix
$\mathcal F_k$. Lemma~\ref{lem:lp_sampling_validity}-(a) gives
coordinate probabilities $v_{{\rm eng},i}/L_6$, where
$v_{{\rm eng},i}:=(\gamma_1+|v_{k,i}|)^2+\delta_{v_k,i}^2$.
These probabilities are legal by Lemma~\ref{lem:lp_normalization}-(c).
Complete this one iteration with the same fresh randomness, temporarily
ignoring the cutoff. Let $w$ be its vector after the mirror update and
shrink, and let $\overline n_k^{\rm shrink}$ count its shrink changes.

We have
\begin{align*}
 \E[\mathscr{J}(w)+\overline n_k^{\rm shrink}/16\mid\mathcal F_k]
 &\le\mathscr{J}(v_k)+\frac{\langle v_k-v_*,\delta_{v_k}\rangle}{2L_6}
 +\frac{3\|\delta_{v_k}\|_2^2}{16L_6}\\
 &\le(1-1/(\gamma_3L_5))\mathscr{J}(v_k)
 -\frac{\|\delta_{v_k}\|_2^2}{16L_6}\\
 &\le(1-1/(2L_5))\mathscr{J}(v_k)
 -\frac{\|\delta_{v_k}\|_2^2}{16L_6}.
\end{align*}
The first step averages the coordinate bound in
Lemma~\ref{lem:lp_coordinate_update}-(b), with $c=2$ and probabilities $v_{{\rm eng},i}/L_6$,
then applies the shrink bound in Lemma~\ref{lem:lp_shrink}-(b).
A $\mathsf{null}$ outcome leaves the vector unchanged and makes no shrink changes.
The second step uses Lemma~\ref{lem:lp_potential_contraction}-(c),
$L_6=\gamma_1^2L_5/4$, and $-1/4+3/16=-1/16$.
The third step uses $\gamma_3\le2$ and $\mathscr{J}(v_k)\ge0$.

We have
\[
 J_{k+1}\mathscr{J}(v_{k+1})+n_k^{\rm shrink}/16
 \le\mathscr{J}(w)+\overline n_k^{\rm shrink}/16.
\]
If $J_{k+1}=1$, the completed iteration has vector $w$ and the same
shrink changes. Otherwise the left potential is zero, the right
potential is nonnegative, and the completed shrink changes form a
prefix of those in the full iteration. This includes a cutoff during
sampling, maintenance, shrink, or the terminal test. Taking conditional
expectations proves Part~(a). The same argument holds for every fixed
choice of sketches.

{\bf Proof of Part (b).}
We have
\begin{equation}\label{eq:lp_expected_potential_decay}
 \E[J_k\mathscr{J}(v_k)]\le(1-1/(2L_5))^k\mathscr{J}({\bf 0}_n)
 \le ne^{-k/(2L_5)}.
\end{equation}
The first step takes expectations in Part~(a), drops its nonpositive
residual and shrink terms, uses $J_0\le1$, and iterates the contraction. The second step uses
$1-t\le e^{-t}$ and $\mathscr{J}({\bf 0}_n)\le3\gamma_3 n/8\le n$, using Lemma~\ref{lem:lp_potential_contraction}-(b) and $\gamma_3\le2$.

We have
\begin{equation}\label{eq:lp_terminal_iteration_budget}
 \log(2^{21}n^5(13n)^2)
 =\log(2^{21}\cdot13^2)+7L_1\log2<32L_1=L_7/(2L_5).
\end{equation}
The first step uses $(13n)^2=13^2n^2$ and $\log n=L_1\log2$.
The second step uses $\log(2^{21}\cdot13^2)<20$, $7\log2<5$, and
$20+5L_1\le25L_1<32L_1$ for $L_1\ge1$.
The third step follows from the definition $L_7:=64L_5 L_1$ in
Algorithm~\ref{alg:iterative_solve}.

We have
\begin{equation}\label{eq:lp_terminal_expected_potential}
 \E[J_{L_7}\mathscr{J}(v_{L_7})]\le\frac{n}{2^{21}n^5(13n)^2}
 =\frac1{2^{21}n^4(13n)^2}.
\end{equation}
Here the first step uses Eqs.~\eqref{eq:lp_expected_potential_decay}
and~\eqref{eq:lp_terminal_iteration_budget}; the second step cancels $n$.

We have
\begin{equation}
 \E[J_{L_7}\|\delta_{v_{L_7}}\|_2^2]\le4\E[J_{L_7}\|v_{L_7}-v_*\|_2^2]\le2\gamma_1^2(13n)^2\E[J_{L_7}\mathscr{J}(v_{L_7})]\le\frac{\gamma_1^2}{2^{20}n^4},\label{eq:lp_terminal_residual_second_moment}
\end{equation}
where the first step follows from $\delta_{v_{L_7}}=\mathscr{A}(\beta_1(v_{L_7})-\beta_1(v_*))-(v_{L_7}-v_*)$
and nonexpansiveness of $\beta_1$ and $\mathscr{A}$;
the second step integrates the lower curvature bound in
Definition~\ref{def:lp_beta_2}-(b) in the Bregman divergence from
Definition~\ref{def:bregman_divergence}.
On the event $J_{L_7}=1$, we have $\|v_*\|_\infty\le\gamma_1 n/2$ and $\|v_{L_7}\|_\infty\le(41/8)\gamma_1 n$,
so every scalar $\zeta$ on the relevant coordinate segments satisfies
\[
 \beta_2''(\zeta)\ge\frac{\gamma_2}{\gamma_1^2(1+41n/8)^2}
 \ge\frac{16}{675\gamma_1^2n^2}
 \ge\frac4{\gamma_1^2(13n)^2},
\]
where the first step uses Definition~\ref{def:lp_beta_2}-(b) and the coordinate bounds;
the second step uses $\gamma_2\ge3/4$ and $1+41n/8\le45n/8$ for $n\ge2$;
and the third step uses $4\cdot13^2=676\ge675$.
The third step in Eq.~\eqref{eq:lp_terminal_residual_second_moment} substitutes
Eq.~\eqref{eq:lp_terminal_expected_potential}.
We have
\begin{equation}\label{eq:lp_terminal_failure_before_stop}
 \Pr[J_{L_7}=1\text{ and }\|\delta_{v_{L_7}}\|_2>\gamma_1/(256n^2)]
 \le\frac{2^{16}n^4}{\gamma_1^2}\E[J_{L_7}\|\delta_{v_{L_7}}\|_2^2]
 \le\frac1{16},
\end{equation}
where the first step applies Markov's inequality to
$J_{L_7}\|\delta_{v_{L_7}}\|_2^2$;
and the second step substitutes Eq.~\eqref{eq:lp_terminal_residual_second_moment} and
uses $2^{16}/2^{20}=1/16$.
A trial that completes its terminal test without a cutoff either has
$J_{L_7}=1$ or has reached a bad state. The latter event has probability
at most $\epsilon_{\rm ds}$. Adding this probability proves Part~(b).

{\bf Proof of Part (c).}
Taking expectations in Part~(a) and dropping the nonpositive
potential-contraction and shrink terms gives
\[
 \E[J_k\|\delta_{v_k}\|_2^2]
 \le16L_6(\E[J_k\mathscr{J}(v_k)]-\E[J_{k+1}\mathscr{J}(v_{k+1})]).
\]
Summing over $k=0,\ldots,L_7-1$ telescopes the right-hand side.
Since $J_0\le1$ and $v_0={\bf 0}_n$ whenever $J_0=1$, we obtain
\begin{align*}
 \sum_{k=0}^{L_7-1}\E[J_k\|\delta_{v_k}\|_2^2]
 &\le16L_6(\mathscr{J}({\bf 0}_n)-\E[J_{L_7}\mathscr{J}(v_{L_7})])\\
 &\le16L_6\mathscr{J}({\bf 0}_n)\le6\gamma_3L_6n\le12L_6n.
\end{align*}
The second step uses nonnegativity of the potential from
Lemma~\ref{lem:lp_potential_contraction}-(a).
The third step uses $\mathscr{J}({\bf 0}_n)\le3\gamma_3n/8$ from
Lemma~\ref{lem:lp_potential_contraction}-(b).
The fourth step uses $\gamma_3\le2$. This proves Part~(c).

{\bf Proof of Part (d).}
We have
\[
 \frac1{16}\E[\sum_{k=0}^{L_7-1}n_k^{\rm shrink}]
 \le \mathscr{J}({\bf 0}_n)-\E[J_{L_7}\mathscr{J}(v_{L_7})]\le \mathscr{J}({\bf 0}_n)\le n,
\]
where the first step takes expectations in Part~(a),
drops the nonpositive potential-contraction and residual terms, uses $J_0\le1$, and
sums over $k=0,\ldots,L_7-1$; the second step uses $J_{L_7}\mathscr{J}(v_{L_7})\ge0$;
and the third step uses $\mathscr{J}({\bf 0}_n)\le3\gamma_3 n/8$ and $\gamma_3\le2$. The potential bounds follow from
Lemma~\ref{lem:lp_potential_contraction}. Multiplying by $16$ proves
Part~(d).
\end{proof}

\subsection{Implementation and base sampling}\label{sec:lp_implementation_and_base_sampling}
Algorithm~\ref{alg:iterative_solve} uses the
\textsc{EntryState} representation in Definition~\ref{def:lp_entry_state_invariant}.

\begin{algorithm}[!ht]
\small
\caption{\textsc{IterativeSolve}: implementation}\label{alg:iterative_solve}
\begin{algorithmic}[1]
\Procedure{IterativeSolve}{$\mathscr{A}\in\R^{n\times n},b\in\R^n$} \Comment{Lemma~\ref{lem:lp_iterative_solve_time}}
\State $n'\gets2^{\lceil\log_2 n\rceil}$; pad using $\mathrm{blockDiag}(\mathscr{A},0)$ and $(b,{\bf 0}_{n'-n})$; set $n\gets n'$.
\State $L_1\gets\log_2n$; $L_2\gets L_1+4$; $L_3\gets2^{\lceil\log_2\max\{2,L_1\}\rceil}$; $L_{4,j}\gets12\gamma_1^2n2^{j-1}$ for $j\in[L_2]$.
\State $L_4\gets\sum_{j=1}^{L_2}L_{4,j}$; $L_5\gets2^{12}n^2$; $L_6\gets\gamma_1^2L_5/4$; $L_{6,4}\gets L_6-L_4$; $L_7\gets64L_5 L_1$.
\State \textsc{EntryState} $\mathrm{entryState}$. \Comment{Algorithm~\ref{alg:entry_state}}
\Loop \Comment{$O(1)$ expected iterations}
\State \Call{\ensuremath{\mathrm{entryState}}.Init}{$\mathscr{A},b,L_2,L_3$}. \Comment{Algorithm~\ref{alg:entry_state}}
\For{$k=0,\ldots,L_7-1$} \Comment{$O(n^2 \log n)$ iterations}
\State $j\gets\lfloor k/(32L_5)\rfloor$; $T\gets\max\{2,L_3 2^{-j}\}$.
\State \Call{\ensuremath{\mathrm{entryState}.\mathrm{exactL2Sampler}}.UpdateT}{$T$}. \Comment{Algorithm~\ref{alg:exact_l2_sampler}}
\If{a coin with success probability $L_4/L_6$ succeeds}
\State Choose $j\in[L_2]$ with probability $L_{4,j}/L_4$. \Comment{Lemma~\ref{lem:lp_base_sampling}-(d)}
\If{$S_j=\emptyset$}
\State $\iota\gets\mathsf{null}$.
\Else
\State Choose $i$ uniformly from $S_j$.
\State $v_i\gets\Call{\ensuremath{\mathrm{entryState}}.Query}{i}$. \Comment{Algorithm~\ref{alg:entry_state}}
\If{a coin with success probability $|S_j|(\gamma_1+|v_i|)^2/L_{4,j}$ succeeds}
\State $\iota\gets i$.
\Else
\State $\iota\gets\mathsf{null}$.
\EndIf
\EndIf
\Else
\State $\iota\gets\Call{\ensuremath{\mathrm{entryState}.\mathrm{exactL2Sampler}}.L2Sample}{L_{6,4}}$. \Comment{Algorithm~\ref{alg:exact_l2_sampler_sample}}
\EndIf
\If{$\iota\in\{\mathsf{null},\mathsf{fail}\}$}
\State Continue to the next iteration.
\EndIf
\State $v_\iota\gets\Call{\ensuremath{\mathrm{entryState}}.Query}{\iota}$. \Comment{Algorithm~\ref{alg:entry_state}}
\State $\delta_{v,\iota}\gets\Call{\ensuremath{\mathrm{entryState}.\mathrm{exactIP}}.Query}{\iota}+b_\iota-v_\iota$. \Comment{Algorithm~\ref{alg:exact_ip}}
\State $v_{{\rm drift},\iota}\gets\gamma_1+|v_\iota|$; $v_{{\rm eng},\iota}\gets v_{{\rm drift},\iota}^2+\delta_{v,\iota}^2$; $v_{{\rm inv},\iota}\gets \delta_{v,\iota}/(2v_{{\rm eng},\iota})$. \Comment{Lemma~\ref{lem:lp_coordinate_update}}
\State $v_{{\rm tra},\iota}\gets(\beta_2')^{-1}(\beta_2'(v_\iota)+v_{{\rm inv},\iota})$; \Call{\ensuremath{\mathrm{entryState}}.UpdateE}{$\iota,v_{{\rm tra},\iota}$}. \Comment{Algorithm~\ref{alg:entry_state}}
\State \Call{\ensuremath{\mathrm{entryState}}.Shrink}{}. \Comment{Algorithm~\ref{alg:entry_state}}
\EndFor
\State $v_i\gets\Call{\ensuremath{\mathrm{entryState}}.Query}{i}$ for every $i\in[n]$. \Comment{Algorithm~\ref{alg:entry_state}}
\If{$\|\mathscr{A}\beta_1(v)+b-v\|_2\le\gamma_1/(256n^2)$}
\State \Return the original coordinates of $v$.
\EndIf
\State Restart with fresh randomness; also restart upon reaching $O(n^2\log n)$ operations, including initialization, all coin calls, and certification.
\EndLoop
\EndProcedure
\end{algorithmic}
\end{algorithm}

Sampling a level before sampling a coordinate within it lets the base branch realize its squared drift weights in constant expected time.

\begin{lemma}[Base sampling]\label{lem:lp_base_sampling}
Use the hypotheses of Lemma~\ref{lem:lp_potential_contraction}. Let $n=2^{L_1}$ with
$L_1\ge1$, and consider an iteration-start state of
Algorithm~\ref{alg:iterative_solve} satisfying
Definition~\ref{def:lp_entry_state_invariant} and
$Z\le(41/8)\gamma_1 n$. Set $L_2:=L_1+4$,
$v_{{\rm drift},i}:=\gamma_1+|v_i|$, $q_j:=|S_j|$, and
$\mathcal N_2:=\|\mathscr{A}\beta_1(v)+b-v\|_2^2$. Define
\[
 L_{4,j}:=12\gamma_1^2n2^{j-1}\quad(j\in[L_2]),\qquad
 L_4:=\sum_{j=1}^{L_2}L_{4,j},\qquad
 L_6:=2^{10}\gamma_1^2n^2,\qquad L_{6,4}:=L_6-L_4.
\]
For a complete base proposal, before applying the operation cutoff,
the following guarantees hold for every fixed history and choice of sketches.
\begin{itemize}
\item[(a)] The weights satisfy $q_jv_{{\rm drift},i}^2\le L_{4,j}$ for
$j\in[L_2]$ and $i\in S_j$, and $L_4<192\gamma_1^2n^2$.
\item[(b)] The residual denominator satisfies $L_{6,4}>6\mathcal N_2$.
\item[(c)] The probability that the base branch selects coordinate
$i\in[n]$ is $v_{{\rm drift},i}^2/L_6$. An empty bucket or a rejected proposal
returns $\mathsf{null}$.
\item[(d)] The base proposal takes $O(1)$ expected time, including
the branch coin, bucket selection, uniform coordinate selection, and
acceptance coin.
\end{itemize}
\end{lemma}
\begin{proof}
{\bf Proof of Part (a).}
By Definition~\ref{def:lp_entry_state_invariant},
$v_{{\rm drift},i}<2\gamma_1$ for $i\in S_1$, and
$v_{{\rm drift},i}<\gamma_1(1+2^{j-1})$ for $i\in S_j$ with $j\ge2$.
For $j=1$, the bound $q_1\le n$ gives
$q_1v_{{\rm drift},i}^2<4\gamma_1^2n\le L_{4,1}$.
For $2\le j\le4$, we have
\[
 \frac{q_jv_{{\rm drift},i}^2}{\gamma_1^2n2^{j-1}}
 \le\frac{(1+2^{j-1})^2}{2^{j-1}}\le\frac{81}{8}<12.
\]
The first step uses $q_j\le n$ and the bound on $v_{{\rm drift},i}$.
The second step evaluates the expression at $j=2,3,4$, whose maximum is $81/8$.
For $j\ge5$, the lower bucket boundary gives
$q_j\gamma_1 2^{j-2}\le\sum_{i\in S_j}|v_i|\le Z$.
The assumed bound on $Z$ gives $q_j\le41n/(4\cdot2^{j-1})$.

We have
\[
 \frac{q_jv_{{\rm drift},i}^2}{\gamma_1^2n2^{j-1}}
 \le\frac{41}{4}(1+2^{1-j})^2
 \le\frac{41}{4}(\frac{17}{16})^2=\frac{11849}{1024}<12.
\]
The first step substitutes the bounds on $q_j$ and $v_{{\rm drift},i}$.
The second step uses $j\ge5$, and the third step simplifies the constants.
These cases prove the coordinate-weight bound.

We have
\[
 L_4=12\gamma_1^2n(2^{L_2}-1)<192\gamma_1^2n^2.
\]
The first step sums the geometric series. The second step uses
$2^{L_2}=16n$. This proves Part~(a).

{\bf Proof of Part (b).}
We have
\[
 L_4+6\mathcal N_2
 <(192+6\cdot2^{7})\gamma_1^2n^2
 <2^{10}\gamma_1^2n^2=L_6.
\]
The first step uses Part~(a) and Lemma~\ref{lem:lp_normalization}-(b).
The second step uses $192+6\cdot2^{7}=960<2^{10}$, and the third step
substitutes the definition of $L_6$. Rearranging proves Part~(b).
In particular, $0<L_4/L_6<1$, so the branch coin is well defined.

{\bf Proof of Part (c).}
Fix $i\in S_j$, so $q_j>0$. We have
\[
 \Pr[\text{the base branch selects }i]
 =\frac{L_4}{L_6}\frac{L_{4,j}}{L_4}\frac1{q_j}
   \frac{q_jv_{{\rm drift},i}^2}{L_{4,j}}
 =\frac{v_{{\rm drift},i}^2}{L_6}.
\]
The first step multiplies the probabilities of entering the base
branch, choosing bucket $j$, choosing $i$ uniformly from $S_j$, and
accepting $i$. The acceptance probability lies in $[0,1]$ by Part~(a).
The second step cancels the common factors. The algorithm returns
$\mathsf{null}$ when the bucket is empty or the proposal is rejected.
This proves Part~(c).

{\bf Proof of Part (d).}
To choose the bucket, generate an integer $k\ge0$ with
$\Pr[k=t]=2^{-t-1}$ using fair coins. Reject and repeat if $k\ge L_2$,
and otherwise set $j:=L_2-k$. We have
\[
 \Pr[j=s\mid k<L_2]
 =\frac{2^{-(L_2-s)-1}}{1-2^{-L_2}}
 =\frac{2^{s-1}}{2^{L_2}-1}
 =\frac{L_{4,s}}{L_4}\qquad(s\in[L_2]).
\]
The first step conditions the geometric distribution on $k<L_2$.
The second step multiplies the numerator and denominator by $2^{L_2}$.
The third step uses the weight definitions. Each attempt uses two
fair coins in expectation, and it is accepted with probability
$1-2^{-L_2}\ge1/2$. Thus bucket selection takes $O(1)$ expected time.

For a nonempty bucket with $j\ge2$, we have
\[
 q_j\le\frac{(41/8)n}{2^{j-2}}=\frac{41}{32}2^k\le2^{k+1}.
\]
The first step uses $q_j\gamma_1 2^{j-2}\le Z\le(41/8)\gamma_1 n$.
The second step substitutes $j=L_2-k$, $L_2=L_1+4$, and $n=2^{L_1}$.
The third step uses $41/32\le2$. For $j=1$, we instead have
$q_1\le n=2^{k-3}\le2^{k+1}$.
Find the smallest power of two at least $q_j$ by repeated doubling,
then generate a uniform index by fair bits and rejection. Each attempt
is accepted with probability at least $1/2$, so its expected time,
including finding the power of two, is $O(k+1)$.

We have
\[
 \E[k+1\mid k<L_2]
 =\frac{\sum_{t=0}^{L_2-1}(t+1)2^{-t-1}}{1-2^{-L_2}}
 \le\frac{2}{1-2^{-L_2}}\le4.
\]
The first step uses the conditional geometric distribution.
The second step extends the sum to all $t\ge0$, whose value is $2$.
The third step uses $1-2^{-L_2}\ge1/2$.
The branch and acceptance coins each require one uniform draw and
comparison. The remaining operations read a bucket size, an array
entry, and the selected coordinate weight, and take $O(1)$ time.
The coordinate query takes $O(1)$ time by
Lemma~\ref{lem:lp_entry_state_query}.
Combining these bounds proves Part~(d).
\end{proof}

\subsection{Exact sampling and adaptive validity}\label{sec:lp_sampling_validity}

The next lemma proves the sampling law used in
Theorem~\ref{thm:lp_convergence} and bounds the probability of a bad
data-structure state. Its proof uses the pointwise sampling and
concentration guarantees from Section~\ref{sec:lp_exact_l2_sampler}.

\begin{lemma}[Sampling and data-structure failure]\label{lem:lp_sampling_validity}
Use the hypotheses of Lemma~\ref{lem:lp_potential_contraction} and consider one trial of
Algorithm~\ref{alg:iterative_solve}, with its padded dimension $n$,
$L_5=2^{12}n^2$, $L_6=\gamma_1^2L_5/4$, and $L_7=64L_5L_1$.
In the sketched branch, call a state good when it satisfies
Lemma~\ref{lem:lp_ensemble}; in the exact branch, call every state good.
Define $\delta_v:=\mathscr{A}\beta_1(v)+b-v$, $v_{{\rm drift},i}:=\gamma_1+|v_i|$, and
$v_{{\rm eng},i}:=v_{{\rm drift},i}^2+\delta_{v,i}^2$. The following guarantees hold.
\begin{itemize}
\item[(a)] At every reached good history, the complete coordinate-selection
step, before applying the operation cutoff, selects $i$ with probability
$v_{{\rm eng},i}/L_6$ and returns $\mathsf{null}$ otherwise. This law holds conditional
on every fixed collection of sketches and the entire past.
\item[(b)] In the sketched branch, the probability of reaching a bad
iteration-start state before the cutoff is at most
$L_7n^{-14}\le n^{-6}$. Thus one may take $\epsilon_{\rm ds}=n^{-6}$
in Theorem~\ref{thm:lp_convergence}.
\item[(c)] In the exact branch, Part~(a) holds at every reached history,
and one may take $\epsilon_{\rm ds}=0$.
\end{itemize}
\end{lemma}
\begin{proof}
{\bf Proof of Part (a).}
We first verify that the maintained data represent the current vector.
The \textsc{Init} call establishes Definition~\ref{def:lp_entry_state_invariant}
by Theorem~\ref{thm:lp_entry_state_interface}. In particular,
$\mathrm{exactIP}.b=\beta_1(v)$, and
Lemma~\ref{lem:lp_entry_state_query} makes each
$\mathrm{entryState}.\textsc{Query}(i)$ return the current $v_i$
without changing the state. Thus
Theorem~\ref{thm:lp_exact_ip_interface} makes each computed
$\delta_{v,i}=\mathrm{exactIP}.\textsc{Query}(i)+b_i-v_i$ exact.
The caller changes $v$ only through \textsc{UpdateE} and \textsc{Shrink}.
Lemma~\ref{lem:lp_normalization}-(d) keeps their coordinate arguments in
$(-8\gamma_1 n,8\gamma_1 n)$, and
Theorem~\ref{thm:lp_entry_state_interface} proves that these calls
preserve the invariant. The level and bucket arrays are shared
references, so the base sampler reads their updated values.
Threshold updates and sampling calls leave the represented vector
unchanged. Induction over completed calls therefore proves the
maintenance claim for every fixed collection of sketches.
If the cutoff interrupts a call, the trial discards that state.

The initial vector is zero, and each completed coordinate update is
followed by shrink.

Lemma~\ref{lem:lp_shrink}-(c) gives
$Z\le(41/8)\gamma_1 n$ at each reached iteration start. A
$\mathsf{null}$ or $\mathsf{fail}$ outcome leaves the vector unchanged.
Thus Lemma~\ref{lem:lp_base_sampling} applies; its Part~(b) gives
$L_{6,4}>6\mathcal N_2$.

We have
\[
 \Pr[\iota=i\mid\text{sketches and past}]
 =\frac{v_{{\rm drift},i}^2}{L_6}
  +\frac{L_{6,4}}{L_6}\frac{\delta_{v,i}^2}{L_{6,4}}
 =\frac{v_{{\rm eng},i}}{L_6}.
\]
The first step uses the base-selection probability from
Lemma~\ref{lem:lp_base_sampling}-(c), the probability of entering the residual
branch, and Lemma~\ref{lem:lp_exact_l2_sampler_sample} at a good state.
The second step cancels $L_{6,4}$ and substitutes the definition of $v_{{\rm eng},i}$.
All other outcomes are $\mathsf{null}$. This proves Part~(a).

{\bf Proof of Part (b).}
Let $S$ denote all initialization sketches. Let $K(v,B)$ be the
probability that a complete coordinate-selection step with the law in
Part~(a), followed by its coordinate update and shrink, ends in a
measurable set $B$. The kernel $K$ does not depend on $S$: coordinate
updates are exact, and shrink uses fresh randomness depending only on
the current vector. The normalization and shrink lemmas make $K$ a
probability kernel on cleaned states.

For each fixed $S=s$, let $G_s$ be its set of good states. Let
$\mu_k^s(B)$ be the conditional probability that iteration $k$ is
reached before the cutoff, its vector belongs to $B$, and all earlier
iteration-start states are good. The current state need not be good
in this definition. Thus $\mu_k^s$ is a measure of total mass at most
one. Let $\nu_0$ be the point mass at ${\bf 0}_n$ and define
$\nu_{k+1}(B):=\int K(v,B)\nu_k(\d v)$; these measures are independent
of $s$.

We have
\[
 \mu_{k+1}^s(B)
 \le\int_{G_s}K(v,B)\mu_k^s(\d v)
 \le\int K(v,B)\nu_k(\d v)=\nu_{k+1}(B).
\]
The first step applies Part~(a) at every reached good history. A cutoff,
including one inside a call, only removes outcomes from the completed
transition. The second step uses the induction hypothesis
$\mu_k^s\le\nu_k$ and nonnegativity of $K(v,B)$.
The third step is the definition of $\nu_{k+1}$.
The induction starts from $\mu_0^s\le\nu_0$.

We have
\begin{align*}
 \Pr[\text{first bad state is reached at iteration }k]
 &=\int\mu_k^s(G_s^c)\Pr[S\in\d s]\\
 &\le\int\nu_k(G_s^c)\Pr[S\in\d s]\\
 &=\int\Pr_S[v\notin G_S]\nu_k(\d v)\le n^{-14}.
\end{align*}
The first step conditions on the initialization sketches. The second
step uses the measure domination just proved for each fixed $s$.
The third step exchanges the two nonnegative integrals. The fourth
step uses Lemma~\ref{lem:lp_ensemble} for each fixed $v$, since $\nu_k$
is independent of the sketches. Summing over $k=0,\ldots,L_7-1$ gives
the bound $L_7n^{-14}$.

In the sketched branch, $L_{s,3}\le n$ and $L_{s,3}\ge2^{16}L_1^2$ imply
$n\ge100$. We have
\[
 L_7=2^{18}n^2\log_2n\le2^{18}n^3\le n^8.
\]
The first step substitutes the definitions of $L_5$ and $L_7$.
The second step uses $\log_2n\le n$. The third step uses
$n\ge100$, which gives $n^5\ge100^5>2^{18}$.
Consequently $L_7n^{-14}\le n^{-6}$, proving Part~(b).

{\bf Proof of Part (c).}
Lemma~\ref{lem:lp_exact_l2_sampler_fallback} gives the exact residual
sampling law for every state. Combining it with the base branch proves
Part~(a) in this case. Every state is good by definition, so the
probability of a bad state is zero. This proves Part~(c).
\end{proof}

\subsection{Expected work of IterativeSolve}\label{sec:lp_iterative_solve_work}
Theorem~\ref{thm:lp_convergence} and Lemmas~\ref{lem:lp_entry_state_exact_ip_cost} and~\ref{lem:lp_entry_state_cache_cost} bound the work before a bad state or cutoff. Combining these bounds with the failure probability in Lemma~\ref{lem:lp_sampling_validity} accounts for the cost of restarting.

\begin{lemma}[Expected work of the implemented IterativeSolve]\label{lem:lp_iterative_solve_time}
Let $n\ge2$, $\mathscr{A}\in\R^{n\times n}$, and $b\in\R^n$ satisfy the hypotheses
of Lemma~\ref{lem:lp_potential_contraction}, using $\beta_1,\beta_2$ with the same fixed
parameters $\gamma_1\ge2$ and $3/4\le\gamma_2\le1\le\gamma_3\le2$.
Algorithm~\ref{alg:iterative_solve} has the following guarantees.
\begin{itemize}
\item[(a)] It terminates almost surely and returns a vector
$v$ satisfying $\|\mathscr{A}\beta_1(v)+b-v\|_2\le\gamma_1/(256n^2)$.
\item[(b)] Its expected work, including initialization, rejected proposals, shrink,
restarts, and certification, is $\E[T_{\rm IterativeSolve}]=O(n^2\log n)$.
\item[(c)] Its space usage is $O(n^2)$.
\end{itemize}
\end{lemma}
\begin{proof}
{\bf Proof of Parts (a) and (b).}
Within this proof, $n$ denotes the padded dimension, which is less than
twice the input dimension. This changes the stated bounds by only a
constant factor.
Zero padding extends $v_*$ by zeros, preserves $v_*=\mathscr{A}\beta_1(v_*)+b$ and its $\ell_1$ norm,
and does not decrease the dimension, so the bound $\|v_*\|_1\le3\gamma_1 n/8$ is preserved. Let $L_1:=\log_2n$, and use $L_5,L_6,L_7$ from
Theorem~\ref{thm:lp_convergence}, so $L_2=L_1+4=O(L_1)$ and $L_7=O(n^2L_1)$.
We count initialization, rejected proposals, shrink, restarts, and final
residual certification. In one trial, let $T_{\rm stop}$ count operations
up to the first bad iteration-start state, the operation cutoff, or
completion, whichever occurs first. Initialization and terminal
verification are included when reached. Use the indicators $J_k$ from
Theorem~\ref{thm:lp_convergence}.

\paragraph{Initialization.}
Lemma~\ref{lem:lp_entry_state_init} initializes the complete
\textsc{EntryState} instance in $O(n^2L_1)$ work and $O(n^2)$ space.
In the exact branch its initialization costs $O(n^2)$.

\paragraph{Phase-by-phase work.}
Partition the $L_7=64L_5 L_1$ iterations into exactly $2L_1$ phases,
each of length $32L_5$. Let
\begin{equation}\label{eq:lp_phase_thresholds}
L_3:=2^{\lceil\log_2\max\{2,L_1\}\rceil},\qquad
T_j:=\max\{2,L_3 2^{-j}\}\quad(0\le j<2L_1).
\end{equation}
Use threshold $T_j$ throughout phase $j$. When the threshold changes,
call \textsc{\ensuremath{\mathrm{exactL2Sampler}}.UpdateT}, which costs $O(n^2)$ by
Lemma~\ref{lem:lp_exact_l2_sampler_update_t}.
There are $O(1+\log L_1)$ changes, giving $O(n^2L_1)$ total rebuilding work.
The threshold schedule depends only on the iteration number.

Use the retained actual iterates and residuals of Theorem~\ref{thm:lp_convergence}.
For $0\le j<2L_1$, set $t_j:=32L_5j$. We first show
\begin{equation}\label{eq:lp_phase_potential_bound}
 \E[J_{t_j}\mathscr{J}(v_{t_j})]\le(3/4)n4^{-j}.
\end{equation}
We have
\[
 \E[J_{t_j}\mathscr{J}(v_{t_j})] \le(1-1/(2L_5))^{t_j}\mathscr{J}({\bf 0}_n) \le e^{-16j}\mathscr{J}({\bf 0}_n) \le4^{-j}\mathscr{J}({\bf 0}_n)\le(3/4)n4^{-j}.
\]
The first step takes expectations in Theorem~\ref{thm:lp_convergence}-(a),
drops the nonpositive residual and shrink terms, and iterates the resulting
contraction from $v_0={\bf 0}_n$, using $J_0\le1$.
The second step uses $1-x\le e^{-x}$ and $t_j=32L_5j$.
The third step uses $e^{-16}\le1/4$ and $j\ge0$.
The fourth step uses $\mathscr{J}({\bf 0}_n)\le3\gamma_3 n/8$ from
Lemma~\ref{lem:lp_potential_contraction}-(b) and $\gamma_3\le2$.

We next show
\begin{equation}\label{eq:lp_phase_residual_bound}
 \sum_{k=t_j}^{t_j+32L_5-1}\frac{\E[J_k\|\delta_{v_k}\|_2^2]}{L_6}
 \le12n4^{-j}.
\end{equation}
We have
\[
 \sum_{k=t_j}^{t_j+32L_5-1}\frac{\E[J_k\|\delta_{v_k}\|_2^2]}{L_6} \le16(\E[J_{t_j}\mathscr{J}(v_{t_j})]-\E[J_{t_j+32L_5}\mathscr{J}(v_{t_j+32L_5})]) \le16\E[J_{t_j}\mathscr{J}(v_{t_j})]\le12n4^{-j}.
\]
The first step takes expectations in Theorem~\ref{thm:lp_convergence}-(a),
drops the nonpositive potential-contraction and shrink terms, and sums
over $k=t_j,\ldots,t_j+32L_5-1$, then multiplies by $16$.
The second step uses $\mathscr{J}(v_{t_j+32L_5})\ge0$ from
Lemma~\ref{lem:lp_potential_contraction}-(a).
The third step uses Eq.~\eqref{eq:lp_phase_potential_bound} and $16\cdot(3/4)=12$.

At a good state, the gate probability for a complete call satisfies
\begin{equation}\label{eq:lp_residual_gate_probability}
\frac{L_{6,4}}{L_6}\frac{2M}{L_{6,4}}
=\frac{4\widetilde{\mathcal{N}}_2}{L_6}\le\frac{6\mathcal{N}_2}{L_6}.
\end{equation}
The left-hand side is the unconditional gate probability. The first step
cancels $L_{6,4}$ and uses $M=2\widetilde{\mathcal{N}}_2$; the second step uses
$\widetilde{\mathcal{N}}_2\le3\mathcal{N}_2/2$ from Eq.~\eqref{eq:lp_sketch_and_proposal_bounds}.
Multiplying the conditional gate bound by $J_k$ and summing
Eq.~\eqref{eq:lp_phase_residual_bound} gives at most $72n4^{-j}$ expected
gate entries in phase $j$ before stopping.

Let $T_{\mathrm{prop}}$ denote the total time spent on proposal operations in the
actual trial, stopped at the first bad state, operation cutoff, or completion.
We can show
\begin{align}
\E[T_{\mathrm{prop}}]
&\le O(nL_1^2)\sum_{j=0}^{2L_1-1}4^{-j}
 +O(n^2L_1^2)\sum_{j=0}^{2L_1-1}\frac{4^{-j}}{T_j}\notag\\
&\le O(nL_1^2)+O(n^2L_1^2/L_3)
=O(n^2L_1).\label{eq:lp_total_proposal_work}
\end{align}
The first step sums Lemma~\ref{lem:lp_entry_state_cache_cost}-(b) over
the expected gate entries in each phase. If the cutoff interrupts a
proposal, its work is at most that of the complete call using the same randomness.
The second step uses the geometric sum $\sum_{j=0}^{2L_1-1}4^{-j}\le4/3$.
Also, $T_j\ge L_3 2^{-j}$ bounds the sum involving $T_j$ by
$L_3^{-1}\sum_{j=0}^{2L_1-1}2^{-j}\le2/L_3$, using the geometric series with ratio $1/2$.
The third step uses $L_3\ge L_1$ and $L_1\le n$.

We have
\begin{equation}\label{eq:lp_total_threshold_cache_work}
O(L_5)\sum_{j=0}^{2L_1-1}T_j=O(n^2L_1),
\end{equation}
where the step uses $T_j\le2+L_3 2^{-j}$ and $L_3\le2L_1$ for $L_1\ge1$.
By Lemma~\ref{lem:lp_entry_state_cache_cost}, the left-hand side bounds
the expected work of the base-selected threshold-cache changes.

Each iteration makes at most two calls to
$\mathrm{entryState}.\textsc{Query}$, which take $O(1)$ total time and
leave the state unchanged by Lemma~\ref{lem:lp_entry_state_query}.
All other per-iteration costs are constant in expectation by
Lemma~\ref{lem:lp_entry_state_exact_ip_cost} and Lemma~\ref{lem:lp_base_sampling}-(d).
The lemma includes the internal root cache of $\mathrm{exactL2Sampler}$.
Over $L_7=O(n^2L_1)$ iterations these costs total $O(n^2L_1)$.
Lemmas~\ref{lem:lp_entry_state_exact_ip_cost} and~\ref{lem:lp_entry_state_cache_cost}
bound residual-selected and shrink work before stopping by $O(n^2)$ in expectation.
Corollary~\ref{cor:lp_entry_state_shrink_work} also accounts for
the constant-time tests in shrink calls making no changes.
The $n$ terminal coordinate queries reconstruct the current vector in
$O(n)$ time and $O(n)$ space by Lemma~\ref{lem:lp_entry_state_query}.
No update occurs during these queries or the ensuing residual test.
Terminal verification therefore still costs $O(n^2)$.

Each conditional cost bound is applied at a reached good history and
multiplied by $J_k$. The phase estimates in
Eq.~\eqref{eq:lp_phase_potential_bound} and Eq.~\eqref{eq:lp_phase_residual_bound}
already concern these retained histories. A cutoff can only reduce the
work of a call relative to its complete execution with the same
randomness. The bounds above include one additional $O(n)$ partial
shrink scan or update when needed. Thus $\E[T_{\rm stop}]=O(n^2L_1)$
in the sketched branch, uniformly in the operation cutoff.

\paragraph{The exact fallback.}
If $L_{s,3}>n$, use the exact implementation in Section~\ref{sec:lp_exact_l2_sampler_fallback}.
Its initialization costs $O(n^2)$, and each sampling or coordinate-update
call costs $O(n)$. Thus $L_7=O(n^2L_1)$ iterations and the shrink budget of
Theorem~\ref{thm:lp_convergence}-(d) give $O(n^3L_1)$ expected work before stopping.
Lemma~\ref{lem:lp_sampling_validity}-(c) gives the exact sampling law,
so Theorem~\ref{thm:lp_convergence} applies with $\epsilon_{\rm ds}=0$. Since this branch has $n\le2^{27}$ by
Section~\ref{sec:lp_exact_l2_sampler_fallback}, its expected work is also $O(n^2L_1)$
with a universal constant. Here $T_{\rm stop}$ counts the entire capped
trial, including a partial call if the cutoff is reached.

\paragraph{Restarts.}
Choose a universal constant $C_0>0$ so that
$B(n):=C_0n^2\log n$ bounds $\E[T_{\rm stop}]$ for all $n\ge2$,
including the exact fallback. Choose the constant in the operation cutoff so that Algorithm~\ref{alg:iterative_solve} caps each run at
$32B(n)=32C_0n^2\log n$ operations, including setup, cache rebuilding,
shrink, coin calls, and terminal verification. Reaching this cutoff discards
the run and restarts from zero with fresh randomness.

We have
\[
 \Pr[\text{cutoff before a bad state}]
 \le\Pr[T_{\rm stop}\ge32B(n)]
 \le\frac{\E[T_{\rm stop}]}{32B(n)}\le\frac1{32}.
\]
The first step uses the definition of the operation cutoff. The second
step applies Markov's inequality. The third step uses
$\E[T_{\rm stop}]\le B(n)$.

Every unsuccessful trial either reaches a bad state, reaches the cutoff
before a bad state, or completes on good states and fails the terminal
test. Eq.~\eqref{eq:lp_terminal_failure_before_stop} bounds the last
event by $1/16$. Lemma~\ref{lem:lp_sampling_validity} gives
$\epsilon_{\rm ds}=n^{-6}$ in the sketched branch, where $n\ge100$,
and $\epsilon_{\rm ds}=0$ in the exact branch. A union bound gives
success probability at least $1-1/16-1/32-\epsilon_{\rm ds}>3/4$.

Let $T_{\rm run}$ denote the actual work of one capped trial. We have
\[
 \E[T_{\rm run}]
 \le\E[T_{\rm stop}]+32B(n)\epsilon_{\rm ds}
 \le(1+32n^{-6})B(n)=O(n^2\log n).
\]
The first step charges the full cutoff if a bad state is reached; on
the complementary event, $T_{\rm run}=T_{\rm stop}$. The second step
uses $\E[T_{\rm stop}]\le B(n)$ and $\epsilon_{\rm ds}\le n^{-6}$.
The third step uses $B(n)=C_0n^2\log n$ and $n\ge2$.

Fresh runs are independent and succeed with probability greater than
$3/4$. Summing each run's expected work times the probability of reaching
that run multiplies the bound by at most $4/3$. Every returned answer
passes the exact residual test, and termination is almost sure.
This proves Parts~(a) and~(b).

{\bf Proof of Part (c).}
All stored matrices and caches use $O(n^2)$ registers; the reciprocal
coin uses only counts and running sums.
\end{proof}

\subsection{Initialization and the LP guarantee}\label{sec:lp_reduction}

The input LP need not come with a strictly feasible primal-dual pair. We augment its standard form so that \textsc{OneStep} can start from an explicit pair.

\begin{lemma}[Strictly feasible initialization]\label{lem:lp_initialization}
Let $\bar{A}\in\R^{d\times n}$ have full row rank, $\bar{b}\in\R^d$, and $\bar{c}\in\R^n$.
Assume that $\{x\ge{\bf 0}_n:\bar{A}x=\bar{b}\}$ is nonempty and contained in $[0,B_{\mathrm{box}}]^n$,
where $B_{\mathrm{box}}>0$, and let $0<\epsilon<1$. Let
$\epsilon_1:=\epsilon/(20\max\{n,2\})$, $v_{\rm obj}:=\|\bar{c}\|_\infty$ if $\bar{c}\ne{\bf 0}_n$, and
$v_{\rm obj}:=1$ if $\bar{c}={\bf 0}_n$. Define the augmented data for the standard-form
initialization used in Algorithm~\ref{alg:main}:
\[
 {\widehat{A}}:=\begin{bmatrix}\bar{A}&{\bf 0}_d&\bar{b}/B_{\mathrm{box}}-\bar{A}{\bf 1}_n\\{\bf 1}_n^\top&1&0\end{bmatrix},
 \quad {\widehat{b}}:=\begin{bmatrix}\bar{b}/B_{\mathrm{box}}\\n+1\end{bmatrix},\quad
 {\widehat{c}}:=\begin{bmatrix}\epsilon_1 \bar{c}/v_{\rm obj}\\0\\1\end{bmatrix}.
\]
The following guarantees hold.
\begin{itemize}
\item[(a)] The matrix ${\widehat{A}}$ has full row rank.
\item[(b)] The state $x:={\bf 1}_{n+2}$, $y:=({\bf 0}_d,-1)$,
$s:=({\bf 1}_n+\epsilon_1 \bar{c}/v_{\rm obj},1,1)$, and $t:=1-\epsilon_1$ satisfies the input conditions of
Algorithm~\ref{alg:one_step} in dimension $N:=n+2$.
\end{itemize}
\end{lemma}
\begin{proof}
{\bf Proof of Part (a).}
The added row is independent of the original rows because of its entry
in column $n+1$; hence ${\widehat{A}}$ has full row rank.

{\bf Proof of Part (b).}
Direct substitution
shows that $x={\bf 1}_{n+2}$, $y=({\bf 0}_d,-1)$, and
$s=({\bf 1}_n+\epsilon_1 \bar{c}/v_{\rm obj},1,1)$ are strictly feasible. Their products belong
to $[1-\epsilon_1,1+\epsilon_1]$. Recall $\epsilon_{1,t}=0.1$ from
Lemma~\ref{lem:lp_projection}. We have
\[
 \epsilon_1<1/40<\frac{\epsilon_{1,t}}{2+\epsilon_{1,t}}.
\]
The first step uses $\epsilon<1$ and $\max\{n,2\}\ge2$;
the second substitutes $\epsilon_{1,t}=0.1$.
Thus $t=1-\epsilon_1>0$, and
\[
 t=1-\epsilon_1\le x_is_i\le1+\epsilon_1
 \le(1+\epsilon_{1,t})(1-\epsilon_1)=(1+\epsilon_{1,t})t.
\]
The first and last steps use the definition of $t$.
The second and third steps use the product bounds.
The fourth step rearranges $\epsilon_1\le\epsilon_{1,t}/(2+\epsilon_{1,t})$.
This establishes the initial neighborhood.
\end{proof}

Algorithm~\ref{alg:main} first converts the box-constrained LP to a
standard-form LP with $m=2n$ variables and $d+n$ equality constraints.
It applies Lemma~\ref{lem:lp_initialization} to $({\bar{A}},{\bar{b}},{\bar{c}})$
with $B_{\mathrm{box}}=1$ and accuracy $\eta$, and normalizes the final vector to lie
in the original box.

\begin{algorithm}[!ht]
\caption{Main algorithm}\label{alg:main}
\begin{algorithmic}[1]
\Procedure{Main}{$A\in\R^{d\times n},b\in\R^d,c,\ell,u\in\R^n,\epsilon\in(0,1]$} \Comment{Theorem~\ref{thm:lp_reduction}}
\State Let $d$ and $n$ be the number of rows and columns of $A$, respectively.
\State $W\gets\max\{2,\|A\|_1,\|c\|_\infty,\|\ell\|_\infty,\|u\|_\infty\}$; $\eta\gets\epsilon/(2nW^2)$.
\State $D\gets\operatorname{diag}(u-\ell)$; $A_0\gets AD$; $b_0\gets b-A\ell$; $c_0\gets Dc$.
\State ${\bar{A}}\gets\begin{bmatrix}A_0&{\bf 0}_{d\times n}\\I_n&I_n\end{bmatrix}$, ${\bar{b}}\gets\begin{bmatrix}b_0\\{\bf 1}_n\end{bmatrix}$, ${\bar{c}}\gets\begin{bmatrix}c_0\\{\bf 0}_n\end{bmatrix}$.
\State $m\gets2n$; $N\gets m+2$; $\epsilon_{1,t}\gets0.1$.
\State $\epsilon_1\gets\eta/(20m)$; $v_{\rm obj}\gets\|{\bar{c}}\|_\infty$ (set $v_{\rm obj}\gets1$ if ${\bar{c}}={\bf 0}_m$).
\State ${\widehat{A}}\gets\begin{bmatrix}{\bar{A}}&{\bf 0}_{d+n}&{\bar{b}}-{\bar{A}}{\bf 1}_m\\{\bf 1}_m^\top&1&0\end{bmatrix}$.
\State ${\widehat{b}}\gets\begin{bmatrix}{\bar{b}}\\m+1\end{bmatrix}$; ${\widehat{c}}\gets\begin{bmatrix}\epsilon_1{\bar{c}}/v_{\rm obj}\\0\\1\end{bmatrix}$.
\State $x\gets{\bf 1}_N$; $y\gets({\bf 0}_{d+n},-1)$.
\State $s\gets({\bf 1}_m+\epsilon_1{\bar{c}}/v_{\rm obj},1,1)$; $t\gets1-\epsilon_1$. \Comment{Lemma~\ref{lem:lp_initialization}}
\While{$(1+\epsilon_{1,t})Nt>\epsilon_1^2$}
\State $(x,y,s,t)\gets\Call{OneStep}{{\widehat{A}},x,y,s,t}$. \Comment{Algorithm~\ref{alg:one_step}}
\EndWhile
\State $\widehat y\gets(x_1,\ldots,x_n)$; $\zeta\gets x_N$.
\State $y_{\rm out}\gets\widehat y/(1+\zeta)$; $x_{\rm out}\gets\ell+Dy_{\rm out}$.
\State \Return $x_{\rm out}$.
\EndProcedure
\end{algorithmic}
\end{algorithm}

Combining this initialization with the per-step guarantee gives bounds on feasibility, objective error, and expected work in the original box coordinates.

\begin{theorem}[Box-constrained LP guarantee and arithmetic complexity, formal version of Theorem~\ref{thm:main_informal}]\label{thm:lp_reduction}
Let $A\in\R^{d\times n}$ have full row rank, with $1\leq d\leq n$,
and let $b\in\R^d$ and $c,\ell,u\in\R^n$.
Suppose $\ell<u$ and the bounded-box linear program
\[
 \OPT:=\min\{c^\top x:Ax=b,\ \ell\leq x\leq u\}
\]
is feasible. Define
\[
 W:=\max\{2,\|A\|_1,\|c\|_\infty,\|\ell\|_\infty,\|u\|_\infty\},
\]
where $\|A\|_1:=\sum_{i=1}^d\sum_{j=1}^n|A_{ij}|$ is the entrywise $\ell_1$ norm.
For $0<\epsilon\leq1$, Algorithm~\ref{alg:main} has the following guarantees.
\begin{itemize}
\item[(a)] It terminates almost surely and returns $x_{\rm out}\in\R^n$ with $\ell\leq x_{\rm out}\leq u$.
\item[(b)] The output satisfies $\|Ax_{\rm out}-b\|_\infty\leq\epsilon$.
\item[(c)] The output satisfies $c^\top x_{\rm out}\leq\OPT+\epsilon$.
\item[(d)] Its expected number of real arithmetic operations is $O((n^\omega+n^2\log n)\log(nW/\epsilon))$.
\item[(e)] The memory requirement is $O(n^2)$.
\end{itemize}
\end{theorem}
\begin{proof}
{\bf Standard-form reduction.}
We first analyze the standard-form stage of Algorithm~\ref{alg:main}
using Lemma~\ref{lem:lp_initialization}. In this part of the proof,
the data $\bar{A},\bar{b},\bar{c},n,B_{\mathrm{box}}$ and parameters $\epsilon,\epsilon_1,v_{\rm obj}$ are those of
that lemma. Write $\OPT:=\min\{\bar{c}^\top x:\bar{A}x=\bar{b},\ x\geq{\bf 0}_n\}$.

Initialize the augmented data and state as in that lemma, repeatedly
call \textsc{OneStep} until $t\leq\epsilon_1^2/((1+\epsilon_{1,t})(n+2))$, and let
$\bar{x}:=B_{\mathrm{box}}(x_1,\ldots,x_n)$ for the final augmented primal vector $x$.
Here $\|\bar{A}\|_1:=\sum_{i=1}^d\sum_{j=1}^n|\bar{A}_{ij}|$ is the entrywise $\ell_1$ norm.
With $N:=n+2$, the expected number of real arithmetic operations is
\begin{equation}\label{eq:lp_standard_runtime}
 O((N^\omega+N^2\log N)\log(2n/\epsilon)).
\end{equation}
The memory requirement is $O(N^2)$.

Lemma~\ref{lem:lp_initialization} gives a valid initial primal-dual state.
The feasible set is nonempty, closed, and bounded, so an optimal original
point $x^*$ exists. It lifts to the nonnegative augmented feasible point
\[
 x^{\rm cmp}:=(x^*/B_{\mathrm{box}},\ n+1-{\bf 1}_n^\top x^*/B_{\mathrm{box}},\ 0).
\]
The middle coordinate is at least one.

At any strictly feasible augmented primal-dual state in the maintained neighborhood, we have
\begin{equation}\label{eq:lp_augmented_gap}
 (\widehat{c})^\top x-(\widehat{c})^\top x^{\rm cmp}
 \le(\widehat{c})^\top x-(\widehat{b})^\top y
 =x^\top s\le(1+\epsilon_{1,t})Nt,
\end{equation}
where the first step follows from weak duality applied to the feasible
comparator $x^{\rm cmp}$;
the second step follows from ${\widehat{A}} x={\widehat{b}}$ and
$(\widehat{A})^\top y+s={\widehat{c}}$;
and the third step sums $x_is_i\le(1+\epsilon_{1,t})t$ over the $N$ coordinates.

Theorem~\ref{thm:lp_one_step} preserves all prerequisites, and each step
multiplies $t$ by $1-\epsilon_{2,t}$. Therefore the stopping condition
$(1+\epsilon_{1,t})Nt\le\epsilon_1^2$ is reached after $O(\log(2n/\epsilon))$ calls to \textsc{OneStep}.
Thus this stage terminates almost surely.

Let $(x_{\rm core},q,\zeta)\ge{\bf 0}_{n+2}$ be the final augmented primal vector,
with artificial variable $\zeta$. Then $\bar{x}=B_{\mathrm{box}}x_{\rm core}\ge{\bf 0}_n$ and
${\bf 1}_n^\top x_{\rm core}+q=n+1$, so ${\bf 1}_n^\top x_{\rm core}\le n+1$.

We have
\begin{equation}\label{eq:lp_final_augmented_objective}
 \epsilon_1 \bar{c}^\top x_{\rm core}/v_{\rm obj}+\zeta
 \le\epsilon_1\OPT/(B_{\mathrm{box}}v_{\rm obj})+\epsilon_1^2,
\end{equation}
where the step substitutes ${\widehat{c}}=(\epsilon_1 \bar{c}/v_{\rm obj},0,1)$ and
$(\widehat{c})^\top x^{\rm cmp}=\epsilon_1\OPT/(B_{\mathrm{box}}v_{\rm obj})$ into
Eq.~\eqref{eq:lp_augmented_gap}, and uses the stopping condition
$(1+\epsilon_{1,t})Nt\le\epsilon_1^2$.

Dropping the nonnegative $\zeta$, dividing by $\epsilon_1$, and multiplying by
$B_{\mathrm{box}}v_{\rm obj}$ gives
\begin{equation}\label{eq:lp_output_objective}
 \bar{c}^\top\bar{x}\le\OPT+B_{\mathrm{box}}v_{\rm obj}\epsilon_1.
\end{equation}

We have
\begin{equation}\label{eq:lp_objective_comparison_bounds}
 \OPT/(B_{\mathrm{box}}v_{\rm obj})\le n,\qquad \bar{c}^\top x_{\rm core}/v_{\rm obj}\ge-(n+1).
\end{equation}
The choice of $v_{\rm obj}$ gives $v_{\rm obj}>0$ and $|\bar{c}_i|\le v_{\rm obj}$ for every $i$, including
when $\bar{c}={\bf 0}_n$. For the first bound, the optimal feasible point $x^*$ satisfies
\[
 \frac{\OPT}{B_{\mathrm{box}}v_{\rm obj}}
 =\frac{\bar{c}^\top x^*}{B_{\mathrm{box}}v_{\rm obj}}
 \le\frac{{\bf 1}_n^\top x^*}{B_{\mathrm{box}}}\le n.
\]
The first step uses $\OPT=\bar{c}^\top x^*$.
The second step uses $\bar{c}_i/v_{\rm obj}\le1$ and $x_i^*\ge0$ term by term.
The third step uses $x_i^*\le B_{\mathrm{box}}$ for every $i$, since the original feasible
set is contained in $[0,B_{\mathrm{box}}]^n$ and $B_{\mathrm{box}}>0$.

For the second bound, we have
\[
 \frac{\bar{c}^\top x_{\rm core}}{v_{\rm obj}}\ge-{\bf 1}_n^\top x_{\rm core}=q-(n+1)\ge-(n+1).
\]
The first step uses $\bar{c}_i/v_{\rm obj}\ge-1$ and $x_{\rm core,i}\ge0$ term by term.
The second step uses the augmented feasibility constraint
${\bf 1}_n^\top x_{\rm core}+q=n+1$. The third step uses $q\ge0$.

We have
\begin{equation}\label{eq:lp_artificial_variable_bound}
 \zeta\le(2n+1)\epsilon_1+\epsilon_1^2<4n\epsilon_1\le\epsilon/5,
\end{equation}
where the first step follows by rearranging
Eq.~\eqref{eq:lp_final_augmented_objective} and using
Eq.~\eqref{eq:lp_objective_comparison_bounds};
the second step uses $n\ge1$ and $0<\epsilon_1<1$;
and the third step uses $\epsilon_1\le\epsilon/(20n)$.

The first augmented equality is
$\bar{A}x_{\rm core}+(\bar{b}/B_{\mathrm{box}}-\bar{A}{\bf 1}_n)\zeta=\bar{b}/B_{\mathrm{box}}$.
Multiplying by $B_{\mathrm{box}}$ and using $\bar{x}=B_{\mathrm{box}}x_{\rm core}$ gives
\begin{equation}\label{eq:lp_output_feasibility}
 \bar{A}\bar{x}-\bar{b}=(B_{\mathrm{box}}\bar{A}{\bf 1}_n-\bar{b})\zeta.
\end{equation}

For any original feasible $x^*$,
$B_{\mathrm{box}}\bar{A}{\bf 1}_n-\bar{b}=\bar{A}(B_{\mathrm{box}}{\bf 1}_n-x^*)$, with ${\bf 0}_n\le B_{\mathrm{box}}{\bf 1}_n-x^*\le B_{\mathrm{box}}{\bf 1}_n$.

We have
\[
 \|\bar{A}\bar{x}-\bar{b}\|_\infty
 \le \zeta\sum_{i=1}^d\sum_{j=1}^n|\bar{A}_{ij}|(B_{\mathrm{box}}-x_j^*)
 \le \zeta B_{\mathrm{box}}\|\bar{A}\|_1
 \le(\epsilon/5)B_{\mathrm{box}}\|\bar{A}\|_1,
\]
where the first step uses the triangle inequality and bounds the maximum row
sum by the sum over all rows, the second step uses $0\le B_{\mathrm{box}}-x_j^*\le B_{\mathrm{box}}$
and the definition of $\|\bar{A}\|_1$, and the third step uses the bound on $\zeta$
in Eq.~\eqref{eq:lp_artificial_variable_bound}.

Each call to \textsc{OneStep} computes ${\widehat{A}}X^2(\widehat{A})^\top$, its inverse, $H$, and $\overline{H}-H$
in $O(N^\omega)$ operations, padding rectangular products if needed.
The remaining dense-vector products cost $O(N^2)$.

For each call to \textsc{IterativeSolve} in Algorithm~\ref{alg:one_step},
and Lemma~\ref{lem:lp_root}-(d) provides a root with the required $\ell_1$ norm bound.
Thus the matrix, right-hand-side, and root assumptions of Lemma~\ref{lem:lp_potential_contraction} hold.
Lemma~\ref{lem:lp_iterative_solve_time}-(b), including padding by less than a factor of
two, proves Eq.~\eqref{eq:lp_standard_runtime}. The sampler tree stores
$O(N^2)$ numbers when $L_{s,3}\le n$, and the exact fallback also uses
$O(N^2)$ memory. Repeated calls to \textsc{OneStep} reuse this memory.

{\bf Proof of Part (a).}
We now return to the bounded-box data in the theorem. Set
$D:=\operatorname{diag}(u-\ell)$, $A_0:=AD$, $b_0:=b-A\ell$, and $c_0:=Dc$.
The substitution $x=\ell+Dy$ maps the original box to $[0,1]^n$.
Introduce a nonnegative slack vector $w\in\R^n$ and define
\[
 {\bar{A}}:=\begin{bmatrix}A_0&{\bf 0}_{d\times n}\\I_n&I_n\end{bmatrix},
 \qquad {\bar{b}}:=\begin{bmatrix}b_0\\{\bf 1}_n\end{bmatrix},
 \qquad {\bar{c}}:=\begin{bmatrix}c_0\\{\bf 0}_n\end{bmatrix}.
\]
The standard-form LP with data $({\bar{A}},{\bar{b}},{\bar{c}})$ has $2n$ variables
and $d+n$ equality constraints. Its feasible points are exactly
$(y,{\bf 1}_n-y)$ with $A_0y=b_0$ and ${\bf 0}_n\leq y\leq{\bf 1}_n$.
Hence its feasible set is nonempty and contained in $[0,1]^{2n}$,
and its optimum is $\OPT_0:=\OPT-c^\top\ell$.
The matrix $D$ is invertible since $\ell<u$, so $A_0$ has full row rank.
The last $n$ columns of ${\bar{A}}$ force the coefficients of the last
$n$ rows in any row dependence to vanish; full row rank of $A_0$
then forces the remaining coefficients to vanish. Thus ${\bar{A}}$ also
has full row rank.

The input bounds give $0<u_i-\ell_i\leq2W$ for every $i$.
We have
\[
 \|A_0\|_1
 =\sum_{i=1}^d\sum_{j=1}^n|A_{ij}|(u_j-\ell_j)
 \leq2W\|A\|_1\leq2W^2,
 \qquad \|c_0\|_\infty\leq2W\|c\|_\infty\leq2W^2.
\]
Each inequality uses the bound on the coordinate widths and the
definition of $W$.

Algorithm~\ref{alg:main} sets $m=2n$, $\eta=\epsilon/(2nW^2)$,
and $\epsilon_1=\eta/(20m)=\eta/(40n)$.
Its augmented data and initial state are precisely those of
Lemma~\ref{lem:lp_initialization} applied to $({\bar{A}},{\bar{b}},{\bar{c}})$,
with $m$ variables, $d+n$ rows, $B_{\mathrm{box}}=1$, and accuracy $\eta<1$.
In particular, ${\widehat{A}}\in\R^{(d+n+1)\times(2n+2)}$, and
the initialized state satisfies the prerequisites of \textsc{OneStep}.
The algorithm uses $v_{\rm obj}=\|{\bar{c}}\|_\infty=\|c_0\|_\infty$ when
$c_0\ne{\bf 0}_n$, and $v_{\rm obj}=1$ otherwise; in both cases $v_{\rm obj}\leq2W^2$.
The standard-form guarantee proved above therefore applies to its loop.
It gives almost-sure termination and a final augmented primal vector
$x=(\widehat y,\widehat w,q,\zeta)$ with all coordinates nonnegative.
We have
\begin{equation}\label{eq:lp_box_intermediate}
 A_0\widehat y-b_0=(A_0{\bf 1}_n-b_0)\zeta,\qquad 0\leq \zeta<\eta/5,\quad \widehat y+\widehat w=(1+\zeta){\bf 1}_n,\quad c_0^\top\widehat y\leq\OPT_0+v_{\rm obj}\epsilon_1,
\end{equation}
where the first two identities follow from Eq.~\eqref{eq:lp_output_feasibility},
the bound on $\zeta$ follows from Eq.~\eqref{eq:lp_artificial_variable_bound},
and the objective bound follows from Eq.~\eqref{eq:lp_output_objective} for the reduced LP.
Algorithm~\ref{alg:main} reads $\zeta=x_{m+2}$ from the final augmented
primal vector and returns
\[
 y_{\rm out}:=\widehat y/(1+\zeta),\qquad
 x_{\rm out}:=\ell+Dy_{\rm out}.
\]
Nonnegativity of $\widehat y,\widehat w$ and
Eq.~\eqref{eq:lp_box_intermediate} imply
${\bf 0}_n\leq y_{\rm out}\leq{\bf 1}_n$, so
$\ell\leq x_{\rm out}\leq u$ exactly.

{\bf Proof of Part (b).}
The original feasible set is nonempty and compact. Fix an optimal
point $x^*$ and put $y^*:=D^{-1}(x^*-\ell)$, so
$A_0y^*=b_0$ and ${\bf 0}_n\leq y^*\leq{\bf 1}_n$.
We have
\[
 Ax_{\rm out}-b
 =A_0y_{\rm out}-b_0
 =\frac{\zeta}{1+\zeta}(A_0{\bf 1}_n-2b_0)
 =\frac{\zeta}{1+\zeta}A_0({\bf 1}_n-2y^*).
\]
The first step substitutes $x_{\rm out}=\ell+Dy_{\rm out}$,
$A_0=AD$, and $b_0=b-A\ell$.
The second step uses $y_{\rm out}=\widehat y/(1+\zeta)$ and
Eq.~\eqref{eq:lp_box_intermediate}.
The third step uses $A_0y^*=b_0$.

We have
\[
 \|Ax_{\rm out}-b\|_\infty
 \leq\frac{\zeta}{1+\zeta}\|A_0\|_1
 \leq\frac{2\eta W^2}{5}
 =\frac{\epsilon}{5n}\leq\epsilon.
\]
The first step uses the residual identity above, $|1-2y_i^*|\leq1$,
the triangle inequality, and the definition of the entrywise $\ell_1$ norm.
The second step uses $0\leq\zeta<\eta/5$, $1+\zeta\geq1$, and
$\|A_0\|_1\leq2W^2$.
The third step substitutes $\eta=\epsilon/(2nW^2)$.
The fourth step uses $n\geq1$.

{\bf Proof of Part (c).}
The reduced optimum satisfies
$|\OPT_0|=|c_0^\top y^*|\leq n\|c_0\|_\infty\leq nv_{\rm obj}$.
We have
\[
 c^\top x_{\rm out}-\OPT =c_0^\top y_{\rm out}-\OPT_0 \leq\frac{v_{\rm obj}\epsilon_1-\zeta\OPT_0}{1+\zeta} \leq v_{\rm obj}(\epsilon_1+n\zeta) \leq nv_{\rm obj}\eta \leq2nW^2\eta=\epsilon.
\]
The equality cancels the constant $c^\top\ell$ in the objective.
The first inequality uses Eq.~\eqref{eq:lp_box_intermediate};
the second uses $\zeta\geq0$ and $|\OPT_0|\leq nv_{\rm obj}$;
the third uses $\epsilon_1=\eta/(40n)$, $\zeta<\eta/5$, and $n\geq1$;
and the last uses $v_{\rm obj}\leq2W^2$ and the choice of $\eta$.
This proves Part~(c).

{\bf Proof of Part (d).}
Computing $W,A_0,b_0,c_0$ and forming the reduced LP cost $O(n^2)$ time,
and recovering $\zeta$ and $x_{\rm out}$ costs $O(n)$ time.
The standard-form stage of Algorithm~\ref{alg:main} has $2n$ variables
and $d+n\leq2n$ equalities;
its internal augmented dimension is $2n+2=O(n)$.
We have
\[
 \log(4n/\eta)=\log(8n^2W^2/\epsilon)
 =O(\log(nW/\epsilon)),
\]
where the equality substitutes $\eta=\epsilon/(2nW^2)$ and the bound
uses $n\geq1$, $W\geq2$, and $0<\epsilon\leq1$.
Eq.~\eqref{eq:lp_standard_runtime} therefore proves
Part~(d).

{\bf Proof of Part (e).}
The reduced data and all subroutine
storage use $O(n^2)$ space. Restricting to $d\leq n\leq C_0d$
gives Theorem~\ref{thm:main_informal}.

\paragraph{Arithmetic model.}
The guarantee counts exact real-arithmetic operations, including square roots
and the indicated random draws. A finite-precision implementation additionally
requires bounds on errors in projection formation, mirror inversion, residual
certification, and rejection probabilities. The theorem does not assert a
bit-complexity bound.
\end{proof}

\section{Exact inner-product data structure}\label{sec:lp_exact_ip}

In Sections~\ref{subsec:lp_exact_ip_main_data_structure} and~\ref{subsec:lp_exact_ip_initialization}, we describe and initialize the partial inner products stored by \textsc{ExactIP}.
In Section~\ref{subsec:lp_exact_ip_inner_product_query}, we use these sums to compute a requested entry of a matrix-vector product.
In Sections~\ref{subsec:lp_exact_ip_vector_entry_update} and~\ref{subsec:lp_exact_ip_level_update}, we update the stored sums when a vector entry or its level changes.

\subsection{Main data structure}
\label{subsec:lp_exact_ip_main_data_structure}
\textsc{ExactIP} answers queries for individual entries of $Ab$ while $A$ stays fixed. The levels of the coordinates determine the cost of queries and updates to $b$.

\begin{theorem}[ExactIP]\label{thm:lp_exact_ip_interface}
There is a deterministic data structure (\textnormal{\textsc{ExactIP}} in Algorithms~\ref{alg:exact_ip} and~\ref{alg:exact_ip_update_l}) that uses $O(n^2+nL_2)$ space and maintains a fixed matrix
$A\in\R^{n\times n}$, a vector $b\in\R^n$, and a level array
$\lev\in[L_2]^n$. It supports the following procedures.
\begin{enumerate}[label=(\alph*)]
\item \textnormal{\textsc{Init}}$(A \in \R^{n \times n},L_2 \in \mathbb{N}_{>0})$ It takes $A$ and $L_2$ as inputs, sets $b={\bf 0}_n$ and $\lev={\bf 1}_n$, and runs in
$O(n^2+nL_2)$ time.
\item \textnormal{\textsc{Query}}$(i \in [n])$ This procedure takes $i$ as input and outputs $(Ab)_i$ exactly in
$O(1+\cnt[\lev[i]])$ time, where $\cnt[j]:=|\{i\in[n]:\lev[i]\ge j\}|$ for $j\in[L_2+1]$.
\item \textnormal{\textsc{UpdateV}}$(i \in [n],\zeta \in \R)$. This procedure takes $i$ and $\zeta$ as inputs, update $b_i$ by $\zeta$. This procedure runs in  $O(1+\cnt[\lev[i]+1])$ time.
\item \textnormal{\textsc{UpdateL}}$(i \in [n],k \in [L_2])$. This procedure takes $i$ and $k$ as inputs and updates the old level $k^{\old}:=\lev[i]$ by $k$. This procedure runs in $O(1+|k-k^{\old}|+\cnt[\min\{k^{\old},k\}])$ time.
\end{enumerate}
All counts and levels in these bounds are measured before the procedure call; in particular, $\cnt[L_2+1]=0$.
The counts describe the running times and need not be stored or computed by the data structure.
The guarantees hold for arbitrary sequences of valid procedure calls.
\end{theorem}

\begin{proof}
Lemma~\ref{lem:lp_exact_ip_init} establishes the partial inner product
invariant in Definition~\ref{def:lp_partial_inner_product_invariant}, the
bucket representation described below, and the initialization and space bounds
for Algorithm~\ref{alg:exact_ip}.
Lemma~\ref{lem:lp_exact_ip_query} proves the inner-product guarantee and leaves the state unchanged.
Lemmas~\ref{lem:lp_exact_ip_update_v} and~\ref{lem:lp_exact_ip_update_l}
prove that value and level updates preserve both the invariant and the bucket
representation, with the stated running times. Induction on the sequence of
procedure calls proves all correctness guarantees. Each procedure uses the
stored arrays and constant additional space, so the space bound continues to hold.
\end{proof}

For each row, \textsc{ExactIP} stores only the contribution from coordinates at strictly lower levels. A query supplies the remaining terms.

\begin{definition}[Partial inner product invariant]\label{def:lp_partial_inner_product_invariant}
Given $A\in\R^{n\times n}$, $b\in\R^n$, and a level array
$\lev\in[L_2]^n$, a vector $c\in\R^n$ satisfies the
\emph{partial inner product invariant} if, for every $i\in[n]$,
\[
 c_i=\sum_{\substack{j=1\\\lev[j]<\lev[i]}}^nA_{ij}b_j.
\]
\end{definition}
For each level $r\in[L_2]$, store the coordinates with $\lev[i]=r$ in an
unordered bucket array $S_r$ of capacity $n$, together with its length.
Store an inverse position array $\mathrm{pos}\in[n]^n$, so that
$S_{\lev[i]}[\mathrm{pos}[i]]=i$ for every $i\in[n]$.
The nonempty levels are stored in increasing order in a doubly linked list,
with a head pointer and predecessor and successor arrays
$\mathrm{prev},\mathrm{next}\in\{0,\ldots,L_2\}^{L_2}$; the value zero denotes
a missing predecessor or successor. Thus each nonempty level has a direct
array entry for its list node. Algorithms~\ref{alg:exact_ip} and~\ref{alg:exact_ip_update_l} give the implementation.
Both blocks use the same \textsc{ExactIP} members.

\begin{figure}[!htbp]
\centering
\begin{tikzpicture}[lpfig/base]
\fill[lpOrange!12] (1,0) rectangle (5,4);
\foreach \r/\lpwidth in {2/1,3/2,4/3}{
  \fill[lpTeal!18] (1,4-\r) rectangle +(\lpwidth,1);
}
\foreach \r in {1,...,4}{\foreach \c in {1,...,4}{
  \draw[white,line width=1pt] (\c,4-\r) rectangle +(1,1);
}}
\draw[lpGray] (1,0) rectangle (5,4);
\foreach \k in {1,...,4}{\node[above] at (\k+.5,4) {$\k$};\node[left] at (1,4.5-\k) {$\k$};}
\node[above] at (3,4.65) {column level};
\node[rotate=90] at (.1,2) {row level};
\draw[lpBlue,line width=1.1pt] (.9,1) rectangle (5.1,2);
\node[lpfig/box,lpfig/teal,text width=6.3cm] (cache) at (10.4,3.2)
  {Stored contribution\\[4pt]
   $c_i=\sum_{j:\lev[j]<\lev[i]}A_{ij}b_j$};
\node[lpfig/box,lpfig/orange,text width=6.3cm] (tail) at (10.4,1.15)
  {Computed by \textsc{Query}$(i)$\\[4pt]
   $\sum_{j:\lev[j]\ge\lev[i]}A_{ij}b_j$};
\draw[lpfig/arrow] (4.5,1.5)--(tail.west);
\end{tikzpicture}
\caption{The partial inner product in Definition~\ref{def:lp_partial_inner_product_invariant}, shown in level order. Each square is a level block, and the highlighted row has level $3$. Equal-level terms, including $A_{ii}b_i$, belong to the computed part. This ordering is illustrative; the data structure need not reorder $A$.}
\label{fig:lp_partial_inner_product}
\end{figure}
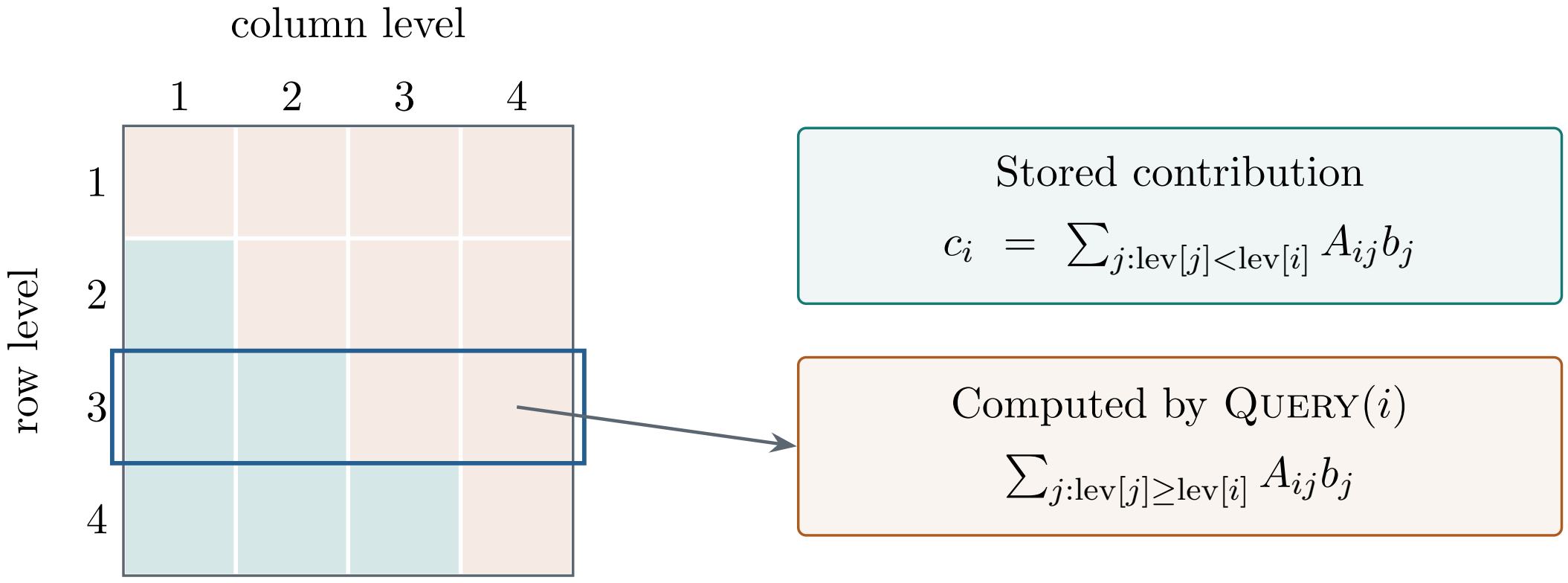

\begin{algorithm}[!ht]
\caption{\textsc{ExactIP} data structure: initialization, queries, and value updates}\label{alg:exact_ip}
\begin{algorithmic}[1]
\DataStructure{ExactIP} \Comment{Theorem~\ref{thm:lp_exact_ip_interface}}
\Members
\State $A\in\R^{n\times n}$; $L_2\in\mathbb{N}_{>0}$; $b,c\in\R^n$; level array $\lev\in[L_2]^n$.
\State Bucket arrays $S_1,\ldots,S_{L_2}$ and their lengths; $\mathrm{pos}$; the head pointer and arrays $\mathrm{prev},\mathrm{next}$.
\State $c_i=\sum_{\substack{j=1\\\lev[j]<\lev[i]}}^nA_{ij}b_j$. \Comment{Definition~\ref{def:lp_partial_inner_product_invariant}}
\EndMembers
\State
\Procedure{Init}{$A \in \R^{n \times n},L_2 \in \mathbb{N}_{>0}$} \Comment{Lemma~\ref{lem:lp_exact_ip_init}}
\State Store $A,L_2$; $b\gets{\bf 0}_n$; $c\gets{\bf 0}_n$; $\lev\gets{\bf 1}_n$.
\State Set $S_1\gets(1,\ldots,n)$, all other buckets empty, and $\mathrm{pos}[i]\gets i$ for every $i$; initialize the list with level $1$.
\EndProcedure
\State
\Procedure{Query}{$i\in[n]$} \Comment{Lemma~\ref{lem:lp_exact_ip_query}}
\State \Return $c_i+\sum_{\substack{j=1\\\lev[j]\ge\lev[i]}}^nA_{ij}b_j$.
\EndProcedure
\State
\Procedure{UpdateV}{$i\in[n],\zeta\in\R$} \Comment{Lemma~\ref{lem:lp_exact_ip_update_v}}
\For{every $j$ with $\lev[j]>\lev[i]$}
\State $c_j\gets c_j+A_{ji}(\zeta-b_i)$.
\EndFor
\State $b_i\gets \zeta$.
\EndProcedure
\EndDataStructure
\end{algorithmic}
\end{algorithm}

\begin{figure}[!htbp]
\centering
\begin{tikzpicture}[lpfig/base]
\foreach \r/\x in {1/0,2/3,3/6,4/9,5/12}{
 \node[lpfig/tree] (b\r) at (\x,1.8) {$S_{\r}$};
}
\node[below=.2cm of b1,align=center] {$(2,5,7)$};
\node[below=.2cm of b2] {$\emptyset$};
\node[below=.2cm of b3] {$(1,4)$};
\node[below=.2cm of b4] {$\emptyset$};
\node[below=.2cm of b5] {$(3,6)$};
\draw[<->,lpBlue,line width=1pt] (b1.north) to[bend left=25] (b3.north);
\draw[<->,lpBlue,line width=1pt] (b3.north) to[bend left=25] (b5.north);
\node[lpfig/note] at (3,3.25) {skip empty levels};
\node[lpfig/note] at (9,3.25) {skip empty levels};
\draw[lpfig/arrow] (-1.2,2.6)--(b1.north west) node[pos=0,above] {head};
\node[lpfig/box,text width=5.2cm] at (2.3,-1.15)
  {Direct location of coordinate $4$\\[3pt]
   $\lev[4]=3$, $\mathrm{pos}[4]=2$\\$S_3[2]=4$};
\node[lpfig/box,lpfig/teal,text width=6.3cm] at (10,-1.15)
  {Tail traversal from level $3$\\[3pt]
   $S_3\longrightarrow S_5$ visits $1,4,3,6$\\
   $\cnt[3]=4$};
\end{tikzpicture}
\caption{The bucket representation used by \textsc{ExactIP}, with $n=7$ and $L_2=5$. Bucket arrays and inverse positions give direct access to coordinates; predecessor and successor pointers skip empty levels. Traversal time is charged to the enumerated entries; the counts $\cnt[j]$ need not be maintained.}
\label{fig:lp_exact_ip_buckets}
\end{figure}
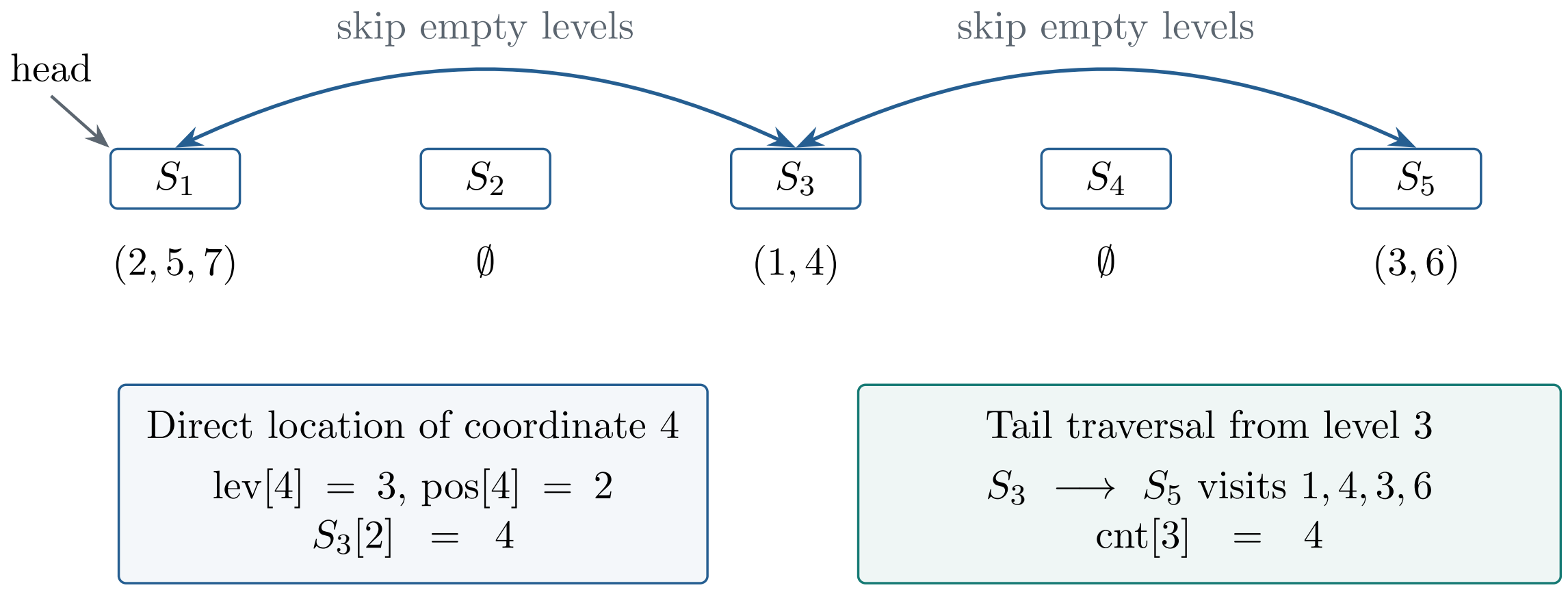

\subsection{Initialization}
\label{subsec:lp_exact_ip_initialization}
We begin by establishing a valid initial state for \textsc{ExactIP}.
The following lemma verifies the initialization procedure and bounds its time and space.

\begin{lemma}[Initialization]\label{lem:lp_exact_ip_init}
The procedure \textnormal{\textsc{Init}}$(A\in\R^{n\times n},L_2\in\mathbb{N}_{>0})$ in Algorithm~\ref{alg:exact_ip} sets $b={\bf 0}_n$ and $\lev={\bf 1}_n$,
establishes the invariant in Definition~\ref{def:lp_partial_inner_product_invariant}, and costs
$O(n^2+nL_2)$ time and space.
\end{lemma}
\begin{proof}
Store $A$ and set $b=c={\bf 0}_n$ and $\lev={\bf 1}_n$.
Allocate $n$ slots for each of the $L_2$ bucket arrays. Set
$S_1=(1,\ldots,n)$ and $\mathrm{pos}[i]=i$ for every $i\in[n]$;
all other buckets have length zero. Set the list head to $1$ and
all predecessor and successor entries to zero. This establishes the bucket
representation, including when $L_2=1$. Storing $A$ costs $O(n^2)$,
the bucket allocation costs $O(nL_2)$, and the remaining arrays cost $O(n+L_2)$
time and space. Since $n,L_2\ge1$, the total is $O(n^2+nL_2)$.
Every partial inner product is zero. This satisfies the invariant in
Definition~\ref{def:lp_partial_inner_product_invariant}, since all coordinates have the same level.
\end{proof}

\subsection{Inner-product query}
\label{subsec:lp_exact_ip_inner_product_query}
The \textsc{Query} procedure combines the stored partial inner product with contributions
from coordinates at the same or higher levels. The following lemma gives the
correctness and running time of this procedure.

\begin{lemma}[Query]\label{lem:lp_exact_ip_query}
Assume the invariant in Definition~\ref{def:lp_partial_inner_product_invariant} and the bucket representation above hold. The procedure
\textnormal{\textsc{Query}}$(i\in[n])$ in Algorithm~\ref{alg:exact_ip} returns $(Ab)_i$ exactly in
$O(1+\cnt[\lev[i]])$ time and leaves the stored state unchanged.
\end{lemma}
\begin{proof}
We have
\[
 c_i+\sum_{\substack{j=1\\\lev[j]\ge\lev[i]}}^nA_{ij}b_j
 =\sum_{\substack{j=1\\\lev[j]<\lev[i]}}^nA_{ij}b_j
  +\sum_{\substack{j=1\\\lev[j]\ge\lev[i]}}^nA_{ij}b_j
 =\sum_{j=1}^n A_{ij}b_j=(Ab)_i.
\]
The first step uses the invariant in Definition~\ref{def:lp_partial_inner_product_invariant}. The second step
partitions the coordinates by their levels, including the diagonal term
in the second sum. The third step is the definition of matrix-vector
multiplication.

The bucket containing $i$ is nonempty. Starting at this bucket, follow
the nonempty-level list and enumerate all coordinates with level at least
$\lev[i]$. Each visited bucket contains a coordinate, so the traversal
and the sum cost $O(1+\cnt[\lev[i]])$ time. No stored value is changed.
\end{proof}

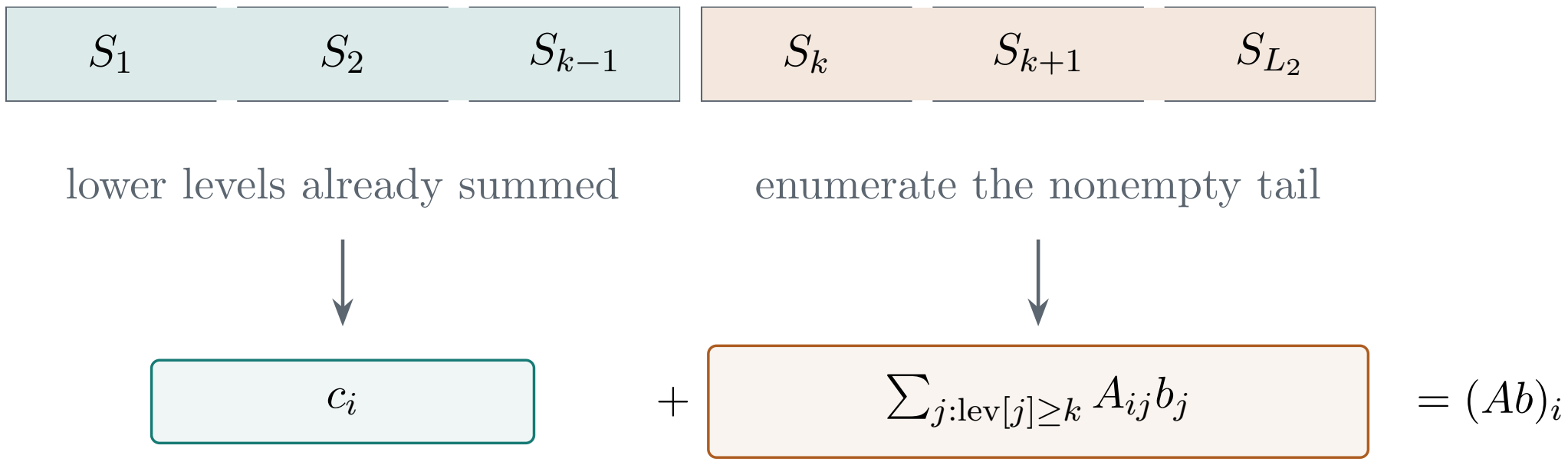
\begin{figure}[!htbp]
\centering
\begin{tikzpicture}[lpfig/base]
\foreach \j/\x/\lab in {1/0/1,2/2/2,3/4/{k-1},4/6/k,5/8/{k+1},6/10/{L_2}}{
 \node[lpfig/cell,minimum width=1.8cm,minimum height=.8cm] (q\j) at (\x,1.7) {$S_{\lab}$};
}
\fill[lpTeal!15] (-.9,1.3) rectangle (4.9,2.1);
\fill[lpOrange!15] (5.1,1.3) rectangle (10.9,2.1);
\foreach \j/\x/\lab in {1/0/1,2/2/2,3/4/{k-1},4/6/k,5/8/{k+1},6/10/{L_2}}{
 \node at (\x,1.7) {$S_{\lab}$};}
\node[lpfig/note] at (2,.55) {lower levels already summed};
\node[lpfig/note] at (8,.55) {enumerate the nonempty tail};
\draw[lpfig/arrow] (2,.1)--(2,-.65);
\draw[lpfig/arrow] (8,.1)--(8,-.65);
\node[lpfig/box,lpfig/teal,minimum width=3.3cm] (stored) at (2,-1.3) {$c_i$};
\node[lpfig/box,lpfig/orange,text width=5.2cm] (scan) at (8,-1.3)
  {$\sum_{j:\lev[j]\ge k}A_{ij}b_j$};
\node at (4.85,-1.3) {$+$};
\node at (11.9,-1.3) {$=(Ab)_i$};
\end{tikzpicture}
\caption{The \textsc{Query} procedure in Lemma~\ref{lem:lp_exact_ip_query}, with $k=\lev[i]$. The stored lower-level contribution and the scanned tail partition $(Ab)_i$ exactly; the tail includes the diagonal term. Intermediate buckets are suppressed. The query takes $O(1+\cnt[k])$ time and leaves the state unchanged.}
\label{fig:lp_exact_ip_query}
\end{figure}

\subsection{Vector entry update}
\label{subsec:lp_exact_ip_vector_entry_update}
We next analyze how \textsc{UpdateV} maintains the partial inner product invariant
when one entry of $b$ changes. The levels remain fixed during this operation.

\begin{lemma}[Value update]\label{lem:lp_exact_ip_update_v}
Assume the invariant in Definition~\ref{def:lp_partial_inner_product_invariant} and the bucket representation above hold.
\textnormal{\textsc{UpdateV}}$(i\in[n],\zeta\in\R)$ in Algorithm~\ref{alg:exact_ip} replaces $b_i$ by $\zeta$,
preserves the invariant and all levels, and costs
$O(1+\cnt[\lev[i]+1])$ time.
\end{lemma}
\begin{proof}
In the cache of row $j$, column $i$ appears
exactly when $\lev[i]<\lev[j]$. The procedure adds $A_{ji}(\zeta-b_i)$
to precisely these caches and then assigns $b_i\gets \zeta$. Every
other contribution is unchanged, proving the invariant in Definition~\ref{def:lp_partial_inner_product_invariant}
for the new vector. In particular, row $i$ has no self-contribution in
its cache.

Start at the successor of bucket $\lev[i]$ in the nonempty-level list.
Enumerating the remaining buckets visits exactly
$\cnt[\lev[i]+1]$ coordinates, with constant time per coordinate.
If there is no successor, the enumeration is empty; in particular,
this holds when $\lev[i]=L_2$, for which $\cnt[L_2+1]=0$.
Assigning $b_i\gets \zeta$ costs $O(1)$, giving the claimed bound. All levels and bucket memberships stay unchanged.
\end{proof}

\subsection{Level update}
\label{subsec:lp_exact_ip_level_update}
Finally, we analyze \textsc{UpdateL}, which changes a coordinate's level while
keeping $b$ fixed.

\begin{algorithm}[!ht]
\caption{\textsc{ExactIP} data structure: level updates}\label{alg:exact_ip_update_l}
\begin{algorithmic}[1]
\DataStructure{ExactIP} \Comment{Theorem~\ref{thm:lp_exact_ip_interface}}
\Procedure{UpdateL}{$i\in[n],k\in[L_2]$} \Comment{Lemma~\ref{lem:lp_exact_ip_update_l}}
\State $k^{\old}\gets\lev[i]$.
\If{$k=k^{\old}$}
\State \Return
\EndIf
\If{$k>k^{\old}$}
\For{every $j\ne i$ with $k^{\old}<\lev[j]\le k$}
\State $c_j\gets c_j-A_{ji}b_i$.
\EndFor
\State $c_i\gets c_i+\sum_{\substack{j=1\\j\ne i\\k^{\old}\le\lev[j]<k}}^nA_{ij}b_j$.
\ElsIf{$k<k^{\old}$}
\For{every $j\ne i$ with $k<\lev[j]\le k^{\old}$}
\State $c_j\gets c_j+A_{ji}b_i$.
\EndFor
\State $c_i\gets c_i-\sum_{\substack{j=1\\j\ne i\\k\le\lev[j]<k^{\old}}}^nA_{ij}b_j$.
\EndIf
\State If $S_k$ is empty, find its list position by scanning from $k^{\old}$ toward $k$, and insert level $k$.
\State Delete $i$ from $S_{k^{\old}}$ by swapping with its last entry; update $\mathrm{pos}$.
\State Append $i$ to $S_k$; set $\mathrm{pos}[i]$ to its new position and $\lev[i]\gets k$.
\State If $S_{k^{\old}}$ is empty, remove level $k^{\old}$ from the list.
\EndProcedure
\EndDataStructure
\end{algorithmic}
\end{algorithm}

When a coordinate changes levels, its own cached sum changes, as can the sums of rows whose levels it crosses.

\begin{lemma}[Level update]\label{lem:lp_exact_ip_update_l}
Assume the invariant in Definition~\ref{def:lp_partial_inner_product_invariant} and the bucket representation above hold. The procedure
\textnormal{\textsc{UpdateL}}$(i\in[n],k\in[L_2])$ in Algorithm~\ref{alg:exact_ip_update_l} changes $k^{\old}:=\lev[i]$ to $k$,
preserves the invariant and the vector $b$, and costs
$O(1+|k-k^{\old}|+\cnt[\min\{k^{\old},k\}])$ time. Counts and levels in this bound
are measured before the procedure call.

\end{lemma}
\begin{proof}
Suppose first that $k>k^{\old}$. The contribution of column $i$ leaves exactly
the caches of rows $j\ne i$ with $k^{\old}<\lev[j]\le k$. The cache of row
$i$ gains exactly the columns $j\ne i$ with $k^{\old}\le\lev[j]<k$.
These are the two updates in the increasing-level branch of
Algorithm~\ref{alg:exact_ip_update_l}. If $k<k^{\old}$, column $i$ enters exactly the
caches with $k<\lev[j]\le k^{\old}$, and row $i$ loses exactly the columns
with $k\le\lev[j]<k^{\old}$. These are the decreasing-level updates.
If $k=k^{\old}$, no cache changes. The diagonal term is excluded in each case,
so the invariant in Definition~\ref{def:lp_partial_inner_product_invariant} is preserved and $b$ is unchanged.

The case $k=k^{\old}$ returns immediately in $O(1)$ time. Otherwise,
evaluate the two sums by scanning the bucket levels from
$\min\{k^{\old},k\}$ through $\max\{k^{\old},k\}$, before changing any bucket
or level. There are $1+|k-k^{\old}|$ such levels, and each coordinate
is inspected at most twice. Every inspected coordinate belongs to the old
tail with level at least $\min\{k^{\old},k\}$. Thus these scans and the
partial inner product updates cost
$O(1+|k-k^{\old}|+\cnt[\min\{k^{\old},k\}])$.

If $S_k$ is empty, locate its position in the ordered nonempty-level list
while $S_{k^{\old}}$ is still nonempty. For $k>k^{\old}$, scan the intervening
levels in increasing order and remember the last nonempty level, starting
with $k^{\old}$; insert $k$ immediately after this level. For $k<k^{\old}$,
scan in decreasing order and insert $k$ immediately before the last nonempty
level encountered, again starting with $k^{\old}$. The saved level and its
successor or predecessor identify both neighbors, so the scan costs
$O(1+|k-k^{\old}|)$ and the list insertion costs $O(1)$, including at the head
or tail. If $S_k$ is already nonempty, no list insertion is needed.

Use $\mathrm{pos}[i]$ to find $i$ in its old bucket, replace it by the last
entry, update the position of that entry, and decrease the bucket length.
Append $i$ to $S_k$, record its new position, and set $\lev[i]=k$.
If the old bucket becomes empty, unlink its level using its predecessor and
successor, updating the head when necessary. These steps cost $O(1)$ and
restore the bucket representation. Inserting the new level before removing
the old one also covers the case where the old bucket contains only $i$.
The total cost is therefore
$O(1+|k-k^{\old}|+\cnt[\min\{k^{\old},k\}])$, with the count evaluated before
the procedure call.

\end{proof}

\section{Dyadic tree data structure}\label{sec:lp_proposal_tree}

Sketching is a powerful tool for improving the running time in the numerical linear algebra \cite{cw13,nn13}. For more details of different sketching matrices and their applications, we refer the readers to Woodruff's survey \cite{w14}. In this section, we apply sketching matrices to create our data-structure.

In Sections~\ref{subsec:dyadic_tree_structure} and~\ref{subsec:dyadic_tree_init}, we describe the shared tree representation and initialize a tree for a given sketch.
In Section~\ref{subsec:dyadic_tree_entry}, we show how to query an entry using the cache and the high coordinates.
In Sections~\ref{subsec:dyadic_tree_norms} and~\ref{subsec:dyadic_tree_branch}, we compute squared sketch norms at tree nodes and use them to determine branch probabilities.
In Sections~\ref{subsec:dyadic_tree_sample} and~\ref{subsec:dyadic_tree_probability}, we sample a coordinate and evaluate the proposal probability of a specified coordinate.

\subsection{Main data structure}
\label{subsec:dyadic_tree_structure}
The data structure \textsc{ProposalTree} samples a coordinate by descending
a dyadic tree and then sampling within a leaf. It also evaluates the
probability of proposing a specified coordinate.

Let $n=2^{L_1}$ with $L_1\ge1$, let $L_{s,1},L_{s,2},q$ be positive integers, and let
$L_{s,3}\le n$ be a power of two. Let $\gamma_1$ and $\beta_1$ be as in
Definition~\ref{def:lp_beta_1}. All arithmetic and random draws use the
model of Section~\ref{sec:preliminaries}.

All tree instances use the same partition and cache; the node rows they query depend on the sketch index.

\begin{definition}[Shared tree representation]\label{def:lp_proposal_tree_representation}
The object $\mathcal T$ stores $n,L_{s,3}$ and the common dyadic partition of
$[n]$ into leaves of size $L_{s,3}$. Each node is a contiguous interval;
its endpoints, children, and cardinality are accessible in $O(1)$ time.
For $t\in[L_{s,2}]$, let $S_{{\rm sk},t}\in\R^{L_{s,1}\times n}$ be fixed.
Let $C\in\R^{q\times2n}$ and $\mathrm{bias}\in\R^q$ be fixed, and let
$v,\delta_v\in\R^n$ and $T\ge2$ describe the current state. Write $\delta_{v,i}:=(\delta_v)_i$ and $\delta_{v,\mathcal{I}}:=(\delta_v)_{\mathcal{I}}$ for coordinates and subvectors. Define
\[
w:=\begin{bmatrix}\beta_1(v)\\v\end{bmatrix},\qquad
\mathrm{val}:=Cw+\mathrm{bias},\qquad
H:=\{j\in[n]:2|v_j|/\gamma_1\ge T\},\qquad h_T:=|H|.
\]
The array $\mathrm{High}$ lists each element of $H$ once.
For each tree node, $\mathrm{NodeRows}[t,\mathcal{I}]$ is an ordered list of $L_{s,1}$
indices in $[q]$; for $i\in[n]$, $\mathrm{EntryIndex}[i]\in[q]$.
The shared record $\mathcal C$ contains references to
$C,c^{\rm low},v,\mathrm{High},\mathrm{NodeRows},\mathrm{EntryIndex}$,
where $c^{\rm low}\in\R^q$. These arrays satisfy
\[
c^{\rm low}=\mathrm{bias}+\sum_{j\in[n]\setminus H}
(C_{:,j}\beta_1(v_j)+C_{:,n+j}v_j),
\]
\[
\mathrm{val}_{\mathrm{NodeRows}[t,\mathcal{I}]}=(S_{{\rm sk},t})_{:,\mathcal{I}}\delta_{v,\mathcal{I}},
\qquad \mathrm{val}_{\mathrm{EntryIndex}[i]}=\delta_{v,i}.
\]
Array access and enumeration of $\mathrm{High}$ take constant time per entry.
The shared state remains fixed during each procedure call.
\end{definition}

The two children determine each branch probability, while the exact
entries of $\delta_v$ determine the conditional probabilities within a leaf.
The following definition also specifies the probabilities when a
normalizing denominator is zero.

\begin{definition}[Tree proposal probabilities]\label{def:lp_proposal_tree_probabilities}
Use the representation in Definition~\ref{def:lp_proposal_tree_representation}.
For a node $\mathcal{I}$, define
$e_{t,\mathcal{I}}:=\|(S_{{\rm sk},t})_{:,\mathcal{I}}\delta_{v,\mathcal{I}}\|_2^2$ and $\mathcal{N}_{2,\mathcal{I}}:=\|\delta_{v,\mathcal{I}}\|_2^2$.
For an internal node $\mathcal{I}$ with children $L,R$ and $K\in\{L,R\}$, define
\[
b_t(\mathcal{I},K):=
\begin{cases}
e_{t,K}/(e_{t,L}+e_{t,R}),&e_{t,L}+e_{t,R}>0,\\
0,&e_{t,L}+e_{t,R}=0.
\end{cases}
\]
For $i\in[n]$, denote its leaf by $\mathcal{I}(i)$ and the ordered edges on the
path from the root to that leaf by $\mathcal P_i$. Define
\[
\lambda_i:=
\begin{cases}
\delta_{v,i}^2/\mathcal{N}_{2,\mathcal{I}(i)},&\mathcal{N}_{2,\mathcal{I}(i)}>0,\\
0,&\mathcal{N}_{2,\mathcal{I}(i)}=0,
\end{cases}
\qquad
p^{\rm prop}_{t,i}:=\lambda_i\prod_{(\mathcal{I},K)\in\mathcal P_i}b_t(\mathcal{I},K).
\]
The empty product is one, including when the root is a leaf.
\end{definition}

The following theorem gives the interface of one tree instance. Its
initialization takes references to an already prepared shared representation.

\begin{theorem}[ProposalTree]\label{thm:lp_proposal_tree_interface}
Assume Definition~\ref{def:lp_proposal_tree_representation} and use the
probabilities in Definition~\ref{def:lp_proposal_tree_probabilities}.
There is a randomized data structure
(\textnormal{\textsc{ProposalTree}} in
Algorithms~\ref{alg:lp_proposal_tree_queries}--\ref{alg:lp_proposal_tree_probability})
supporting the following procedures.
\begin{enumerate}[label=(\alph*)]
\item \textnormal{\textsc{Init}}$(t_{\mathrm{in}},\mathcal T',\mathcal C')$ binds the
instance to tree $t_{\mathrm{in}}\in[L_{s,2}]$ and the supplied topology and cache in $O(1)$ time.
\item \textnormal{\textsc{Query}}$(k\in[q])$ returns $\mathrm{val}_k$
exactly in $O(1+h_T)$ time.
\item \textnormal{\textsc{NodeSquaredNorm}}$(\mathcal{I})$ returns $e_{t,\mathcal{I}}$ exactly
in $O(L_{s,1}(1+h_T))$ time.
\item \textnormal{\textsc{LeftProbability}}$(\mathcal{I})$, for an internal node
with children $L,R$, returns $b_t(\mathcal{I},L)$ when $e_{t,L}+e_{t,R}>0$ and
$\mathsf{fail}$ otherwise. It takes $O(L_{s,1}(1+h_T))$ time.
\item \textnormal{\textsc{Propose}}$()$ returns a tuple
$(i,p^{\rm prop}_{t,i},\delta_{v,i}^2,\lambda_i)$ whose first entry is $i$ with probability
$p^{\rm prop}_{t,i}$, and returns $\mathsf{fail}$ with the remaining probability.
Every returned tuple has $\delta_{v,i}^2>0$ and $p^{\rm prop}_{t,i}>0$.
The procedure takes $O((L_{s,1}L_1+L_{s,3})(1+h_T))$ time.
\item \textnormal{\textsc{Probability}}$(i,\lambda)$, for $\delta_{v,i}\ne0$ and
$\lambda=\lambda_i$, returns $p^{\rm prop}_{t,i}$ exactly in $O(L_{s,1}L_1(1+h_T))$ time.
\end{enumerate}
Each instance uses $O(1)$ storage in addition to the shared representation.
The procedure \textnormal{\textsc{Propose}} uses $O(L_{s,3})$ temporary storage;
every other procedure uses $O(1)$ temporary storage. All procedures leave
the shared representation unchanged. The guarantees hold for every fixed
choice of sketches and every state satisfying
Definition~\ref{def:lp_proposal_tree_representation}.
\end{theorem}
\begin{proof}
Lemma~\ref{lem:lp_proposal_tree_init} proves that \textsc{Init} binds the
three stored members in constant time and space.
Lemma~\ref{lem:lp_proposal_tree_query} proves the exact coordinate
evaluation and its running time. Lemmas~\ref{lem:lp_proposal_tree_node_norm}
and~\ref{lem:lp_proposal_tree_left_probability} give the node and branch
guarantees. Lemma~\ref{lem:lp_proposal_tree_propose} proves the proposal
distribution, returned values, and running time, and
Lemma~\ref{lem:lp_proposal_tree_probability} proves the probability-query
guarantee. Each lemma proves that the shared representation is unchanged.
The proposal procedure stores at most $L_{s,3}$ leaf weights, while the other
procedures use a constant number of local values. These bounds give all
six guarantees and the stated persistent and temporary space bounds.
\end{proof}

\subsection{Initialization}
\label{subsec:dyadic_tree_init}
The topology and cache are shared by all tree instances. Initialization
records references to these objects and the index of the chosen sketch.

\begin{lemma}[Initialization]\label{lem:lp_proposal_tree_init}
Suppose $(\mathcal T',\mathcal C')$ satisfies
Definition~\ref{def:lp_proposal_tree_representation} and $t_{\mathrm{in}}\in[L_{s,2}]$.
The procedure \textnormal{\textsc{Init}}$(t_{\mathrm{in}},\mathcal T',\mathcal C')$ in
Algorithm~\ref{alg:lp_proposal_tree_queries} binds the instance to these
objects and tree index $t_{\mathrm{in}}$ in $O(1)$ time and additional space.
It leaves the shared representation unchanged.
\end{lemma}
\begin{proof}
The procedure assigns the index $t_{\mathrm{in}}$ and stores one reference to each
supplied object. These three assignments take $O(1)$ time and space.
Every later access through the stored references reads the corresponding
shared object. No shared array is copied or changed, so the supplied
representation remains valid and the instance has the stated members.
\end{proof}

\subsection{Entry query}
\label{subsec:dyadic_tree_entry}
The cached low-coordinate contribution and the explicitly enumerated
high-coordinate contributions together determine a coordinate of $\mathrm{val}$.

\begin{lemma}[Query]\label{lem:lp_proposal_tree_query}
Assume Definition~\ref{def:lp_proposal_tree_representation} and an initialized
tree instance. The procedure \textnormal{\textsc{Query}}$(k\in[q])$
in Algorithm~\ref{alg:lp_proposal_tree_queries} returns $\mathrm{val}_k$ exactly in
$O(1+h_T)$ time and $O(1)$ temporary space. Evaluating any $s$ requested
coordinates takes $O(s(1+h_T))$ time. These calls leave the shared
representation unchanged.
\end{lemma}
\begin{proof}
Fix $k\in[q]$ and define
$d_j:=C_{k,j}\beta_1(v_j)+C_{k,n+j}v_j$ for $j\in[n]$.
Let $x$ denote the returned value.
We have
\[
x=c^{\rm low}_k+\sum_{j\in H}d_j
=\mathrm{bias}_k+\sum_{j\in[n]\setminus H}d_j+\sum_{j\in H}d_j
=\mathrm{bias}_k+\sum_{j=1}^n d_j.
\]
The first step follows from the loop in \textsc{Query}.
The second step uses the cache identity in
Definition~\ref{def:lp_proposal_tree_representation}.
The third step partitions $[n]$ into its low and high coordinates.

We have
\[
\mathrm{bias}_k+\sum_{j=1}^n d_j
=\mathrm{bias}_k+\sum_{j=1}^n(C_{k,j}\beta_1(v_j)+C_{k,n+j}v_j)
=\mathrm{bias}_k+\sum_{j=1}^{2n}C_{k,j}w_j
=(Cw+\mathrm{bias})_k=\mathrm{val}_k.
\]
The first step uses the definition of $d_j$.
The second step uses the two blocks of $w$ in
Definition~\ref{def:lp_proposal_tree_representation}.
The third step is the definition of matrix-vector multiplication, and
the fourth step uses the definition of $\mathrm{val}$.

The initial cache access takes $O(1)$ time. Each of the $h_T$ loop
iterations uses a constant number of arithmetic operations, including
one evaluation of $\beta_1$ by Definition~\ref{def:lp_beta_1}(c).
Thus one call takes $O(1+h_T)$ time, and $s$ calls take $O(s(1+h_T))$ time.
The accumulator and loop index use $O(1)$ space. The procedure only reads
the shared arrays, so they remain unchanged.
\end{proof}

\subsection{Squared norms of node sketches}
\label{subsec:dyadic_tree_norms}
The row map for a node identifies the $L_{s,1}$ coordinates of its sketch.
Evaluating these coordinates gives the squared norm used during descent.

\begin{lemma}[Node squared norm]\label{lem:lp_proposal_tree_node_norm}
Assume Definition~\ref{def:lp_proposal_tree_representation} and an initialized
tree instance with index $t$. For every node $\mathcal{I}$, the procedure
\textnormal{\textsc{NodeSquaredNorm}}$(\mathcal{I})$ in
Algorithm~\ref{alg:lp_proposal_tree_queries} returns $e_{t,\mathcal{I}}$ exactly
in $O(L_{s,1}(1+h_T))$ time and $O(1)$ temporary space. It leaves the shared
representation unchanged.
\end{lemma}
\begin{proof}
Write the ordered list $\mathrm{NodeRows}[t,\mathcal{I}]$ as $(k_1,\ldots,k_{L_{s,1}})$,
and let $e$ denote the returned value.
We have
\[
e=\sum_{r=1}^{L_{s,1}} \mathrm{val}_{k_r}^2
=\|\mathrm{val}_{\mathrm{NodeRows}[t,\mathcal{I}]}\|_2^2
=\|(S_{{\rm sk},t})_{:,\mathcal{I}}\delta_{v,\mathcal{I}}\|_2^2=e_{t,\mathcal{I}}.
\]
The first step uses the coordinate-evaluation guarantee in
Lemma~\ref{lem:lp_proposal_tree_query} for the values accumulated by
the loop. The second step is the definition of the squared Euclidean norm.
The third step uses the node-coordinate identity in
Definition~\ref{def:lp_proposal_tree_representation}, and the fourth step
uses the definition of $e_{t,\mathcal{I}}$.

There are $L_{s,1}$ coordinate evaluations, each costing $O(1+h_T)$ by
Lemma~\ref{lem:lp_proposal_tree_query}. The squaring and accumulation
take $O(L_{s,1})$ additional time, giving $O(L_{s,1}(1+h_T))$ total time.
Only the current coordinate, accumulator, and loop index are stored.
The coordinate evaluations leave the shared representation unchanged,
so the entire procedure does as well and uses $O(1)$ temporary space.
\end{proof}

\subsection{Branch probability}
\label{subsec:dyadic_tree_branch}
At an internal node, the squared norms of its two child sketches determine
the probability of choosing the left child.

\begin{lemma}[Branch probability]\label{lem:lp_proposal_tree_left_probability}
Assume Definition~\ref{def:lp_proposal_tree_representation} and an initialized
tree instance. For an internal node $\mathcal{I}$ with children $L,R$, the procedure
\textnormal{\textsc{LeftProbability}}$(\mathcal{I})$ in
Algorithm~\ref{alg:lp_proposal_tree_queries} returns $\mathsf{fail}$
if $e_{t,L}+e_{t,R}=0$, and otherwise returns $b_t(\mathcal{I},L)\in[0,1]$.
It takes $O(L_{s,1}(1+h_T))$ time and $O(1)$ temporary space and leaves the
shared representation unchanged.
\end{lemma}
\begin{proof}
Lemma~\ref{lem:lp_proposal_tree_node_norm} gives
$e_L=e_{t,L}$ and $e_R=e_{t,R}$ for the two computed values. Thus the
zero-denominator test returns $\mathsf{fail}$ exactly in the stated case.
When the denominator is positive, the returned value $p$ satisfies
\[
p=\frac{e_L}{e_L+e_R}
=\frac{e_{t,L}}{e_{t,L}+e_{t,R}}=b_t(\mathcal{I},L).
\]
The first step is the assignment to $p$ in \textsc{LeftProbability}.
The second step uses Lemma~\ref{lem:lp_proposal_tree_node_norm}.
The third step uses Definition~\ref{def:lp_proposal_tree_probabilities}.
Both squared norms are nonnegative, so $0\le e_L\le e_L+e_R$ and
$p\in[0,1]$.

The two node evaluations take $O(L_{s,1}(1+h_T))$ time by
Lemma~\ref{lem:lp_proposal_tree_node_norm}. The remaining operations
take constant time and space. Both evaluations leave the shared
representation unchanged, which proves the remaining claims.
\end{proof}

Algorithm~\ref{alg:lp_proposal_tree_queries} implements initialization,
coordinate evaluation, node squared norms, and branch probabilities.
Within the tree algorithms, $n,L_{s,3}$ are fields of $\mathcal T$, and
$C$, $c^{\rm low}$, $v$, $\mathrm{High}$, $\mathrm{NodeRows}$, and $\mathrm{EntryIndex}$ are
the current fields of $\mathcal C$. The operations \textsc{Left} and
\textsc{Right} return the child intervals stored in $\mathcal T$.

\begin{algorithm}[!htb]
\caption{\textsc{ProposalTree} data structure: initialization and queries}\label{alg:lp_proposal_tree_queries}
\begin{algorithmic}[1]
\DataStructure{ProposalTree} \Comment{Theorem~\ref{thm:lp_proposal_tree_interface}}
\Members
\State Tree index $t$; references $\mathcal T,\mathcal C$ to the common topology and shared cache.
\EndMembers
\Procedure{Init}{$t_{\mathrm{in}}\in[L_{s,2}]$, topology $\mathcal T'$, cache $\mathcal C'$} \Comment{Lemma~\ref{lem:lp_proposal_tree_init}}
\State $t\gets t_{\mathrm{in}}$; $\mathcal T\gets\operatorname{ref}(\mathcal T')$; $\mathcal C\gets\operatorname{ref}(\mathcal C')$.
\EndProcedure
\State
\Procedure{Query}{$k\in[q]$} \Comment{Lemma~\ref{lem:lp_proposal_tree_query}}
\State $x\gets c^{\rm low}_k$.
\For{$j\in\mathrm{High}$}
\State $x\gets x+C_{k,j}\beta_1(v_j)+C_{k,n+j}v_j$.
\EndFor
\State \Return $x$.
\EndProcedure
\State
\Procedure{NodeSquaredNorm}{$\mathcal{I}$: node of $\mathcal T$} \Comment{Lemma~\ref{lem:lp_proposal_tree_node_norm}}
\State $e\gets0$.
\For{$k\in\mathrm{NodeRows}[t,\mathcal{I}]$}
\State $x\gets\Call{Query}{k}$; $e\gets e+x^2$.
\EndFor
\State \Return $e$.
\EndProcedure
\State
\Procedure{LeftProbability}{$\mathcal{I}$: internal node of $\mathcal T$} \Comment{Lemma~\ref{lem:lp_proposal_tree_left_probability}}
\State $L\gets\Call{Left}{\mathcal{I}}$; $R\gets\Call{Right}{\mathcal{I}}$.
\State $e_L\gets\Call{NodeSquaredNorm}{L}$; $e_R\gets\Call{NodeSquaredNorm}{R}$.
\If{$e_L+e_R=0$}
\State \Return $\mathsf{fail}$.
\EndIf
\State $p\gets e_L/(e_L+e_R)$.
\State \Return $p$.
\EndProcedure
\EndDataStructure
\end{algorithmic}
\end{algorithm}

\subsection{Proposal sampling}
\label{subsec:dyadic_tree_sample}
The proposal procedure records its branch probabilities and then uses
exact squared residual entries to select a coordinate within the reached leaf.

\begin{lemma}[Proposal]\label{lem:lp_proposal_tree_propose}
Assume Definition~\ref{def:lp_proposal_tree_representation} and an initialized
tree instance with index $t$. The procedure \textnormal{\textsc{Propose}}$()$
in Algorithm~\ref{alg:lp_proposal_tree_propose} returns a tuple
$(i,p^{\rm prop}_{t,i},\delta_{v,i}^2,\lambda_i)$ whose first entry is $i$ with probability
$p^{\rm prop}_{t,i}$, and returns $\mathsf{fail}$ with the remaining probability.
Every returned tuple has $\delta_{v,i}^2>0$ and $p^{\rm prop}_{t,i}>0$.
The procedure takes $O((L_{s,1}L_1+L_{s,3})(1+h_T))$ time and $O(L_{s,3})$ temporary
space and leaves the shared representation unchanged.
\end{lemma}
\begin{proof}
At an internal node with a positive denominator,
Lemma~\ref{lem:lp_proposal_tree_left_probability} and the Bernoulli draw
give branch probability $b_t(\mathcal{I},K)$ for either child $K$.
A zero denominator returns $\mathsf{fail}$ before a child is selected.
Induction along the path therefore shows that the probability of reaching
the leaf $\mathcal{I}(i)$ is $\prod_{(\mathcal{I},K)\in\mathcal P_i}b_t(\mathcal{I},K)$.
The variable $p_{\rm prop}$ stores the same product along the sampled path.

At a reached leaf $\mathcal{I}$, Lemma~\ref{lem:lp_proposal_tree_query} and
the entry-coordinate identity in
Definition~\ref{def:lp_proposal_tree_representation} give $\omega_i=\delta_{v,i}^2$
for each $i\in \mathcal{I}$. Hence the accumulated value is $\mathcal{N}_{2,\mathcal{I}}$.
If this value is zero, the procedure returns $\mathsf{fail}$ as specified.
Otherwise define $S_i:=\sum_{j\in \mathcal{I},\,j\le i}\omega_j$ and
$S_{\min \mathcal{I}-1}:=0$. The scan selects $i$ exactly when
$\xi\in[S_{i-1},S_i)$. It terminates within the leaf because
$\xi<\mathcal{N}_{2,\mathcal{I}}=S_{\max \mathcal{I}}$. Conditional on reaching $\mathcal{I}$, we have
\[
\Pr[\text{select }i\mid \mathcal{I}]
=\frac{S_i-S_{i-1}}{\mathcal{N}_{2,\mathcal{I}}}
=\frac{\omega_i}{\mathcal{N}_{2,\mathcal{I}}}
=\frac{\delta_{v,i}^2}{\mathcal{N}_{2,\mathcal{I}}}=\lambda_i.
\]
The first step uses the uniform draw on $[0,\mathcal{N}_{2,\mathcal{I}})$.
The second step uses the definition of $S_i$.
The third step uses the exact leaf weights, and the fourth step uses
Definition~\ref{def:lp_proposal_tree_probabilities}.

For any coordinate whose path and leaf have positive denominators, we have
\[
\Pr[\textsc{Propose}\text{ returns coordinate }i]
=\lambda_i\prod_{(\mathcal{I},K)\in\mathcal P_i}b_t(\mathcal{I},K)
=p^{\rm prop}_{t,i}.
\]
The first step multiplies the probability of reaching the leaf by its
conditional selection probability. The second step uses
Definition~\ref{def:lp_proposal_tree_probabilities}.
If a path denominator or branch probability is zero, the coordinate is
unreachable and the same definition gives $p^{\rm prop}_{t,i}=0$.
If its leaf has zero squared norm, the procedure returns $\mathsf{fail}$
there and again $p^{\rm prop}_{t,i}=0$. These cases prove the distribution for
every $i$, with all remaining outcomes equal to $\mathsf{fail}$.
The scan skips zero weights, and every selected branch has positive
probability. Thus a successful call has $s=\delta_{v,i}^2>0$,
$\lambda=\lambda_i$, and the final product $p_{\rm prop}=p^{\rm prop}_{t,i}>0$.

There are at most $L_1$ calls to \textsc{LeftProbability}. Each uses
$O(L_{s,1}(1+h_T))$ time by Lemma~\ref{lem:lp_proposal_tree_left_probability}.
The $L_{s,3}$ leaf evaluations take $O(L_{s,3}(1+h_T))$ time by
Lemma~\ref{lem:lp_proposal_tree_query}, and the scan takes $O(L_{s,3})$ time.
The other operations take $O(L_1+1)$ time. Since $L_{s,1},L_1\ge1$, these costs
sum to $O((L_{s,1}L_1+L_{s,3})(1+h_T))$.
The leaf weights occupy an array of length $L_{s,3}$, indexed by their
offsets in $\mathcal{I}$; all other local values use $O(1)$ space.
Every helper call reads the shared representation without changing it.
Thus the total temporary space is $O(L_{s,3})$ and the shared state is unchanged.
\end{proof}

Algorithm~\ref{alg:lp_proposal_tree_propose} uses a cumulative-weight scan
in the natural order of the leaf interval. Each random draw is fresh.

\begin{algorithm}[!htb]
\caption{\textsc{ProposalTree} data structure: proposing a coordinate}\label{alg:lp_proposal_tree_propose}
\begin{algorithmic}[1]
\DataStructure{ProposalTree} \Comment{Theorem~\ref{thm:lp_proposal_tree_interface}}
\Procedure{Propose}{} \Comment{Lemma~\ref{lem:lp_proposal_tree_propose}}
\State $\mathcal{I}\gets[n]$; $p_{\rm prop}\gets1$.
\While{$|\mathcal{I}|>L_{s,3}$}
\State $p\gets\Call{LeftProbability}{\mathcal{I}}$. \Comment{Lemma~\ref{lem:lp_proposal_tree_left_probability}}
\If{$p=\mathsf{fail}$}
\State \Return $\mathsf{fail}$.
\EndIf
\State $\eta\gets\Call{Bernoulli}{p}$.
\If{$\eta=1$}
\State $p_{\rm prop}\gets p_{\rm prop} p$; $\mathcal{I}\gets\Call{Left}{\mathcal{I}}$.
\Else
\State $p_{\rm prop}\gets p_{\rm prop}(1-p)$; $\mathcal{I}\gets\Call{Right}{\mathcal{I}}$.
\EndIf
\EndWhile
\State $\mathcal{N}_{2,\mathcal{I}}\gets0$.
\For{$i\in \mathcal{I}$}
\State $x\gets\Call{Query}{\mathrm{EntryIndex}[i]}$. \Comment{Algorithm~\ref{alg:lp_proposal_tree_queries}}
\State $\omega_i\gets x^2$; $\mathcal{N}_{2,\mathcal{I}}\gets\mathcal{N}_{2,\mathcal{I}}+\omega_i$.
\EndFor
\If{$\mathcal{N}_{2,\mathcal{I}}=0$}
\State \Return $\mathsf{fail}$.
\EndIf
\State $\xi\gets\Call{Uniform}{[0,\mathcal{N}_{2,\mathcal{I}})}$; $i\gets\min \mathcal{I}$; $c\gets\omega_i$.
\While{$\xi\ge c$}
\State $i\gets i+1$; $c\gets c+\omega_i$.
\EndWhile
\State $s\gets\omega_i$; $\lambda\gets s/\mathcal{N}_{2,\mathcal{I}}$; $p_{\rm prop}\gets p_{\rm prop}\lambda$.
\State \Return $(i,p_{\rm prop},s,\lambda)$.
\EndProcedure
\EndDataStructure
\end{algorithmic}
\end{algorithm}

\subsection{Proposal probability query}
\label{subsec:dyadic_tree_probability}
Algorithm~\ref{alg:lp_proposal_tree_probability} takes the exact leaf probability $\lambda_i$ as input. This value can be reused across tree indices because all trees share the leaf partition and residual vector.

\begin{algorithm}[!htb]
\caption{\textsc{ProposalTree} data structure: evaluating a proposal probability}\label{alg:lp_proposal_tree_probability}
\begin{algorithmic}[1]
\DataStructure{ProposalTree} \Comment{Theorem~\ref{thm:lp_proposal_tree_interface}}
\Procedure{Probability}{$i\in[n],\lambda\in(0,1]$} \Comment{Lemma~\ref{lem:lp_proposal_tree_probability}}
\State $\mathcal{I}\gets[n]$; $p_{\rm prop}\gets1$.
\While{$|\mathcal{I}|>L_{s,3}$}
\State $p\gets\Call{LeftProbability}{\mathcal{I}}$. \Comment{Lemma~\ref{lem:lp_proposal_tree_left_probability}}
\If{$p=\mathsf{fail}$}
\State \Return $0$.
\EndIf
\State $L\gets\Call{Left}{\mathcal{I}}$.
\If{$i\in L$}
\State $q\gets p$; $\mathcal{I}\gets L$.
\Else
\State $q\gets1-p$; $\mathcal{I}\gets\Call{Right}{\mathcal{I}}$.
\EndIf
\If{$q=0$}
\State \Return $0$.
\EndIf
\State $p_{\rm prop}\gets p_{\rm prop} q$.
\EndWhile
\State $p_{\rm prop}\gets p_{\rm prop}\lambda$.
\State \Return $p_{\rm prop}$.
\EndProcedure
\EndDataStructure
\end{algorithmic}
\end{algorithm}

To evaluate the probability of a given coordinate, the query multiplies the probabilities along its unique path by the supplied within-leaf probability.

\begin{lemma}[Probability query]\label{lem:lp_proposal_tree_probability}
Assume Definition~\ref{def:lp_proposal_tree_representation} and an initialized
tree instance with index $t$. For $i\in[n]$ with $\delta_{v,i}\ne0$ and
$\lambda=\lambda_i$, the procedure
\textnormal{\textsc{Probability}}$(i,\lambda)$ in
Algorithm~\ref{alg:lp_proposal_tree_probability} returns $p^{\rm prop}_{t,i}$ exactly.
It takes $O(L_{s,1}L_1(1+h_T))$ time and $O(1)$ temporary space and leaves the
shared representation unchanged.
\end{lemma}
\begin{proof}
Since $\delta_{v,i}\ne0$, the leaf containing $i$ has positive squared norm and
$\lambda_i\in(0,1]$. At each internal node, membership in the left
child interval identifies the unique next child on $\mathcal P_i$.
Lemma~\ref{lem:lp_proposal_tree_left_probability} gives the probability
of that branch. A zero denominator or zero branch probability makes the
product in Definition~\ref{def:lp_proposal_tree_probabilities} zero,
so either early return is correct.

Otherwise the loop multiplies precisely the branch probabilities on
$\mathcal P_i$. If $p_{\rm out}$ denotes the returned value, we have
\[
p_{\rm out}
=\lambda\prod_{(\mathcal{I},K)\in\mathcal P_i}b_t(\mathcal{I},K)
=\lambda_i\prod_{(\mathcal{I},K)\in\mathcal P_i}b_t(\mathcal{I},K)
=p^{\rm prop}_{t,i}.
\]
The first step follows from the loop and final multiplication in
\textsc{Probability}. The second step uses the input condition
$\lambda=\lambda_i$. The third step uses
Definition~\ref{def:lp_proposal_tree_probabilities}.
This also covers a root that is a leaf, when the loop is empty.

There are at most $L_1$ calls to \textsc{LeftProbability}. Each uses
$O(L_{s,1}(1+h_T))$ time by Lemma~\ref{lem:lp_proposal_tree_left_probability}.
Interval membership and the remaining arithmetic cost $O(L_1+1)$ time,
so the total is $O(L_{s,1}L_1(1+h_T))$ because $L_{s,1},L_1\ge1$.
The procedure stores only a constant number of local values and its
helper calls use $O(1)$ temporary space. All accesses to the shared
representation are reads, proving the remaining claims.
\end{proof}

\section{Exact \texorpdfstring{$\ell_2$}{L2}-sampling data structure}\label{sec:lp_exact_l2_sampler}

In Sections~\ref{subsec:lp_exact_l2_sampler_main_data_structure} and~\ref{subsec:lp_exact_l2_sampler_initialization}, we specify the sampling guarantee and initialize the sketches and trees.
In Sections~\ref{subsec:lp_exact_l2_sampler_threshold_update} and~\ref{sec:lp_exact_l2_sampler_entry_updates}, we update the sampler when the threshold or a coordinate changes.
In Section~\ref{subsec:lp_exact_l2_sampler_sampling_procedure}, we prove that the sampler gives each coordinate its prescribed probability at a good state.
In Sections~\ref{subsec:lp_exact_l2_sampler_concentration_along_a_tree_path} and~\ref{subsec:lp_exact_l2_sampler_simultaneous_sketch_and_proposal_bounds}, we bound proposal probabilities along one tree path and then across the ensemble.
In Section~\ref{sec:reciprocal_coin}, we construct the reciprocal coin used for acceptance. In Section~\ref{sec:lp_exact_l2_sampler_fallback}, we give an exact sampler for small dimensions.

\subsection{Main data structure}
\label{subsec:lp_exact_l2_sampler_main_data_structure}
The sampler keeps $\mathscr{A}\in\R^{n\times n}$ and $b\in\R^n$ fixed while $v\in\R^n$ and the threshold $T\ge2$ may change. Its sampling law depends on the squared residual entries, and its query cost depends on how many coordinates are classified as high.

\begin{definition}[\textsc{ExactL2Sampler} notation]\label{def:lp_exact_l2_sampler_notation}
Let $n=2^{L_1}$ with $L_1\ge1$, $\mathscr{A}\in\R^{n\times n}$, $b,v\in\R^n$, and $T\ge2$.
Let $\gamma_1\ge2$ and $\beta_1$ be as in Definition~\ref{def:lp_beta_1}.
Define the mapped vector, the residual, and its squared norm by
\[
v^{\new}:=\mathscr{A}\beta_1(v)+b,\qquad \delta_v:=v^{\new}-v,\qquad \mathcal{N}_2:=\|\delta_v\|_2^2.
\]
The threshold $T$ is measured in normalized magnitudes $2|v_i|/\gamma_1$.
Define the number of high coordinates by
\[
h_T:=|\{i\in[n]:2|v_i|/\gamma_1\ge T\}|.
\]
\end{definition}

The data structure maintains the residual under coordinate updates and uses tree proposals to sample from it. At a good state, the correction step gives each coordinate exactly its prescribed probability.

\begin{theorem}[ExactL2Sampler]\label{thm:lp_exact_l2_sampler_interface}
Use the notation in Definition~\ref{def:lp_exact_l2_sampler_notation}
and the parameters in Definition~\ref{def:lp_exact_l2_sampler_parameters}.
For $L_{s,3}\le n$, there is a randomized data structure
(\textnormal{\textsc{ExactL2Sampler}} in Algorithms~\ref{alg:exact_l2_sampler}
and~\ref{alg:exact_l2_sampler_sample}) that uses $O(n^2)$ space and maintains
fixed $\mathscr{A}\in\R^{n\times n}$ and $b\in\R^n$, a vector $v\in\R^n$, and
a threshold $T\ge2$. The maintenance guarantees hold for arbitrary sequences of valid procedure calls, with old and new levels in $[L_2]$. It supports the following procedures.
\begin{enumerate}[label=(\alph*)]
\item \textnormal{\textsc{Init}}$(\mathscr{A}\in\R^{n\times n},b\in\R^n,T\in\R_{\ge2})$.
This procedure takes $\mathscr{A},b,T$ as inputs, sets $v={\bf 0}_n$, and chooses
independent Gaussian sketches. It runs in $O(n^2L_1)$ time.
\item \textnormal{\textsc{UpdateT}}$(T^{\new}\in\R_{\ge2})$.
This procedure takes $T^{\new}$ as input and changes the threshold to
$T^{\new}$. It runs in $O(1)$ time if the threshold is unchanged and
$O(n^2)$ time otherwise.
\item \textnormal{\textsc{UpdateVL}}$(i\in[n],\zeta\in\R,j\in[L_2])$.
This procedure takes $i,\zeta,j$ as inputs and updates $v_i$ to $\zeta$
and its level to $j$. It preserves the invariant in
Definition~\ref{def:lp_exact_l2_sampler_invariant} and runs in $O(n)$ time.
\item \textnormal{\textsc{L2Sample}}$(L_{6,4}\in\R_{>0})$.
This procedure takes $L_{6,4}$ as input. At a good state in the sense of
Lemma~\ref{lem:lp_ensemble}, if $L_{6,4}\ge6\mathcal{N}_2$, it outputs $i$ with probability
$\delta_{v,i}^2/L_{6,4}$ and $\mathsf{null}$ otherwise. Each fixed residual independent of the
sketches is good with probability at least $1-n^{-14}$. This procedure runs in $O(L_1^2(1+h_T))$ expected time.
\end{enumerate}
\end{theorem}

\begin{proof}
Lemma~\ref{lem:lp_exact_l2_sampler_init} establishes the invariant in
Definition~\ref{def:lp_exact_l2_sampler_invariant}, the stored representation
described below, and the initialization and space bounds.
Lemmas~\ref{lem:lp_exact_l2_sampler_update_t} and~\ref{lem:lp_exact_l2_sampler_update_vl}
prove that threshold and coordinate updates preserve this invariant and
representation with the stated running times. Induction on the sequence
of procedure calls proves the maintenance guarantees.
Lemma~\ref{lem:lp_exact_l2_sampler_sample} proves the sampling law and expected running time,
and Lemma~\ref{lem:lp_ensemble} gives the good-state probability.
Sampling leaves the stored state unchanged, and the procedure bounds
include the temporary storage, so the space bound continues to hold.
\end{proof}

We first specify the parameters used by the Gaussian sketches and the dyadic trees.
These parameters remain fixed throughout the sequence of procedure calls.

\begin{definition}[Parameters for \textsc{ExactL2Sampler}]\label{def:lp_exact_l2_sampler_parameters}
For $n=2^{L_1}$ and $L_1\ge1$, define
\[
L_{s,1}:=2^{\lceil\log_2(1024L_1)\rceil},\qquad
L_{s,2}:=2^{\lceil\log_2(64L_1)\rceil},\qquad L_{s,3}:=L_{s,1}L_{s,2},
\qquad L_{s,4}:=4096L_1.
\]
Define $L_2:=L_1+4$.
\end{definition}

For $L_{s,3}\le n$, store independent standard Gaussian matrices
$S_{{\rm sk},0}\in\R^{L_{s,4}\times n}$ and $S_{{\rm sk},t}\in\R^{L_{s,1}\times n}$ for $t\in[L_{s,2}]$.
Use $L_{s,2}$ trees with a common dyadic partition of $[n]$ and leaf size $L_{s,3}$.
For every node $\mathcal{I}$ of tree $t$, store
\[
\mathscr{A}^{\rm sk}_{t,\mathcal{I}}:=(S_{{\rm sk},t})_{:,\mathcal{I}}\mathscr{A}_{\mathcal{I},:},\qquad
 \mathrm{bias}_{t,\mathcal{I}}:=(S_{{\rm sk},t})_{:,\mathcal{I}}b_{\mathcal{I}}.
\]
The representation of the root sketch's squared norm uses
\[
B:=\begin{bmatrix}S_{{\rm sk},0}\mathscr{A}&-S_{{\rm sk},0}\end{bmatrix},\qquad c:=S_{{\rm sk},0}b,
\qquad w:=\begin{bmatrix}\beta_1(v)\\v\end{bmatrix},
\qquad Q_0:=B^\top B,\qquad d:=B^\top c.
\]
Store a separate \textsc{ExactIP} instance $\mathrm{exactIP}_0$ for $Q_0w$,
with dimension $2n$ and $L_2$ levels. Coordinates $i$ and $n+i$ share
an auxiliary level in $[L_2]$, initially one.

Assemble a fixed matrix $C$ with $2n$ columns as follows. For every tree
node, include the $L_{s,1}$ rows
$\begin{bmatrix}\mathscr{A}^{\rm sk}_{t,\mathcal{I}}&-(S_{{\rm sk},t})_{:,\mathcal{I}}\end{bmatrix}$, extending the second
block by zeros off $\mathcal{I}$, with constant term $\mathrm{bias}_{t,\mathcal{I}}$.
Also include the $n$ rows $\begin{bmatrix}\mathscr{A}&-I\end{bmatrix}$ with constant
term $b$. Define $\mathrm{bias}$ as the vector of these constant terms.
Store the high coordinates in an unordered array with inverse membership
pointers, using a last-entry swap for deletions.
Denote this array by $\mathrm{High}$. While assembling $C$, record the $L_{s,1}$
row indices for node $\mathcal{I}$ of tree $t$ in $\mathrm{NodeRows}[t,\mathcal{I}]$ and the
row index for residual coordinate $i$ in $\mathrm{EntryIndex}[i]$.
These arrays form the shared record $\mathcal C$ in
Definition~\ref{def:lp_proposal_tree_representation}; the object $\mathcal T$
stores $n,L_{s,3}$ and the common dyadic topology.
The objects $\mathrm{Tree}[1],\ldots,\mathrm{Tree}[L_{s,2}]$ are the instances
of Theorem~\ref{thm:lp_proposal_tree_interface} that read these shared objects.

The following invariant specifies the squared norm of the root sketch and the low-coordinate
contributions maintained by the data structure.

\begin{definition}[\textsc{ExactL2Sampler} invariant]\label{def:lp_exact_l2_sampler_invariant}
For the stored matrices above, $\delta_v=\mathscr{A}\beta_1(v)+b-v$, and $T\ge2$, the
\textnormal{\textsc{ExactL2Sampler}} invariant consists of the following identities:
\[
\mathrm{exactIP}_0.b=w,\qquad
\mathcal{N}_{2,0}=w^\top Q_0w+2d^\top w+\|c\|_2^2=\|S_{{\rm sk},0}\delta_v\|_2^2,
\]
\[
c^{\rm low}=\mathrm{bias}+\sum_{\substack{i=1\\2|v_i|/\gamma_1<T}}^n
(C_{:,i}\beta_1(v_i)+C_{:,n+i}v_i).
\]
The shared records $\mathcal T,\mathcal C$ also satisfy
Definition~\ref{def:lp_proposal_tree_representation} for this residual $\delta_v$.
\end{definition}

Algorithms~\ref{alg:exact_l2_sampler} and~\ref{alg:exact_l2_sampler_sample}
give the implementation for $L_{s,3}\le n$. Both blocks use the same
\textsc{ExactL2Sampler} members.

\begin{algorithm}[!ht]
\caption{\textsc{ExactL2Sampler} data structure: initialization and updates}\label{alg:exact_l2_sampler}
\begin{algorithmic}[1]
\DataStructure{ExactL2Sampler} \Comment{Theorem~\ref{thm:lp_exact_l2_sampler_interface}}
\Members
\State $\mathscr{A}\in\R^{n\times n}$; $b,v\in\R^n$; threshold $T$; parameters $L_{s,1},L_{s,2},L_{s,3},L_{s,4},L_2$.
\State Gaussian matrices $S_{{\rm sk},0},S_{{\rm sk},t}$; a common dyadic tree and tables $\mathscr{A}^{\rm sk}_{t,\mathcal{I}},\mathrm{bias}_{t,\mathcal{I}}$.
\State Shared records $\mathcal T,\mathcal C$; \textsc{ProposalTree} instances $\mathrm{Tree}[1],\ldots,\mathrm{Tree}[L_{s,2}]$.
\State $Q_0,d,\mathcal{N}_{2,0}$ and \textsc{ExactIP} $\mathrm{exactIP}_0$; $C,\mathrm{bias},c^{\rm low}$ and the high-coordinate array.
\EndMembers
\State
\Procedure{Init}{$\mathscr{A}\in\R^{n\times n},b\in\R^n,T\in\R_{\ge2}$} \Comment{Lemma~\ref{lem:lp_exact_l2_sampler_init}}
\State Store $\mathscr{A},b,T$ and the parameters; $L_2\gets L_1+4$; draw independent standard Gaussian $S_{{\rm sk},0}\in\R^{L_{s,4}\times n}$ and $S_{{\rm sk},t}\in\R^{L_{s,1}\times n}$ for $1\le t\le L_{s,2}$; $v\gets{\bf 0}_n$.
\State Build the dyadic tree with leaves of size $L_{s,3}$. For each leaf $\mathcal{I}$, set $\mathscr{A}^{\rm sk}_{t,\mathcal{I}}\gets(S_{{\rm sk},t})_{:,\mathcal{I}}\mathscr{A}_{\mathcal{I},:}$ and $\mathrm{bias}_{t,\mathcal{I}}\gets(S_{{\rm sk},t})_{:,\mathcal{I}}b_{\mathcal{I}}$; sum children to form parent tables.
\State Form $Q_0,d$; $\mathcal{N}_{2,0}\gets\|S_{{\rm sk},0}b\|_2^2$.
\State \Call{\ensuremath{\mathrm{exactIP}_0}.Init}{$Q_0,L_2$}. \Comment{Algorithm~\ref{alg:exact_ip}}
\State Assemble $C,\mathrm{bias}$ and the row maps; $c^{\rm low}\gets \mathrm{bias}$; $\mathrm{High}\gets()$. \Comment{Definition~\ref{def:lp_exact_l2_sampler_invariant}}
\State $\mathcal T\gets$ the common topology with $n,L_{s,3}$; $\mathcal C\gets$ references to the shared arrays and row maps.
\For{$t=1,\ldots,L_{s,2}$}
\State $\mathrm{Tree}[t].\Call{Init}{t,\mathcal T,\mathcal C}$. \Comment{Algorithm~\ref{alg:lp_proposal_tree_queries}}
\EndFor
\EndProcedure
\State
\Procedure{UpdateT}{$T^{\new}\in\R_{\ge2}$} \Comment{Lemma~\ref{lem:lp_exact_l2_sampler_update_t}}
\If{$T^{\new}\ne T$}
\State $T\gets T^{\new}$; rebuild $c^{\rm low}$ and the high-coordinate array. \Comment{Definition~\ref{def:lp_exact_l2_sampler_invariant}}
\EndIf
\EndProcedure
\State
\Procedure{UpdateVL}{$i\in[n],\zeta\in\R,j\in[L_2]$} \Comment{Lemma~\ref{lem:lp_exact_l2_sampler_update_vl}}
\State Update $\mathcal{N}_{2,0}$ by Eq.~\eqref{eq:lp_root_sketch_update}.
\State \Call{\ensuremath{\mathrm{exactIP}_0}.UpdateV}{$i,\beta_1(\zeta)$}. \Comment{Algorithm~\ref{alg:exact_ip}}
\State \Call{\ensuremath{\mathrm{exactIP}_0}.UpdateV}{$n+i,\zeta$}. \Comment{Algorithm~\ref{alg:exact_ip}}
\State \Call{\ensuremath{\mathrm{exactIP}_0}.UpdateL}{$i,j$}. \Comment{Algorithm~\ref{alg:exact_ip_update_l}}
\State \Call{\ensuremath{\mathrm{exactIP}_0}.UpdateL}{$n+i,j$}. \Comment{Algorithm~\ref{alg:exact_ip_update_l}}
\State Subtract the old contribution to $c^{\rm low}$ if $2|v_i|/\gamma_1<T$.
\State Add the new contribution if $2|\zeta|/\gamma_1<T$; update the high-coordinate array; $v_i\gets\zeta$.
\EndProcedure
\EndDataStructure
\end{algorithmic}
\end{algorithm}

\subsection{Initialization}
\label{subsec:lp_exact_l2_sampler_initialization}
We begin by constructing the sketch tables and a valid initial state for
\textsc{ExactL2Sampler}. The following lemma gives the correctness,
running time, and space bound of the initialization procedure.

\begin{lemma}[Initialization]\label{lem:lp_exact_l2_sampler_init}
For $L_{s,3}\le n$, the procedure
\textnormal{\textsc{Init}}$(\mathscr{A}\in\R^{n\times n},b\in\R^n,T\in\R_{\ge2})$
in Algorithm~\ref{alg:exact_l2_sampler} sets $v={\bf 0}_n$, establishes
the invariant in Definition~\ref{def:lp_exact_l2_sampler_invariant} and
the stored representation above, and takes $O(n^2L_1)$ time and $O(n^2)$ space.
\end{lemma}
\begin{proof}
Choose a fixed square matrix-multiplication algorithm costing $O(s^\gamma)$
for some $2<\gamma<5/2$. The choice $\gamma=12/5$ is available from
\cite{g14}. No subpolynomial factor is hidden in this fixed algorithm.
At each leaf, stack its $L_{s,2}$ sketches into a $L_{s,3}$ by $L_{s,3}$ matrix and multiply
by $\mathscr{A}_{\mathcal{I},:}$. Splitting the output columns into blocks of size $L_{s,3}$ costs
$O(nL_{s,3}^{\gamma-1})$ at that leaf.

We have
\begin{equation}\label{eq:lp_leaf_sketch_setup_work}
\text{leaf setup}=O(n^2L_{s,3}^{\gamma-2})
=O(n^2L_1^{2\gamma-4})=O(n^2L_1).
\end{equation}
The first step sums over the $n/L_{s,3}$ leaves; the second step uses $L_{s,3}=O(L_1^2)$;
and the third step uses $\gamma<5/2$ and $L_1\ge1$.

All internal $M$ and $\mathrm{bias}$ tables are obtained by summing children in
$O(n^2)$ additional time. Generating the Gaussian matrices costs
$O(L_{s,3}n)\le O(n^2)$. Forming $S_{{\rm sk},0}\mathscr{A}$, $Q_0=B^\top B$, and $d=B^\top c$ by ordinary
multiplication takes $O(n^2L_1)$ time and $O(n^2)$ space.
Initializing $\mathrm{exactIP}_0$ takes $O(n^2+nL_2)=O(n^2)$ time and space
by Lemma~\ref{lem:lp_exact_ip_init} and $L_2=L_1+4=O(n)$.

There are $2n/L_{s,3}-1$ nodes per tree. Thus all node matrices have
$L_{s,2}L_{s,1}(2n/L_{s,3}-1)n=O(n^2)$ entries, and the Gaussian arrays have $L_{s,3}n\le n^2$
entries. The matrix $C$ has
\[
q=L_{s,2}L_{s,1}(2n/L_{s,3}-1)+n=3n-L_{s,3}\le3n
\]
rows. The first step counts $L_{s,1}$ rows per node in each of the $L_{s,2}$ trees
and the $n$ exact residual rows. The second step uses $L_{s,3}=L_{s,1}L_{s,2}$,
and the third step uses $L_{s,3}>0$. Thus assembling $C,\mathrm{bias}$ takes $O(n^2)$
time and space. Recording the row maps takes $O(n)$ time and space.
Binding the $L_{s,2}$ tree instances to the common topology and cache takes
$O(L_{s,2})\le O(n)$ time and space because $L_{s,3}=L_{s,1}L_{s,2}\le n$.
The remaining matrices and caches also use $O(n^2)$ space.
Compute one leaf block at a time, recursively evaluating and accumulating
the fixed bilinear multiplication algorithm's subproducts sequentially;
the temporary storage is $O(L_{s,3}^2)$.

Parent tables are sums of their child tables because the children
partition the parent node. This establishes the definitions of all
$\mathscr{A}^{\rm sk}_{t,\mathcal{I}}$ and $\mathrm{bias}_{t,\mathcal{I}}$.

We have
\begin{equation}\label{eq:lp_root_sketch_quadratic}
\|S_{{\rm sk},0}\delta_v\|_2^2=\|Bw+c\|_2^2
=w^\top Q_0w+2d^\top w+\|c\|_2^2.
\end{equation}
The first step substitutes $\delta_v=\mathscr{A}\beta_1(v)+b-v$ and the definitions of
$B,w,c$. The second step expands the squared norm and uses
$Q_0=B^\top B$ and $d=B^\top c$.

The assembled blocks also give the node-coordinate identity in
Definition~\ref{def:lp_proposal_tree_representation}. For every $v$, we have
\[
\begin{aligned}
(Cw+\mathrm{bias})_{\mathrm{NodeRows}[t,\mathcal{I}]}
&=\mathscr{A}^{\rm sk}_{t,\mathcal{I}}\beta_1(v)+\mathrm{bias}_{t,\mathcal{I}}-(S_{{\rm sk},t})_{:,\mathcal{I}}v_{\mathcal{I}}\\
&=(S_{{\rm sk},t})_{:,\mathcal{I}}(\mathscr{A}_{\mathcal{I},:}\beta_1(v)+b_{\mathcal{I}}-v_{\mathcal{I}})
=(S_{{\rm sk},t})_{:,\mathcal{I}}\delta_{v,\mathcal{I}}.
\end{aligned}
\]
The first step uses the block of $C$ and constant term indexed by
$\mathrm{NodeRows}[t,\mathcal{I}]$. The second step substitutes the definitions of
$\mathscr{A}^{\rm sk}_{t,\mathcal{I}}$ and $\mathrm{bias}_{t,\mathcal{I}}$. The third step uses the definition of $\delta_v$.

For the exact residual coordinates, we have
\[
(Cw+\mathrm{bias})_{\mathrm{EntryIndex}[i]}
=(\mathscr{A}\beta_1(v))_i+b_i-v_i=\delta_{v,i}.
\]
The first step uses the exact residual block of $C$ and its constant term.
The second step uses the definition of $\delta_v$.
Thus the row maps have the meanings required by
Definition~\ref{def:lp_proposal_tree_representation} for every $v$.

Initially $v={\bf 0}_n$ and $\beta_1(0)=0$, so $w={\bf 0}_{2n}$ and
$\mathcal{N}_{2,0}=\|S_{{\rm sk},0}b\|_2^2$. Lemma~\ref{lem:lp_exact_ip_init} initializes
$\mathrm{exactIP}_0.b={\bf 0}_{2n}$ and all paired levels to one.
All coordinates are low, so setting $c^{\rm low}=\mathrm{bias}$ and the
high-coordinate array to empty establishes the cache and membership
conditions in Definition~\ref{def:lp_proposal_tree_representation}.
Lemma~\ref{lem:lp_proposal_tree_init} binds each tree instance to these
shared records. Together with the root representation, this establishes
Definition~\ref{def:lp_exact_l2_sampler_invariant}.
\end{proof}

\subsection{Threshold update}
\label{subsec:lp_exact_l2_sampler_threshold_update}
The threshold determines which coordinate contributions are stored in the
low-coordinate cache. Changing it rebuilds this cache and the high-coordinate
array while keeping the residual fixed.

\begin{lemma}[Threshold update]\label{lem:lp_exact_l2_sampler_update_t}
Assume the invariant in Definition~\ref{def:lp_exact_l2_sampler_invariant}
and the stored representation above hold. The procedure
\textnormal{\textsc{UpdateT}}$(T^{\new}\in\R_{\ge2})$ in
Algorithm~\ref{alg:exact_l2_sampler} changes $T$ to $T^{\new}$ and
preserves the invariant, $v$, and all levels. It takes $O(1)$ time if
the threshold is unchanged and $O(n^2)$ time otherwise.
\end{lemma}
\begin{proof}
An unchanged threshold requires only a comparison. Otherwise classify
the $n$ coordinates under the new threshold and rebuild $c^{\rm low}$
by summing the low contributions in
Definition~\ref{def:lp_exact_l2_sampler_invariant}.
The matrix $C$ has at most $3n$ rows, as counted in the proof of
Lemma~\ref{lem:lp_exact_l2_sampler_init}, so this takes $O(n^2)$ time.
Rebuild the high-coordinate array and its membership pointers in $O(n)$
time. The rebuilt cache and membership array satisfy
Definition~\ref{def:lp_proposal_tree_representation} for $T^{\new}$.
The row maps and their represented values are unchanged because $\delta_v,v$
are unchanged. Thus these assignments establish the full invariant for
$T^{\new}$. The vector, levels, sketches, and squared norm of the root sketch are unchanged.
The procedure reuses the stored arrays, so the space bound is unchanged.
\end{proof}

\subsection{Entry and level updates}\label{sec:lp_exact_l2_sampler_entry_updates}
We next analyze how \textsc{UpdateVL} maintains the squared norm of the root sketch and cached
contributions when one coordinate changes. These updates also preserve
the shared representation used by \textsc{ProposalTree}.

\begin{lemma}[Coordinate update]\label{lem:lp_exact_l2_sampler_update_vl}
Assume the invariant in Definition~\ref{def:lp_exact_l2_sampler_invariant}
and the stored representation above hold. The procedure
\textnormal{\textsc{UpdateVL}}$(i\in[n],\zeta\in\R,j\in[L_2])$ in
Algorithm~\ref{alg:exact_l2_sampler} replaces $v_i$ by $\zeta$, changes
both paired levels to $j$, preserves the invariant and threshold, and
takes $O(n)$ time. Its node and residual-row cache update takes $O(n)$
time if coordinate $i$ is low before or after the change, and $O(1)$
time otherwise.
\end{lemma}
\begin{proof}
Define $I:=(i,n+i)$ and
$\Delta:=\begin{bmatrix}\beta_1(\zeta)-\beta_1(v_i)\\\zeta-v_i\end{bmatrix}$.
Before changing $\mathrm{exactIP}_0$, compute $(Q_0w+d)_I$ and define
\begin{equation}\label{eq:lp_root_sketch_update}
\mathcal{N}_{2,0}^{\new}:=\mathcal{N}_{2,0}+2\Delta^\top(Q_0w+d)_I
+\Delta^\top(Q_0)_{I,I}\Delta.
\end{equation}
To verify this update, define $\delta\in\R^{2n}$ to be zero outside $I$ with
$\delta_I=\Delta$. We have
\[(w+\delta)^\top Q_0(w+\delta)+2d^\top(w+\delta)+\|c\|_2^2=\mathcal{N}_{2,0}+2\Delta^\top(Q_0w+d)_I+\Delta^\top(Q_0)_{I,I}\Delta=\mathcal{N}_{2,0}^{\new}.\]
The first step expands the quadratic form, uses the symmetry of $Q_0$,
and substitutes the old invariant. The second step uses
Eq.~\eqref{eq:lp_root_sketch_update}. Updating the two stored values and their
levels in $\mathrm{exactIP}_0$ therefore preserves the identity for the squared norm of the root sketch
and the paired-level condition, by Theorem~\ref{thm:lp_exact_ip_interface}.
There are a constant number of \textsc{ExactIP} calls, each taking
$O(n+L_2)=O(n)$ time because $L_2=L_1+4$ and $n=2^{L_1}$.

For the low-coordinate cache, subtract the old contribution of $i$ if
$2|v_i|/\gamma_1<T$ and add its new contribution if
$2|\zeta|/\gamma_1<T$. Then set $v_i=\zeta$. These assignments preserve
the cache identity in Definition~\ref{def:lp_exact_l2_sampler_invariant},
including at threshold crossings. Since $C$ has at most $3n$ rows,
this takes $O(n)$ time when either test holds and $O(1)$ time otherwise.
If the coordinate changes membership, delete or append it in the
high-coordinate array and update its inverse pointer in $O(1)$ time.
All other members and the threshold are unchanged. The updates reuse
the stored arrays and constant additional space.

The fixed blocks of $C$ and their constant terms represent
$(S_{{\rm sk},t})_{:,\mathcal{I}}(\mathscr{A}_{\mathcal{I},:}\beta_1(v)+b_{\mathcal{I}}-v_{\mathcal{I}})$ at each node and
$\mathscr{A}\beta_1(v)+b-v$ at the exact residual entries for every $v$.
Their values after the update therefore represent the new residual and
its node sketches. Together with the updated cache and membership array,
this preserves Definition~\ref{def:lp_proposal_tree_representation}.
\end{proof}

\subsection{Sampling procedure}
\label{subsec:lp_exact_l2_sampler_sampling_procedure}
The sampling procedure uses the maintained squared norm of the root sketch to gate a tree proposal.
Rejection sampling and a reciprocal coin then give the exact squared-residual
probabilities at a good state.

Algorithm~\ref{alg:exact_l2_sampler_sample} calls the \textsc{ProposalTree}
procedures in Algorithms~\ref{alg:lp_proposal_tree_queries}--\ref{alg:lp_proposal_tree_probability}.
A successful proposal returns $(i,p_{\rm prop},s,\lambda)$, where $s=\delta_{v,i}^2>0$,
$\lambda=\delta_{v,i}^2/\mathcal{N}_{2,\mathcal{I}}$ for the leaf $\mathcal{I}$ containing $i$, and $p_{\rm prop}$
is the full probability of proposing $i$. Each call to \textsc{Bernoulli}
or \textsc{Uniform} uses fresh randomness. The stored state remains fixed
throughout each call to \textsc{L2Sample}.

Each call to $\mathsf{Coin}$ runs \textsc{InputCoin} with fresh randomness. When its unknown bias satisfies $p_i\ge3/4$, \textsc{ReciprocalCoin} (Lemma~\ref{lem:reciprocal_coin}) accepts with probability $1/(2p_i)$, correcting the provisional acceptance probability.

\begin{lemma}[Sampling]\label{lem:lp_exact_l2_sampler_sample}
Assume the invariant in Definition~\ref{def:lp_exact_l2_sampler_invariant}
and the stored representation above hold. At a good state in the sense of
Lemma~\ref{lem:lp_ensemble}, the procedure
\textnormal{\textsc{L2Sample}}$(L_{6,4}\in\R_{>0})$ in
Algorithm~\ref{alg:exact_l2_sampler_sample}, with $L_{6,4}\ge6\mathcal{N}_2$, returns $i$ with
probability $\delta_{v,i}^2/L_{6,4}$ and $\mathsf{null}$ otherwise. At any fixed state and for any
fixed sketches, computing and testing the gate take $O(1)$ time, and
the conditional expected time after entering the gate is
$O(L_1^2(1+h_T))$, including all rejections and reciprocal-coin calls.
The procedure leaves the stored state unchanged and uses $O(n^2)$ total space.
\end{lemma}
\begin{proof}
By Lemma~\ref{lem:lp_proposal_tree_propose}, the proposal returns
$p^{\rm prop}_{t,i}$ and the exact leaf probability $\lambda_i$.
Lemma~\ref{lem:lp_proposal_tree_probability} evaluates $p^{\rm prop}_{t,i}$ for
each fresh tree using the same $\lambda_i$ at the unchanged state.
Consequently, each call to \textsc{InputCoin} has the bias $p_i$ below.

For $\mathcal{N}_2>0$ and $\delta_{v,i}\ne0$, we have
\[
p_i:=\frac1{L_{s,2}}\sum_{t=1}^{L_{s,2}}\min\{1,Mp^{\rm prop}_{t,i}/\delta_{v,i}^2\}\ge3/4,
\]
where $p_i$ is the input-coin bias; the inequality holds because at least
three quarters of the trees obey the envelope and
$M=2\widetilde{\mathcal{N}}_2\ge2\mathcal{N}_2$.

We have
\begin{equation}\label{eq:lp_exact_sampling}
\Pr[\text{return }i\mid \delta_v,(S_{{\rm sk},t}),S_{{\rm sk},0}]
=\frac{2M}{L_{6,4}}\frac{\delta_{v,i}^2p_i}{M}\frac1{2p_i}
=\frac{\delta_{v,i}^2}{L_{6,4}}.
\end{equation}
The first step uses the provisional proposal probability $\delta_{v,i}^2p_i/M$
conditional on every fixed collection of sketches, the reciprocal-coin
factor $1/(2p_i)$, and the gate probability $2M/L_{6,4}\le6\mathcal{N}_2/L_{6,4}\le1$.
The second step cancels these factors.

All acceptance probabilities are legal. Nodes with positive squared residual norm have
nonzero sketches almost surely at a fixed good state; branches with zero squared residual norm have
zero probability. If $\mathcal{N}_2=0$, the maintained squared norm of the root sketch is zero, so the gate returns $\mathsf{null}$
with probability one.

Reading $\mathcal{N}_{2,0}$ and testing the sampling gate take $O(1)$ time.
The following bound does not require the good-state event.

For each proposed residual sample, descent queries at most $2L_{s,1}L_1$ rows;
the leaf uses $L_{s,3}$ exact residual rows. A reciprocal-coin input evaluates
at most $2L_{s,1}L_1$ more rows, reusing the same squared residual norm at the leaf. For every fixed state and collection of sketches, we have
\begin{equation}\label{eq:lp_sampler_proposal_work}
O((L_{s,1}L_1+L_{s,3})(1+h_T))=O(L_1^2(1+h_T)).
\end{equation}
The expression on the left multiplies the $O(L_{s,1}L_1+L_{s,3})$ expected number
of queried rows by the $O(1+h_T)$ coordinate-evaluation cost from
Lemma~\ref{lem:lp_proposal_tree_query}; the expected number of reciprocal-coin
inputs is $O(1)$ by Lemma~\ref{lem:reciprocal_coin}.
The step uses the parameter choices in Definition~\ref{def:lp_exact_l2_sampler_parameters}.
Power-of-two rounding satisfies $x\le2^{\lceil\log_2x\rceil}<2x$ for $x>0$,
so $1024L_1\le L_{s,1}<2048L_1$ and $64L_1\le L_{s,2}<128L_1$.
Thus $L_{s,1}L_1=\Theta(L_1^2)$ and $L_{s,3}=L_{s,1}L_{s,2}=\Theta(L_1^2)$, which give
$L_{s,1}L_1+L_{s,3}=\Theta(L_1^2)$ with absolute constants. The factor $1+h_T$ is unchanged.
All randomness and arithmetic inside rejection and reciprocal-coin
procedures are included.

The queried rows and the leaf residuals require $O(L_{s,1}+L_{s,3}+L_1)$ temporary
space, and the reciprocal coin uses constant additional arithmetic storage
by Lemma~\ref{lem:reciprocal_coin}. Since $L_{s,3}\le n$, this is within the
$O(n^2)$ space bound. The procedure reads the maintained arrays without
changing them.
\end{proof}

Algorithm~\ref{alg:exact_l2_sampler_sample} implements the sampling procedure
using explicit gate, proposal, and acceptance steps.

\begin{algorithm}[!htb]
\caption{\textsc{ExactL2Sampler} data structure: sampling}\label{alg:exact_l2_sampler_sample}
\begin{algorithmic}[1]
\DataStructure{ExactL2Sampler} \Comment{Theorem~\ref{thm:lp_exact_l2_sampler_interface}}
\Procedure{L2Sample}{$L_{6,4}\in\R_{>0}$} \Comment{Lemma~\ref{lem:lp_exact_l2_sampler_sample}}
\State $\widetilde{\mathcal{N}}_2\gets4\mathcal{N}_{2,0}/(3L_{s,4})$; $M\gets2\widetilde{\mathcal{N}}_2$.
\If{$2M>L_{6,4}$}
\State \Return $\mathsf{fail}$.
\EndIf
\State $\eta\gets\Call{Bernoulli}{2M/L_{6,4}}$.
\If{$\eta=0$}
\State \Return $\mathsf{null}$.
\EndIf
\State $t\gets\Call{Uniform}{[L_{s,2}]}$; $P\gets\mathrm{Tree}[t].\Call{Propose}{}$. \Comment{Algorithm~\ref{alg:lp_proposal_tree_propose}}
\If{$P=\mathsf{fail}$}
\State \Return $\mathsf{fail}$.
\EndIf
\State $(i,p_{\rm prop},s,\lambda)\gets P$; $\alpha\gets\min\{1,s/(Mp_{\rm prop})\}$.
\State $\eta\gets\Call{Bernoulli}{\alpha}$.
\If{$\eta=0$}
\State \Return $\mathsf{null}$.
\EndIf
\State $\mathsf{Coin}\gets()\mapsto\Call{InputCoin}{i,s,\lambda,M}$.
\State $\eta\gets\Call{ReciprocalCoin}{\mathsf{Coin}}$. \Comment{Lemma~\ref{lem:reciprocal_coin}}
\If{$\eta=0$}
\State \Return $\mathsf{null}$.
\EndIf
\State \Return $i$.
\EndProcedure
\State
\Procedure{InputCoin}{$i\in[n],s\in\R_{>0},\lambda\in(0,1],M\in\R_{>0}$}
\State $t\gets\Call{Uniform}{[L_{s,2}]}$.
\State $p_{\rm prop}\gets\mathrm{Tree}[t].\Call{Probability}{i,\lambda}$. \Comment{Algorithm~\ref{alg:lp_proposal_tree_probability}}
\State $\alpha\gets\min\{1,Mp_{\rm prop}/s\}$; $\eta\gets\Call{Bernoulli}{\alpha}$.
\State \Return $\eta$.
\EndProcedure
\EndDataStructure
\end{algorithmic}
\end{algorithm}

\subsection{Concentration along a tree path}
\label{subsec:lp_exact_l2_sampler_concentration_along_a_tree_path}
The tree proposal of Definition~\ref{def:lp_proposal_tree_probabilities} and Lemma~\ref{lem:lp_proposal_tree_propose} uses dependent sketches along a single path. To bound the chance of undersampling a coordinate, we compare its proposal probability with its share of the squared residual norm.

\begin{lemma}[Shared-tree moment bound]\label{lem:lp_path_concentration}
Fix a residual independent of the Gaussian matrices and define $\mathcal{N}_2:=\|\delta_v\|_2^2$.
For $\delta_{v,i}\ne0$, define $P_i:=p^{\rm prop}_{t,i}/(\delta_{v,i}^2/\mathcal{N}_2)$. Then
\[
\E[P_i^{-8}]\le e,\qquad
\Pr[p^{\rm prop}_{t,i}<\delta_{v,i}^2/(2\mathcal{N}_2)]\le e/256<1/64.
\]
\end{lemma}
\begin{proof}
Suppress the tree index and define $g_{\mathcal{I}}:=(S_{\rm sk})_{:,\mathcal{I}}\delta_{v,\mathcal{I}}$. At an internal node
$\mathcal{I}$ with children $\mathcal{I}_1,\mathcal{I}_2$, define
\[
R_{\mathcal{I}}:=\frac{\|g_{\mathcal{I}_1}+g_{\mathcal{I}_2}\|_2^2}
{\|g_{\mathcal{I}_1}\|_2^2+\|g_{\mathcal{I}_2}\|_2^2}.
\]
For a path to a nonzero coordinate, every selected child has positive
squared residual norm and a nonzero sketch almost surely. The branch probability to a
selected child $C$ equals $\|g_C\|_2^2R_{\mathcal{I}}/\|g_{\mathcal{I}}\|_2^2$.

Define $L_i$ as the leaf containing $i$ and $\mathcal{N}_{2,L_i}:=\|\delta_{v,L_i}\|_2^2$.
We have
\begin{equation}\label{eq:lp_proposal_path_telescope}
P_i=\frac{\|g_{L_i}\|_2^2/(L_{s,1}\mathcal{N}_{2,L_i})}{\|g_{[n]}\|_2^2/(L_{s,1}\mathcal{N}_2)}
\prod_{\mathcal{I}\text{ on the path}}R_{\mathcal{I}}.
\end{equation}
This follows by telescoping the branch probabilities along the path.

We first bound a product of inverse powers of the $R_{\mathcal{I}}$. Conditional on
all inner products among Gaussian column vectors inside each of the two
child subtrees, the directions of their nonzero sum vectors are independent
and uniform on the unit sphere. Indeed the two collections are independent,
and their laws are unchanged by separate orthogonal transformations;
conditioning on within-subtree inner products preserves this invariance.
All descendant ratios and child norms are functions of these inner products.

We have
\[
R_{\mathcal{I}}=1+\theta T,\qquad 0\le\theta\le1,
\]
where, conditional on the child data, $T$ is the inner product of two
independent uniform directions. If a child has zero squared residual norm, $R_{\mathcal{I}}=1$.

For nonzero children, $(1+T)/2$ has the beta distribution with both
parameters $(L_{s,1}-1)/2$, from the spherical density proportional to
$(1-t^2)^{(L_{s,1}-3)/2}$.

For a positive integer $p$ with $L_{s,1}\ge4p+2$, we have
\begin{equation}\label{eq:lp_branch_ratio_inverse_moment}
\E[(1+\theta T)^{-p}\mid\text{child data}]
\le\E[(1+T)^{-p}]
=\prod_{j=1}^p\frac{L_{s,1}-1-j}{L_{s,1}-1-2j}
\le\exp(p(p+1)/L_{s,1}).
\end{equation}
The first step uses symmetry and convexity in $\theta$.
The second step follows by integrating the beta density and applying the
gamma recurrence. For the third step, each factor is
$1+j/(L_{s,1}-1-2j)$, whose logarithm is at most $2j/L_{s,1}$; summing over
$1\le j\le p$ gives $p(p+1)/L_{s,1}$.

For path length $h\le L_1$, we have
\begin{equation}\label{eq:lp_path_ratio_inverse_moment}
\E[\prod_{\mathcal{I}} R_{\mathcal{I}}^{-p}]\le\exp(hp(p+1)/L_{s,1}).
\end{equation}
The bound follows by integrating the root ratio first, conditioning on its
child data, and repeating down the selected path. All remaining path ratios
are measurable in that data, so Eq.~\eqref{eq:lp_branch_ratio_inverse_moment} applies at each step.
This argument does not assert independence of the ratios.

Both normalized endpoint norms in Eq.~\eqref{eq:lp_proposal_path_telescope} have
law $X:=\chi_{L_{s,1}}^2/L_{s,1}$.

For a positive integer $k\le L_{s,1}/4$, we have
\begin{equation}\label{eq:lp_normalized_chi_square_moments}
\E[X^k]=\prod_{j=0}^{k-1}(1+2j/L_{s,1})\le e^{k(k-1)/L_{s,1}},
\qquad
\E[X^{-k}]=\prod_{j=1}^k(1-2j/L_{s,1})^{-1}
\le e^{2k(k+1)/L_{s,1}}.
\end{equation}
For the first step in each moment bound in Eq.~\eqref{eq:lp_normalized_chi_square_moments},
$X$ has density
\[
(L_{s,1}/2)^{L_{s,1}/2}x^{L_{s,1}/2-1}e^{-L_{s,1}x/2}/\Gamma(L_{s,1}/2)\qquad(x>0).
\]
Integrating $x^s$ against this density and substituting $y=L_{s,1}x/2$ gives
\[
 \E[X^s]=(2/L_{s,1})^s\frac{\Gamma(L_{s,1}/2+s)}{\Gamma(L_{s,1}/2)}
 \qquad(s>-L_{s,1}/2),
\]
by the defining integral of the gamma function.
For $s=k$, applying $\Gamma(t+1)=t\Gamma(t)$ exactly $k$ times expands
$\Gamma(L_{s,1}/2+k)/\Gamma(L_{s,1}/2)$ into
$\prod_{j=0}^{k-1}(L_{s,1}/2+j)$.
Multiplication by $(2/L_{s,1})^k$ gives the first product.
For $s=-k$, the same recurrence gives
$\Gamma(L_{s,1}/2-k)/\Gamma(L_{s,1}/2)=\prod_{j=1}^k(L_{s,1}/2-j)^{-1}$;
multiplication by $(L_{s,1}/2)^k$ gives the second product.
This inverse moment is finite because $k\le L_{s,1}/4<L_{s,1}/2$.

For the second step in the bound on $\E[X^k]$, take logarithms of its product and apply
$\log(1+2j/L_{s,1})\le2j/L_{s,1}$ term by term. Summing gives
$\log\E[X^k]\le(2/L_{s,1})\sum_{j=0}^{k-1}j=k(k-1)/L_{s,1}$.
For the second step in the bound on $\E[X^{-k}]$, $1\le j\le k\le L_{s,1}/4$ implies
$0\le2j/L_{s,1}\le1/2$, so $-\log(1-x)\le2x$ with $x:=2j/L_{s,1}$ gives
$-\log(1-2j/L_{s,1})\le4j/L_{s,1}$ for each factor.
Summing gives
$\log\E[X^{-k}]\le(4/L_{s,1})\sum_{j=1}^k j=2k(k+1)/L_{s,1}$.
The second step in each logarithmic calculation evaluates the arithmetic sum; exponentiating
proves both bounds in Eq.~\eqref{eq:lp_normalized_chi_square_moments}.

We can show
\[
\E[P_i^{-8}]
\le\exp((552+1200+600h)/(3L_{s,1}))
\le\exp(784L_1/L_{s,1})\le e.
\]
The first step applies H\"older's inequality with three equal exponents
to the two endpoint factors and the path product in Eq.~\eqref{eq:lp_proposal_path_telescope}.
Their moments are bounded by Eqs.~\eqref{eq:lp_normalized_chi_square_moments}
and~\eqref{eq:lp_path_ratio_inverse_moment}, respectively, with $k=p=24$;
the choice $L_{s,1}\ge1024L_1$ satisfies every moment restriction.
The second step uses $h\le L_1$ and $L_1\ge1$, which give
$552+1200+600h\le2352L_1$.
The third step uses $L_{s,1}\ge1024L_1$ and $784/1024<1$.
For the probability bound in the lemma, the first step applies Markov's
inequality to $P_i^{-8}$ at threshold $2^8$; the second step uses $e<4$.
\end{proof}

\subsection{Simultaneous sketch and proposal bounds}
\label{subsec:lp_exact_l2_sampler_simultaneous_sketch_and_proposal_bounds}
We combine the path bound in Lemma~\ref{lem:lp_path_concentration} across independent trees and control
the separate root sketch. The resulting good-state event is the condition
used in the sampling guarantee.

\begin{lemma}[Simultaneous ensemble envelope]\label{lem:lp_ensemble}
For a fixed residual $\delta_v$ independent of all sketches, define
$\widetilde{\mathcal{N}}_2:=4\|S_{{\rm sk},0}\delta_v\|_2^2/(3L_{s,4})$. Then, with probability at least
$1-n^{-14}$, simultaneously
\begin{equation}\label{eq:lp_sketch_and_proposal_bounds}
\mathcal{N}_2\le\widetilde{\mathcal{N}}_2\le3\mathcal{N}_2/2,
\qquad
|\{t:p^{\rm prop}_{t,i}\ge \delta_{v,i}^2/(2\mathcal{N}_2)\}|\ge3L_{s,2}/4
\quad(\delta_{v,i}\ne0).
\end{equation}
Require also that every node with positive squared residual norm have a nonzero sketch, and
define such a state to be good. This extra condition holds almost surely for a
fixed residual. The zero residual is good as well.
\end{lemma}
\begin{proof} By Lemma~\ref{lem:lp_path_concentration}, for each fixed coordinate $i$ with $\delta_{v,i}\ne0$, each tree $t$ satisfies $\Pr[p^{\rm prop}_{t,i}<\delta_{v,i}^2/(2\mathcal{N}_2)]\le1/64$. We have \[2^{L_{s,2}}(1/64)^{L_{s,2}/4}=2^{-L_{s,2}/2}\le n^{-32}.
\]
The left-hand side bounds the probability of at least $L_{s,2}/4$ bad trees
for a fixed coordinate, by independence and a union bound over subsets
of $L_{s,2}/4$ trees. The first step uses $64=2^6$ and combines the powers
of $2$. The second step uses $L_{s,2}\ge64L_1$ and $n=2^{L_1}$.

A union bound over coordinates gives failure probability at most $n^{-31}$.

For the independent root sketch and $\mathcal{N}_2>0$, define
$Z:=\|S_{{\rm sk},0}\delta_v\|_2^2/\mathcal{N}_2$.
$S_{{\rm sk},0}\delta_v/\sqrt{\mathcal{N}_2}$ has $L_{s,4}$ independent standard Gaussian entries.
For the lower tail, we have
\[
\Pr[Z<3L_{s,4}/4]\le2e^{-L_{s,4}/128}=2e^{-32L_1},
\]
where the first step follows from Corollary~\ref{cor:gaussian_squared_norm}
with $k=L_{s,4}$ and $\delta=1/4$, since $Z<3L_{s,4}/4$ implies $|Z/L_{s,4}-1|>1/4$;
and the second step substitutes $L_{s,4}=4096L_1$.
For the upper tail, first observe that
\begin{equation}\label{eq:lp_root_sketch_log_bound}
\frac18-\log\frac98=\int_0^{1/8}\frac{x}{1+x}\d x
\ge\frac89\int_0^{1/8}x\d x=\frac1{144},
\end{equation}
where the first step follows by integration; the second step uses $1+x\le9/8$
for $0\le x\le1/8$; and the third step evaluates the integral.
We have
\[
\Pr[Z>9L_{s,4}/8]\le e^{-L_{s,4}(1/8-\log(9/8))/2}
\le e^{-L_{s,4}/288}=e^{-128L_1/9},
\]
where the first step applies Lemma~\ref{lem:chernoff_bound} with parameter $1/18$
and Eq.~\eqref{eq:gaussian_squared_norm_mgf}, giving
$e^{-L_{s,4}/16}(1-1/9)^{-L_{s,4}/2}=e^{-L_{s,4}(1/8-\log(9/8))/2}$;
the second step uses Eq.~\eqref{eq:lp_root_sketch_log_bound};
and the third step substitutes $L_{s,4}=4096L_1$.
Outside these two tail events, $3L_{s,4}/4\le Z\le9L_{s,4}/8$, so
$\mathcal{N}_2\le\widetilde{\mathcal{N}}_2=4Z\mathcal{N}_2/(3L_{s,4})\le3\mathcal{N}_2/2$.
A union bound over the tree event and these two root events gives total failure probability at most
\[
n^{-31}+2e^{-32L_1}+e^{-128L_1/9}
\le n^{-31}+3n^{-16}\le4n^{-16}\le n^{-14},
\]
where the first step uses $n=2^{L_1}$ and $32,128/9>16\log2$;
the second step uses $n\ge1$; and the third step uses $n\ge2$.
For $\mathcal{N}_2=0$, the root sketch is zero and the claimed bounds hold deterministically.
\end{proof}

\subsection{Reciprocal coin construction}\label{sec:reciprocal_coin}
The final acceptance step needs a coin of bias $1/(2p)$ when the input bias $p$ is at least $3/4$. Its expected cost must remain bounded even when the sampler is outside the good-state event.

\begin{lemma}[Reciprocal coin]\label{lem:reciprocal_coin}
Given independent Bernoulli coins of unknown bias $p\in[3/4,1]$, a
procedure can generate a Bernoulli coin of bias $1/(2p)$ with a universal
constant expected number of input coins and arithmetic operations, and
constant additional arithmetic storage. It remains defined and has the
same uniform expected running-time bound for every $p\in[0,1]$, although the
reciprocal formula is asserted only for $p\ge3/4$.
\end{lemma}
\begin{proof}
Define $S(t):=126t^5-420t^6+540t^7-315t^8+70t^9$ and
\[
F(x):=\begin{cases}
1/2,&0\le x\le1/2,\\
1/2+S(4x-2)(1/(2x)-1/2),&1/2<x<3/4,\\
1/(2x),&3/4\le x\le1.
\end{cases}
\]
The identity $S'(t)=630t^4(1-t)^4$ shows that $F$ is $C^4$. It satisfies
$1/2\le F\le4/5$: for $x\le5/8$ in the middle interval use
$S(4x-2)\le1/2$ to obtain $F\le3/4$; for $x\ge5/8$, the expression is
a convex combination of $1/2$ and $1/(2x)\le4/5$.

We have
\begin{equation}\label{eq:lp_coin_derivative_bounds}
\|F''\|_\infty\le M_2:=316,\quad
\|F'''\|_\infty\le M_3:=12384,\quad
\|F''''\|_\infty\le M_4:=535200.
\end{equation}
Here is an explicit certificate for the only nontrivial interval. Define
$t:=4x-2\in[0,1]$. The numerator of $F^{(j)}(x)$ over $(t+2)^{j+1}$
is a degree-ten polynomial. Its degree-ten Bernstein coefficient vectors are
\begin{align*}
j=2:\quad &(0,0,0,1344,1536,-608,-2528,-1448,64,64,64),\\
j=3:\quad &(0,0,86016,43008,-112128,-198144,\\
&-73344,144384,144384,-768,-768),\\
j=4:\quad &(0,6193152,-2064384,-10924032,-9781248,\\
&3001344,17126400,15979008,-4632576,-15664128,12288).
\end{align*}
Bernstein basis functions are nonnegative and sum to one. Dividing the
largest absolute coefficients by $2^{j+1}$ gives
Eq.~\eqref{eq:lp_coin_derivative_bounds}. The outer intervals satisfy these bounds
directly.

Define $A(x):=x(1-x)F''(x)$ and
\[
C:=(3M_2+2M_3)/16+3M_4/128+1=14152,\qquad B:=2C=28304.
\]
For an integer $s\ge s_0:=1024$ define
\[
g_s(x):=F(x)-A(x)/(2s),\quad
l_s(j):=g_s(j/s)-B/s^2,\quad
h_s(j):=g_s(j/s)+B/s^2.
\]
Since $M_2/(8s_0)+B/s_0^2\le1/10$, all coefficients lie in $[0,1]$,
with $l_s\le h_s$.

Fix $j\in\{0,\ldots,2s\}$ and define $x:=j/(2s)$. Define $K$ as the success
count in a uniformly chosen subset of $s$ of $2s$ bits containing $j$ ones.
Define $V:=K/s-x$. We have $\E[V]=\E[V^3]=0$ by exchange of the two halves.

We have
\begin{equation}\label{eq:lp_coin_hypergeometric_moments}
\E[V^2]=x(1-x)/(2s-1),\qquad \E[V^4]\le3/(16s^2).
\end{equation}
For the fourth moment, pair the population uniformly and independently
choose one element from each pair. The resulting subset is uniform by
permutation symmetry. Conditional on the pairing, $V$ is a sum of at most
$s$ independent Rademacher variables divided by $2s$. Its fourth moment
is at most $3s^2/(16s^4)$.

We can show
\begin{equation}\label{eq:lp_coin_taylor_f}
|\E[F(K/s)]-F(x)-A(x)/(2(2s-1))|\le M_4/(128s^2).
\end{equation}
This follows from fourth-order Taylor expansion, the even moments in
Eq.~\eqref{eq:lp_coin_hypergeometric_moments}, and $\E[V]=\E[V^3]=0$.

We have
\begin{equation}\label{eq:lp_coin_second_derivative_bound}
 \|A''\|_\infty\le2M_2+2M_3+M_4/4,
\end{equation}
so
\begin{equation}\label{eq:lp_coin_taylor_a}
|\E[A(K/s)]-A(x)|\le\|A''\|_\infty/(8s).
\end{equation}
This follows from second-order Taylor expansion, $\E[V]=0$, and the
second-moment formula in Eq.~\eqref{eq:lp_coin_hypergeometric_moments}.

We have
\begin{equation}\label{eq:lp_coin_coefficient_identity}
\frac1{2(2s-1)}-\frac1{2s}+\frac1{4s}=\frac1{4s(2s-1)},
\end{equation}
by bringing the three fractions to a common denominator.
We first bound the three error terms. Observe that
\[
|A(x)|=x(1-x)|F''(x)|\le\frac{|F''(x)|}{4}\le\frac{M_2}{4}.
\]
The first step uses the definition of $A$.
The second step uses $x(1-x)\le1/4$.
The third step uses Eq.~\eqref{eq:lp_coin_derivative_bounds}.
Consequently,
\begin{align}
\frac{M_4}{128s^2}+\frac{\|A''\|_\infty}{16s^2}
+\frac{|A(x)|}{4s(2s-1)}
&\le\frac{M_4}{128s^2}+\frac{\|A''\|_\infty}{16s^2}
+\frac{M_2}{16s(2s-1)}\notag\\*
&\le\frac{M_4}{128s^2}+\frac{\|A''\|_\infty}{16s^2}
+\frac{M_2}{16s^2}\notag\\*
&\le\frac1{s^2}(\frac{M_4}{128}
+\frac{2M_2+2M_3+M_4/4}{16}+\frac{M_2}{16})\notag\\*
&=\frac{C-1}{s^2}\notag\\*
&\le\frac{C}{s^2}.\label{eq:lp_coin_error_constants}
\end{align}
The first step uses $|A(x)|\le M_2/4$.
The second step uses $2s-1\ge s$.
The third step uses Eq.~\eqref{eq:lp_coin_second_derivative_bound}.
The fourth step uses the definition of $C$.
The fifth step uses $1/s^2>0$.

By the definition of $g_s$,
\[
\E[g_s(K/s)]-g_{2s}(x)
=\E[F(K/s)]-F(x)-\frac{\E[A(K/s)]}{2s}+\frac{A(x)}{4s}.
\]
We have
\begin{align}
&|\E[g_s(K/s)]-g_{2s}(x)|\notag\\*
&=|\E[F(K/s)]-F(x)-\frac{A(x)}{2(2s-1)}
-\frac{\E[A(K/s)]-A(x)}{2s}+\frac{A(x)}{4s(2s-1)}|\notag\\*
&\le |\E[F(K/s)]-F(x)-\frac{A(x)}{2(2s-1)}|
+\frac{|\E[A(K/s)]-A(x)|}{2s}+\frac{|A(x)|}{4s(2s-1)}\notag\\*
&\le \frac{M_4}{128s^2}
+\frac{|\E[A(K/s)]-A(x)|}{2s}+\frac{|A(x)|}{4s(2s-1)}\notag\\*
&\le \frac{M_4}{128s^2}
+\frac{\|A''\|_\infty}{16s^2}+\frac{|A(x)|}{4s(2s-1)}\notag\\*
&\le \frac{C}{s^2}.\label{eq:lp_coin_degree_elevation_error}
\end{align}
The first step follows from Eq.~\eqref{eq:lp_coin_coefficient_identity}.
The second step is the triangle inequality.
The third step follows from Eq.~\eqref{eq:lp_coin_taylor_f}.
The fourth step follows from Eq.~\eqref{eq:lp_coin_taylor_a}.
The fifth step follows from Eq.~\eqref{eq:lp_coin_error_constants}.

We have
\begin{equation}\label{eq:lp_coin_nested_envelopes}
\E[l_s(K)]\le l_{2s}(j)\le h_{2s}(j)\le\E[h_s(K)].
\end{equation}
The first step follows from the first line of Eq.~\eqref{eq:lp_coin_envelope_gaps}.
The second step follows from
$h_{2s}(j)-l_{2s}(j)=2B/(2s)^2=B/(2s^2)>0$.
The third step follows from the second line of Eq.~\eqref{eq:lp_coin_envelope_gaps}.
For these estimates, define $\delta:=\E[g_s(K/s)]-g_{2s}(x)$, so
$|\delta|\le C/s^2$ by Eq.~\eqref{eq:lp_coin_degree_elevation_error}. Then
\begin{equation}\label{eq:lp_coin_envelope_gaps}
\begin{aligned}
l_{2s}(j)-\E[l_s(K)]
&=\frac{3B}{4s^2}-\delta
\ge\frac{3B/4-C}{s^2}\ge0,\\
\E[h_s(K)]-h_{2s}(j)
&=\frac{3B}{4s^2}+\delta
\ge\frac{3B/4-C}{s^2}\ge0.
\end{aligned}
\end{equation}
In each line of Eq.~\eqref{eq:lp_coin_envelope_gaps}, the first step
substitutes the definitions of $l_s$ and $h_s$,
using $x=j/(2s)$ and $1/s^2-1/(2s)^2=3/(4s^2)$.
The second step uses $|\delta|\le C/s^2$, and the third step uses
$B=2C\ge4C/3$.

The coin procedure is as follows. Flip $s_0$ input coins, with success count
$J$. Accept with probability $l_{s_0}(J)$, reject with probability
$1-h_{s_0}(J)$, and otherwise continue. If undecided at size $s$, flip $s$
additional input coins and define $j$ as the total success count at size $2s$.
Compute $\bar l:=\E[l_s(K)]$ and $\bar h:=\E[h_s(K)]$ using the
hypergeometric weights in Eq.~\eqref{eq:lp_coin_hypergeometric_weights}.
Accept with probability
\begin{equation}\label{eq:lp_coin_refinement_probabilities}
\frac{l_{2s}(j)-\bar l}{\bar h-\bar l},
\qquad\text{reject with probability}\qquad
\frac{\bar h-h_{2s}(j)}{\bar h-\bar l},
\end{equation}
and otherwise continue at size $2s$. The denominator equals $2B/s^2>0$,
and Eq.~\eqref{eq:lp_coin_nested_envelopes} makes these legal probabilities.

For correctness, pre-generate all input coins. Conditional on their first
$s$ success count, induction gives probability $l_s$ of having accepted
and probability $h_s$ of not having rejected. Conditioning on the success
count at $2s$ elevates these coefficients by the hypergeometric distribution.
The transition probabilities in Eq.~\eqref{eq:lp_coin_refinement_probabilities} give the same induction at
$2s$. Thus the probability of remaining undecided at size $s$ is exactly
$2B/s^2$, for every $p$. The limiting acceptance probability is $F(p)$,
by the law of large numbers and the uniform convergence of $g_s$ to $F$.
Termination is almost sure, and for $p\ge3/4$ the bias is $1/(2p)$.

Each refinement uses $s$ more input coins and $O(s)$ arithmetic operations.
Indeed, for $k_0:=\max\{0,j-s\}$ through $k_1:=\min\{j,s\}$, the
hypergeometric weights are
\begin{equation}\label{eq:lp_coin_hypergeometric_weights}
q_k:=\frac{\binom jk\binom{2s-j}{s-k}}{\binom{2s}s}.
\end{equation}
The numerator counts subsets containing $k$ of the $j$ ones and $s-k$
of the $2s-j$ zeros; the denominator counts all equally likely subsets
of size $s$. For $k_0\le k<k_1$, we have
\begin{equation}\label{eq:lp_coin_hypergeometric_ratio}
\frac{q_{k+1}}{q_k}
=\frac{\binom{j}{k+1}\binom{2s-j}{s-k-1}}
        {\binom{j}{k}\binom{2s-j}{s-k}}
=\frac{j-k}{k+1}\frac{s-k}{s-j+k+1}
=\frac{(j-k)(s-k)}{(k+1)(s-j+k+1)}.
\end{equation}
The first step substitutes $q_k$ from Eq.~\eqref{eq:lp_coin_hypergeometric_weights}
and cancels the common normalizing factor $\binom{2s}{s}^{-1}$.
The second step writes each binomial coefficient as a quotient of factorials
and cancels the common factorials. The third step multiplies the two fractions.
The support bounds $k_0\le k<k_1$ ensure $q_k>0$.
They also give $k+1\ge1$ and $s-j+k+1\ge1$, since $k\ge\max\{0,j-s\}$,
so both denominator factors are positive.

Initialize $q_{k_0}$ using $O(s)$ products, then sum the coefficients
in order using Eq.~\eqref{eq:lp_coin_hypergeometric_ratio}.
Values of $F$ and $F''$ have constant-cost piecewise rational formulas.
Only counts and running sums are stored.

Choose an absolute constant $c\ge1$ such that initialization costs at most
$cs_0$ and refinement from size $s$ to $2s$ costs at most $cs$, counting
input coins and arithmetic operations. The refinement from size $2^js_0$
is performed exactly when the procedure remains undecided after $2^js_0$
input coins, which has probability $2B/(2^js_0)^2$ for every $p\in[0,1]$.
We have
\[
\E[\text{total time}]
\le cs_0+\sum_{j=0}^{\infty}c(2^js_0)\frac{2B}{(2^js_0)^2}
=cs_0+\frac{2cB}{s_0}\sum_{j=0}^{\infty}2^{-j}
=cs_0+\frac{4cB}{s_0}
=O(s_0+B/s_0)=O(1).
\]
The first step sums the cost of each stage times the probability that it
is executed; monotone convergence justifies taking expectations through
the sum of nonnegative costs. The second step cancels one factor of
$2^js_0$ in each summand. The third step uses
$\sum_{j=0}^{\infty}2^{-j}=1/(1-1/2)=2$.
The fourth step absorbs the absolute constants $c$ and $4$ into the
$O(\cdot)$ notation. The fifth step uses that $s_0=1024$ and $B=28304$
are fixed constants. None of these bounds depends on $p$, so the
expected running-time bound is uniform over $p\in[0,1]$.
\end{proof}

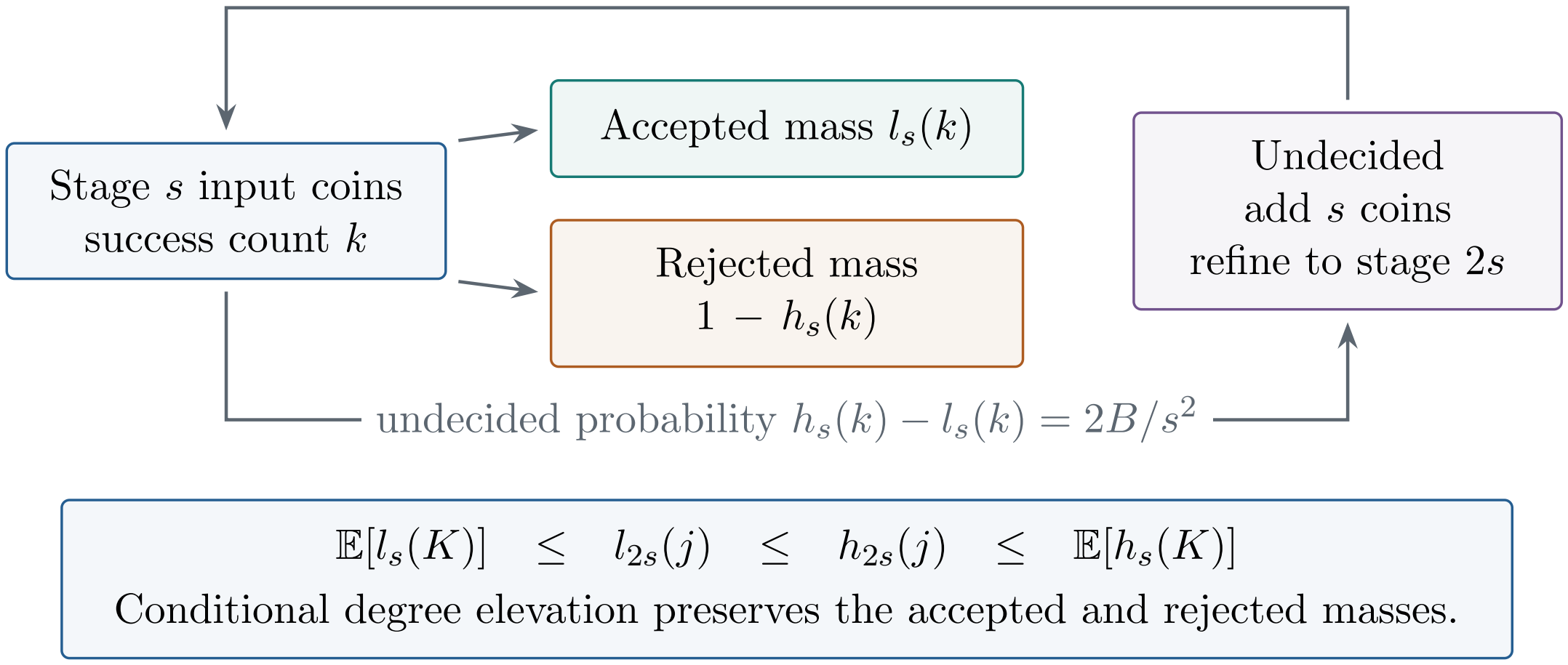
\begin{figure}[!htbp]
\centering
\begin{tikzpicture}[lpfig/base]
\node[lpfig/box,text width=3.5cm] (s) at (1.9,1.8)
  {Stage $s$ input coins\\success count $k$};
\node[lpfig/box,lpfig/teal,text width=3.8cm] (accept) at (7,2.55)
  {Accepted mass $l_s(k)$};
\node[lpfig/box,lpfig/orange,text width=3.8cm] (reject) at (7,1.05)
  {Rejected mass $1-h_s(k)$};
\node[lpfig/box,lpfig/purple,text width=3.4cm] (next) at (12.1,1.8)
  {Undecided\\add $s$ coins\\refine to stage $2s$};
\draw[lpfig/arrow,shorten <=3pt,shorten >=3pt] (s.north east)--(accept.west);
\draw[lpfig/arrow,shorten <=3pt,shorten >=3pt] (s.south east)--(reject.west);
\draw[lpfig/arrow,shorten <=3pt,shorten >=3pt] (s.south)--(1.9,-.1)--(12.1,-.1)--(next.south);
\node[fill=white,lpfig/note] at (7,-.1) {undecided probability $h_s(k)-l_s(k)=2B/s^2$};
\draw[lpfig/arrow,shorten <=3pt,shorten >=3pt] (next.north)--(12.1,3.65)--(1.9,3.65)--(s.north);
\node[lpfig/box,text width=12.7cm] at (7,-1.55)
  {$\E[l_s(K)]\le l_{2s}(j)\le h_{2s}(j)\le\E[h_s(K)]$\\[3pt]
   Conditional degree elevation preserves the accepted and rejected masses.};
\end{tikzpicture}
\caption{The refinement argument in Lemma~\ref{lem:reciprocal_coin} for constructing a coin of bias $1/(2p)$ from coins of bias $p$. At size $s$, the boxes give the cumulative accepted and rejected masses for an observed success count $k$, using the proof's envelopes $l_s,h_s$ and constant $B$. In the refinement inequality, $K$ counts successes in the first $s$ coins conditional on $j$ successes among all $2s$. The reciprocal formula holds for $p\ge3/4$, while the expected-time bound is uniform over $p\in[0,1]$.}
\label{fig:lp_reciprocal_coin}
\end{figure}

\subsection{Exact fallback for small dimensions}\label{sec:lp_exact_l2_sampler_fallback}
When the sketch leaf size exceeds the dimension, the data structure stores
the residual explicitly. This gives the same interface without Gaussian sketches.

\begin{lemma}[Exact fallback]\label{lem:lp_exact_l2_sampler_fallback}
For $L_{s,3}>n$, the four procedures of
Theorem~\ref{thm:lp_exact_l2_sampler_interface} have a deterministic
maintenance implementation using $O(n^2)$ space, with $O(n^2)$
initialization time, $O(1)$ time per threshold change, and $O(n)$ time
per coordinate update or sample. For $L_{6,4}>0$ and $L_{6,4}\ge6\mathcal{N}_2$,
\textnormal{\textsc{L2Sample}}$(L_{6,4})$ returns $i$ with probability $\delta_{v,i}^2/L_{6,4}$
and $\mathsf{null}$ otherwise, without a sketch event. This branch has $n\le2^{27}$.
\end{lemma}
\begin{proof}
The procedure \textsc{Init} stores $\mathscr{A},b$, sets $v={\bf 0}_n$, and sets
$\delta_v=b$ in $O(n^2)$ time and space. This is the correct residual because
$\beta_1(0)=0$. The procedure \textsc{UpdateT} only stores the new
threshold, taking $O(1)$ time.

For \textsc{UpdateVL}$(i,\zeta,j)$, define
\[
\delta_{v_{\rm tra}}:=\delta_v+\mathscr{A}_{:,i}(\beta_1(\zeta)-\beta_1(v_i))
-e_i(\zeta-v_i),\qquad v_{{\rm tra},i}:=\zeta.
\]
Substituting $\delta_v=\mathscr{A}\beta_1(v)+b-v$ shows that
$\delta_{v_{\rm tra}}=\mathscr{A}\beta_1(v_{\rm tra})+b-v_{\rm tra}$, since only coordinate $i$
changes. The update takes $O(n)$ time. The level argument is unnecessary
in this branch.

The procedure \textsc{L2Sample}$(L_{6,4})$ scans the residual entries to compute
$\mathcal{N}_2$. If $\mathcal{N}_2=0$, it returns $\mathsf{null}$. Otherwise it passes a gate with probability
$\mathcal{N}_2/L_{6,4}$ and, conditional on passing, chooses $i$ with probability $\delta_{v,i}^2/\mathcal{N}_2$.
The product is $\delta_{v,i}^2/L_{6,4}$, and the gate probability is at most one because
$L_{6,4}\ge6\mathcal{N}_2$. A scan or rebuilt sum tree implements this in $O(n)$ time
and $O(n)$ additional space. Sampling leaves the stored residual unchanged.

Finally, $n=2^{L_1}<L_{s,3}<2^{18}L_1^2$ implies $L_1\le27$:
$2^{28}>2^{18}28^2$, and $2^{L_1}/L_1^2$ increases for $L_1\ge3$.
Thus $n\le2^{27}$, as claimed.
\end{proof}

\section{Entry state data structure}\label{sec:lp_entry_state}

In Sections~\ref{subsec:lp_entry_state_main_data_structure} and~\ref{subsec:lp_entry_state_initialization}, we describe and initialize the state that \textsc{IterativeSolve} maintains for its current vector.
In Sections~\ref{subsec:lp_entry_state_entry_query} and~\ref{subsec:lp_entry_state_entry_update}, we show how to read and update a single entry.
In Section~\ref{subsec:lp_entry_state_shrinking_large_entries}, we shrink large entries and bound the expected decrease in potential.
In Sections~\ref{subsec:lp_entry_state_exact_inner_product_costs}, \ref{subsec:lp_entry_state_residual_cache_costs}, and~\ref{subsec:lp_entry_state_total_shrink_work}, we bound the work spent on exact inner products, residual-cache updates, and shrink calls, respectively.

\subsection{Main data structure}
\label{subsec:lp_entry_state_main_data_structure}
The data structure \textsc{EntryState} maintains the current vector,
its shrink statistic, and the two data structures used by
\textsc{IterativeSolve}. Use the sampler parameters in
Definition~\ref{def:lp_exact_l2_sampler_parameters} and the arithmetic
model of Section~\ref{sec:preliminaries}.

\begin{theorem}[EntryState]\label{thm:lp_entry_state_interface}
Let $n=2^{L_1}$ with $L_1\in\mathbb{N}_{>0}$, and set
$L_2:=L_1+4$. Fix $\gamma_1\ge2$ and $\beta_1$ as in
Definition~\ref{def:lp_beta_1}, and set $c_{{\rm sh},1}:=41/8$ and
$c_{{\rm sh},2}:=4$. For $v\in\R^n$, define
$Z(v):=\sum_{i=1}^n|v_i|\mathbbm{1}[|v_i|\ge\gamma_1]$.
There is a randomized data structure
(\textnormal{\textsc{EntryState}} in Algorithm~\ref{alg:entry_state})
that uses $O(n^2)$ space and maintains fixed $\mathscr{A}\in\R^{n\times n}$ and
$b\in\R^n$, a vector $v\in\R^n$ with $\|v\|_\infty<8\gamma_1 n$,
the scalar $Z=Z(v)$, and the invariant in
Definition~\ref{def:lp_entry_state_invariant}.
It supports the following procedures.
\begin{enumerate}[label=(\alph*)]
\item \textnormal{\textsc{Init}}$(\mathscr{A}\in\R^{n\times n},b\in\R^n,L_2\in\mathbb{N}_{>0},L_3\in\R_{\ge2})$, where
$L_2=L_1+4$, sets $v={\bf 0}_n$, $Z=0$, and the sampler threshold to
$L_3$. It takes $O(n^2L_1)$ time, or $O(n^2)$ time in the exact fallback.
\item \textnormal{\textsc{Query}}$(i\in[n])$ takes $i$ as input and returns the
current stored coordinate $v_i$ exactly in $O(1)$ time. It leaves the
stored state unchanged.
\item \textnormal{\textsc{UpdateE}}$(i\in[n],\zeta\in(-8\gamma_1 n,8\gamma_1 n))$
replaces $v_i$ by $\zeta$, preserves the
invariant and sampler threshold, and takes $O(n)$ time.
\item \textnormal{\textsc{Shrink}}$()$ takes no explicit arguments and updates the stored vector $v\in\R^n$. It repeatedly samples a coordinate
with $|v_i|\ge\gamma_1$ with probability $|v_i|/Z$ and divides it by
$c_{{\rm sh},2}$ while $Z>c_{{\rm sh},1}\gamma_1 n$.
It terminates, preserves the invariant and sampler threshold, and takes
$O(1+n\cdot n^{\rm shrink})$ time if it makes $n^{\rm shrink}$ coordinate changes. Its output $v_{\rm tra}$ satisfies
$Z(v_{\rm tra})\le c_{{\rm sh},1}\gamma_1 n$,
$\|v_{\rm tra}\|_1\le(c_{{\rm sh},1}+1)\gamma_1 n$, and
$\|v_{\rm tra}\|_\infty\le c_{{\rm sh},1}\gamma_1 n$.
Under the hypotheses of Lemma~\ref{lem:lp_potential_contraction}, it also satisfies
$\E[\mathscr{J}(v_{\rm tra})+n^{\rm shrink}/16\mid v]\le\mathscr{J}(v)$.
\end{enumerate}
The maintenance guarantees hold for arbitrary sequences of valid calls,
for every fixed choice of the sampler sketches. The stored members
$\mathrm{exactIP}$ and $\mathrm{exactL2Sampler}$ support the inner-product evaluations,
threshold updates, and sampling procedures of
Theorems~\ref{thm:lp_exact_ip_interface} and~\ref{thm:lp_exact_l2_sampler_interface},
with their stated hypotheses; the exact fallback uses
Lemma~\ref{lem:lp_exact_l2_sampler_fallback}. All changes to $v$ are made
through \textnormal{\textsc{UpdateE}} or \textnormal{\textsc{Shrink}}.
\end{theorem}
\begin{proof}
Lemma~\ref{lem:lp_entry_state_init} establishes the invariant and
initialization and space bounds. Lemma~\ref{lem:lp_entry_state_query}
proves that \textsc{Query} returns the current coordinate in constant time
without changing the state. Lemma~\ref{lem:lp_entry_state_update}
proves preservation of the invariant and the coordinate-update bound.
Lemma~\ref{lem:lp_entry_state_shrink} proves that \textsc{Shrink}
implements the stated process, preserves the invariant, and has the
claimed running time. Its termination and norm bounds follow from the
same lemma, and its potential guarantee follows from
Lemma~\ref{lem:lp_shrink}. Induction on the sequence of calls proves
all maintenance guarantees. The underlying inner-product evaluations and sampling calls
leave the stored vectors unchanged. Threshold updates preserve their
vectors and levels by Lemma~\ref{lem:lp_exact_l2_sampler_update_t}, or by
Lemma~\ref{lem:lp_exact_l2_sampler_fallback} in the exact branch. Thus these
calls also preserve the entry-state invariant. The proofs reuse
the stored arrays and $O(n)$ additional space, so the space bound remains
$O(n^2)$ throughout.
\end{proof}

A coordinate change must keep the shrink statistic, the inner-product buckets, and the sampler consistent with the same vector. The invariant records what each component stores.

\begin{definition}[Entry-state invariant]\label{def:lp_entry_state_invariant}
For $v\in\R^n$ with $\|v\|_\infty<8\gamma_1 n$, define
$S_1:=\{i\in[n]:|v_i|<\gamma_1\}$ and
$S_j:=\{i\in[n]:\gamma_1 2^{j-2}\le|v_i|<\gamma_1 2^{j-1}\}$
for $2\le j\le L_2$. Set $\lev[i]=j$ for $i\in S_j$ and
$\cnt[j]:=|\{i\in[n]:\lev[i]\ge j\}|$ for $j\in[L_2+1]$.
The entry-state invariant consists of the following conditions.
\begin{enumerate}[label=(\alph*)]
\item The stored scalar satisfies $Z=Z(v)$, where $Z(v)$ is defined
in Theorem~\ref{thm:lp_entry_state_interface}.
\item The instance $\mathrm{exactIP}$ stores the fixed matrix $\mathscr{A}$,
the vector $\mathrm{exactIP}.b=\beta_1(v)$, and the level array $\lev$.
It satisfies Definition~\ref{def:lp_partial_inner_product_invariant}
and its bucket arrays represent $S_1,\ldots,S_{L_2}$.
\item The instance $\mathrm{exactL2Sampler}$ stores the same fixed
$\mathscr{A},b$ and a separate vector equal to $v$. In the sketched branch it
satisfies Definition~\ref{def:lp_exact_l2_sampler_invariant}, with both
paired levels of coordinate $i$ equal to $\lev[i]$ for every $i\in[n]$.
In the exact branch it stores the residual $\mathscr{A}\beta_1(v)+b-v$ as in
Lemma~\ref{lem:lp_exact_l2_sampler_fallback}.
\end{enumerate}
In the analysis of Algorithm~\ref{alg:iterative_solve}, $v$ denotes the
current vector stored in $\mathrm{entryState}$. The algorithm reads each
required coordinate through \textnormal{\textsc{Query}} and reconstructs the full vector
from these queries before its terminal residual test. The symbols
$Z,\mathrm{exactIP},\mathrm{exactL2Sampler}$ denote the corresponding
members of $\mathrm{entryState}$. The symbols $\lev,S_1,\ldots,S_{L_2}$
denote its level array and bucket arrays, read through shared references
to $\mathrm{exactIP}$ without additional copies.
\end{definition}

\begin{algorithm}[!ht]
\caption{\textsc{EntryState} data structure}\label{alg:entry_state}
\begin{algorithmic}[1]
\DataStructure{EntryState} \Comment{Theorem~\ref{thm:lp_entry_state_interface}}
\Members
\State $n=2^{L_1}$ with $L_1\in\mathbb{N}_{>0}$; $L_2=L_1+4$; $v\in\R^n$; $Z\in\R_{\ge0}$.
\State Boundary table $(\gamma_1 2^{j-1})_{j=1}^{L_2}$.
\State Fixed parameter $\gamma_1\ge2$ and shrink constants $c_{{\rm sh},1},c_{{\rm sh},2}$ from Lemma~\ref{lem:lp_shrink}.
\State \textsc{ExactIP} $\mathrm{exactIP}$. \Comment{Algorithms~\ref{alg:exact_ip} and~\ref{alg:exact_ip_update_l}}
\State \textsc{ExactL2Sampler} $\mathrm{exactL2Sampler}$. \Comment{Algorithm~\ref{alg:exact_l2_sampler}}
\EndMembers
\State
\Procedure{Init}{$\mathscr{A}\in\R^{n\times n},b\in\R^n,L_2\in\mathbb{N}_{>0},L_3\in\R_{\ge2}$} \Comment{Lemma~\ref{lem:lp_entry_state_init}}
\State Store $n,L_2$; $v\gets{\bf 0}_n$; $Z\gets0$; form the boundary table by repeated doubling.
\State \Call{\ensuremath{\mathrm{exactIP}}.Init}{$\mathscr{A},L_2$}. \Comment{Algorithm~\ref{alg:exact_ip}}
\State \Call{\ensuremath{\mathrm{exactL2Sampler}}.Init}{$\mathscr{A},b,L_3$}. \Comment{Algorithm~\ref{alg:exact_l2_sampler}}
\EndProcedure
\State
\Procedure{Query}{$i\in[n]$} \Comment{Lemma~\ref{lem:lp_entry_state_query}}
\State \Return $v_i$.
\EndProcedure
\State
\Procedure{UpdateE}{$i\in[n],\zeta\in(-8\gamma_1 n,8\gamma_1 n)$} \Comment{Lemma~\ref{lem:lp_entry_state_update}}
\State Find the level $j$ of $\zeta$ in Definition~\ref{def:lp_entry_state_invariant}, scanning from $\mathrm{exactIP}.\lev[i]$.
\State \Call{\ensuremath{\mathrm{exactIP}}.UpdateV}{$i,\beta_1(\zeta)$}.
\State \Call{\ensuremath{\mathrm{exactL2Sampler}}.UpdateVL}{$i,\zeta,j$}. \Comment{Algorithms~\ref{alg:exact_ip} and~\ref{alg:exact_l2_sampler}}
\State $Z\gets Z-|v_i|\mathbbm{1}[|v_i|\ge\gamma_1]+|\zeta|\mathbbm{1}[|\zeta|\ge\gamma_1]$; $v_i\gets\zeta$.
\If{$j\ne\mathrm{exactIP}.\lev[i]$}
\State \Call{\ensuremath{\mathrm{exactIP}}.UpdateL}{$i,j$}. \Comment{Algorithm~\ref{alg:exact_ip_update_l}}
\EndIf
\EndProcedure
\State
\Procedure{Shrink}{$ $} \Comment{Stored $v\in\R^n$; Lemma~\ref{lem:lp_entry_state_shrink}}
\While{$Z>c_{{\rm sh},1}\gamma_1 n$}
\State Choose $i$ with $|v_i|\ge\gamma_1$ with probability $|v_i|/Z$. 
\State \Call{UpdateE}{$i,v_i/c_{{\rm sh},2}$}. \Comment{Algorithm~\ref{alg:entry_state}}
\EndWhile
\EndProcedure
\EndDataStructure
\end{algorithmic}
\end{algorithm}

\subsection{Initialization}
\label{subsec:lp_entry_state_initialization}

At the zero vector every coordinate belongs to the lowest level. This gives the inner-product structure and the sampler a common initial state.

\begin{lemma}[Initialization]\label{lem:lp_entry_state_init}
For $\mathscr{A}\in\R^{n\times n}$, $b\in\R^n$, $L_2=L_1+4\in\mathbb{N}_{>0}$, and $L_3\in\R_{\ge2}$,
the procedure \textnormal{\textsc{Init}} in Algorithm~\ref{alg:entry_state}
sets $v={\bf 0}_n$ and $Z=0$, establishes
Definition~\ref{def:lp_entry_state_invariant}, and sets the sampler
threshold to $L_3$. It takes $O(n^2L_1)$ time and $O(n^2)$ space.
For $L_{s,3}>n$, its running time is $O(n^2)$.
\end{lemma}
\begin{proof}
Storing $n,L_2$ and initializing $v,Z$ costs $O(n)$ time and space.
The boundary table is formed by repeated doubling in $O(L_2)=O(n)$
time and space.
Lemma~\ref{lem:lp_exact_ip_init} initializes $\mathrm{exactIP}$ with its
stored vector zero, all levels one, and bucket $S_1=[n]$.
Since $\beta_1(0)=0$, this vector equals $\beta_1(v)$, and these levels
are the magnitude levels of $v={\bf 0}_n$.
The scalar $Z=0$ equals $Z({\bf 0}_n)$, establishing
Definition~\ref{def:lp_entry_state_invariant}-(a).
For $L_{s,3}\le n$, Lemma~\ref{lem:lp_exact_l2_sampler_init} initializes the
sampler with a separate zero vector, both paired levels one, and
threshold $L_3$, and establishes its invariant. For $L_{s,3}>n$,
Lemma~\ref{lem:lp_exact_l2_sampler_fallback} initializes its vector to zero
and its residual to $b=\mathscr{A}\beta_1({\bf 0}_n)+b-{\bf 0}_n$.
This proves the entry-state invariant in both branches.

The first instance costs $O(n^2+nL_2)=O(n^2)$ time and space, since
$L_2=L_1+4=O(n)$. The sketched sampler costs $O(n^2L_1)$ time and
$O(n^2)$ space; the exact sampler costs $O(n^2)$ time and space.
Adding the $O(n)$ wrapper storage and initialization work proves the
stated bounds. All members are reinitialized when \textsc{Init} is
called again, and the sketched branch draws fresh independent sketches.
\end{proof}

\subsection{Entry query}
\label{subsec:lp_entry_state_entry_query}

We next describe how to read a coordinate of the maintained vector.
The query accesses one stored entry and leaves the state unchanged.

\begin{lemma}[Coordinate query]\label{lem:lp_entry_state_query}
Assume Definition~\ref{def:lp_entry_state_invariant}.
The procedure \textnormal{\textsc{Query}}$(i\in[n])$ in
Algorithm~\ref{alg:entry_state} returns the current coordinate $v_i$
exactly in $O(1)$ time and $O(1)$ additional space. It leaves the stored
state unchanged.
\end{lemma}
\begin{proof}
The vector $v$ is stored as an array. The procedure reads its $i$th entry
and returns that scalar, which takes $O(1)$ time and $O(1)$ additional
space. It performs no writes, so the vector, the invariant, and all
other stored members remain unchanged.
\end{proof}

\subsection{Entry update}
\label{subsec:lp_entry_state_entry_update}

Changing one entry can change its level as well as its contribution to the shrink statistic. The update propagates the new value and level to the two data structures.

\begin{lemma}[Coordinate update]\label{lem:lp_entry_state_update}
Assume Definition~\ref{def:lp_entry_state_invariant}.
For $i\in[n]$ and $\zeta\in(-8\gamma_1 n,8\gamma_1 n)$, the procedure
\textnormal{\textsc{UpdateE}}$(i,\zeta)$ in
Algorithm~\ref{alg:entry_state} replaces $v_i$ by $\zeta$, preserves
the invariant and sampler threshold, and takes $O(n)$ time and $O(1)$
additional space. The update of $Z$ itself takes $O(1)$ time.
\end{lemma}
\begin{proof}
Let $v_{\rm tra}$ agree with $v$ except that $v_{{\rm tra},i}=\zeta$, and let
$k=\lev[i]$ be the old level. The buckets in
Definition~\ref{def:lp_entry_state_invariant} partition all magnitudes
below $8\gamma_1 n=\gamma_1 2^{L_2-1}$, so the new level $j\in[L_2]$
exists and is unique. Scanning the stored boundary table from level $k$ finds $j$ in
$O(1+|j-k|)$ operations.

The call $\mathrm{exactIP}.\textsc{UpdateV}(i,\beta_1(\zeta))$
sets its vector to $\beta_1(v_{\rm tra})$ and preserves its partial inner
product invariant at the old levels, by
Lemma~\ref{lem:lp_exact_ip_update_v}. The sampler call replaces its
separate coordinate by $\zeta$ and changes its paired levels to $j$,
while preserving its invariant and threshold, by
Lemma~\ref{lem:lp_exact_l2_sampler_update_vl}. The exact branch instead
uses Lemma~\ref{lem:lp_exact_l2_sampler_fallback}.
Since the sampler vector is stored separately, the wrapper's $v_i$
still has its old value when $Z$ is updated. We have
\[
 Z^{\new}=Z-|v_i|\mathbbm{1}[|v_i|\ge\gamma_1]
 +|\zeta|\mathbbm{1}[|\zeta|\ge\gamma_1]
 =\sum_{r=1}^n|v_{{\rm tra},r}|\mathbbm{1}[|v_{{\rm tra},r}|\ge\gamma_1].
\]
The first step is the assignment to $Z$ in Algorithm~\ref{alg:entry_state};
the second step uses Definition~\ref{def:lp_entry_state_invariant}-(a)
and the fact that only coordinate $i$ changes. Assigning $v_i\gets\zeta$ therefore
preserves the scalar identity. If $j\ne k$,
Lemma~\ref{lem:lp_exact_ip_update_l} shows that the final level update
preserves the partial inner product invariant and moves $i$ to the
correct bucket. If $j=k$, these memberships are already correct.
All other coordinates are unchanged, proving the full invariant.

The value update costs $O(n)$, and the level update costs
$O(n+|j-k|)=O(n)$ because $L_2=O(n)$, by
Theorem~\ref{thm:lp_exact_ip_interface}. The sampler update costs $O(n)$
in either branch. The level search costs $O(L_2)=O(n)$; the scalar
update and assignments cost $O(1)$. These calls reuse their arrays and
constant additional space, which proves the claimed bounds.
\end{proof}

\subsection{Shrinking large entries}
\label{subsec:lp_entry_state_shrinking_large_entries}

When large entries carry too much total mass, the solver shrinks an entry chosen in proportion to its magnitude. The expected drop in potential pays for the number of such changes.

\begin{lemma}[Shrink decreases the potential in expectation]\label{lem:lp_shrink}
Use the hypotheses and potential of Lemma~\ref{lem:lp_potential_contraction}.
Set $c_{{\rm sh},1}:=41/8$ and $c_{{\rm sh},2}:=4$. Define
$Z(v):=\sum_{i=1}^n|v_i| \mathbbm{1}[|v_i| \geq \gamma_1]$.
Starting from $v$, consider the following shrink process: while
$Z(v)>c_{{\rm sh},1}\gamma_1 n$, choose $i$ among $|v_i|\ge\gamma_1$
with probability $|v_i|/Z(v)$ and replace $v_i$ by $v_i/c_{{\rm sh},2}$.
Let $v_{\rm tra}$ be the resulting vector. The following guarantees hold.
\begin{itemize}
\item[(a)] Each coordinate change decreases $\mathscr{J}$ by at least $1/16$ in conditional expectation.
\item[(b)] If $n^{\rm shrink}$ counts the coordinate changes in this call, then
$\E[\mathscr{J}(v_{\rm tra})+n^{\rm shrink}/16\mid v]\le\mathscr{J}(v)$.
\item[(c)] Shrink terminates with
$Z(v_{\rm tra})\le c_{{\rm sh},1}\gamma_1 n$, $\|v_{\rm tra}\|_1\le(c_{{\rm sh},1}+1)\gamma_1 n$, and
$\|v_{\rm tra}\|_\infty\le c_{{\rm sh},1}\gamma_1 n$.
\end{itemize}
\end{lemma}
\begin{proof}
{\bf Proof of Part (a).}
Let $I:=\{i\in[n]:|v_i|\ge\gamma_1\}$, $m_i:=|v_i|$, and $\rho_i:=m_i/Z$ for $i\in I$.
Let $\mathscr{J}_i$ denote the contribution of coordinate $i$ to $\mathscr{J}$, and define
$F(r):=\int_{1/c_{{\rm sh},2}}^1 s/(r+s)^2\d s$ for $r\ge0$.
For $i\in I$, we have
\begin{align}
 \mathscr{J}_i(v_i)-\mathscr{J}_i(v_i/c_{{\rm sh},2})
 &\ge\gamma_2\int_{m_i/c_{{\rm sh},2}}^{m_i}\frac{t\d t}{(\gamma_1+t)^2}
 -\gamma_3|v_{*,i}|\int_{m_i/c_{{\rm sh},2}}^{m_i}\frac{\d t}{(\gamma_1+t)^2}\notag\\*
 &\ge\gamma_2F(\gamma_1/m_i)-\frac{(c_{{\rm sh},2}-1)\gamma_3|v_{*,i}|}{m_i}.
 \label{eq:lp_shrink_coordinate_decrease}
\end{align}
where the first step integrates $\mathscr{J}_i'(s)=(s-v_{*,i})\beta_2''(s)$
from $v_i/c_{{\rm sh},2}$ to $v_i$, uses $t=|s|$, bounds the term containing $v_{*,i}$
by its absolute value, and applies Definition~\ref{def:lp_beta_2}-(b);
and the second step substitutes $t=m_i s$ in the first integral and bounds
the second integral by $\int_{m_i/c_{{\rm sh},2}}^{m_i}t^{-2}\d t=(c_{{\rm sh},2}-1)/m_i$.

Differentiating under the integral gives
$F'(r)=-2\int_{1/c_{{\rm sh},2}}^1s/(r+s)^3\d s<0$, so $F$ is decreasing.
Differentiating again gives
$F''(r)=6\int_{1/c_{{\rm sh},2}}^1s/(r+s)^4\d s>0$, so $F$ is convex.
We have
\begin{equation}\label{eq:lp_shrink_mean_scale}
 \sum_{i\in I}\rho_i\frac{\gamma_1}{m_i}
 =\frac{\gamma_1|I|}{Z}\le\frac{\gamma_1 n}{Z}<\frac1{c_{{\rm sh},1}},
\end{equation}
where the first step substitutes $\rho_i=m_i/Z$; the second step uses
$|I|\le n$; and the third step uses $Z>c_{{\rm sh},1}\gamma_1 n$.

Then, we can show
\begin{equation}\label{eq:lp_shrink_average_f}
 \sum_{i\in I}\rho_iF(\gamma_1/m_i)
 \ge F(\sum_{i\in I}\rho_i\gamma_1/m_i)\ge F(1/c_{{\rm sh},1}),
\end{equation}
where the first step follows from Jensen's inequality and
$\sum_{i\in I}\rho_i=1$; and the second step follows from
Eq.~\eqref{eq:lp_shrink_mean_scale} and monotonicity of $F$.

Put $y:=123/269$. We have
\begin{equation}\label{eq:lp_shrink_log_bound}
 \log(196/73)=2\int_0^y\frac{\d t}{1-t^2}\ge2\int_0^y(1+t^2)\d t=2(y+y^3/3)>19/20,
\end{equation}
where the first step uses $(1+y)/(1-y)=196/73$ and integrates
$2/(1-t^2)$; the second step uses $1/(1-t^2)\ge1+t^2$ for $0\le t<1$;
the third step integrates the polynomial; and the fourth step substitutes
$y=123/269$ and uses $2(y+y^3/3)-19/20=10990609/389302180>0$.
It follows that
\begin{equation}\label{eq:lp_shrink_f_lower}
 F(1/c_{{\rm sh},1})=\log(196/73)-984/3577>19/20-984/3577,
\end{equation}
where the first step evaluates the defining integral of $F$ using $c_{{\rm sh},1}=41/8$
and $c_{{\rm sh},2}=4$, with rational term $8/49-32/73=-984/3577$;
and the second step uses Eq.~\eqref{eq:lp_shrink_log_bound}.

Let $v'$ be the state after one shrink change. We have
\begin{align*}
 \E[\mathscr{J}(v)-\mathscr{J}(v')\mid v]
 &\ge\gamma_2\sum_{i\in I}\rho_iF(\gamma_1/m_i)
 -\frac{(c_{{\rm sh},2}-1)\gamma_3\|v_*\|_1}{Z}\\
 &\ge\gamma_2F(1/c_{{\rm sh},1})-\frac{(c_{{\rm sh},2}-1)\gamma_3\|v_*\|_1}{Z}\\
 &>\frac34(19/20-984/3577)-\frac{18}{41}
 =\frac{787929}{11732560}>\frac1{16},
\end{align*}
where the first step averages Eq.~\eqref{eq:lp_shrink_coordinate_decrease}
with probabilities $\rho_i$ and uses $\sum_{i\in I}|v_{*,i}|\le\|v_*\|_1$;
the second step uses Eq.~\eqref{eq:lp_shrink_average_f};
the third step uses Eq.~\eqref{eq:lp_shrink_f_lower}, $\gamma_2\ge3/4$,
$\gamma_3\le2$, $\|v_*\|_1\le3\gamma_1 n/8$, $Z>c_{{\rm sh},1}\gamma_1 n$, and
$c_{{\rm sh},1}=41/8$, $c_{{\rm sh},2}=4$;
the fourth step simplifies the rational expression; and the fifth step uses
$787929/11732560-1/16=13661/2933140>0$.
This proves Part~(a).

{\bf Proof of Part (c).}
Each change at coordinate $i$ reduces $Z$ by at least
$(1-1/c_{{\rm sh},2})|v_i|\ge(1-1/c_{{\rm sh},2})\gamma_1>0$, so the number of
changes is bounded by a finite deterministic function of the starting $Z$.
At termination, $Z\le c_{{\rm sh},1}\gamma_1 n$. The coordinates below magnitude $\gamma_1$ contribute less than
$\gamma_1 n$ to the one-norm; every other coordinate has magnitude at most $Z$.
Since $c_{{\rm sh},1}n\ge1$, this proves Part~(c).

{\bf Proof of Part (b).}
The number of changes has the deterministic bound established in Part~(c).
Telescoping the conditional inequalities from Part~(a) therefore proves
Part~(b). Individual shrink changes need not decrease $\mathscr{J}$.
\end{proof}

The shrink statistic gives the stopping test, and the entry-update procedure records each change. This implements the shrink rule on the maintained vector.

\begin{lemma}[Shrink implementation]\label{lem:lp_entry_state_shrink}
Assume Definition~\ref{def:lp_entry_state_invariant}.
The procedure \textnormal{\textsc{Shrink}} in
Algorithm~\ref{alg:entry_state} implements the shrink process in
Lemma~\ref{lem:lp_shrink}, preserves the invariant and sampler threshold,
and terminates with the norm and scalar bounds in that lemma's Part~(c).
A call making $n^{\rm shrink}$ coordinate changes costs $O(1+n\cdot n^{\rm shrink})$ time, including
all coordinate updates, and uses $O(1)$ additional space.
These maintenance, termination, and time guarantees do not require
the root or potential hypotheses of Lemma~\ref{lem:lp_potential_contraction}.
\end{lemma}
\begin{proof}
The stored value of $Z$ is exact. If its stopping condition holds, the
procedure returns in $O(1)$ time. Otherwise $Z>0$, and an $O(n)$ scan
samples $i$ with weights $|v_i|\mathbbm{1}[|v_i|\ge\gamma_1]$,
so its conditional distribution is exactly $|v_i|/Z$ on the selected
set. A scan accumulates these weights and compares against one uniform
draw on $[0,Z)$, using constant additional storage.
The call to \textsc{UpdateE} divides $v_i$ by four.
It is valid because division reduces its magnitude, and it preserves
the invariant and threshold by Lemma~\ref{lem:lp_entry_state_update}.
Induction over the changes proves that the implemented vector sequence
has precisely the law of the shrink process.

Every change decreases $Z$ by at least $3|v_i|/4\ge3\gamma_1/4$.
Indeed, when the coordinate remains above the cutoff, its contribution
is divided by four; when it falls below the cutoff, its contribution
becomes zero. Thus the number of changes is at most
$4Z_{\rm initial}/(3\gamma_1)$ and termination is finite.
At termination, $Z\le c_{{\rm sh},1}\gamma_1 n$.
The coordinates below magnitude $\gamma_1$ contribute less than
$\gamma_1 n$ to the one-norm, and each remaining magnitude is at most
$Z$. Since $c_{{\rm sh},1}n\ge1$, these observations give the bounds
in Lemma~\ref{lem:lp_shrink}-(c) without a potential assumption.
There are $n^{\rm shrink}$ scans and $n^{\rm shrink}$ coordinate updates, each costing $O(n)$,
and $n^{\rm shrink}+1$ constant-time stopping tests. Their total is $O(1+n\cdot n^{\rm shrink})$.
The scans and updates use constant additional space.
\end{proof}

\subsection{Exact inner-product costs}
\label{subsec:lp_entry_state_exact_inner_product_costs}

The cost of an \textsc{ExactIP} operation depends on how many coordinates lie at or above the relevant level. To bound the expected work, we weight these costs by the coordinate-selection probabilities.

\begin{lemma}[Exact inner-product costs in IterativeSolve]\label{lem:lp_entry_state_exact_ip_cost}
Use the hypotheses and parameters of Theorem~\ref{thm:lp_convergence} and set $L_2:=L_1+4$.
Use the buckets $S_j$, levels $\lev$, and counts $\cnt$ in
Definition~\ref{def:lp_entry_state_invariant} for a state of
Algorithm~\ref{alg:iterative_solve}. A cleaned state means
$Z\le c_{{\rm sh},1}\gamma_1 n$.

The following bounds hold.
\begin{itemize}
\item[(a)] The \textnormal{\textsc{ExactIP}} operations for a regular mirror update, including
its inner-product evaluation, value update, and level update, cost $O(1+\cnt[\max\{\lev[i]-1,1\}])$, with the old level used
in the bound. A shrink coordinate update costs $O(n)$.
\item[(b)] At a cleaned state, $\cnt[j]\le49n/2^{j+1}$ for $j\ge2$ and $\cnt[1]=n$.
Writing $C_i$ for the cost of these exact inner-product operations at coordinate $i$, we have
\[
 \sum_{i=1}^n(\gamma_1+|v_i|)^2C_i=O(\gamma_1^2n^2).
\]
\item[(c)] The bounds in Parts~(a) and~(b) hold for a constant number of fixed matrices on duplicated
coordinates, with each copy assigned the level of its original coordinate.
\item[(d)] In one trial, stopped at the first bad state, operation cutoff,
or completion, the expected base-update work is $O(1)$ per iteration,
and the total expected residual-selected and shrink work is $O(n^2)$.
\end{itemize}
\end{lemma}
\begin{proof}
{\bf Proof of Part (a).}
A regular update changes a level by at most one. To see this, put
$t:=|v_i|$ and $t^{\new}:=|v_{{\rm tra},i}|$. Lemma~\ref{lem:lp_coordinate_update}-(c)
gives
\[
 \frac34(\gamma_1+t)-\gamma_1
 \le t^{\new}\le\frac32(\gamma_1+t)-\gamma_1.
\]
The first step follows by multiplying the lower bound in
Lemma~\ref{lem:lp_coordinate_update}-(c) by $\gamma_1+t$ and subtracting
$\gamma_1$ from both sides. The second step applies the same operations
to the upper bound in Lemma~\ref{lem:lp_coordinate_update}-(c).
Both steps preserve the inequality direction because $\gamma_1+t>0$,
as $\gamma_1\ge2$ and $t\ge0$.

If $t\ge\gamma_1$, these bounds imply
$t/2\le t^{\new}\le2t$.
If $t<\gamma_1$, the upper bound gives $t^{\new}<2\gamma_1$.
The dyadic buckets are closed at their lower endpoints and open at their
upper endpoints, so these bounds change the level by at most one.
Thus the new level is determined by the neighboring powers of two.
Lemmas~\ref{lem:lp_exact_ip_query} and~\ref{lem:lp_exact_ip_update_v},
together with Lemma~\ref{lem:lp_exact_ip_update_l}, give total time
$O(1+\cnt[\max\{\lev[i]-1,1\}])$. A division by $c_{{\rm sh},2}=4$ in shrink crosses
at most two bucket boundaries. Its value and level changes therefore
cost $O(n)$, and determining its new level costs $O(1)$.

{\bf Proof of Part (b).}
At a cleaned state, we have
\[
 \cnt[j]\le49n/2^{j+1}\quad(j\ge2),\qquad \cnt[1]=n.
\]
Each coordinate counted by $\cnt[j]$ has magnitude
at least $\gamma_1 2^{j-2}$, and
Lemma~\ref{lem:lp_shrink}-(c) gives $\|v\|_1\le(49/8)\gamma_1 n$.
The second bound counts all coordinates.

Write $v_{{\rm drift},i}:=\gamma_1+|v_i|$. For an old level $j\ge3$,
the tail-count bound just established and Part~(a) give
$C_i=O(1+\gamma_1 n/|v_i|)$, since
$\gamma_1 2^{j-2}\le|v_i|<\gamma_1 2^{j-1}$.
For $j=1,2$, we have $C_i=O(n)$ and $v_{{\rm drift},i}^2=O(\gamma_1^2)$,
so these coordinates contribute $O(\gamma_1^2n^2)$ to
the weighted sum in Part~(b). For the remaining coordinates, we have
\[
 \sum_{\substack{i=1\\\lev[i]\ge3}}^nv_{{\rm drift},i}^2C_i
 =O(\sum_{i=1}^n|v_i|^2+\gamma_1 n\sum_{i=1}^n|v_i|)
 =O(\gamma_1^2n^2).
\]
The first step uses $v_{{\rm drift},i}^2=O(|v_i|^2)$ and the bound on $C_i$.
The second step uses $\sum_{i=1}^n|v_i|^2\le\|v\|_1^2$ and
$\|v\|_1\le(49/8)\gamma_1 n$. This proves Part~(b).

{\bf Proof of Part (c).}
The same argument holds for any constant number of fixed matrices on
duplicated coordinates, with each copy given the level of the original
$v_i$. Duplication multiplies the tail counts and number of calls by
constant factors. Updating the copies sequentially changes each intermediate
tail count by at most a constant. Matrix entries and vector signs do not
affect the operation counts. In particular, this covers both $\mathrm{exactIP}$ and the
paired instance $\mathrm{exactIP}_0$ for the squared norm of the root sketch in
Section~\ref{sec:lp_exact_l2_sampler_entry_updates}.

{\bf Proof of Part (d).}
Use the indicators $J_k$ of Theorem~\ref{thm:lp_convergence}. At each
reached good history, the complete base selection chooses $i$ with
probability $v_{{\rm drift},i}^2/L_6$. Part~(b) and
$L_6=2^{10}\gamma_1^2n^2$ give $O(1)$ expected base time for that
iteration. A cutoff can only reduce the work relative to the complete call.
Let $N_{\rm res}$ count the residual-selected updates begun before
stopping, including any interrupted update.

We have
\[
 \E[N_{\rm res}]
 \le\sum_{k=0}^{L_7-1}\frac{\E[J_k\|\delta_{v_k}\|_2^2]}{L_6}
 \le12n.
\]
The first step uses the exact conditional sampling law in
Lemma~\ref{lem:lp_sampling_validity} and includes any selected update
interrupted by the cutoff. The second step uses
Theorem~\ref{thm:lp_convergence}-(c). Each update costs $O(n)$, giving
$O(n^2)$ total expected work.
Theorem~\ref{thm:lp_convergence}-(d) bounds the expected number of
completed shrink changes by $16n$. Each costs $O(n)$, and at most
one interrupted shrink update adds $O(n)$ work. This proves Part~(d).
\end{proof}

\subsection{Residual-cache costs}
\label{subsec:lp_entry_state_residual_cache_costs}

The threshold trades the cost of querying uncached coordinates against the cost of maintaining cached contributions. We bound both costs for a stopped trial.

\begin{lemma}[Residual-cache costs in IterativeSolve]\label{lem:lp_entry_state_cache_cost}
Use the hypotheses and parameters of Theorem~\ref{thm:lp_convergence}.
In the sketched branch of Algorithm~\ref{alg:iterative_solve}, fix a threshold $T\ge2$.
Define $v_{{\rm drift},i}:=\gamma_1+|v_i|$ for $i\in[n]$. The following bounds hold.
\begin{itemize}
\item[(a)] At a cleaned state, the number
$h_T:=|\{i\in[n]:2|v_i|/\gamma_1\ge T\}|$ of high coordinates satisfies
$h_T\le49n/(4T)$, and querying $s$ node-sketch or exact residual rows costs $O(s(1+n/T))$.
\item[(b)] Conditional on any fixed sketches and on entering the sampling gate,
the expected sampling work is $O(L_1^2(1+n/T))$.
\item[(c)] The low-coordinate cache changes caused by the base branch cost $O(T)$
in expectation per iteration before stopping. In one trial, stopped at
the first bad state, operation cutoff, or completion, all residual-selected
and shrink changes to this cache cost $O(n^2)$ in expectation.
\end{itemize}
\end{lemma}
\begin{proof}
{\bf Proof of Part (a).}
At a cleaned state, Lemma~\ref{lem:lp_shrink}-(c) gives
$\sum_{i=1}^n2|v_i|/\gamma_1\le49n/4$. Every high coordinate contributes at
least $T$ to this sum, so $h_TT\le49n/4$.
Substituting this bound into the coordinate-evaluation bound of
Lemma~\ref{lem:lp_proposal_tree_query} proves
Part~(a).

{\bf Proof of Part (b).}
Substitution into the sampling bound
of Lemma~\ref{lem:lp_exact_l2_sampler_sample} proves
Part~(b), including all rejections and
reciprocal-coin calls.

{\bf Proof of Part (c).}
If a regular update requires an $O(n)$ cache change, its old normalized magnitude
$2|v_i|/\gamma_1$ is less than $2T$. This is immediate if it was low before. If it becomes low,
Lemma~\ref{lem:lp_coordinate_update}-(c) gives
$\gamma_1+|v_i|\le(4/3)(\gamma_1+|v_{{\rm tra},i}|)<(4\gamma_1/3)(1+T/2)$,
which gives $2|v_i|/\gamma_1<(4T+2)/3\le2T$ for $T\ge2$.

For $m_i:=2|v_i|/\gamma_1$, we have
\begin{equation}\label{eq:lp_truncated_weights}
\sum_{\substack{i=1\\m_i<2T}}^nv_{{\rm drift},i}^2=\frac{\gamma_1^2}{4}\sum_{\substack{i=1\\m_i<2T}}^n(2+m_i)^2
\le\frac{\gamma_1^2}{4}(4n+(4+2T)\sum_{i=1}^n m_i)=O(\gamma_1^2nT).
\end{equation}
The first step substitutes $v_{{\rm drift},i}=\gamma_1+|v_i|=(\gamma_1/2)(2+m_i)$.
For the second step, expanding the square gives
$(2+m_i)^2=4+4m_i+m_i^2$. On the summation set, $0\le m_i<2T$ implies
$m_i^2\le2Tm_i$, so each summand is at most $4+(4+2T)m_i$.
There are at most $n$ summands, and extending the sum of $m_i$ to all
coordinates increases it because $m_i\ge0$.
The third step uses $\sum_{i=1}^n m_i=2\|v\|_1/\gamma_1\le49n/4$, by
Lemma~\ref{lem:lp_shrink}. Since $T\ge2$, we have $4+2T\le4T$ and
$4n\le2nT$. Thus the expression before $O(\cdot)$ is at most
$(51/4)\gamma_1^2nT$, which proves the stated order bound.

By Lemma~\ref{lem:lp_exact_l2_sampler_update_vl}, a base-selected change costs $O(n)$
only on this set, and $O(1)$ otherwise.

Lemma~\ref{lem:lp_base_sampling}-(c)
gives unconditional selection probability $v_{{\rm drift},i}^2/L_6$.
Consequently its expected cache work per iteration is
$O(1+(n/L_6)\sum_{\substack{i=1\\m_i<2T}}^nv_{{\rm drift},i}^2)=O(T)$, using
$L_6=2^{10}\gamma_1^2n^2$ and $T\ge2$.
Each residual-selected or shrink change costs $O(n)$ by
Lemma~\ref{lem:lp_exact_l2_sampler_update_vl}. Apply the conditional
selection probabilities at the reached good histories, with indicators
$J_k$ from Theorem~\ref{thm:lp_convergence}. Parts~(c) and~(d) of that
theorem bound the expected numbers of residual-selected updates and
completed shrink changes before stopping by $12n$ and $16n$,
respectively. One interrupted shrink change adds at most $O(n)$ work.
Thus the total expected work is $O(n^2)$.

\end{proof}

\subsection{Total shrink work}
\label{subsec:lp_entry_state_total_shrink_work}

A shrink call still performs a stopping test when it changes no entries. We count these tests separately from the work of the coordinate changes.

\begin{corollary}[Total shrink work]\label{cor:lp_entry_state_shrink_work}
Use the hypotheses and parameters of Theorem~\ref{thm:lp_convergence}.
In one trial of Algorithm~\ref{alg:iterative_solve}, stopped at the
first bad state, operation cutoff, or completion, all shrink calls with
\textnormal{\textsc{EntryState}} take
$O(L_7+n^2)=O(n^2\log n)$ expected work, including calls that make no
coordinate changes. The bound holds for every fixed choice of sketches
and also for the exact fallback.
\end{corollary}
\begin{proof}
There are at most $L_7$ calls. Use the completed shrink counts
$n_k^{\rm shrink}$ of Theorem~\ref{thm:lp_convergence}.
At most one additional shrink scan or update is interrupted, and it
costs $O(n)$ time. Lemma~\ref{lem:lp_normalization} ensures that every intermediate
coordinate has magnitude below $8\gamma_1 n$, so the invariant's range
condition holds. Lemma~\ref{lem:lp_entry_state_shrink} and
Theorem~\ref{thm:lp_convergence}-(d) give
\[
 \E[\text{total shrink work}]
 =O(L_7+n+n\E[\sum_{k=0}^{L_7-1}n_k^{\rm shrink}])
 =O(L_7+n^2)=O(n^2\log n).
\]
The first step sums $O(1+n\cdot n^{\rm shrink})$ over completed calls and includes the partial call; the second step
uses $\E[\sum_{k=0}^{L_7-1}n_k^{\rm shrink}]\le16n$; and the third
step uses $L_7=O(n^2\log n)$. The maintenance bounds are deterministic
for fixed sketches and hold in the exact branch. The proof of
Theorem~\ref{thm:lp_convergence}-(a), and hence its shrink budget,
also holds for every fixed collection of sketches, with the execution
stopped at the first bad state or cutoff. Thus the bound applies
directly to the retained actual execution in both branches.
\end{proof}

\section{More potential functions}\label{sec:lp_more_potentials}

We record an alternative choice of $\beta_2$ and its analytic properties.
Throughout this section, $\beta_1$ and $\gamma_1\ge2$ are as in
Definition~\ref{def:lp_beta_1}.

In Lemma~\ref{lem:lp_arctan_potential}, we derive the curvature and contraction bounds for the arctangent potential.
In Remark~\ref{rem:lp_arctan_evaluation}, we identify the operations needed to evaluate its mirror map and inverse exactly.

\begin{lemma}[An arctangent potential]\label{lem:lp_arctan_potential}
For $\zeta\in\R$, define
\begin{equation}\label{eq:lp_arctan_potential}
 \beta_2(\zeta):=\frac{\zeta}{\gamma_1}\arctan(\zeta/\gamma_1)
 -\frac12\log(1+\zeta^2/\gamma_1^2).
\end{equation}
Then the following properties hold.
\begin{enumerate}[label=(\alph*)]
\item The function $\beta_2$ is infinitely differentiable and strictly convex,
with $\beta_2(0)=\beta_2'(0)=0$. Its derivatives are
\begin{equation}\label{eq:lp_arctan_derivatives}
 \beta_2'(\zeta)=\frac1{\gamma_1}\arctan(\zeta/\gamma_1),\qquad
 \beta_2''(\zeta)=\frac1{\gamma_1^2+\zeta^2}.
\end{equation}
The map $\beta_2'$ is a bijection from $\R$ onto
$I_{\beta_2}=(-\pi/(2\gamma_1),\pi/(2\gamma_1))$, with inverse
\begin{equation}\label{eq:lp_arctan_inverse}
 (\beta_2')^{-1}(\theta)=\gamma_1\tan(\gamma_1\theta)
 \qquad (\theta\in I_{\beta_2}).
\end{equation}
\item For every $\zeta\in\R$, the curvature satisfies
\begin{equation}\label{eq:lp_arctan_curvature}
 \frac1{(\gamma_1+|\zeta|)^2}\le\beta_2''(\zeta)
 \le\frac2{(\gamma_1+|\zeta|)^2}.
\end{equation}
Moreover, $0\le\beta_2(\zeta)\le\pi|\zeta|/(2\gamma_1)$.
\item For every $\zeta\in\R$, the compatibility inequality is
\begin{equation}\label{eq:lp_arctan_compatibility}
 1-|\beta_1'(\zeta)|\ge\frac{\gamma_1^2}{2}\beta_2''(\zeta).
\end{equation}
Consequently, for every $\zeta_1,\zeta_2\in\R$,
\begin{equation}\label{eq:lp_arctan_scalar_contraction}
 |\beta_1(\zeta_1)-\beta_1(\zeta_2)|
 \le |\zeta_1-\zeta_2|-\frac{\gamma_1^2}{2}|\beta_2'(\zeta_1)-\beta_2'(\zeta_2)|.
\end{equation}
\end{enumerate}
\end{lemma}
\begin{proof}
For (a), both terms in Eq.~\eqref{eq:lp_arctan_potential} are infinitely
differentiable on $\R$, and substitution gives $\beta_2(0)=0$.
We have
\[
 \beta_2'(\zeta)
 =\frac1{\gamma_1}\arctan(\zeta/\gamma_1)
   +\frac{\zeta}{\gamma_1^2+\zeta^2}-\frac{\zeta}{\gamma_1^2+\zeta^2}
 =\frac1{\gamma_1}\arctan(\zeta/\gamma_1),
\]
where the first step follows by differentiating the product and logarithm;
and the second step cancels the last two terms.
Differentiating once more gives Eq.~\eqref{eq:lp_arctan_derivatives}.
In particular, $\beta_2'(0)=0$ and $\beta_2''(\zeta)>0$, so $\beta_2$ is strictly
convex and $\beta_2'$ is strictly increasing.
The limits of $\beta_2'(\zeta)$ at $\pm\infty$ are $\pm\pi/(2\gamma_1)$.
Continuity and strict monotonicity therefore give the asserted range and
bijection. Solving $\gamma_1\theta=\arctan(\zeta/\gamma_1)$ gives
Eq.~\eqref{eq:lp_arctan_inverse}, since $|\gamma_1\theta|<\pi/2$.

For (b), we have
\[
 \gamma_1^2+\zeta^2\le(\gamma_1+|\zeta|)^2\le2(\gamma_1^2+\zeta^2),
\]
where the first step uses $2\gamma_1|\zeta|\ge0$; and the second step uses
$2\gamma_1|\zeta|\le\gamma_1^2+\zeta^2$.
Taking reciprocals and using Eq.~\eqref{eq:lp_arctan_derivatives} proves
Eq.~\eqref{eq:lp_arctan_curvature}.
Convexity and $\beta_2'(0)=0$ show that zero is a minimizer, so
$\beta_2(\zeta)\ge0$. Integrating $|\beta_2'(\zeta_3)|\le\pi/(2\gamma_1)$ from zero
to $\zeta$ gives the upper bound on $\beta_2(\zeta)$.

For (c), we have
\[
 1-|\beta_1'(\zeta)|
 =1-\frac{|\zeta|}{\sqrt{\zeta^2+\gamma_1^2}}
 =\frac{\gamma_1^2}
 {\sqrt{\zeta^2+\gamma_1^2}(\sqrt{\zeta^2+\gamma_1^2}+|\zeta|)}
 \ge\frac{\gamma_1^2}{2(\zeta^2+\gamma_1^2)}
 =\frac{\gamma_1^2}{2}\beta_2''(\zeta),
\]
where the first step differentiates $\beta_1$ from
Definition~\ref{def:lp_beta_1}; the second step rationalizes the numerator;
the third step uses $|\zeta|\le\sqrt{\zeta^2+\gamma_1^2}$; and the fourth step uses
Eq.~\eqref{eq:lp_arctan_derivatives}.
This proves Eq.~\eqref{eq:lp_arctan_compatibility}.

For $\zeta_1\ge \zeta_2$, we have
\[
 |\beta_1(\zeta_1)-\beta_1(\zeta_2)|
 \le\int_{\zeta_2}^{\zeta_1}|\beta_1'(\zeta_3)|\d \zeta_3
 \le \zeta_1-\zeta_2-\frac{\gamma_1^2}{2}\int_{\zeta_2}^{\zeta_1}\beta_2''(\zeta_3)\d \zeta_3
 =|\zeta_1-\zeta_2|-\frac{\gamma_1^2}{2}|\beta_2'(\zeta_1)-\beta_2'(\zeta_2)|,
\]
where the first step follows from the fundamental theorem of calculus
and the triangle inequality; the second step uses
Eq.~\eqref{eq:lp_arctan_compatibility}; and the third step integrates
$\beta_2''$ and uses that $\beta_2'$ is increasing.
Interchanging $\zeta_1,\zeta_2$ proves the remaining case, establishing
Eq.~\eqref{eq:lp_arctan_scalar_contraction}.
\end{proof}

The arctangent choice meets the curvature requirements. Whether its mirror map can be evaluated exactly at unit cost depends on the arithmetic model.

\begin{remark}[Relation to the assumptions in the main text]
\label{rem:lp_arctan_evaluation}
The arctangent choice satisfies properties (a) and (b) of
Definition~\ref{def:lp_beta_2} with any $\gamma_2\in[3/4,1]$ and
$\gamma_3=2$, by Eq.~\eqref{eq:lp_arctan_curvature}.
The value $\gamma_3=2$ is necessary within the permitted range, since
$\beta_2''(\gamma_1)=1/(2\gamma_1^2)$.
Its compatibility inequality in Eq.~\eqref{eq:lp_arctan_compatibility}
is stronger than the bound used in Lemma~\ref{lem:lp_potential_contraction}.
For the potential $\mathscr{J}$ in Eq.~\eqref{eq:lp_potential} and a root with
$\|v_*\|_1\le3\gamma_1 n/8$, part (b) gives
$\mathscr{J}({\bf 0}_n)=\sum_{i=1}^n\beta_2(v_{*,i})\le3\pi n/16\le n$.

The formulas for $\beta_2'$ and its inverse require evaluations of
$\arctan$ and $\tan$. Section~\ref{sec:preliminaries} does not include
these functions among the unit-cost primitives, so the formulas alone
do not establish the exact $O(1)$ evaluation requirement in
Definition~\ref{def:lp_beta_2}(c).
That requirement holds in a model which additionally permits unit-cost
exact evaluation of $\arctan$ and $\tan$.
An implementation by numerical approximation in the original model
requires additional error and cost bounds.
\end{remark}

\clearpage
\part{Linear programming with sparse constraint matrices}\label{part:sparse_lp}

This part establishes the sparse-input guarantee of
Theorem~\ref{thm:tall_sparse_bounded_box_lp}. For bounded-box linear programs
with general $d\leq n$ satisfying its assumptions, the expected running time
is $\widetilde O((\sqrt{\nnz(A)nd}+d^{5/2})\poly\log(nW/\epsilon))$.
The first term accounts for the sparsity of the constraint matrix.

Section~\ref{sec:tall_sparse_lp} presents the algorithm and proves its
correctness and running-time bound. Sections~\ref{sec:tall_inverse_maintenance}
and~\ref{sec:tall_lewis_maintenance} develop the inverse-maintenance and
Lewis weight maintenance data structures. Section~\ref{sec:tall_proportional_sampling}
gives the proportional sampling data structure, and
Sections~\ref{sec:tall_primal_gradient} and~\ref{sec:tall_dual_slack} develop
the data structures for primal and gradient quantities and dual slacks.

\section{Sparse-input LP algorithm and analysis}\label{sec:tall_sparse_lp}
This section proves a sparse-input bound for bounded-box linear programming.
The matrix notation in this section is $A\in\R^{d\times n}$, with $n$ variables
and $d$ equality constraints. We write $a_i:=A_{:,i}\in\R^d$ for
column $i$ of $A$. In particular, $A$ here is the input constraint
matrix, rather than the coordinate-system matrix of Section~\ref{sec:lp_proofs}.
Consider the feasible program
\begin{equation}\label{eq:tall_bounded_box_lp}
 \OPT:=\min\{c^\top x:Ax=b,\ \ell\leq x\leq u\}.
\end{equation}
Write $\nnz(A)$ for the number of nonzero entries of $A$, and define
\begin{equation}\label{eq:tall_input_scale}
 W:=\max\{2,\|A\|_{\max},\|b\|_\infty,\|c\|_\infty,\|\ell\|_\infty,\|u\|_\infty,\max_i(u_i-\ell_i)/\min_i(u_i-\ell_i)\}.
\end{equation}
Here $\|A\|_{\max}:=\max_{i,j}|A_{ij}|$.
All running times count exact real-arithmetic operations, comparisons,
evaluations of the fixed scalar barrier and potential functions, positive
square roots, and independent standard Gaussian draws at unit cost.
We use OpenAI's bound $\omega\leq9/4$~\cite{o26a}. For every fixed
$\xi>0$, it gives square matrix multiplication in
$O(d^{9/4+\xi})$ arithmetic operations.
The notation $\widetilde O$ suppresses polylogarithmic dimension and
failure-probability factors. The theorem gives an arithmetic bound;
it does not assert a bit-complexity bound.

The bound below separates the work that depends on the number of nonzeros from the work that depends only on the number of equality constraints.

\begin{theorem}[Sparse-input bounded-box LP, formal version of Theorem~\ref{thm:tall_main_informal}]
\label{thm:tall_sparse_bounded_box_lp}
Let $A\in\mathbb R^{d\times n}$ have full row rank and no zero
columns, with $n\geq d\geq2$.
Suppose $\ell<u$ and Eq.~\eqref{eq:tall_bounded_box_lp} is feasible. Let $W$ be given by Eq.~\eqref{eq:tall_input_scale}.
For $0<\epsilon\leq1$, there is a randomized algorithm returning,
with probability at least $2/3$, a vector satisfying
\[
 \ell\leq x_{\rm out}\leq u,\qquad
 \|Ax_{\rm out}-b\|_\infty\leq\epsilon,\qquad
 c^\top x_{\rm out}\leq\OPT+\epsilon
\]
in expected
\[
 \widetilde O((\sqrt{\nnz(A)nd}+d^{5/2})\poly \log(nW/\epsilon))
\]
operations in the arithmetic model specified above, using the
matrix-multiplication bound $\omega\leq9/4$ of OpenAI~\cite{o26a}.
\end{theorem}

The proof combines the two-sided stochastic interior-point method
of~\cite{bllsssw21} with the stable inverse maintenance of~\cite{blss20}.

The data structures used by the algorithm are proved in
Sections~\ref{sec:tall_inverse_maintenance}--\ref{sec:tall_dual_slack}.
This section verifies their input conditions and combines their
procedure costs along the LP trajectory.

In Section~\ref{subsec:tall_setup}, we introduce the barrier notation and recall the short-step guarantees.
In Sections~\ref{subsec:tall_sparse_lp_main_algorithm} and~\ref{subsec:tall_sparse_lp_initialization_and_approximate_output}, we present the sparse LP algorithm, construct a centered starting point, and certify its output.
In Section~\ref{subsec:tall_maintenance_assumptions}, we verify the data-structure assumptions at the iterates of the sparse LP algorithm.
In Section~\ref{subsec:tall_sparse_lp_entry_changes_and_total_work}, we bound the work between observer resets and choose the epoch length.
In Section~\ref{subsec:tall_success_probability}, we combine these bounds and control the failure probability over the adaptive execution.

\subsection{Barrier notation and imported short-step guarantees}\label{subsec:tall_setup}

We use the arithmetic model specified at the beginning of this section.
The definitions below translate the source notation to a constraint matrix
with $d$ rows and $n$ columns.

\begin{definition}[Barrier and Lewis-weight notation]\label{def:tall_barrier_notation}
Let $A\in\R^{d\times n}$ have full row rank and no zero columns, with
$n\geq d\geq2$, and let $\ell<u$. Write $a_i:=A_{:,i}$.
For $x\in(\ell,u)$, define
$$
 \phi_i(x_i):=-\log(x_i-\ell_i)-\log(u_i-x_i),\qquad
 \phi(x):=\sum_{i=1}^n\phi_i(x_i),\qquad
 \Phi''(x):=\operatorname{Diag}(\phi_i''(x_i)).
$$
For $H\succ0$, write $\|v\|_H:=(v^\top Hv)^{1/2}$; a positive
vector in the subscript denotes its diagonal matrix. For a
full-column-rank matrix $B$, define its leverage scores by
$\sigma_i(B):=e_i^\top B(B^\top B)^{-1}B^\top e_i$.
Set
$$
 \rho:=d/n,\qquad p:=1-\frac1{4\log(4n/d)},\qquad
 z:=2\rho{\bf 1}_n,\qquad g(x)_i:=\phi_i''(x_i)^{-1/2}.
$$
For $g\in\R_{>0}^n$, let $\tau(g)$ be the regularized Lewis vector
characterized by
$$
 \tau(g)=\sigma(\operatorname{Diag}(\tau(g)^{1/2-1/p}g)A^\top)+z.
$$
Write $\tau(x):=\tau(g(x))$ when the argument is a point.
For positive vectors, $u\approx_\epsilon v$ means
$e^{-\epsilon}v_i\leq u_i\leq e^\epsilon v_i$ for every $i$;
for positive definite matrices it has the analogous Loewner meaning.
Let $C$ be the sufficiently large absolute constant in the source
parameter choices, and define
$$
 C_{\rm norm}:=\frac{C}{1-p},\qquad
 \|h\|_{\theta+\infty}:=\|h\|_\infty+C_{\rm norm}\|h\|_\theta
 \quad\text{for }\theta\in\R_{>0}^n.
$$
The source method is the two-sided-barrier algorithm of~\cite{bllsssw21}.
Its constraint matrix is $A^\top$, and its row and column dimensions
$(m,n)$ are $(n,d)$ here. An input row in that paper corresponds to a
column of $A$ here. Its internal parameters are denoted by
$\epsilon_c,\gamma,\lambda$ in this section, with $\epsilon_c$ replacing
the source centrality parameter $\epsilon$.
\end{definition}

The cost analysis uses two size estimates: the range of possible input sparsities and the total mass of the regularized Lewis weights.

\begin{lemma}[Sparsity and regularized-weight bounds]\label{lem:tall_weight_bounds}
Under Definition~\ref{def:tall_barrier_notation},
$n\leq\nnz(A)\leq nd$. For every $g\in\R_{>0}^n$, the vector
$\tau(g)$ exists uniquely and satisfies
$$
 2\rho\leq\tau(g)_i\leq1+2\rho\quad\text{for every }i\in[n],
 \qquad \sum_{i=1}^n\tau(g)_i=3d.
$$
\end{lemma}
\begin{proof}
Each nonzero column has at least one and at most $d$ nonzero entries,
which proves the sparsity bounds. Our choice satisfies $1/2\leq p<1$.
Apply
Lemma~\ref{lem:tall_static_regularized_lewis_initialization} to
$B=\operatorname{Diag}(g)A^\top$, which has full column rank, and to
$z=2\rho{\bf 1}_n$. This gives existence and uniqueness. The leverage
scores of a full-column-rank $n\times d$ matrix lie in $[0,1]$ and
sum to $d$, because they are the diagonal entries of its rank-$d$
orthogonal projection. Adding $z$ in the defining equation for
$\tau(g)$ gives the coordinate bounds and
$\sum_i\tau(g)_i=d+2\rho n=3d$.
\end{proof}

The short step needs a notion of centrality that allows a small primal residual. We measure that residual in the metric induced by the barrier Hessian and the Lewis weights.

\begin{definition}[Centered state]\label{def:tall_centered_state}
Let $x\in(\ell,u)$, $s_{\rm dual}\in\R^n$, and $\mu>0$.
For $0<\epsilon_c\leq1/80$, call $(x,s_{\rm dual},\mu)$
$\epsilon_c$-centered if there exists $y\in\R^d$ such that
\begin{align}
 s_{\rm dual}&=c-A^\top y,\notag\\*
 |\frac{s_{{\rm dual},i}+\mu\tau(x)_i\phi_i'(x_i)}
 {\mu\tau(x)_i\sqrt{\phi_i''(x_i)}}|&\leq\epsilon_c
 \quad\text{for every }i\in[n],\notag\\*
 \|Ax-b\|_{(A(\operatorname{Diag}(\tau(x))\Phi''(x))^{-1}A^\top)^{-1}}
 &\leq\epsilon_c\gamma/C_{\rm norm}.
\label{eq:tall_source_invariants}
\end{align}
The weights in this definition are the exact vector $\tau(x)$.
The dual vector is not required to be positive.
\end{definition}

We use the short-step guarantee of~\cite{bllsssw21} to reduce the barrier parameter while preserving centrality. Its hypotheses specify the approximation and sampling guarantees that our implementation must meet.

\begin{lemma}[Imported short-step guarantee]\label{lem:tall_source_short_steps}
Use the parameter choices and approximation tolerances of
Algorithms 1--2 of~\cite{bllsssw21}, with the dimension convention
in Definition~\ref{def:tall_barrier_notation}. Start from an interior
feasible primal point and an $\epsilon_c/C_{\rm start}$-centered state,
where $C_{\rm start}$ is the source sufficiently large constant.
Let $0<\mu_{\rm end}<\mu_0$.
Consider the mathematical source PathFollowing execution in which
all approximations satisfy its stated requirements, the Newton
actions within each short step use one common symmetric approximate
inverse, and the conditional sampling law given the history before
each sample is $C_{\rm valid}$-valid in
\cite[Definition 4.13]{bllsssw21}.
It makes $\widetilde O(\sqrt d\log(\mu_0/\mu_{\rm end}))$ short steps.
With probability at least $1-n^{-5}$, at the start and end of every
step the primal point belongs to $(\ell,u)$ and
Eq.~\eqref{eq:tall_source_invariants} holds.
Dual feasibility is preserved by every well-defined update.
An implementation satisfying the state-approximation and
data-structure contracts of~\cite[Definition 6.5]{bllsssw21}
realizes these mathematical updates.
\end{lemma}
\begin{proof}
Apply Lemma 4.12 of~\cite{bllsssw21} with its constraint matrix
$A^\top$ and its dimensions $(m,n)$ replaced by $(n,d)$.
Its centered-state definition is Definition~\ref{def:tall_centered_state}.
Its centered-state and feasibility conclusions give
Eq.~\eqref{eq:tall_source_invariants} at every step endpoint with
probability at least $1-n^{-5}$, and its call bound gives the stated
number of steps. Every dual increment belongs to
$\operatorname{range}(A^\top)$, so a suitable update of $y$ preserves
$s_{\rm dual}=c-A^\top y$ exactly. The sign of this multiplier is
chosen to match our convention.
Lemma 6.4 of~\cite{bllsssw21} proves the implementation assertion
under the specified contracts. Its two Newton actions use the same
approximate inverse; the source analysis does not require unbiased
inverse answers. Lemma 4.42 of~\cite{bllsssw21} establishes the five
sampling conditions for its proportional law. The validity of the
balanced law used here is proved separately in
Lemma~\ref{lem:tall_sparsity_balanced_sampler}.
This lemma concerns the mathematical source execution. The proof of
Theorem~\ref{thm:tall_sparse_bounded_box_lp} verifies the contracts
for our implementation and separately accounts for primitive
failures and transcript-coupling errors.
\end{proof}

To bound maintenance work, we compare the iterates with a nearby path whose barrier scales and Lewis weights change slowly.

\begin{lemma}[Nearby stable path]\label{lem:tall_nearby_stability}
Let $T$ be the declared horizon of a source short-step execution
satisfying the hypotheses of~\cite[Lemma 4.44]{bllsssw21}.
Choose $0<\beta_{\rm stab}<\gamma$ and
$C_{\rm valid}\geq\beta_{\rm stab}^{-2}\log(nT)$, together with
the source lower bounds on $C_{\rm valid}$.
With probability at least $1-n^{-10}$, there are nearby interior
points $\widehat x_t$ such that, writing
$\widehat g_t:=g(\widehat x_t)$,
$\widehat\tau_t:=\tau(\widehat g_t)$, and
$f_t:=\widehat\tau_t^{1/2-1/p}\widehat g_t$, the following hold
for every applicable $t$:
\begin{equation}\label{eq:tall_stable_reference_changes}
 \|f_{t+1}/f_t-{\bf 1}_n\|_{\widehat\tau_t}
 +\|\widehat\tau_{t+1}/\widehat\tau_t-{\bf 1}_n\|_{\widehat\tau_t}
 +\|f_{t+1}/f_t-{\bf 1}_n\|_\infty=O(\gamma).
\end{equation}
On this event,
$g(x_t)\approx_{O(\beta_{\rm stab})}\widehat g_t$ and
$\tau(x_t)\approx_{O(\beta_{\rm stab})}\widehat\tau_t$.
If the stored approximations satisfy
\cite[Definition 6.5]{bllsssw21}, their barrier scales and Lewis
vectors are $e^{O(\gamma)}$-close to
$\widehat g_t$ and $\widehat\tau_t$, respectively.
\end{lemma}
\begin{proof}
Lemma 4.44 of~\cite{bllsssw21} supplies the nearby path with the
stated probability after the dimension substitution $(m,n)=(n,d)$.
Parts 3 and 4 of its Lemma 4.45 bound the relative changes of
$\widehat\tau_t$ and $f_t$ in the stronger
$\widehat\tau_t+\infty$ norm. Since $C_{\rm norm}\geq1$, that norm
bounds both the weighted Euclidean norm and the infinity norm.
Adding the resulting three inequalities proves
Eq.~\eqref{eq:tall_stable_reference_changes}.
Part 1 of Lemma 4.45 gives a relative scale error at most
$\beta_{\rm stab}$, which implies the stated multiplicative
comparison because the source parameters make $\beta_{\rm stab}$
sufficiently small. Lemma 4.22 of~\cite{bllsssw21} transfers this
comparison to the regularized Lewis weights.
Finally, the stored primal-point accuracy in Definition 6.5 and
self-concordance give an $e^{O(\gamma)}$ comparison of the stored
and actual barrier scales. The stored-weight accuracy in that
definition gives the same comparison for the Lewis vectors.
Combining these bounds with $\beta_{\rm stab}<\gamma$ proves the
last assertion. These conclusions hold on the nearby-path event;
no conditioning on future source or data-structure success is used.
\end{proof}

The reported entry changes in a freshly initialized epoch are bounded
in Lemma~\ref{lem:tall_reported_entry_charge}.

\subsection{Main algorithm}
\label{subsec:tall_sparse_lp_main_algorithm}
Algorithm~\ref{alg:tall_sparse_lp}, its short step, and its observer reset share five data structures from Sections~\ref{sec:tall_inverse_maintenance}--\ref{sec:tall_dual_slack}. We specify which vectors and scalars these procedures store in common.

\begin{definition}[State of the sparse LP solver]\label{def:tall_algorithm_state}
Use the augmented instance of Lemma~\ref{lem:tall_sparse_centered_augmentation},
and write $A,b,c,\ell,u,n,d$ for its data and dimensions within the algorithms.
The procedures share their data-structure instances and the stored vectors
$\bar x,\bar s,\theta\in\R^n$, the scalars $\mu,\bar\mu>0$, the exact
residual $\Delta\in\R^d$, the multiplier $y\in\R^d$, and the private
archive handle $h_{\rm inv}$. Define, coordinatewise,
\[
 g_i:=\phi_i''(\bar x_i)^{-1/2},\qquad
 r_i:=\frac{\bar s_i+\bar\mu\theta_i\phi_i'(\bar x_i)}
 {\bar\mu\theta_i\sqrt{\phi_i''(\bar x_i)}},\qquad
 e^x_i:=g_i,\qquad e^s_i:=\bar\mu\theta_i/g_i.
\]
Write $C:=\operatorname{Diag}(g)$,
$\widetilde w:=\theta^{1-2/p}g^2$, and
$\widetilde\lambda:=\theta-\rho\mathbf1$.
These expressions are evaluated only at changed coordinates, except at
bulk initialization or a displayed full-vector inverse update.
The primal structure stores signed scales $-\gamma g$, whereas the
sampler and Lewis structure use the positive scales $g$.

Use the source parameters $\gamma,\lambda$ and relative decrease
$\kappa_\mu$ from Algorithms 4--5 of~\cite{bllsssw21}, with the
accuracy reductions in Section~\ref{subsec:tall_setup}.
Every primal-gradient instance uses the source-compatible mixed-norm
constant $C_{\rm norm}$ specified in that section. In particular,
the same constant is used for its unit-ball maximization and the
source feasibility tolerance. Take $\gamma\leq1/4$.
Take observer accuracy $\alpha\leq\gamma/2^{16}$, and choose
$\delta_{\rm inv},\nu_{\rm inv},\beta_{\rm inv}$ and the declared
horizons $J,U$ as in Section~\ref{subsec:tall_success_probability}.
Set $\eta:=\sqrt{n/\nnz(A)}$, $q_0:=\lceil\sqrt d\rceil$, and
$\mu_{\rm end}:=\epsilon/(C_{\rm end}d)$, where $C_{\rm end}$ is the
constant of Lemma~\ref{lem:tall_sparse_terminal_correction}.
Choose $L_{\rm epoch}$ by Corollary~\ref{cor:tall_warm_epoch_optimization},
including its cached-$\mu$ restriction.
\end{definition}

The helper procedures use the shared state by reference. Returned
approximate vectors are accessed through their pointers and changed-index
lists; they are not copied at each short step. Every Lewis-layer solve
uses \textnormal{\textsc{InverseMaintain.Solve}} with the private archive
handle for its own endpoint. Supplying an archived preconditioner to
\textnormal{\textsc{LewisMaintain.Init}} passes a private reference.

\begin{algorithm}[!ht]
\caption{\textsc{SparseMain}}\label{alg:tall_sparse_lp}
\begin{algorithmic}[1]
\Procedure{SparseMain}{$A\in\R^{d\times n},b\in\R^d,c,\ell,u\in\R^n,\epsilon\in(0,1]$}
\Comment{Theorem~\ref{thm:tall_sparse_bounded_box_lp}}
\State Retain the input; form the augmented LP and centered start $(x,s_{\rm dual},\mu_0)$. \Comment{Lemma~\ref{lem:tall_sparse_centered_augmentation}}
\State Use the augmented notation and parameters of Definition~\ref{def:tall_algorithm_state}.
\State \textsc{InverseMaintain} $\mathrm{inverse}$. \Comment{Algorithms~\ref{alg:tall_inverse_maintain} and~\ref{alg:tall_inverse_solve}}
\State \textsc{LewisMaintain} $\mathrm{lewis}$. \Comment{Algorithm~\ref{alg:tall_lewis_maintain}}
\State \textsc{SparseSampler} $\mathrm{sampler}$. \Comment{Algorithms~\ref{alg:tall_sparse_sampler_update} and~\ref{alg:tall_sparse_sampler_sample}}
\State \textsc{PrimalGradient} $\mathrm{primal}$. \Comment{Algorithms~\ref{alg:tall_primal_gradient_updates} and~\ref{alg:tall_primal_gradient_queries}}
\State \textsc{DualSlack} $\mathrm{dual}$. \Comment{Algorithm~\ref{alg:tall_dual_slack}}
\State $\mu\gets\mu_0$; $\Delta\gets{\bf0}_d$; $y\gets{\bf0}_d$.
\While{$\mu>\mu_{\rm end}$}
  \State $\bar x\gets x$; $g\gets g(x)$; compute a numerical Lewis seed $\theta$ at $g$. \Comment{Lemma~\ref{lem:tall_sparse_static_lewis_initialization}}
  \State \Call{\ensuremath{\mathrm{inverse}}.Init}{$A,\widetilde w,\widetilde\lambda$}. \Comment{Algorithm~\ref{alg:tall_inverse_maintain}}
  \State \Call{ResetSparseObservers}{$x,s_{\rm dual},\theta$}. \Comment{Algorithm~\ref{alg:tall_observer_reset}}
  \For{$k=1,\ldots,q_0$}
    \If{$\mu\leq\mu_{\rm end}$}
      \State \textbf{break}.
    \EndIf
    \If{$k>1$ and $k-1$ is a multiple of $L_{\rm epoch}$}
      \State $x\gets\Call{\ensuremath{\mathrm{primal}}.ComputeExactSum}{}$. \Comment{Algorithm~\ref{alg:tall_primal_gradient_queries}}
      \State $s_{\rm dual}\gets\Call{\ensuremath{\mathrm{dual}}.ComputeExact}{}$. \Comment{Algorithm~\ref{alg:tall_dual_slack}}
      \State \Call{ResetSparseObservers}{$x,s_{\rm dual},\theta$}. \Comment{Algorithm~\ref{alg:tall_observer_reset}}
    \EndIf
    \State \Call{SparseShortStep}{}. \Comment{Algorithm~\ref{alg:tall_sparse_short_step}}
  \EndFor
  \State $x\gets\Call{\ensuremath{\mathrm{primal}}.ComputeExactSum}{}$. \Comment{Algorithm~\ref{alg:tall_primal_gradient_queries}}
  \State $s_{\rm dual}\gets\Call{\ensuremath{\mathrm{dual}}.ComputeExact}{}$. \Comment{Algorithm~\ref{alg:tall_dual_slack}}
\EndWhile
\State Recover and clip $x_{\rm out}$ from $x,s_{\rm dual}$. \Comment{Lemma~\ref{lem:tall_sparse_terminal_correction}}
\If{Eq.~\eqref{eq:tall_output_certificate} holds for the original input, $x_{\rm out}$, and $y$}
  \State \Return $x_{\rm out}$.
\Else
  \State \Return $\mathsf{fail}$.
\EndIf
\EndProcedure
\end{algorithmic}
\end{algorithm}

\begin{algorithm}[!ht]
\caption{\textsc{ResetSparseObservers}}\label{alg:tall_observer_reset}
\begin{algorithmic}[1]
\Procedure{ResetSparseObservers}{$x,s_{\rm dual}\in\R^n,\theta^{\rm seed}\in\R_{>0}^n$}
\Comment{Lemma~\ref{lem:tall_residual_preserving_restart}}
\State $\bar x\gets x$; $\bar s\gets s_{\rm dual}$; $\bar\mu\gets\mu$; $\theta\gets\theta^{\rm seed}$.
\State Recompute $g,\widetilde w,\widetilde\lambda$ from Definition~\ref{def:tall_algorithm_state}.
\State \Call{\ensuremath{\mathrm{inverse}}.Update}{$\widetilde w,\widetilde\lambda$}. \Comment{Algorithm~\ref{alg:tall_inverse_maintain}}
\State $h_{\rm inv}\gets\Call{\ensuremath{\mathrm{inverse}}.Archive}{}$. \Comment{Algorithm~\ref{alg:tall_inverse_maintain}}
\State \Call{\ensuremath{\mathrm{lewis}}.Init}{$A,g,\theta,\mathrm{inverse}.P_{h_{\rm inv}}$}. \Comment{Algorithm~\ref{alg:tall_lewis_maintain}}
\State $(\theta^{\rm tmp},I_\theta)\gets\Call{\ensuremath{\mathrm{lewis}}.Query}{}$; copy $\theta^{\rm tmp}$ into $\theta$. \Comment{Algorithm~\ref{alg:tall_lewis_maintain}}
\State Recompute $r,e^x,e^s,C$ from Definition~\ref{def:tall_algorithm_state}.
\State \Call{\ensuremath{\mathrm{primal}}.Init}{$A,x,-\gamma g,\theta,r,e^x,\alpha$}. \Comment{Algorithm~\ref{alg:tall_primal_gradient_updates}}
\State \Call{\ensuremath{\mathrm{dual}}.Init}{$A,s_{\rm dual},e^s,\alpha,L_{\rm epoch}$}. \Comment{Algorithm~\ref{alg:tall_dual_slack}}
\State \Call{\ensuremath{\mathrm{sampler}}.Init}{$A,C,\theta$}. \Comment{Algorithm~\ref{alg:tall_sparse_sampler_update}}
\State Start fresh observer clocks; retain $\Delta,y,\mu$ and the persistent inverse.
\EndProcedure
\end{algorithmic}
\end{algorithm}

\begin{algorithm}[!ht]
\caption{\textsc{SparseShortStep}}\label{alg:tall_sparse_short_step}
\begin{algorithmic}[1]
\Procedure{SparseShortStep}{} \Comment{Shared state in Definition~\ref{def:tall_algorithm_state}}
\State $h'\gets\Call{\ensuremath{\mathrm{primal}}.QueryProduct}{}$. \Comment{Algorithm~\ref{alg:tall_primal_gradient_queries}}
\State $q_i\gets\min\{1,C_{\rm sp}\theta_i\gamma^{-2}\log(nJU)\}$ for every $i\in[n]$.
\State Form $F$ by independently including row $g_i a_i^\top/\sqrt{\theta_i q_i}$ with probability $q_i$. \Comment{Lemma~\ref{lem:tall_layer_target_coverage}}
\If{$h'=0$ and $\Delta=0$}
  \State $v_0\gets{\bf0}_d$; $v_\Delta\gets{\bf0}_d$.
\Else
  \State Compute $B,V$ with $\begin{bmatrix}h'&h'+\Delta\end{bmatrix}=BV$ and $B$ of full column rank.
  \State $Y\gets\Call{\ensuremath{\mathrm{inverse}}.Solve}{h_{\rm inv},F,B,\delta_{\rm inv},\nu_{\rm inv},\beta_{\rm inv}}$. \Comment{Algorithm~\ref{alg:tall_inverse_solve}}
  \State $\begin{bmatrix}v_0&v_\Delta\end{bmatrix}\gets YV$.
\EndIf
\State $(R,h_{\rm sp},r_{\rm sp})\gets\Call{\ensuremath{\mathrm{sampler}}.Sample}{v_\Delta}$. \Comment{Algorithm~\ref{alg:tall_sparse_sampler_sample}}
\State $(x^{\rm tmp},I_x)\gets\Call{\ensuremath{\mathrm{primal}}.QuerySum}{h_{\rm sp}}$. \Comment{Algorithm~\ref{alg:tall_primal_gradient_queries}}
\State $\Delta\gets\Delta+h'+r_{\rm sp}$.
\State \Call{FinishSparseStep}{$x^{\rm tmp},I_x,v_0$}. \Comment{Algorithm~\ref{alg:tall_sparse_step_updates}}
\State $\mu\gets(1-\kappa_\mu)\mu$.
\EndProcedure
\end{algorithmic}
\end{algorithm}

\begin{algorithm}[!ht]
\caption{\textsc{FinishSparseStep}: reported coordinate changes}\label{alg:tall_sparse_step_updates}
\begin{algorithmic}[1]
\Procedure{FinishSparseStep}{$x^{\rm tmp}\in\R^n,I_x\subseteq[n],v_0\in\R^d$}
\For{$i\in I_x$ with $\sqrt{\phi_i''(x_i^{\rm tmp})}|x_i^{\rm tmp}-\bar x_i|>\gamma/2^{12}$}
  \State $\bar x_i\gets x_i^{\rm tmp}$; $g_i\gets\phi_i''(\bar x_i)^{-1/2}$.
  \State \Call{\ensuremath{\mathrm{lewis}}.UpdateG}{$i,g_i$}. \Comment{Algorithm~\ref{alg:tall_lewis_maintain}}
  \State \Call{RefreshSparseCoordinate}{$i$}. \Comment{Procedure below}
\EndFor
\State \Call{\ensuremath{\mathrm{inverse}}.Update}{$\widetilde w,\widetilde\lambda$}. \Comment{Algorithm~\ref{alg:tall_inverse_maintain}}
\State $h_{\rm inv}\gets\Call{\ensuremath{\mathrm{inverse}}.Archive}{}$. \Comment{Algorithm~\ref{alg:tall_inverse_maintain}}
\State Bind the current Lewis-layer endpoints to $h_{\rm inv}$; retain their historical handles.
\State $(\theta^{\rm tmp},I_\theta)\gets\Call{\ensuremath{\mathrm{lewis}}.Query}{}$. \Comment{Algorithm~\ref{alg:tall_lewis_maintain}}
\For{$i\in I_\theta$ with $\theta_i^{\rm tmp}\not\approx_{\gamma/2^{10}}\theta_i$}
  \State $\theta_i\gets\theta_i^{\rm tmp}$; \Call{RefreshSparseCoordinate}{$i$}.
\EndFor
\State $(s^{\rm tmp},I_s)\gets\Call{\ensuremath{\mathrm{dual}}.Add}{\mu v_0}$; $y\gets y-\mu v_0$. \Comment{Algorithm~\ref{alg:tall_dual_slack}}
\For{$i\in I_s$ with $|s_i^{\rm tmp}-\bar s_i|>\gamma e_i^s/2^{10}$}
  \State $\bar s_i\gets s_i^{\rm tmp}$; recompute $r_i$ from Definition~\ref{def:tall_algorithm_state}.
  \State \Call{\ensuremath{\mathrm{primal}}.Update}{$i,-\gamma g_i,\theta_i,r_i$}. \Comment{Algorithm~\ref{alg:tall_primal_gradient_updates}}
\EndFor
\EndProcedure
\State
\Procedure{RefreshSparseCoordinate}{$i\in[n]$}
\State Recompute $r_i,e_i^x,e_i^s$ from Definition~\ref{def:tall_algorithm_state}.
\State \Call{\ensuremath{\mathrm{primal}}.Update}{$i,-\gamma g_i,\theta_i,r_i$}. \Comment{Algorithm~\ref{alg:tall_primal_gradient_updates}}
\State \Call{\ensuremath{\mathrm{primal}}.SetAccuracy}{$i,e_i^x$}. \Comment{Algorithm~\ref{alg:tall_primal_gradient_updates}}
\State \Call{\ensuremath{\mathrm{dual}}.SetAccuracy}{$i,e_i^s$}. \Comment{Algorithm~\ref{alg:tall_dual_slack}}
\State \Call{\ensuremath{\mathrm{sampler}}.Update}{$i,g_i,\theta_i$}. \Comment{Algorithm~\ref{alg:tall_sparse_sampler_update}}
\EndProcedure
\end{algorithmic}
\end{algorithm}

The constants and failure budgets are those used in the proof of
Theorem~\ref{thm:tall_sparse_bounded_box_lp}. In particular, $C_{\rm sp}$
is sufficiently large, and the sampled factor is processed sparsely.
All private solves and observer calls use the parameters in their
interface theorems. An undefined operation or an exceeded work cap returns $\mathsf{fail}$ from the trial.
Internal histories, inverse products, and data-dependent timings remain
private. The algorithm is one trial; independent trials may be repeated
until the output certificate passes, as proved below.

\subsection{Initialization and approximate output}
\label{subsec:tall_sparse_lp_initialization_and_approximate_output}

To start path following at a feasible box midpoint, we introduce two nonnegative artificial variables per equality. This adds only two nonzeros per row.

\begin{lemma}[Sparse centered augmentation]
\label{lem:tall_sparse_centered_augmentation}
Consider a feasible bounded-box LP
$\min\{c^\top x:Ax=b,\ \ell\leq x\leq u\}$ with
$\ell_i<u_i$ and $0<\epsilon\leq1$.
Let $W$ be as in Eq.~\eqref{eq:tall_input_scale}.
In $O(\nnz(A)+n+d)$ operations one can form a bounded-box LP with
$N:=n+2d$ variables, $d$ equalities, and $\nnz(A)+2d$ nonzeros, together
with an exactly feasible box-midpoint start and dual slack equal
to its objective vector. This start is source-centered at a
parameter $\mu_0=\poly(nW/\epsilon)$.
If an exactly feasible point of the augmented LP has additive
objective error at most $\epsilon/8$, its original-variable part
has objective error at most $\epsilon/8$ and equality residual at
most $\epsilon/8$ in infinity norm. The latter point may then be
perturbed by a sufficiently small inverse-polynomial coordinate
error and clipped to retain the same guarantees with $\epsilon$
in place of $\epsilon/8$.
\end{lemma}

\begin{proof}
Let $x_0:=(\ell+u)/2$, $\Xi:=\max_i(u_i-\ell_i)$, and write
$x=x_0+\Xi y$. Put
\[
 B:=\Xi A,\quad r:=b-Ax_0,\quad c':=\Xi c,
 \quad a_i:=(u_i-\ell_i)/(2\Xi).
\]
Then $1/(2W)\leq a_i\leq1/2$ and the normalized original LP is
$\min\{(c')^\top y:By=r,-a\leq y\leq a\}$, up to the
constant $c^\top x_0$. Set
\[
 K:=1+\|r\|_\infty,\quad
 u_0:=K\mathbf1+r/2,\quad v_0:=K\mathbf1-r/2,
 \quad F:=\sum_{i=1}^n2a_i|c'_i|,\quad M:=8(F+1)/\epsilon.
\]
The augmented constraint matrix is $\mathcal A:=[B\ I_d\ {-I_d}]\in\R^{d\times N}$.
Its equality is $By+u-v=r$, its boxes are
$-a\leq y\leq a$, $0\leq u\leq2u_0$, $0\leq v\leq2v_0$,
and its objective is
$(c')^\top y+M\mathbf1^\top(u+v)$.
In particular the artificial penalty cannot be made
negative by moving an artificial variable below zero.
The point $(0,u_0,v_0)$ is strictly feasible and is the midpoint
of every box. All barrier first derivatives vanish there.
Let $B_{\max}$ denote the largest box width and
$C_{\max}:=\max\{\|c'\|_\infty,M\}$.
Since $\tau_i\geq2d/N$ and
$\sqrt{\phi_i''}\geq1/B_{\max}$, the choice
\[
 \mu_0:=\frac{C_{\max}B_{\max}N}{d\epsilon_{\mathrm{start}}}
\]
gives normalized centrality at most $\epsilon_{\mathrm{start}}/2$.
Here $\epsilon_{\mathrm{start}}$ is the source starting tolerance.
Take the dual multiplier zero. The equality residual is zero.
All displayed quantities and the augmented width ratio are bounded
by polynomials in $nW/\epsilon$, so this is an admissible source
start with only logarithmic scale dependence.

Every feasible original $y$ embeds with $u=v=0$. Thus the augmented
optimum is at most the original optimum. If $(y,u,v)$ is augmented
feasible and within $\epsilon/8$ of its optimum, nonnegativity gives
\[
 (c')^\top y\leq\OPT_{\rm original}+\epsilon/8,
 \qquad M\mathbf1^\top(u+v)\leq F+\epsilon/8.
\]
The second inequality uses that the range of $(c')^\top y$ over
the original box is $F$. Consequently
$\|By-r\|_\infty\leq\mathbf1^\top(u+v)\leq\epsilon/8$.
Clipping an approximate $y$ to its box cannot increase its
coordinatewise distance from this feasible $y$. An error at most
\begin{equation}\label{eq:tall_terminal_coordinate_accuracy}
 \delta_y:=\frac{\epsilon}{16(1+\|c'\|_1+
                  \max_j\sum_{i=1}^n|B_{ji}|)}
\end{equation}
changes its objective and equality residual by at most
$\epsilon/16$ each. Finally return $x_0+\Xi y$.
The equality residual and objective difference in original
coordinates are exactly the ones just bounded.
\end{proof}

At termination the augmented iterate can still have a small equality residual. We correct it to recover a point that meets the tolerances of the original LP.

\begin{lemma}[Sparse approximate terminal correction]
\label{lem:tall_sparse_terminal_correction}
For the augmented instance of Lemma~\ref{lem:tall_sparse_centered_augmentation}, a valid source terminal point
with $\mu\leq\epsilon/(C d)$ can be converted to the accepted
bounded-box output in
$\widetilde O((\nnz(A)+n+d^{\omega+o(1)})\log(nW/\epsilon))$ operations.
\end{lemma}

\begin{proof}
In this proof write $\mathcal A\in\R^{d\times N}$ for the augmented matrix,
$\mathfrak b$ for its right-hand side, $\mathfrak c$ for its cost vector,
$X$ for its current primal
point, and $S$ for its dual slack. Materialize $X,S$ at sparse
cost. Let $\theta$ be a constant-accurate numerical Lewis vector
at the exact $X$; it can be obtained by warm contraction, or by
the cold sparse projected-sign procedure. Define
\[
 D:=\operatorname{Diag}(\theta\phi''(X))^{-1},\quad
 H:=\mathcal A D\mathcal A^\top,\quad
 r_X:=\mathcal A X-\mathfrak b,\quad
 h:=-D\mathcal A^\top H^{-1}r_X.
\]
These are analytic definitions; the inverse is not formed.
The source feasibility invariant and constant comparison of
$\theta$ to $\tau(X)$ give $\|r_X\|_{H^{-1}}=O(\epsilon_c\gamma)$,
where $\epsilon_c$ is the source centrality tolerance.
Lemma~\ref{lem:tall_layer_target_coverage} implies the Newton leverage
bound $\sigma_i(D^{1/2}\mathcal A^\top)\leq C\theta_i$.
Rowwise Cauchy--Schwarz therefore gives
\[
 \sqrt{\phi_i''(X_i)}|h_i|
 \leq\sqrt{\sigma_i(D^{1/2}\mathcal A^\top)/\theta_i}
       \|r_X\|_{H^{-1}}\leq1/4.
\]
Hence $X^*:=X+h$ stays in the box and is exactly equality feasible.

Here is an explicit objective bound. The source centrality
condition says
$S_i=\mu\tau_i(-\phi_i'(X_i)+e_i\sqrt{\phi_i''(X_i)})$ with
$|e_i|\leq\epsilon_c<1/80$. Put
$t_i:=X_i-\ell_i$ and $v_i:=u_i-X_i$.
Since $\phi_i'=-1/t_i+1/v_i$ and
$\sqrt{\phi_i''}\leq1/t_i+1/v_i$, we have
\[
 \frac{S_i t_i}{\mu\tau_i}
 \leq1-t_i/v_i+\epsilon_c(1+t_i/v_i)
 \leq1+\epsilon_c.
\]
Interchanging $t_i,v_i$ gives
$(-S_i)v_i\leq(1+\epsilon_c)\mu\tau_i$.
Use the first inequality when $S_i\geq0$ and the second when $S_i<0$.
Thus for every box point $Y$,
$S_i(X_i-Y_i)\leq(1+\epsilon_c)\mu\tau_i$.
Also $|\phi_i'|\leq\sqrt{\phi_i''}$ gives
$|S_i h_i|\leq(1+\epsilon_c)\mu\tau_i/4$.
For an augmented optimal point $Y$, dual feasibility and the
equalities for $X^*,Y$ yield
\[
 \mathfrak c^\top(X^*-Y)=S^\top(X^*-Y)
 \leq\tfrac54(1+\epsilon_c)\mu\sum_{i=1}^N\tau_i
 =\tfrac{15}4(1+\epsilon_c)d\mu.
\]
The first equality eliminates the dual multiplier; the inequality
uses the two coordinate bounds; the last equality uses
$\sum_{i=1}^N\tau_i=3d$. Choose $C$ to make this at most $\epsilon/8$.

Use the input-sparsity solver in Lemma 8.1 of~\cite{bllsssw21} to approximate
$H^{-1}r_X$ with relative $H$-energy error $\eta$.
The same rowwise argument bounds the resulting coordinate error
in $h$ by $C B_{\max}\eta$, where $B_{\max}$ is the largest
augmented box width and the feasibility invariant bounds
$\|r_X\|_{H^{-1}}$ by an absolute constant.
Choose $\eta:=\delta_y/(C B_{\max})$, with $\delta_y$ from
Eq.~\eqref{eq:tall_terminal_coordinate_accuracy}. If $r_X=0$, take $h=0$ exactly instead.
The solve, sparse products, and clipping cost the claimed amount,
since $\eta^{-1}$ is polynomial in $nW/\epsilon$.
Lemma~\ref{lem:tall_sparse_centered_augmentation} then proves the required output guarantee.
\end{proof}

The solver must decide whether to accept its candidate without knowing $\OPT$. A dual lower bound makes the objective test computable.

\begin{lemma}[A computable output certificate]
\label{lem:tall_output_certificate}
For $y\in\R^d$, define
\begin{equation}\label{eq:tall_dual_lower_bound}
 D(y):=b^\top y+\sum_{i=1}^n
 \min\{\ell_i(c-A^\top y)_i,u_i(c-A^\top y)_i\}.
\end{equation}
Given $x,y$, the three tests
\begin{equation}\label{eq:tall_output_certificate}
 \ell\leq x\leq u,\qquad
 \|Ax-b\|_\infty\leq\epsilon,\qquad
 c^\top x-D(y)\leq\epsilon
\end{equation}
take $O(n+\nnz(A)+d)$ work. Every accepted $x$ satisfies the
output guarantees of Theorem~\ref{thm:tall_sparse_bounded_box_lp}.
On the valid terminal event of Lemma~\ref{lem:tall_sparse_terminal_correction},
the returned point passes these tests with the maintained multiplier.
\end{lemma}

\begin{proof}
For any equality-feasible box point $z$, we have
\[
 c^\top z=b^\top y+(c-A^\top y)^\top z\geq D(y).
\]
Thus $D(y)\leq\OPT$ for every $y$, with no probabilistic assumption.
Sparse products evaluate the tests in the claimed time.

For completeness, the terminal proof gives a gap to a computable
dual bound, not only to an unknown optimum. In its notation put
$S=\mathfrak c-\mathcal A^\top y$ and
$D_{\rm aug}(y):=\mathfrak b^\top y+
\min_{Z\text{ in the augmented box}}S^\top Z$.
The coordinate bounds in that proof hold for every box point $Z$.
Since $\mathcal A X^*=\mathfrak b$, they imply
\[
 \mathfrak c^\top X^*-D_{\rm aug}(y)
 \leq\tfrac{15}4(1+\epsilon_c)d\mu\leq\epsilon/8.
\]
For the augmentation, the two artificial-coordinate contributions
to $D_{\rm aug}(y)$ are
$2u_{0,j}\min\{M-y_j,0\}$ and
$2v_{0,j}\min\{M+y_j,0\}$. Both are nonpositive.
Consequently the normalized original dual bound is at least
$D_{\rm aug}(y)$. The artificial cost at $X^*$ is nonnegative,
so its original-variable objective is no larger than
$\mathfrak c^\top X^*$. The approximation and clipping in
Lemma~\ref{lem:tall_sparse_centered_augmentation} add at most
$\epsilon/16$ to that objective. Undoing $x=x_0+\Xi y_{\rm primal}$
adds the same constant $c^\top x_0$ to the primal objective and
dual bound. The final gap is at most $3\epsilon/16<\epsilon$.
The box and residual tests follow from the same terminal lemma.
Maintain the dual multiplier throughout: an update
$s_{\rm dual}\gets s_{\rm dual}+A^\top h$ is accompanied by
$y\gets y-h$, which costs $O(d)$ and preserves dual feasibility.
\end{proof}

\subsection{Correctness and maintenance assumptions}\label{subsec:tall_maintenance_assumptions}

Inverse maintenance needs both matrix weights and estimates of shifted leverage scores. The stable Lewis path supplies both from the same reference sequence.

\begin{lemma}[Stable reference and shifted leverage hints]
\label{lem:tall_stable_lewis_reference}
Put
\[
 w_t:=f_t^2,\qquad
 \lambda_t:=\sigma(\operatorname{Diag}(\sqrt{w_t})A^\top)+\rho\mathbf1.
\]
Then
\[
 \lambda_t=\widehat\tau_t-\rho\mathbf1,
 \qquad \tfrac12\widehat\tau_t\leq\lambda_t\leq\widehat\tau_t,
\]
and, for any constant inverse-maintenance accuracy $\epsilon_I\in(0,1/10)$,
\[
 \epsilon_I^{-2}\|w_{t+1}/w_t-\mathbf1\|_{\lambda_t}^2
 +\|\lambda_{t+1}/\lambda_t-\mathbf1\|_{\lambda_t}^2=O(1).
\]
At time $t$, let $\bar g_t$ be the stored barrier scale and let
$\theta_{t-1}$ be the previously returned Lewis vector. After making
the source internal accuracy sufficiently small, the inputs
\[
 \widetilde w_t:=\theta_{t-1}^{1-2/p}\bar g_t^2,
 \qquad \widetilde\lambda_t:=\theta_{t-1}-\rho\mathbf1
\]
satisfy the weight-accuracy and shifted-leverage-hint requirements of
Theorem 10 of~\cite{blss20}. A reset that changes only stored representations
can be inserted as a repeated point of the stable sequence.
\end{lemma}

\begin{proof}
The defining identity for the regularized Lewis weights gives the formula for $\lambda_t$.
Since $\widehat\tau_{t,i}\geq2\rho$, subtracting $\rho$ loses at
most a factor two in relative-error bounds. More explicitly,
\[
 \sum_{i=1}^n\lambda_{t,i}
 \frac{(\lambda_{t+1,i}-\lambda_{t,i})^2}{\lambda_{t,i}^2}
 \leq2\sum_{i=1}^n\widehat\tau_{t,i}
 \frac{(\widehat\tau_{t+1,i}-\widehat\tau_{t,i})^2}
 {\widehat\tau_{t,i}^2}.
\]
Writing $a_i:=f_{t+1,i}/f_{t,i}-1$, one has
$w_{t+1,i}/w_{t,i}-1=2a_i+a_i^2$. Since $\|a\|_\infty=O(\gamma)$,
the squared-weight change has weighted norm $O(\gamma)$.
This proves the stability assertion.

The source approximation invariants and one-step Lewis stability give
$\theta_{t-1}\approx_{C\gamma}\widehat\tau_t$ and
$\bar g_t\approx_{C\gamma}\widehat g_t$ for an absolute $C$.
Since $1/2<p<1$, the exponent $1-2/p$ is bounded in absolute value.
Consequently $\widetilde w_t\approx_{C'\gamma}w_t$.
Moreover,
\[
 |\frac{\widetilde\lambda_{t,i}}{\lambda_{t,i}}-1|
 \leq2(e^{C\gamma}-1).
\]
Choose $\gamma$ so that
\begin{equation}\label{eq:tall_observer_error_budgets}
 C'\gamma\le\frac{\epsilon_I}{16\lceil\log d\rceil},\qquad
 2(e^{C\gamma}-1)\le\frac1{16\lceil\log d\rceil}.
\end{equation}
These are the actual tolerances in
Theorem 10 of~\cite{blss20}.
If the leverage hint is compared to the supplied weight rather than
the nearby witness, use the following additional comparison.
For $\widetilde w\approx_a w$, the variational formula for leverage
scores gives $\sigma(\sqrt{\widetilde W}A^\top)\approx_{2a}
\sigma(\sqrt W A^\top)$ on nonzero columns of $A$. Adding the same positive shift
preserves this comparison. Decrease the right-hand sides of Eq.~\eqref{eq:tall_observer_error_budgets}
by another factor four to obtain the required hint accuracy for
both matrices.
The same bounds hold when $\bar g_t$ is replaced by the exact
current $g(x_t)$ at an observer reset. The underlying $x_t$ is not
changed, so its stable witness can simply be repeated. This does
not require a newly centered iterate or an infeasibility correction.
\end{proof}

The Lewis layers and the Newton step use different Gram matrices. A single preconditioner suffices at each time, and an archived copy serves later queries to the same matrices.

\begin{lemma}[Coverage of layer and Newton targets]
\label{lem:tall_layer_target_coverage}
Let $P_t$ be a constant-quality preconditioner for
$A\operatorname{Diag}(\widetilde w_t)A^\top$ from
Lemma~\ref{lem:tall_stable_lewis_reference}.
It is a constant-quality preconditioner for every full Gram matrix
in the source Lewis layers at time $t$, and for the Newton Gram
matrix at that time. It also preconditions every constant spectral
sample of these matrices. A copy of $P_t$ covers a later query to
the same historical target, irrespective of when its right-hand
side becomes known.
\end{lemma}

\begin{proof}
Lemma 5.6 of~\cite{bllsssw21} gives
$v_t^{(j)}\approx_{O(\gamma)}\theta_{t-1}$ for every layer $j$.
The layer has column weights $(v_t^{(j)})^{1-2/p}\bar g_t^2$.
Their coordinatewise comparison to $\widetilde w_t$ gives the
full-Gram comparison by summing rank-one matrices.

For the Newton weights $\theta_{t-1}^{-1}\bar g_t^2$, the ratio
of the Lewis weights to Newton weights is $\theta_{t-1}^{2-2/p}$.
Writing $L:=\log(4n/d)$ gives
$|2-2/p|=2/(4L-1)$. The bounds
$\theta_{t-1,i}\in[e^{-O(\gamma)}2d/n,e^{O(\gamma)}3]$ imply
$|(2-2/p)\log\theta_{t-1,i}|=O(1)$ uniformly in $n/d$.
Thus this ratio lies between two absolute constants. A spectral
sample introduces only another constant. No comparison between
different times is needed: an old query uses its own stored $P_t$.
Copying $P_t$ costs $O(d^2)$; copying public endpoint vectors costs
$O(n)$ per time and per constant number of layers. Polylogarithmically
many copies are harmless under soft notation.
\end{proof}

The short step tracks changes to primal coordinates, gradients, and dual slacks at different accuracy scales. We account for the corresponding observer operations separately.

\begin{lemma}[Observers for centered two-sided variables]
\label{lem:tall_centered_observers}
Let $r\in[-1,1]^n$, $\rho\leq\theta_i\leq4$, and let all coordinate
accuracy scales be positive. The primal/gradient and dual observers
needed by the source short step admit bulk initialization in
$\widetilde O(n+\nnz(A))$ time. A primal/gradient coordinate change
takes $\widetilde O(1+\nnz(a_i))$ amortized time, and a dual accuracy
change takes $\widetilde O(d)$ amortized time with the general-matrix
HeavyHitter of~\cite[Lemma B.1]{bllsssw21}. For a horizon of at most $n$ short steps, their query bounds are
those of Lemma D.3 and Lemma E.5 of~\cite{bllsssw21}, with an additional
$\widetilde O(d)$ per gradient query. The hidden factors are polynomial
in the reciprocal internal accuracy and logarithmic in $n$.
\end{lemma}

\begin{proof}
The primal and gradient guarantees follow from
Theorem~\ref{thm:tall_primal_gradient}. The dual guarantees follow
from Theorem~\ref{thm:tall_dual_slack}. Their initialization
procedures retain the exact current vectors as additive bases and
start new increment histories. The procedure cost bounds in those
two theorems give the stated bounds.
\end{proof}

At an epoch boundary, the solver rebuilds the observers while retaining the reference inverse. The actual iterate and its equality residual remain unchanged.

\begin{lemma}[Residual-preserving observer restart]
\label{lem:tall_residual_preserving_restart}
At a valid source IPM state, suppose the persistent reference inverse
of Lemma~\ref{lem:tall_stable_lewis_reference} is available. The primal,
dual, gradient, proportional sampler, and all recursive Lewis-layer
observers can be reset in $\widetilde O(\nnz(A)+n+d^2)$ work,
apart from one update of the persistent inverse. That update is
charged to the complete inverse-maintenance block in
Corollary~\ref{cor:tall_warm_epoch_optimization}. The reset does not
change the actual iterate, its parameter $\mu$, or its residual
$\Delta:=Ax-b$.
\end{lemma}

\begin{proof}
Before calling Algorithm~\ref{alg:tall_observer_reset},
Algorithm~\ref{alg:tall_sparse_lp} materializes the exact current primal
and dual vectors using the exact-query procedures in
Algorithms~\ref{alg:tall_primal_gradient_queries} and~\ref{alg:tall_dual_slack}.
Their costs
are $\widetilde O(n)$ and $\widetilde O(\nnz(A)+n)$, respectively. Retain
the length-$d$ residual $\Delta$; do not execute the assignment
$\Delta\gets0$ in the cold PathFollowing initializer.
Set stored $\bar x,\bar s,\bar\mu$ to the exact current values.
The previous Lewis vector is still a constant-accurate warm seed,
since the old stored barrier scale and the exact one are close.

Use Lemma~\ref{lem:tall_stable_lewis_reference} to update the persistent
inverse with the new public scale and old Lewis hint.
Keep this update on the same inverse-maintenance clock; its
actual cost is not charged separately to the current epoch.
Apply Lemma~\ref{lem:tall_warm_lewis_initialization}, then bulk initialize
the primal/gradient and dual observers by
Lemma~\ref{lem:tall_centered_observers}, and the sampler by
Lemma~\ref{lem:tall_sparsity_balanced_sampler}, with the exact
current vectors and new numerical Lewis vector. Each bulk
initialization costs $\widetilde O(\nnz(A)+n)$. The initial Lewis query
in Algorithm~\ref{alg:tall_observer_reset} has no intervening scale changes;
its $\widetilde O(n+d^2)$ cost is included in the reset bound by
Lemma~\ref{lem:tall_lewis_epoch_cost}. The returned vector is copied only
at this bulk reset. Discard all obsolete observer histories and
start their dyadic clocks together. The inverse-maintenance state
is retained; only its old snapshots no longer needed by any layer
may be discarded. Initializing the newly computed arrays does not
invoke their per-coordinate Scale procedures.

For the gradient observer use the centered residual
$r_i=(s_{{\rm dual},i}+\mu\theta_i\phi_i'(x_i))/
(\mu\theta_i\sqrt{\phi_i''(x_i)})$ and weight $\theta_i$.
Centrality gives $|r_i|<1$, and the Lewis estimates satisfy
$\rho\leq\theta_i\leq4$. Its primal accuracy scales are
$\phi_i''(x_i)^{-1/2}$; the dual scales are
$\mu\theta_i\sqrt{\phi_i''(x_i)}$. These scales need only be
positive, by Lemma~\ref{lem:tall_centered_observers}.
All new stored quantities satisfy the required approximation
invariants. No geometric IPM step or feasibility projection has
occurred. The cost and preservation assertions follow.
\end{proof}

\subsection{Entry changes and total work}
\label{subsec:tall_sparse_lp_entry_changes_and_total_work}

An observer reports a coordinate only after it moves by a threshold amount. We charge the weight at each report to the motion since initialization or the previous report, allowing the weight itself to change.

\begin{lemma}[Weighted threshold changes after a reset]
\label{lem:tall_local_threshold_charge}
Let $w_{t,i}>0$ and $z_{t,i}\in\R^k$ for $0\leq t\leq T$.
For each coordinate $i$, let $(s,t]$ be pairwise disjoint detection
intervals, each satisfying $\|z_{t,i}-z_{s,i}\|_2\geq a>0$.
Then the sum of $w_{t,i}$ over all detections is at most
\begin{equation}\label{eq:tall_local_threshold_charge}
 4T\sum_{t=1}^T\sum_i
 (a^{-2}w_{t-1,i}\|z_{t,i}-z_{t-1,i}\|_2^2
       +(\sqrt{w_{t,i}}-\sqrt{w_{t-1,i}})^2).
\end{equation}
The same conclusion holds separately on every freshly initialized
epoch. Initialization assignments are excluded from the detections.
\end{lemma}

\begin{proof}
Fix a detection interval $(s,t]$ for coordinate $i$.
If $\sqrt{w_{j-1,i}}\geq\sqrt{w_{t,i}}/2$ for every
$s<j\leq t$, the triangle inequality and Cauchy--Schwarz give
\[
 a^2w_{t,i}\leq
 w_{t,i}(\sum_{j=s+1}^t\|z_{j,i}-z_{j-1,i}\|_2)^2
 \leq4T\sum_{j=s+1}^t w_{j-1,i}\|z_{j,i}-z_{j-1,i}\|_2^2.
\]
Otherwise choose $j$ with
$\sqrt{w_{j-1,i}}<\sqrt{w_{t,i}}/2$. Telescoping the square roots
from $j-1$ to $t$ and applying Cauchy--Schwarz gives
\[
 w_{t,i}\leq4T\sum_{h=j}^t
 (\sqrt{w_{h,i}}-\sqrt{w_{h-1,i}})^2.
\]
Sum the applicable bound over the disjoint intervals for each
coordinate, and then over coordinates. This proves
Eq.~\eqref{eq:tall_local_threshold_charge}. A reset starts new
intervals and introduces no detection term.
\end{proof}

We apply the threshold bound to the primal, dual, and Lewis-weight reports between two resets. The nearby path bounds the total weight of those reports within one epoch.

\begin{lemma}[Reported entry changes within an epoch]
\label{lem:tall_reported_entry_charge}
Consider $T$ valid source short steps between synchronized observer resets,
with the reporting rules and approximation accuracies of
Algorithms~\ref{alg:tall_sparse_short_step} and~\ref{alg:tall_sparse_step_updates}.
Work on the nearby-path event of Lemma~\ref{lem:tall_nearby_stability}.
Assume that the observer and nearby-path errors are smaller than a fixed
sufficiently small fraction of the reporting thresholds, and that the cached
scalar $\bar\mu$ remains within its source approximation tolerance throughout
the epoch. Write $\tau_t:=\tau(x_t)$ for the exact Lewis vector and
$\bar\tau_t$ for the reported Lewis vector stored as $\theta$ by the algorithms.
The reported primal, dual, and Lewis entries satisfy
\begin{equation}\label{eq:tall_source_coordinate_charge}
 \sum_{t=1}^{T}\sum_{i=1}^{n}\tau_{t,i}
 (\mathbbm{1}[\bar x_{t,i}\ne\bar x_{t-1,i}]
 +\mathbbm{1}[\bar s_{t,i}\ne\bar s_{t-1,i}]
 +\mathbbm{1}[\bar\tau_{t,i}\ne\bar\tau_{t-1,i}])
 =\widetilde O(T^2).
\end{equation}
Only thresholded changes after initialization are counted.
The dual reporting rule uses the positive scale
$\mu\tau_i\sqrt{\phi_i''(x_i)}$.
\end{lemma}

\begin{proof}
We apply Lemma~\ref{lem:tall_local_threshold_charge} within this
epoch, starting its detection intervals at the reset. Let
$w_{t,i}:=\widehat\tau_{t,i}$ and put
\[
 z^x_{t,i}:=\log\frac{\widehat x_{t,i}-\ell_i}{u_i-\widehat x_{t,i}},
 \qquad z^\tau_{t,i}:=\log\widehat\tau_{t,i}.
\]
The derivative of the first scalar map is
$1/(x_i-\ell_i)+1/(u_i-x_i)$, between
$\sqrt{\phi_i''(x_i)}$ and $\sqrt{2\phi_i''(x_i)}$.
Lemmas 4.44--4.45 of~\cite{bllsssw21}, and their one-step infinity
bounds, therefore imply
\begin{equation}\label{eq:tall_epoch_signal_energy}
 \sum_{t=1}^T\sum_i
 (w_{t-1,i}(|z^x_{t,i}-z^x_{t-1,i}|^2
             +|z^\tau_{t,i}-z^\tau_{t-1,i}|^2)
       +(\sqrt{w_{t,i}}-\sqrt{w_{t-1,i}})^2)
 =O(T\gamma^2).
\end{equation}
For the square-root term use
$(\sqrt{u}-\sqrt v)^2\leq(u-v)^2/v$ for $u,v>0$.
The actual primal points and numerical Lewis estimates are closer
to these stable quantities than a fixed fraction of their reporting
thresholds. Choose the nearby-path tolerance and all estimation
errors accordingly. Each reported primal or Lewis-coordinate
change then requires an inverse-polylogarithmic displacement of
$z^x_i$ or $z^\tau_i$ since its last reported value, including the
first change after reset. Actual and stable endpoint weights are
within a constant factor. Lemma~\ref{lem:tall_local_threshold_charge}
and Eq.~\eqref{eq:tall_epoch_signal_energy} give total charge
$\widetilde O(T^2)$ for these changes.

For the dual changes define the positive scale
$a_{t,i}:=\mu_t\widehat\tau_{t,i}\sqrt{\phi_i''(\widehat x_{t,i})}$
and the two-dimensional signal
$z^s_{t,i}:=(s_{{\rm dual},t,i}/a_{t,i},\log a_{t,i})$.
Centrality and the nearby-path comparison imply
$|s_{{\rm dual},t,i}/a_{t,i}|=O(1)$.
Also $|d\log\sqrt{\phi_i''}/dz^x_i|\leq1$.
Thus Eq.~\eqref{eq:tall_epoch_signal_energy}, the source dual-step
bound in Corollary 4.30, and
$\sum_i\widehat\tau_{t,i}=3d$ show
\[
 \sum_{t=1}^T\sum_i w_{t-1,i}
 \|z^s_{t,i}-z^s_{t-1,i}\|_2^2=\widetilde O(T).
\]
Here the scalar $\log\mu$ contributes $O(dr^2)$ per step, where
$r=\widetilde O(d^{-1/2})$ is the relative step size.
To verify the normalized-slack term, write its change as
$(s_t-s_{t-1})/a_t+(s_{t-1}/a_{t-1})(a_{t-1}/a_t-1)$;
the first term is covered by the dual-step bound and the second by
the scale bound. At a reported dual change between $s$ and $t$,
the source threshold gives
$|s_{{\rm dual},t,i}-s_{{\rm dual},s,i}|/a_{t,i}\geq a_0$,
with inverse-polylogarithmic $a_0>0$, after allowing for observer
errors. If $|\log(a_{t,i}/a_{s,i})|$ is smaller than a sufficiently
small multiple of $a_0$, the normalized slack changes by at least
$a_0/2$; otherwise the second component of $z^s_i$ changes by that
multiple of $a_0$. Lemma~\ref{lem:tall_local_threshold_charge} again
gives $\widetilde O(T^2)$. This proves
Eq.~\eqref{eq:tall_source_coordinate_charge} for a fresh epoch,
rather than deducing an individual-epoch bound from a block total.
Initialization assignments are paid by the bulk reset.
\end{proof}

To bound the sampling and observer costs of a short step, we need its primal and dual increments in the units used by the corresponding observers.

\begin{lemma}[Short-step normalization]\label{lem:tall_short_step_normalization}
Consider the pre-step state of Algorithm~\ref{alg:tall_sparse_short_step}.
Assume Eq.~\eqref{eq:tall_source_invariants} and the stored-state
approximation requirements of~\cite[Definition 6.5]{bllsssw21}.
Use $\alpha,\gamma\leq1/4$, $\epsilon_c\leq1/80$, and $C_{\rm norm}\geq1$,
and the cached-parameter rule of Definition~\ref{def:tall_algorithm_state}.
Write $T:=\operatorname{Diag}(\theta)$, $C:=\operatorname{Diag}(g)$,
$H:=ACT^{-1}CA^\top$, and $G:=F^\top F$.
Suppose the current sampled Gram and the common inverse $M$ satisfy
$(3/4)H\preceq G\preceq(5/4)H$ and
$(1-\delta_{\rm inv})G^{-1}\preceq M\preceq(1+\delta_{\rm inv})G^{-1}$,
where $\delta_{\rm inv}\leq1/4$.
Let $h$ be the mixed-unit-ball direction returned by the primal structure,
$d_x:=-\gamma Ch$, $h':=Ad_x$, and
$u:=T^{-1}CA^\top M(h'+\Delta)$. The following hold.
\begin{itemize}[leftmargin=2.1em,labelwidth=1.5em,labelsep=0.6em]
\item[(a)] Before drawing the proportional sample,
$\|d_x/g\|_2^2\leq e^\alpha\gamma^2n/(dC_{\rm norm}^2)$,
$\|u\|_T\leq3\gamma/C_{\rm norm}$, and $\|u\|_2^2\leq n/d$.
Also $\sum_i\theta_i\leq6d$.
\item[(b)] On a successful source short step, let $e^{s,\rm new}$
be the positive accuracy scales at the dual Add call and let
$\delta s:=\mu A^\top Mh'$. With $c_i=d$, the dual observer energy
satisfies $\|\delta s/e^{s,\rm new}\|_c^2\leq144n\gamma^2/C_{\rm norm}^2$.
The primal observer energy is $O(\gamma^2n/(dC_{\rm norm}^2))$.
\end{itemize}
Part (a) requires only the pre-step invariants and the stated
sampled-Gram and inverse guarantees.
\end{lemma}

\begin{proof}
{\bf Proof of Part (a).} Put $T:=\operatorname{Diag}(\theta)$,
$C:=\operatorname{Diag}(g)$, and $H:=ACT^{-1}CA^\top$.
Let $h\in\R^n$ be the maximizer used by QueryProduct, so
$d_x=-\gamma Ch$ and
$\|h\|_\infty+C_{\rm norm}\|h\|_{\widehat\theta}\leq1$, where
$\widehat\theta\approx_\alpha\theta$. We use
$\alpha\leq1/4$ and $C_{\rm norm}\geq1$. Consequently,
\begin{equation}\label{eq:tall_dense_direction_normalization}
 \|h\|_T\leq e^{\alpha/2}\|h\|_{\widehat\theta}
 \leq e^{\alpha/2}/C_{\rm norm},\qquad
 \|d_x/g\|_2^2=\gamma^2\|h\|_2^2
 \leq\frac{e^\alpha\gamma^2n}{dC_{\rm norm}^2}.
\end{equation}
The first step uses the coordinatewise comparison of the weights;
the second uses the mixed-unit-ball constraint. The equality uses
$d_x=-\gamma Ch$, and the last step uses $\theta_i\geq\rho=d/n$.
This lower bound follows from $\tau_i\geq2\rho$ and the numerical
Lewis accuracy; the same accuracy gives $\sum_i\theta_i\leq6d$.

Write
$H_{\rm true}:=A(\operatorname{Diag}(\tau(x))\Phi''(x))^{-1}A^\top$.
The stored-scale and Lewis approximation invariants give
$H/2\preceq H_{\rm true}\preceq2H$, after reducing the internal
accuracy constants. The source feasibility invariant therefore gives
\begin{equation}\label{eq:tall_stored_residual_normalization}
 \|\Delta\|_{H^{-1}}
 \leq\sqrt2\|\Delta\|_{H_{\rm true}^{-1}}
 \leq\sqrt2\epsilon_c\gamma/C_{\rm norm}.
\end{equation}
Here the first step is the Loewner comparison of the inverses, and
the second is Eq.~\eqref{eq:tall_source_invariants}.
Let $G=F^\top F$ be the sampled Newton Gram and let $M$ be the
common symmetric inverse returned for the two right-hand sides.
On their valid events, choose the sampling and inverse accuracies so
$(3/4)H\preceq G\preceq(5/4)H$ and
$(1-\delta_{\rm inv})G^{-1}\preceq M\preceq(1+\delta_{\rm inv})G^{-1}$,
where $\delta_{\rm inv}\leq1/4$. Then
$\|H^{1/2}MH^{1/2}\|_{\rm op}\leq2$.
These are consequences of the sampled-Gram and batch guarantees;
no independence of $M$ and the right-hand sides is used.

For $u:=T^{-1}CA^\top M(h'+\Delta)$, we have
\begin{align}
 \|u\|_T
 &=\|H^{1/2}M(h'+\Delta)\|_2\notag\\
 &\leq2(\|h'\|_{H^{-1}}+\|\Delta\|_{H^{-1}})\notag\\
 &\leq2\gamma\|h\|_T+2\sqrt2\epsilon_c\gamma/C_{\rm norm}\notag\\
 &\leq\frac{2\gamma}{C_{\rm norm}}(e^{\alpha/2}+\sqrt2\epsilon_c)
 \leq3\gamma/C_{\rm norm}.
 \label{eq:tall_sampler_weighted_normalization}
\end{align}
The first step follows by squaring both sides and using
$H=ACT^{-1}CA^\top$; the second uses the bound on $M$ and the
triangle inequality. The third uses $h'=-\gamma ACh$,
Eq.~\eqref{eq:tall_stored_residual_normalization}, and
$\|AC h\|_{H^{-1}}\leq\|h\|_T$: the latter holds because
$H^{-1/2}ACT^{-1/2}$ has orthonormal rows. The fourth step uses
Eq.~\eqref{eq:tall_dense_direction_normalization}; the last uses
$\alpha\leq1/4$ and $\epsilon_c<1/80$.
It follows that
\begin{equation}\label{eq:tall_sampler_euclidean_normalization}
 \|u\|_2^2\leq\frac nd\|u\|_T^2
 \leq\frac{9\gamma^2n}{dC_{\rm norm}^2}\leq\frac nd.
\end{equation}
The first step uses $T\succeq(d/n)I$, the second uses
Eq.~\eqref{eq:tall_sampler_weighted_normalization}, and the third
uses $\gamma\leq1/4$ and $C_{\rm norm}\geq1$.
Thus the sampler input $v_\Delta=M(h'+\Delta)$ satisfies its stated
squared-norm hypothesis. In particular the dynamic-branch
denominator $en/d$ is at least $e\|u\|_2^2$.

{\bf Proof of Part (b).} For the dual observer, put $u_0:=T^{-1}CA^\top Mh'$ and let
$e^{s,\rm old}:=\bar\mu\theta/g$ be the accuracy scales before the
step. The same argument, without the residual term, gives
$\|u_0\|_T\leq3\gamma/C_{\rm norm}$, and
\begin{equation}\label{eq:tall_dual_increment_normalization}
 \delta s/e^{s,\rm old}=(\mu/\bar\mu)u_0.
\end{equation}
This equality uses $\delta s=\mu A^\top Mh'$ and the coordinatewise
definition of $e^{s,\rm old}$.
The cached-parameter rule gives $\mu/\bar\mu\leq2$.
On a successful source short step, Lemmas 4.31(a) and 4.22 and
Definition 6.5 of~\cite{bllsssw21} give
$g_i^{\new}/g_i^{\old},\theta_i^{\new}/\theta_i^{\old}\in e^{O(\gamma)}$.
The cached $\bar\mu$ is unchanged within FinishSparseStep, so
$e_i^{s,\new}/e_i^{s,\old}
=(\theta_i^{\new}/\theta_i^{\old})(g_i^{\old}/g_i^{\new})$.
Decreasing the fixed accuracy constants makes this ratio lie in
$[1/2,2]$, including the reported updates before Add. With the coordinate costs $c_i=d$,
\begin{equation}\label{eq:tall_dual_observer_energy}
 \|\delta s/e^{s,\rm new}\|_c^2
 \leq16d\|u_0\|_2^2
 \leq144n\gamma^2/C_{\rm norm}^2.
\end{equation}
The first step uses Eq.~\eqref{eq:tall_dual_increment_normalization}
and the two scale comparisons; the second uses
$T\succeq(d/n)I$ and the bound on $\|u_0\|_T$.
Likewise, Eq.~\eqref{eq:tall_dense_direction_normalization} bounds
the primal observer energy by $O(\gamma^2n/(dC_{\rm norm}^2))$.
The adjacent-scale comparison absorbs the previous-accuracy indexing
in Lemma D.3 of~\cite{bllsssw21}. Thus the primal and dual energies
used below are respectively $\widetilde O(n/d)$ and
$\widetilde O(n)$, with no normalization by a signed dual slack.
\end{proof}

We can now sum the costs of processing reports, sampling columns, and resetting observers over one epoch. The persistent inverse spans several epochs, so its updates are charged separately.

\begin{lemma}[Finite-epoch maintenance accounting]
\label{lem:tall_finite_epoch_accounting}
Consider $T$ valid source short steps between synchronized observer
resets. Use the private batched solves for historical and current
targets, and the sparsity-balanced sampler with
$\eta:=\sqrt{n/\nnz(A)}$. Let all internal tolerances be inverse
polylogarithms. Excluding cold initialization and all actual work
of the persistent inverse-maintenance updates, the expected epoch
cost, stopped at the first failed invariant, is
\begin{equation}\label{eq:tall_epoch_cost}
 \widetilde O(\nnz(A)+n+d^2+T(n+d^2)+nT^2+T\sqrt{\nnz(A)n}).
\end{equation}
This includes recursive Lewis-layer changes, selected-column actions,
the actual primal-update lists, residual products, archived inverse
copies, and all changes caused by the bulk reset other than its
persistent inverse update. The inverse is retained across epochs
and is charged only over its complete block. The displayed
bound assumes that the cached scalar $\bar\mu$ is not changed by a
coordinatewise loop inside the epoch; this is ensured by choosing
the epoch length below a sufficiently small inverse-polylogarithmic
multiple of $\sqrt d$ and setting $\bar\mu:=\mu$ at its start.
\end{lemma}

\begin{proof}
We first check the calls in Algorithms~\ref{alg:tall_sparse_short_step}
and~\ref{alg:tall_sparse_step_updates}. Let $d_x$ be the dense primal
increment stored by $\mathrm{primal}.\textnormal{\textsc{QueryProduct}}$.
The paired query guarantee gives $h'=Ad_x$. The sampler returns the
sparse increment $h_{\rm sp}$ and its exact product $r_{\rm sp}=Ah_{\rm sp}$.
We have
\[
 \Delta^{\new}=\Delta+h'+r_{\rm sp}
 =Ax-b+Ad_x+Ah_{\rm sp}
 =A(x+d_x+h_{\rm sp})-b.
\]
The first equality is the assignment in
Algorithm~\ref{alg:tall_sparse_short_step}; the second uses the two
returned products and the old residual invariant; the third groups
the terms. The procedure QuerySum represents exactly this updated
primal vector. There is no Update or SetAccuracy call between the
paired QueryProduct and QuerySum calls.

For the dual update, we have
\[
 s_{\rm dual}^{\new}=s_{\rm dual}+\mu A^\top v_0
 =c-A^\top y+\mu A^\top v_0
 =c-A^\top(y-\mu v_0).
\]
The first equality is the guarantee of $\mathrm{dual}.\textnormal{\textsc{Add}}$;
the second uses the old dual invariant; the third is algebra.
The multiplier assignment in Algorithm~\ref{alg:tall_sparse_step_updates}
therefore preserves dual feasibility. The rank reduction and
reconstruction of the two-column batch use the convention in
Lemma~\ref{lem:tall_private_preconditioner_batch}; both responses
use the same symmetric approximate inverse. The zero batch is exact.
The reported-coordinate tests and updates are those of
\cite[Algorithm 5]{bllsssw21}, with the five interfaces of this paper
and the residual-preserving reset of Algorithm~\ref{alg:tall_observer_reset}.

Lemma~\ref{lem:tall_short_step_normalization} verifies the sampler
and observer norm bounds used below.

Use the general-matrix HeavyHitter of~\cite[Lemma B.1]{bllsssw21},
with bulk cost
$\widetilde O(\nnz(A)+n)$, coordinate cost $c_i=\widetilde O(d)$,
and query cost $\widetilde O(d)$. Thus $C:=\sum_{i=1}^n c_i=\widetilde O(nd)$.
Since $\tau_i\geq2d/n$, every coordinate operation of cost
$\widetilde O(d)$ can be charged $\widetilde O(n\tau_i)$.
Lemma~\ref{lem:tall_reported_entry_charge} proves
Eq.~\eqref{eq:tall_source_coordinate_charge} for this freshly
initialized epoch.

Thus Eq.~\eqref{eq:tall_source_coordinate_charge} pays
$\widetilde O(nT^2)$ for all externally reported primal, dual,
and Lewis-coordinate changes. This includes changes to barrier
scales, sampler column scales and masses, gradient data, and dual
accuracy scales. The sparse sampler's actual column-scale update cost
$1+\nnz(a_i)$ is no larger than this charge.

Lemma~\ref{lem:tall_lewis_epoch_cost} pays
$\widetilde O(T(n+d^2)+nT^2)$ for all recursive Lewis-layer
updates, queries, forced-column products, and private endpoint copies.

For primal/gradient queries, the dense increment has scaled
Euclidean norm squared $\widetilde O(n/d)$. The bucket construction
in Lemma~\ref{lem:tall_primal_product} and the finite-horizon
bound of Lemma D.3 of~\cite{bllsssw21} therefore cost
$\widetilde O(Td+T^2n/d+L_h)$, where $L_h$ is the actual total
length of the additive sparse lists. For dual Add queries, use
Lemma E.5 of~\cite{bllsssw21}, with fresh zero increment histories
at the start of this epoch, and the source
bound on $\delta s/(\mu\tau\sqrt{\phi''})$. The weighted energy
with costs $c_i=d$ is $\widetilde O(n)$ per step, so the bound is
$\widetilde O(Td+nT^2)$. Neither claim divides by dual slack.

The balanced sampler costs
$\widetilde O(T(n/\eta+\eta(n+\nnz(A))+d^2))$ beyond initialization
and the coordinate changes charged above. This is
$\widetilde O(T(\sqrt{\nnz(A)n}+d^2))$.
Its actual lists and explicit $Ah$ products are included;
in particular $L_h$ is no greater than the already paid work.
The two Newton right-hand sides $h'$ and $h'+\Delta$ are both
fixed before drawing the proportional sample. Solve them in one
batch and cache the response for $h'$ until the dual update. This provides the
common symmetric inverse required within the short step without
requiring one operator for future batches.

Finally Lemma~\ref{lem:tall_residual_preserving_restart} pays the
observer work of the bulk reset. Summing proves
Eq.~\eqref{eq:tall_epoch_cost}. No actual-cost bound for persistent
inverse updates in this individual epoch has been used.
The decrease of $\mu$ per step is an inverse-polylogarithmic
multiple of $1/\sqrt d$. Selecting the epoch cap with a sufficiently
small constant and polylogarithmic slack keeps $\mu/\bar\mu$
within its source threshold throughout the epoch. This avoids
an unaccounted full-array update loop. The explicit procedure calls add
only constant-many procedure calls per reported coordinate, an $O(n)$ scan
per step, and $O(d^2)$ per archive copy, all covered above.
\end{proof}

Short epochs reduce the cost of coordinate changes but require more resets. Balancing these two costs determines the epoch length; the persistent inverse is charged once over the full block.

\begin{corollary}[Epoch-length optimization]
\label{cor:tall_warm_epoch_optimization}
Let $1\leq q=\widetilde O(\sqrt d)$ and choose a common epoch length
$L_{\rm epoch}$ within polylogarithmic factors of $\sqrt{\nnz(A)/n}$, rounded
down and truncated below by one. Over one complete block from
a cold initialization of the persistent inverse, the total expected
work of the observer epochs, the cold setup, all persistent inverse
updates, and snapshot copies is
\[
 \widetilde O(\sqrt{\nnz(A)nd}+d^{5/2})
\]
when $\omega\leq9/4$. In particular the coordinate term is
$\widetilde O(nqL_{\rm epoch})$.
\end{corollary}

\begin{proof}
There are $O(1+q/L_{\rm epoch})$ observer resets and $q$ ordinary
inverse updates, hence $O(q+1)$ inverse updates in all. A reset
is a repeated point of the stable sequence by
Lemma~\ref{lem:tall_stable_lewis_reference}. Apply Theorem 10
of~\cite{blss20} once to this entire sequence starting from its
cold initialization, including the calls made at observer resets.
Let $I_{\rm block}$ denote this inverse work together with its cold
initialization. The constant-accuracy bound is
\begin{align}
 \E[I_{\rm block}]
 &\leq\widetilde O(d^{\omega+o(1)}
       +q(n+d^{\omega-1/2+o(1)}+d^2))\notag\\
 &\leq\widetilde O(d^{\omega+o(1)}+q(n+d^2)).
 \label{eq:tall_inverse_block_work}
\end{align}
The first step applies that theorem to the complete sequence of
$O(q+1)$ calls. The second step uses $\omega\leq9/4$, so a fixed
sufficiently small exponent slack gives
$\omega-1/2+o(1)<2$. An individual epoch may contain a more
expensive update; no uniform actual-cost bound per epoch is asserted. Lemma~\ref{lem:tall_finite_epoch_accounting} bounds
only the freshly reset observer work. Since each epoch has length
at most $L_{\rm epoch}$, the sum of its squared lengths is at most
$qL_{\rm epoch}$. Adding these separate charges gives
\[
 \widetilde O(\nnz(A)+n+d^{\omega+o(1)}
 +(1+q/L_{\rm epoch})(\nnz(A)+n+d^2)+q(n+d^2)+nqL_{\rm epoch}+q\sqrt{\nnz(A)n}).
\]
Here $q\nnz(A)/L_{\rm epoch}$ and $nqL_{\rm epoch}$ are both
$\widetilde O(\sqrt{\nnz(A)nd})$, and $q\sqrt{\nnz(A)n}$ has that same bound.
Also $qd^2/L_{\rm epoch}\leq qd^2=\widetilde O(d^{5/2})$.
Since $n\leq \nnz(A)\leq nd$, one has $\nnz(A)+n\leq2\sqrt{\nnz(A)nd}$ and
$nq\leq\widetilde O(\sqrt{\nnz(A)nd})$. Finally,
$d^{\omega+o(1)}=O(d^{5/2})$ for a fixed sufficiently small
exponent slack. All remaining terms follow.
If rounding or the cache-$\mu$ restriction forces $L_{\rm epoch}=1$,
then $\nnz(A)/n$ is at most polylogarithmic, so $q\nnz(A)$ is still within
the soft target bound.
\end{proof}

\subsection{Adaptive execution and success probability}\label{subsec:tall_success_probability}

\begin{proof}[Proof of Theorem~\ref{thm:tall_sparse_bounded_box_lp}]
Apply Lemma~\ref{lem:tall_sparse_centered_augmentation}. Its $N=n+2d$
columns and $\nnz(A)+2d$ nonzeros are within constant factors of $n$ and $\nnz(A)$.
Use the augmented dimensions in all source parameters, including
$p,z$, and in the choice of the balanced sampler. The starting
tolerance in that lemma is at most the
$\epsilon_{\rm src}/C_{\rm start}$ tolerance required by
Lemma 4.12 of~\cite{bllsssw21}. After fixing the potential parameter $\lambda$, also
make it at most $\log(2N)/(10\lambda)$. Thus the initial potential
is polynomially bounded even when the tail parameters are enlarged.
These are internal inverse-polylogarithmic accuracies, distinct
from the requested output error $\epsilon$.

Run the source two-sided ShortStep algorithm with the balanced
proportional sampler using Algorithms~\ref{alg:tall_sparse_short_step}
and~\ref{alg:tall_sparse_step_updates}. Their exact residual and dual
identities are verified in the proof of
Lemma~\ref{lem:tall_finite_epoch_accounting}.
At the start of a short step, use the previously maintained inverse
to precondition the fresh sampled Newton target. Compute the two
Newton responses in one batch before drawing the proportional
sample. After the primal step and its reported barrier-scale
updates, call the persistent inverse update with the new stored
barrier scale and the old Lewis vector. Only then call Lewis.Query,
and associate this inverse snapshot with the current endpoint of
every Lewis layer. Return the new Lewis vector and use the already
cached Newton response for the dual update. Thus no inverse update
requires the Lewis vector that the pending query is supposed to
compute. Numerical hints from the previous step suffice by
Lemma~\ref{lem:tall_stable_lewis_reference}.

Every Newton target is
a fresh spectral approximation of the publicly specified Newton Gram,
formed by sampling scaled columns of $A$.
It can be formed by scanning the stored Lewis estimates and sampling
columns with probabilities $\min\{1,C\theta_i\gamma^{-2}\log(nJU)\}$.
The Newton leverage domination from
Lemma~\ref{lem:tall_layer_target_coverage} makes this a valid spectral
sample. The scan costs $O(n)$ and the expected sampled-column extraction
cost is $\widetilde O(d^2)$, already included in the per-step budget.
The two Newton right-hand sides are processed together before the
proportional sample; thus one common symmetric approximate inverse
is used in the primal and dual directions. Set its relative error
below the source $\gamma/2$ tolerance. All leverage endpoint solves
instead use the smaller polynomial accuracy required by
Lemma~\ref{lem:tall_pathwise_jl_perturbation}. Those solves use fresh
samples and sign matrices in exactly the source call locations.
Historical queries use the snapshot belonging to their endpoint.

At the beginning, and after each block of at most $\lceil\sqrt d\rceil$
steps, compute a numerical Lewis vector at the current exact barrier
scale and cold-initialize the private reference inverse. A cold
numerical Lewis vector is obtained in $\widetilde O(\nnz(A)+n+d^{\omega+o(1)})$
work by Lemma~\ref{lem:tall_sparse_static_lewis_initialization}.
Initializing the sampled reference
inverse has the same bound. These block boundaries do not change
the iterate, its parameter, or its residual. Between boundaries
retain this inverse and use observer epochs of length
$L_{\rm epoch}=\widetilde\Theta(\sqrt{\nnz(A)/n})$, subject to the cache-$\mu$
cap of Lemma~\ref{lem:tall_finite_epoch_accounting}. Every observer
restart is the residual-preserving restart, not a call to the cold
PathFollowing initializer with zero residual.

Lemma~\ref{lem:tall_short_step_normalization}(a) verifies the squared-norm
bound before the proportional sample is drawn. The sampler in
Lemma~\ref{lem:tall_sparsity_balanced_sampler} uses this normalized
direction and the required leverage domination. Its
Young-inequality mass bound and the proportional-sampling proof
apply verbatim to the two-sided source barrier. They do not use a
positive dual slack. In particular the same five validity conditions
hold conditionally before each sample. The nearby-stability and
coordinate-charge results are consequently applied to this changed
trajectory itself, not transferred from an execution with the old
sampling law.

We justify the source invariants by induction. The augmented
midpoint has the starting invariants. Given a valid state, the
balanced sampler has the source conditional validity properties,
the two Newton responses have the required common-operator
interpretation, the Lewis observers satisfy their specified
accuracies, and Lemma~\ref{lem:tall_centered_observers} supplies
the gradient and coordinate observers on the actual parameter ranges. Lemma 6.4 of~\cite{bllsssw21} therefore identifies
the resulting step with its mathematical ShortStep. Its potential,
feasibility, and nearby-stability conclusions apply to this very
execution. The nearby stable path supplies
Lemma~\ref{lem:tall_stable_lewis_reference}; the latter supplies the
inverse update before the next Lewis query. At a restart the
actual state is unchanged and its stored accuracies improve;
its exact residual and dual multiplier are retained. The observer
clocks restart, while the persistent inverse clock does not.
This closes the induction. In particular, the internal Lewis-layer
charge and the external coordinate charge are consequences of the
source analysis, rather than hypotheses about the new trajectory.

Here is a common probability and work convention for this induction.
Declare the total source step horizon and all dyadic query horizons
in advance. There are polynomially many primitive calls in $n$ and
$M:=1+\lceil\log(nW/\epsilon)\rceil$. Give each batch, column sample,
and JL object failure probability at most $1/(100JU)$, where $J$
is a polynomial upper bound on their total number and $U$ is a
polynomial work cap for the execution. Increase sampling dimensions
and oversampling constants by the corresponding logarithmic
factors. Choose batch and sampler transcript-coupling errors at
most $1/(100JU)$ as well. The inverse-maintenance proof uses the
same failure allocation by increasing its logarithmic oversampling.
Raw inverse matrices, intermediate iterations, node estimates, and
data-dependent timings are never used as public observations.

Set $\beta:=1/(10U)$.
Every inverse state used by Algorithm~\ref{alg:tall_sparse_lp} must remain private. For the nonadaptive
analysis of~\cite{blss20}, replace each released batch
by the public ideal kernel in
Lemma~\ref{lem:tall_private_preconditioner_batch}. Conditional on all
other randomness, that ideal execution is independent of the
inverse-maintenance coins. Lemma~\ref{lem:tall_stable_lewis_reference}
then supplies the stable sequence and sufficiently accurate hints.
Use the stopped transcript comparison in
Corollary~\ref{cor:tall_archived_private_preconditioners} rather than conditioning
on the future event that all inverse updates succeed.
For $J$ batches per macro-phase, allocate coupling error
$\nu\leq\beta/(10J)$ and batch failure at most $\beta/(10J)$.
All historical matrices are private as well; their adaptive future
right-hand sides are allowed by the uniform comparison.
Fixed public iteration counts and a polynomial macro-phase work cap
prevent data-dependent timing and exceptional executions from
invalidating the distribution or expected-work bounds.

Replace inverse responses by their ideal kernels and sampler
responses by their exact kernels. In that execution the update
sequence is independent of the private inverse coins. Apply
Theorem 10 of~\cite{blss20} on each block and the source stable-path estimates,
stopping the analysis at the first failed invariant. The stopped
coupling of Corollary~\ref{cor:tall_archived_private_preconditioners} and
the sampler's exact-output coupling transfer the result to the
implemented execution with total additional error at most $1/(50U)$.
Allocate the same inverse-polynomial error to the source potential,
feasibility, and stable-path estimates: replace their logarithmic
tail parameters by $C\log(nJU)$. Their concentration and potential
bounds then give total invariant-failure probability at most
$1/(100U)$ on the declared horizon. Cap exceptional executions
at $U$ operations and return
failure on undefined arithmetic or an exceeded cap. The cap can be
chosen polynomially larger than the proved expected work, so
Markov's inequality makes a cap-trigger probability below $1/10$.
Bad preprocessing, invariant, and coupling events contribute only $O(1)$ to
expected work. On the valid terminal event,
Lemma~\ref{lem:tall_output_certificate} proves that the certificate
passes. This argument does not condition sampling on future success
and gives acceptance probability at least $2/3$ per trial. On every
other event, an accepted output is still valid by the certificate.

There are $\widetilde O(\log(nW/\epsilon))$ blocks of
$O(\sqrt d)$ steps, by the source decrease of $\mu$:
initialization gives $\mu_0=\poly(nW/\epsilon)$, and we use the terminal
target $\mu_{\rm end}:=\epsilon/(Cd)$, so
$\log(\mu_0/\mu_{\rm end})=O(\log(nW/\epsilon))$.
Corollary~\ref{cor:tall_warm_epoch_optimization} bounds each block by
$\widetilde O(\sqrt{\nnz(A)nd}+d^{5/2})$, including its cold setup.
The final point has $\mu\leq\epsilon/(Cd)$.
Lemma~\ref{lem:tall_sparse_terminal_correction} converts it to the
stated output in
$\widetilde O((\nnz(A)+n+d^{\omega+o(1)})\log(nW/\epsilon))$ additional work.
This cost is absorbed because $n\leq \nnz(A)\leq nd$ and $\omega\leq9/4$.
Including the logarithmic factors from the failure allocation gives total
expected time $\widetilde O((\sqrt{\nnz(A)nd}+d^{5/2})\poly \log(nW/\epsilon))$.
For the independent-restart variant, the probability of reaching
trial $j$ is at most $3^{-(j-1)}$. Its fresh coins are independent
of the preceding failures, so summing expected costs multiplies
the one-trial bound by at most $\sum_{j\geq1}3^{-(j-1)}=3/2$.
The probability of never accepting is zero. No independence between
a trial's cost and its own acceptance event is needed.
\end{proof}

\section{Inverse-maintenance data structure}\label{sec:tall_inverse_maintenance}

In Section~\ref{subsec:tall_inverse_maintenance_main_data_structure}, we describe how to maintain a private preconditioner and archive copies for later use.
In Section~\ref{subsec:tall_inverse_maintenance_batched_solves_with_a_hidden_preconditioner}, we use current and archived preconditioners to solve batches of linear systems with adaptively chosen right-hand sides.

\subsection{Main data structure}
\label{subsec:tall_inverse_maintenance_main_data_structure}

Using the inverse maintenance of~\cite[Theorem 10]{blss20}, \textsc{InverseMaintain} stores the current reference inverse and its historical copies. A batched solve releases only its answer; the stored matrices, intermediate products, and data-dependent timing remain private.

\begin{theorem}[InverseMaintain]\label{thm:tall_inverse_maintain}
Let $A\in\R^{d\times n}$ have full row rank, and fix constants
$\epsilon_I\in(0,1/10)$ and $\chi\ge1$. Assume that the supplied weights and shifted
leverage hints satisfy the hypotheses of \cite[Theorem 10]{blss20}.
There is a data structure
(\textnormal{\textsc{InverseMaintain}} in
Algorithms~\ref{alg:tall_inverse_maintain} and~\ref{alg:tall_inverse_solve})
that maintains the imported private inverse state and an archive of its
symmetric $d\times d$ preconditioners. It supports the following procedures.
\begin{itemize}
\item[(a)] \textnormal{\textsc{Init}}$(A\in\R^{d\times n},\widetilde w\in\R_{>0}^n,\widetilde\lambda\in\R_{>0}^n)$
starts a cold inverse-maintenance block at the supplied weights and hints
and sets the archive to empty. Its guarantee and initialization work are
those of \cite[Theorem 10]{blss20}.
\item[(b)] \textnormal{\textsc{Update}}$(\widetilde w\in\R_{>0}^n,\widetilde\lambda\in\R_{>0}^n)$
updates the same private inverse state under that theorem's contract.
Existing archived matrices are unchanged. The maintenance work is charged
over the complete block starting at \textnormal{\textsc{Init}}.
\item[(c)] \textnormal{\textsc{Archive}}$()$ stores a private copy of the
current preconditioner and returns its handle. It costs $O(d^2)$ time and
$O(d^2)$ additional space per retained copy.
\item[(d)] \textnormal{\textsc{Solve}}$(h,L,B,\delta,\nu,\beta)$ takes an
archive handle $h$, a full-column-rank factor $L\in\R^{m\times d}$, a
full-column-rank matrix $B\in\R^{d\times r}$ with $1\le r\le d$, and
$\delta,\nu,\beta\in(0,1/4)$. Write $G:=L^\top L$ and let $P_h$ be the
archived matrix. Assume $\chi^{-1}G^{-1}\preceq P_h\preceq\chi G^{-1}$
for the fixed constant $\chi$. The procedure returns the matrix $Y$
of Lemma~\ref{lem:tall_private_preconditioner_batch}, with its common
approximate-inverse and total-variation guarantees. If multiplying by
$P_h$ takes $T_{P_h}$ time, its work is
$O(r(T_{P_h}+\nnz(L))\log(dr/(\delta\nu\beta))+mr+dr^2+r^3)$.
It leaves the maintained inverse and the archive unchanged.
\end{itemize}
Every solve uses fresh randomness. For at most $J$ adaptive solves, the
released transcript satisfies Corollary~\ref{cor:tall_archived_private_preconditioners},
including its stopped validity condition. In the LP application,
Lemma~\ref{lem:tall_stable_lewis_reference} verifies the maintenance
hypotheses, and Lemma~\ref{lem:tall_layer_target_coverage} verifies the
spectral promise for each historical query. The total expected inverse
work for a complete cold block is bounded in
Eq.~\eqref{eq:tall_inverse_block_work}; that bound does not assert an
actual-cost bound for an individual observer epoch.
\end{theorem}
\begin{proof}
Parts~(a) and~(b) are the imported procedures of
\cite[Theorem 10]{blss20}, with their state retained between calls.
Part~(c) copies the $d^2$ entries of the current matrix, as in
Lemma~\ref{lem:tall_layer_target_coverage}. Copying does not change the
maintained state or an earlier archive entry.
Lemma~\ref{lem:tall_private_preconditioner_batch} proves Part~(d).
Its computation uses the selected archived matrix without changing it.
The two algorithm blocks use the same private members. The Richardson operator $K$
in Algorithm~\ref{alg:tall_inverse_solve} is applied to vectors; the procedure does not form $L^\top L$ or $K$
densely. The public iteration count depends only on the stated parameters.
Rank-deficient batches are reduced and reconstructed as in
Lemma~\ref{lem:tall_private_preconditioner_batch}; the zero batch is
returned exactly.
Corollary~\ref{cor:tall_archived_private_preconditioners} proves the
adaptive-transcript assertion. The cited application lemmas establish
the required hypotheses without resetting the persistent inverse.
\end{proof}

A solve may need an earlier preconditioner after the weights have changed. An archive handle identifies the copy saved at that endpoint.

\begin{definition}[Private inverse state]\label{def:tall_inverse_state}
The data structure stores the fixed matrix $A$, the current supplied
weights $\widetilde w$ and shifted leverage hints $\widetilde\lambda$,
and the private state of \cite[Theorem 10]{blss20}.
An archive entry $P_h$ is a copy of the symmetric preconditioner at the
endpoint identified by $h$. Later maintenance updates do not change
$P_h$. The handle identifies this private entry; it does not reveal its
matrix. A call to \textnormal{\textsc{Solve}} is valid when that entry
satisfies the spectral promise in Theorem~\ref{thm:tall_inverse_maintain}-(d).
An observer restart retains the persistent inverse state. Historical
copies are retained while a pending query can still use them.
\end{definition}

Algorithm~\ref{alg:tall_inverse_maintain} updates the current inverse state and saves a copy of the preconditioner when a new archive handle is requested.

\begin{algorithm}[!ht]
\caption{\textsc{InverseMaintain} data structure: maintenance and archives}
\label{alg:tall_inverse_maintain}
\begin{algorithmic}[1]
\DataStructure{InverseMaintain} \Comment{Theorem~\ref{thm:tall_inverse_maintain}}
\Members
\State Fixed $A\in\R^{d\times n}$, accuracy $\epsilon_I\in(0,1/10)$, and constant $\chi\ge1$.
\State Current weights $\widetilde w$ and shifted leverage hints $\widetilde\lambda$.
\State Private inverse-maintenance state from \cite[Theorem 10]{blss20}.
\State Private archive $(P_h)_h$ and next unused handle $h_{\rm next}$.
\EndMembers
\State
\Procedure{Init}{$A,\widetilde w,\widetilde\lambda$} \Comment{\cite[Theorem 10]{blss20}}
\State Store $A,\widetilde w,\widetilde\lambda$; cold-initialize the imported inverse state.
\State Set the archive to empty; $h_{\rm next}\gets1$.
\EndProcedure
\State
\Procedure{Update}{$\widetilde w,\widetilde\lambda$} \Comment{\cite[Theorem 10]{blss20}}
\State Apply the imported maintenance update with the supplied weights and hints.
\State Store the supplied $\widetilde w,\widetilde\lambda$; retain all archive entries.
\EndProcedure
\State
\Procedure{Archive}{$ $} \Comment{Lemma~\ref{lem:tall_layer_target_coverage}}
\State $h\gets h_{\rm next}$; copy the current private preconditioner into $P_h$.
\State $h_{\rm next}\gets h_{\rm next}+1$.
\State \Return $h$.
\EndProcedure
\EndDataStructure
\end{algorithmic}
\end{algorithm}

Given a handle, Algorithm~\ref{alg:tall_inverse_solve} applies the same archived preconditioner to every right-hand side in the batch and randomizes the matrix it returns.

\begin{algorithm}[!ht]
\caption{\textsc{InverseMaintain} data structure: a private batched solve}
\label{alg:tall_inverse_solve}
\begin{algorithmic}[1]
\DataStructure{InverseMaintain} \Comment{Theorem~\ref{thm:tall_inverse_maintain}}
\Procedure{Solve}{$h,L,B,\delta,\nu,\beta$} \Comment{Lemma~\ref{lem:tall_private_preconditioner_batch}}
\State $P\gets P_h$; $r\gets\text{number of columns of }B$.
\State Choose the absolute constant $C$ from Lemma~\ref{lem:tall_private_preconditioner_batch}.
\State $\kappa\gets\delta/(C\sqrt{dr+\log(2/\beta)})$; $\zeta\gets\nu\kappa/(C\sqrt{dr})$.
\State $N\gets\lceil\chi^2\log(1/\zeta)\rceil+1$.
\State Apply $K:=\chi^{-1}\sum_{j=0}^{N-1}(I-PL^\top L/\chi)^jP$ by Richardson iteration.
\State $U\gets KB$; $M\gets B^\top U$; compute $R$ with $M=R^\top R$.
\State $X\gets UR^{-1}$.
\State Draw independent standard Gaussian $F\in\R^{m\times r}$.
\State Draw independent symmetric Gaussian $S\in\R^{r\times r}$ with off-diagonal variance $1$ and diagonal variance $2$.
\State $Z\gets KL^\top F$.
\State $Y\gets U+\kappa(Z-UM^{-1}B^\top Z+XS)R$.
\State \Return $Y$.
\EndProcedure
\EndDataStructure
\end{algorithmic}
\end{algorithm}

\subsection{Batched solves with a hidden preconditioner}
\label{subsec:tall_inverse_maintenance_batched_solves_with_a_hidden_preconditioner}

For a matrix $Z$, write $\|Z\|_F$ for its Frobenius norm.
The input factor $L\in\mathbb R^{m\times d}$ has full column rank,
and $G:=L^\top L$. A symmetric matrix $P$ is private: its entries,
intermediate products, and data-dependent timing are not released.
Public inputs below may depend on all previously released responses.

We need one approximate inverse for all columns and a distribution of replies close to one determined by the public inputs alone.

\begin{lemma}[Private-preconditioner batched solve]
\label{lem:tall_private_preconditioner_batch}
Suppose $\chi^{-1}G^{-1}\preceq P\preceq\chi G^{-1}$ for an
absolute constant $\chi\geq1$. Let $B\in\mathbb R^{d\times r}$
have full column rank, where $1\leq r\leq d$, and let
$0<\delta,\nu,\beta<1/4$. There is a randomized procedure returning
$Y\in\mathbb R^{d\times r}$ with the following properties.
\begin{itemize}[leftmargin=2.1em,labelwidth=1.5em,labelsep=0.6em]
\item[(a)]\phantomsection\label{lem:tall_private_preconditioner_batch_a}
With probability at least $1-\beta$, there is one symmetric positive
definite matrix $\widehat M$ such that
\[
 Y=\widehat M B,\qquad
 (1-\delta)G^{-1}\preceq\widehat M\preceq(1+\delta)G^{-1}.
\]
Thus all columns use the same approximate inverse.
\item[(b)]\phantomsection\label{lem:tall_private_preconditioner_batch_b}
For each fixed public pair $(L,B)$, the law of $Y$ is within total
variation distance $\nu$ of a kernel depending only on
$(L,B,\delta,\nu,\beta)$, and not on $P$.
This comparison is uniform over all admissible private $P$.
\item[(c)]\phantomsection\label{lem:tall_private_preconditioner_batch_c}
If multiplication by $P$ costs $T_P$, the work is
\[
 O(r(T_P+\nnz(L))\log(dr/(\delta\nu\beta))
       +mr+dr^2+r^3).
\]
This counts arithmetic operations, square roots, comparisons, and
Gaussian draws; it is not a bit-complexity statement.
\end{itemize}
Only $Y$ is released. The common matrix in part (a) is guaranteed
within this batch, not across later batches. Rank-deficient batches
can be reduced to a basis for their column span and reconstructed
afterward; the zero batch is returned exactly.
\end{lemma}

\begin{proof}
Choose a sufficiently large absolute constant $C$ and put
\[
 \kappa:=\frac{\delta}{C\sqrt{dr+\log(2/\beta)}},\qquad
 \zeta:=\frac{\nu\kappa}{C\sqrt{dr}}.
\]
For $N:=\lceil\chi^2\log(1/\zeta)\rceil+1$, define the symmetric
Richardson operator
\[
 K:=\frac1\chi\sum_{j=0}^{N-1}(I-PG/\chi)^jP.
\]
It is applied to vectors, not formed as a matrix. Indeed, if
$T:=G^{1/2}PG^{1/2}$, then
\[
 G^{1/2}KG^{1/2}=I-(I-T/\chi)^N,
 \qquad
 \|G^{1/2}KG^{1/2}-I\|_{\rm op}\leq\zeta.
\]
All eigenvalues of $T/\chi$ lie in $[\chi^{-2},1]$; this proves
both the displayed bound and positive definiteness of $K$.
The iteration count is fixed from public parameters.

Compute $M:=B^\top KB$, its Cholesky factor $R$ with $M=R^\top R$,
and $X:=KBR^{-1}$. Generate an $m\times r$ matrix $F$ of independent
standard Gaussian variables and an independent symmetric $r\times r$
Gaussian matrix $S$: its entries above the diagonal are independent
$N(0,1)$ variables and its diagonal entries are independent $N(0,2)$
variables. Define
\[
 Z:=KL^\top F,\qquad
 \Pi:=I-KBM^{-1}B^\top,\qquad
 Y:=KB+\kappa(\Pi Z+XS)R.
\]
The matrix $\Pi$ is applied using its low-rank formula. In particular,
the algorithm never forms $G$, $K$, or $\Pi$ densely. It satisfies
the exact reciprocity identity
\[
 B^\top Y=M+\kappa R^\top SR,
\]
which is symmetric. Independent perturbations of the columns would
not, in general, satisfy this identity.

We establish quantitative distribution and accuracy bounds.
These are analytic changes of coordinates, not algorithmic work.
Whiten by $G$ and replace the column basis by one for which
$B^\top G^{-1}B=I$. This reduction preserves the construction's law
up to the corresponding invertible transformation of its output.
To check the column-basis claim, replace $B$ by $BC$ with $C$
invertible. If $R_C$ is the new Cholesky factor, then
$O:=RCR_C^{-1}$ is orthogonal. Right orthogonal invariance of $F$
and conjugation invariance of $S$ show that the new output has the
law of $YC$. Left whitening follows from the covariance identity
$\operatorname{Cov}(KL^\top f)=KGK$.

Consequently it suffices to analyze $G=I$, $B=Q$ with $Q^\top Q=I$,
and $K=I+E$ with $\|E\|_{\rm op}\leq\zeta$. In distribution we may use
the symmetric square root of $M=Q^\top KQ$ instead of its Cholesky
factor, since the two factors differ by an orthogonal matrix.
Let $Z_0$ be a $d\times r$ standard Gaussian matrix. Then
\[
 Y=KQ+\kappa N_K,
\]
where
\[
 N_K:=(I-KQM^{-1}Q^\top)KZ_0M^{1/2}
          +KQM^{-1/2}SM^{1/2}.
\]
Both $N_K$ and $KQ-Q$ lie in the fixed linear space
\[
 \mathcal L:=\{U\in\mathbb R^{d\times r}:Q^\top U
       \text{ is symmetric}\}.
\]
This space has dimension $D:=dr-r(r-1)/2$.
At $K=I$, the noise is
\[
 N_I=(I-QQ^\top)Z_0+QS.
\]
On $\mathcal L$, equipped with its Frobenius inner product, its
covariance operator $\mathcal C_I$ has eigenvalues $1$ or $2$:
the component perpendicular to $Q$ has independent standard
Gaussian entries, whereas the symmetric component has covariance
twice the identity in an orthonormal symmetric-matrix basis.

Here are explicit perturbation estimates needed below. The eigenvalues
of $M$ lie in $[1-\zeta,1+\zeta]$, so
\[
 \|M^{1/2}-I\|_{\rm op}+\|M^{-1/2}-I\|_{\rm op}
       +\|M^{-1}-I\|_{\rm op}=O(\zeta).
\]
Expanding each of the two summands of $N_K-N_I$ by changing one
factor at a time shows that the associated linear maps from the
underlying independent Gaussian coordinates to $\mathcal L$
differ in operator norm by $O(\zeta)$. Left or right multiplication
has Frobenius operator norm at most the spectral norm of its
factor, and every other factor here has bounded spectral norm.
The Gaussian map producing $S$ has operator norm $\sqrt2$.
Thus the covariance $\mathcal C_K$ of $N_K$ satisfies
\[
 \|\mathcal C_K-\mathcal C_I\|_{\rm op}=O(\zeta),\qquad
 \tfrac12 I_{\mathcal L}\preceq\mathcal C_K
       \preceq3I_{\mathcal L}
\]
after increasing $C$. Also $\|(K-I)Q\|_F\leq\zeta\sqrt r$.

{\bf Proof of (b).}
The ideal kernel is the construction above with $K=G^{-1}$.
In the whitened coordinates it is the Gaussian law
$N(Q,\kappa^2\mathcal C_I)$ on the affine space $Q+\mathcal L$.
The actual law is $N(KQ,\kappa^2\mathcal C_K)$ on the same space.
The formula for relative entropy between Gaussians, together with
$x-1-\log x=O((x-1)^2)$ near $1$, gives
\[
 \operatorname{KL}(\text{actual}\|\text{ideal})
       \leq C_0\zeta^2(D+r/\kappa^2)
\]
for an absolute $C_0$. Pinsker's inequality and the chosen $\zeta$
give total variation at most $\nu$. The constants do not depend
on $P$ or on the condition numbers of $G$ and $B$.

{\bf Proof of (a).}
The covariance bound and the Gaussian squared-norm tail bound give
\[
 \|N_K\|_F\leq C_1\sqrt{dr+\log(2/\beta)}
\]
with probability at least $1-\beta$. Hence, writing $D_0:=Y-Q$,
we have $\|D_0\|_{\rm op}\leq\|D_0\|_F\leq\delta/3$ by the choices
of $C,\kappa,\zeta$. Reciprocity says that $Q^\top D_0$ is symmetric.
The symmetric extension
\[
 J:=D_0Q^\top+QD_0^\top-Q(Q^\top D_0)Q^\top
\]
satisfies $JQ=D_0$ and $\|J\|_{\rm op}\leq3\|D_0\|_{\rm op}\leq\delta$.
Thus $Y=(I+J)Q$, and $I+J$ is a common positive definite
approximate inverse. Undoing the analytic coordinate changes
proves part (a).

{\bf Proof of (c).}
Applying $K$ to $O(r)$ vectors takes
$O(rN(T_P+\nnz(L)))$ work: each Richardson step
uses a multiplication by $P$ and two sparse products with $L$.
Generating $F$ costs $O(mr)$, including entries corresponding to
zero rows if these have not been removed. Forming $M$, applying
the low-rank projector, and changing the column basis cost
$O(dr^2+r^3)$. Cholesky factorization uses $O(r^3)$ arithmetic
and square-root operations. No dense factorization of order $d$
is performed. Since $N=O(\log(dr/(\delta\nu\beta)))$, the claimed
bound follows.
\end{proof}

Earlier replies may determine which archive handle is used next. We extend the comparison for one batch to the full sequence of replies.

\begin{corollary}[Adaptive access to archived preconditioners]
\label{cor:tall_archived_private_preconditioners}
Consider at most $J$ public adaptive batch queries. A query specifies
a factor $L$, its right-hand sides $B$, and an archive handle.
Suppose the matrix $P$ at that handle satisfies the spectral
comparison in Lemma~\ref{lem:tall_private_preconditioner_batch}.
All randomness in that lemma is fresh for each query. Then the
released transcript is within total variation distance $J\nu$
of the transcript obtained by replacing each reply by its ideal
kernel. The latter replies do not inspect any archived matrix.
If validity is only known with failure probability $p$ in the ideal
execution, the standard stopped hybrid adds at most $p$ to the
failure bound; it does not justify conditioning on future success.
\end{corollary}

\begin{proof}
At each coupled identical public history, the next public inputs
agree. Condition additionally on the entire private state: the
one-step distance is at most $\nu$ for every admissible state.
Averaging preserves this bound. Couple the two conditional replies
maximally and sum the disagreement probabilities over the $J$
queries. When validity is not a deterministic promise, stop the
coupling at its first failure and include that event separately.
The claim makes no assumption that independently archived matrices
are mutually independent.
\end{proof}

\section{Lewis weight maintenance data structure}\label{sec:tall_lewis_maintenance}

In Section~\ref{subsec:tall_lewis_maintenance_main_data_structure}, we describe how \textsc{LewisMaintain} tracks regularized Lewis weights as the column scales change.
In Sections~\ref{sec:tall_static_regularized_lewis_initialization} and~\ref{subsec:tall_lewis_maintenance_sparse_numerical_lewis_initialization}, we prove contraction of the weight iteration and use it to initialize the weights for a sparse matrix.
In Sections~\ref{subsec:tall_lewis_maintenance_approximate_solves_inside_a_fresh_jl_batch} and~\ref{subsec:tall_lewis_maintenance_warm_initialization}, we bound the error from approximate solves in a fresh sketch and initialize from an existing weight estimate.
In Section~\ref{subsec:tall_lewis_maintenance_recursive_updates_and_query_costs}, we bound the costs of recursive updates and queries.

\subsection{Main data structure}
\label{subsec:tall_lewis_maintenance_main_data_structure}

Algorithm~\ref{alg:tall_sparse_lp} needs updated Lewis weights as the positive column scale $g$ of its fixed matrix $A$ changes. We use the recursive layers of~\cite[Algorithm 3]{bllsssw21}, with the initialization below and the batched solves of Section~\ref{sec:tall_inverse_maintenance}.

\begin{theorem}[LewisMaintain]\label{thm:tall_lewis_maintain}
Use the parameters and stable-input conditions of
Section~\ref{subsec:tall_setup}. Suppose each current or historical Gram target uses the private
preconditioner in
Lemma~\ref{lem:tall_layer_target_coverage}. There is a randomized
data structure (\textnormal{\textsc{LewisMaintain}} in
Algorithm~\ref{alg:tall_lewis_maintain}) that maintains a positive
column scale $g\in\R^n$ and numerical regularized Lewis weights.
It supports the following procedures.
\begin{enumerate}[label=(\alph*)]
\item \textnormal{\textsc{Init}}$(A,g,\theta,P)$ takes a constant-accurate
warm seed $\theta$ and a private preconditioner $P$ satisfying
Lemma~\ref{lem:tall_warm_lewis_initialization}. It initializes all
numerical layers and their histories in
$\widetilde O(n+\nnz(A)+d^2)$ time. A seed for the first initialization
can be computed by Lemma~\ref{lem:tall_sparse_static_lewis_initialization}.
\item \textnormal{\textsc{UpdateG}}$(i\in[n],\zeta\in\R_{>0})$
changes the stored column scale $g_i$ to $\zeta$ and performs the
corresponding scale changes in the source recursive layers.
\item \textnormal{\textsc{Query}}$()$ returns a pointer to the numerical Lewis
vector and the list of changed coordinates supplied by the source
algorithm, with its prescribed multiplicative accuracy. Current
and historical solves use their respective private snapshots.
\end{enumerate}
The procedures use the source accuracy and failure parameters, with
the perturbation allowance in Lemma~\ref{lem:tall_pathwise_jl_perturbation}.
On a freshly initialized epoch of $T$ valid short steps whose
external weighted scale charge is $A_g=\widetilde O(T^2)$, as in
Lemma~\ref{lem:tall_lewis_epoch_cost}, the total maintenance work, excluding persistent inverse updates and the
initialization in Part (a), is
$\widetilde O(T(n+d^2)+nT^2)$, as proved in
Lemma~\ref{lem:tall_lewis_epoch_cost}. This is an aggregate bound;
the costs of internal recursive changes are included.
\end{theorem}

\begin{proof}
Lemma~\ref{lem:tall_warm_lewis_initialization} proves Part (a).
The source Algorithm 3 specifies the scale and query procedures;
its correctness follows from its Theorem 5.1. Lemma~\ref{lem:tall_layer_target_coverage}
supplies each required target preconditioner, and
Lemma~\ref{lem:tall_pathwise_jl_perturbation} proves that the
approximate batched products preserve the source accuracy margins.

Use the notation of Lemma~\ref{lem:tall_pathwise_jl_perturbation}.
For the source leverage queries, draw the entries of
$R\in\R^{m\times r}$ independently from
$\{-1/\sqrt r,1/\sqrt r\}$. The diagonal row samples in
Algorithm 7 of~\cite{bllsssw21} retain and rescale a subset of the
$n$ input rows; a forced-in row is included only once. Thus the
sampled factor $L$ has $m\leq n$ rows, or exactly $n$ rows if zero
rows are retained, and $\|R\|_F^2=m\leq n$.
Choose $\delta$ smaller than
$\epsilon_L/[C_{\rm pert} n^2 r\log(2q)]$ for a sufficiently large absolute
constant $C_{\rm pert}$, where $\epsilon_L$ is the layer
accuracy and $q$ bounds the number of short steps. With fixed target
and row sample of the factor $C$, ordinary JL applies to the exact inverse products.
Lemma~\ref{lem:tall_pathwise_jl_perturbation} then controls the implemented
products. In particular their row-norm error is smaller than the
relative estimation margin, and their normalized endpoint-detection
error is smaller than half the heavy-hitter threshold. Reducing the
source accuracy constants by a factor four leaves these margins.
For an exact endpoint product $U$ and its perturbation $E$, the work estimate uses $\|U+E\|_F^2\leq2\|U\|_F^2+2\|E\|_F^2$;
the additional cost-weighted energy per endpoint query is
$O(\delta^2 n^2)$. After division by its squared detection threshold
and summation over polylogarithmically many endpoint queries per step,
this contributes $\widetilde O(q)$. On a $T$-step epoch its contribution
is $\widetilde O(T)$.
The source argument for charging actual changes of returned scores
also survives: its update threshold is separated from its estimation
error by a constant factor. This applies to Lemmas C.6--C.13 of~\cite{bllsssw21}
with those constant adjustments.

Lemma~\ref{lem:tall_lewis_epoch_cost} gives the total work of the
external and recursive scale changes, queries, and endpoint copies.
These statements prove Parts (b) and (c) with the stated sequence
hypotheses and failure allocation.
\end{proof}

A pending query can refer to an earlier column scale. Each recursive layer therefore keeps a history and a handle for the preconditioner at its endpoint.

\begin{definition}[Lewis-maintenance state]\label{def:tall_lewis_state}
The data structure stores $A,g,p,z$, the numerical vectors of every
recursive Lewis layer, and their leverage and HeavyHitter histories.
A layer at endpoint $t$ uses the full Gram matrix associated with
its scale at that endpoint. Its archive handle refers to a private
preconditioner for that same target, as in
Lemma~\ref{lem:tall_layer_target_coverage}. At initialization and after each Query, the numerical vectors satisfy
the accuracy guarantees of Algorithm 3 of~\cite{bllsssw21}; scale updates
are processed by its next Query.
An initialization starts every layer history at the current scale;
it does not reinitialize the external persistent inverse.
\end{definition}

Algorithm~\ref{alg:tall_lewis_maintain} records changes to individual scales as they arrive and processes the pending changes at the next weight query.

\begin{algorithm}[!ht]
\caption{\textsc{LewisMaintain} data structure}\label{alg:tall_lewis_maintain}
\begin{algorithmic}[1]
\DataStructure{LewisMaintain} \Comment{Theorem~\ref{thm:tall_lewis_maintain}}
\Members
\State $A\in\R^{d\times n}$; $g\in\R_{>0}^n$; parameters $p,z$ and source accuracy parameters.
\State Source recursive Lewis structure $\mathrm{layers}$ and private endpoint handles.
\EndMembers
\State
\Procedure{Init}{$A,g,\theta,P$} \Comment{Lemma~\ref{lem:tall_warm_lewis_initialization}}
\State Store $A,g$ and the parameters.
\State Refine $\theta$ and initialize every numerical layer using $P$ as in Lemma~\ref{lem:tall_warm_lewis_initialization}.
\State Initialize all layer histories at the current scale and record the current endpoint handle.
\EndProcedure
\State
\Procedure{UpdateG}{$i\in[n],\zeta\in\R_{>0}$} \Comment{\cite[Algorithm 3]{bllsssw21}}
\State $\mathrm{layers}.\Call{Scale}{i,\zeta}$; $g_i\gets\zeta$.
\EndProcedure
\State
\Procedure{Query}{$ $} \Comment{Lemma~\ref{lem:tall_lewis_epoch_cost}}
\State Bind each pending current or historical solve to its corresponding private endpoint handle.
\State \Return $\mathrm{layers}.\Call{Query}{}$. \Comment{\cite[Algorithm 3]{bllsssw21}}
\EndProcedure
\EndDataStructure
\end{algorithmic}
\end{algorithm}

\subsection{Regularized Lewis contraction}
\label{sec:tall_static_regularized_lewis_initialization}

To initialize the weights, we express the regularized Lewis equations as a contraction in logarithmic coordinates. The contraction also controls errors in the leverage score estimates used at each iteration.

\begin{lemma}[Regularized Lewis contraction]
\label{lem:tall_static_regularized_lewis_initialization}
Let $B\in\mathbb R^{n\times d}$ have full column rank, let $z>0$, and let $1/2\leq p\leq2$. Write $b_i^\top$ for row $i$ of $B$. For $w>0$, define
\[
 s_p:=2/p-1,\quad W:=\operatorname{Diag}(w),\quad
 H(w):=B^\top W^{-s_p}B,\quad
 \mathcal L_i(w):=w_i^{-s_p}b_i^\top H(w)^{-1}b_i+z_i.
\]
There is a unique positive $\tau=\mathcal L(\tau)$, with
$z\leq\tau\leq z+\mathbf1$ and $\sum_{i=1}^n\tau_i=d+\sum_{i=1}^n z_i$.
The map $F_i(w):=(b_i^\top H(w)^{-1}b_i+z_iw_i^{s_p})^{p/2}$ is a contraction in logarithmic infinity distance with factor $\rho:=1-p/2\leq3/4$.

More precisely, let $0<\epsilon<1/2$, and put
\[
 E_0:=\log(1+1/\min_i z_i),\qquad
 J:=\lceil4\log\max\{e,16E_0/\epsilon\}\rceil.
\]
Starting from $w^{(0)}:=z$, use estimates
$\widehat\ell^{(t)}\approx_{\epsilon/16}\mathcal L(w^{(t)})$
and updates $w_i^{(t+1)}:=((w_i^{(t)})^{s_p}\widehat\ell_i^{(t)})^{p/2}$.
Then $\bar\tau:=w^{(J)}$ satisfies
\[
 \bar\tau\approx_{\epsilon/8}\tau,\qquad
 \bar\tau\approx_\epsilon\mathcal L(\bar\tau).
\]
\end{lemma}
\begin{proof}

If $w\approx_h v$, then $W^{-s_p}\approx_{s_p h}V^{-s_p}$ since $s_p\geq0$.
Congruence by $B$ gives $H(w)\approx_{s_p h}H(v)$ in Loewner order.
Both matrices are positive definite.  Inverting reverses each Loewner
inequality and preserves the same two-sided multiplicative factor.
Consequently their quadratic forms at $b_i$ differ by at most that
factor.  The terms $z_iw_i^{s_p}$ and $z_iv_i^{s_p}$ satisfy the same comparison.
Adding the two nonnegative terms and taking the power $p/2$ therefore
gives
\[
 F(w)\approx_{\rho h}F(v),
 \qquad s_p p/2=1-p/2=\rho.
\]
Equivalently, $u\mapsto\log F(e^u)$ is a $\rho$-contraction of the
complete metric space $(\mathbb R^n,\|\cdot\|_\infty)$.
The contraction mapping theorem gives a unique positive solution $\tau$ to $\tau=F(\tau)$. The equation
$\tau_i=F_i(\tau)$, raised to the power $2/p=s_p+1$ and divided by
$\tau_i^{s_p}$, is precisely $\tau_i=\mathcal L_i(\tau)$; the converse is
the same algebra in reverse.  The first term of $\mathcal L_i(\tau)$ is
a leverage score of $\operatorname{Diag}(\tau)^{-s_p/2}B$.
Such scores lie in $[0,1]$ and sum to its rank $d$.  This proves the existence, uniqueness, and stated bounds for $\tau$.

Put $\eta:=\epsilon/16$.
Initialize $w^{(0)}:=z$ and update coordinatewise by
\[
 w_i^{(t+1)}:=((w_i^{(t)})^{s_p}\widehat\ell_i^{(t)})^{p/2}.
\]
The corresponding exact update is $F(w^{(t)})$, since
$w_i^{s_p}\mathcal L_i(w)=b_i^\top H(w)^{-1}b_i+z_iw_i^{s_p}$.
Write $e_t:=\|\log(w^{(t)}/\tau)\|_\infty$.
The contraction and the assumed estimation error give
\[
 e_{t+1}\leq\rho e_t+(p/2)\eta.
\]
The bounds on $\tau$ give $e_0\leq E_0$.  Summing this recurrence, and
using $1-\rho=p/2$, yields
\[
 e_J\leq\rho^J E_0+\eta(1-\rho^J)
 \leq e^{-J/4}E_0+\eta\leq\epsilon/8.
\]
This argument also covers $p=2$, when $\rho=0$.
Furthermore, $w\approx_h\tau$ implies
$\mathcal L(w)\approx_{2s_p h}\mathcal L(\tau)$: the factor $w_i^{-s_p}$
and the inverse quadratic form each contribute at most $s_p h$, and adding
$z_i$ preserves their comparison.  Hence
\[
 w^{(J)}\approx_{(1+2s_p)e_J}\mathcal L(w^{(J)}).
\]
Here $s_p\leq3$, so $(1+2s_p)e_J\leq7\epsilon/8<\epsilon$, as required.
\end{proof}

\subsection{Sparse numerical Lewis initialization}
\label{subsec:tall_lewis_maintenance_sparse_numerical_lewis_initialization}

Projecting random sign vectors gives the leverage score estimates needed for the weight iteration without forming the full projection matrix.

\begin{lemma}[Sparse static Lewis initialization by projected signs]
\label{lem:tall_sparse_static_lewis_initialization}
Under Lemma~\ref{lem:tall_static_regularized_lewis_initialization}, let $0<\delta<1/2$. Its two approximation guarantees can be obtained with probability at least $1-\delta$ in
\[
 \widetilde O(J\epsilon^{-2}(n+\nnz(B)+d^{\omega+o(1)}))
\]
arithmetic operations. The soft notation also hides logarithms of $1/\epsilon$ and $J/\delta$. At a supplied interior point, applying the lemma with $B=\operatorname{Diag}(g)A^\top$ and the source parameters therefore allows numerical Lewis initialization in $\widetilde O(n+\nnz(A)+d^{\omega+o(1)})$ work. This applies also to the augmentation in Lemma~\ref{lem:tall_sparse_centered_augmentation}, which adds $2d$ columns and $2d$ nonzeros. This is numerical initialization, not initialization of the dynamic observers.
\end{lemma}
\begin{proof}
We implement the approximate updates in Lemma~\ref{lem:tall_static_regularized_lewis_initialization}.
Fix the history of one round and put
\[
 C:=W^{-s_p/2}B,\qquad H:=C^\top C,\qquad
 P:=CH^{-1}C^\top,\qquad \sigma_i:=P_{ii},\qquad
 \theta:=\epsilon/64.
\]
These matrices are analytical notation: neither $H$ nor $P$ is formed.
The matrix $P$ is an orthogonal projection. Draw
$m=O(\theta^{-2}\log(4nJ/\delta))$ independent sign vectors
$r_j\in\{-1,1\}^n$. For each $j$, form $b_j:=C^\top r_j$ and use the
input-sparsity weighted least-squares solver in Lemma 8.1
of~\cite{bllsssw21} to compute $\widetilde v_j$ satisfying
\[
 v_j:=H^{-1}b_j,\qquad
 \|\widetilde v_j-v_j\|_H\leq\eta\|v_j\|_H,
 \qquad \eta:=\theta/(16\sqrt n).
\]
Allocate failure probability $\delta/(2Jm)$ to each solve.
If amplification is needed, repeat independently and retain the candidate
minimizing $\|Cv\|_2^2/2-b_j^\top v$. This objective differs from its
minimum by $\|v-H^{-1}b_j\|_H^2/2$, so selecting it preserves the best
energy-error guarantee; evaluation costs $O(n+\nnz(B)+d)$ per candidate.
The solver costs
$\widetilde O((\nnz(B)+d^{\omega+o(1)})\log(1/\eta))$ per call and does
not require an explicit dense Gram matrix. Form
$Z_{:,j}:=C\widetilde v_j/\sqrt m$ by sparse multiplication.

Since $\|v_j\|_H=\|Pr_j\|_2\leq\sqrt n$, Cauchy--Schwarz in the
$H$-norm gives, for every row $c_i^\top$,
\[
 |c_i^\top(\widetilde v_j-v_j)|
 \leq\sqrt{c_i^\top H^{-1}c_i}\|\widetilde v_j-v_j\|_H
 \leq(\theta/16)\sqrt{\sigma_i}.
\]
Consequently the rowwise Euclidean error between $Z$ and
$P[r_1\ \cdots\ r_m]/\sqrt m$ is at most
$(\theta/16)\sqrt{\sigma_i}$. The fixed-vector sign-JL bound, applied
to all rows of $P$ conditionally on the round's history, gives
\[
 \|e_i^\top P[r_1\ \cdots\ r_m]/\sqrt m\|_2
 \in[(1-\theta/16)\sqrt{\sigma_i},
      (1+\theta/16)\sqrt{\sigma_i}]
\]
simultaneously, except with probability $\delta/(2J)$, after choosing
the absolute constant in $m$ sufficiently large. Thus
$\|Z_{i,:}\|_2\in[(1-\theta/8)\sqrt{\sigma_i},
(1+\theta/8)\sqrt{\sigma_i}]$. Zero rows have zero error and zero
estimate. Squaring, truncating at one, and adding $z_i$ preserves a
logarithmic error at most $\theta$ in $\mathcal L_i(w)$.

The contraction recurrence in
Lemma~\ref{lem:tall_static_regularized_lewis_initialization} therefore
applies with an error no larger than the one used there. Fresh signs and
solver randomness in every round, followed by the tower property and a
union bound, give both claimed output guarantees with probability
$1-\delta$ even though later weights depend on earlier estimates.

Generating the signs and computing row norms cost $O(nm)$. Forming all
$b_j$ and $C\widetilde v_j$ costs $O(\nnz(B)m)$; all $m$ solves have the
stated total cost. Diagonal scaling preserves the nonzero pattern and
needs $O(n+\nnz(B))$ work. Multiplying the per-round cost by $J$ proves the
bound. For source accuracy, $J$ and $\epsilon^{-1}$ are polylogarithmic.
The Phase-I matrix consists of $A$ and two signed diagonal blocks, so it
has $d$ rows, $n+2d$ columns, and $\nnz(A)+2d$ nonzeros. This proves
the initialization specialization without a dense Hadamard pass.
\end{proof}
\subsection{Approximate solves inside a fresh JL batch}
\label{subsec:tall_lewis_maintenance_approximate_solves_inside_a_fresh_jl_batch}

The approximate inverse can depend on the random right-hand sides in the same JL batch. Comparing it with the exact inverse for each draw bounds the error despite this dependence.

\begin{lemma}[Pathwise JL perturbation bound]
\label{lem:tall_pathwise_jl_perturbation}
Let $C=VA^\top\in\mathbb R^{n\times d}$ have full column rank,
$G_0:=C^\top C$, and $G:=L^\top L$ satisfy
$c^{-1}G_0\preceq G\preceq cG_0$ for a constant $c$.
Let $R\in\mathbb R^{m\times r}$ satisfy $\|R\|_F\leq\sqrt n$,
where $L$ has $m$ rows, and put $B:=L^\top R$.
Suppose $Y=MB$ and
$\|G^{1/2}MG^{1/2}-I\|_{\rm op}\leq\delta$, without any independence
assumption between $M$ and $R$. Then
\[
 \|G^{1/2}(Y-G^{-1}B)\|_F\leq\delta\sqrt n,
 \qquad
 \|e_i^\top C(Y-G^{-1}B)\|_2
 \leq\delta\sqrt{cn\sigma_i(C)}.
\]
If $Z\succeq\rho I$ is diagonal, $F$ is a diagonal zero--one mask,
and coordinate costs equal $d$, then
\[
 \|\sqrt d Z^{-1/2}FC(Y-G^{-1}B)\|_F
 \leq\sqrt c\delta n.
\]
For a difference of two endpoint products, the corresponding error
is at most $2\sqrt c\delta n$.
\end{lemma}

\begin{proof}
The matrix $G^{-1/2}L^\top$ has orthonormal rows. Therefore
$\|G^{-1/2}B\|_F\leq\|R\|_F\leq\sqrt n$.
Multiplying by $G^{1/2}MG^{1/2}-I$ proves the first inequality.
For the second, apply Cauchy--Schwarz to each row and use
$e_i^\top CG^{-1}C^\top e_i\leq c\sigma_i(C)$.
Finally,
$\|\sqrt d Z^{-1/2}FCG^{-1/2}\|_{\rm op}\leq\sqrt{cd/\rho}
=\sqrt{cn}$, which proves the Frobenius estimate.
The endpoint assertion follows from the triangle inequality.
All statements are pathwise, including when $M$ is selected after $R$.
\end{proof}

\subsection{Warm initialization}
\label{subsec:tall_lewis_maintenance_warm_initialization}

At an epoch restart, we can initialize the recursive layers from a coarse weight estimate and the existing private preconditioner.

\begin{lemma}[Warm Lewis initialization]
\label{lem:tall_warm_lewis_initialization}
Suppose $\theta\approx_{\epsilon_0}\tau(g)$ with an absolute
$\epsilon_0<1/10$, and a private $P$ is a constant-quality
preconditioner for
$H_0:=A\operatorname{Diag}(\tau(g)^{1-2/p}g^2)A^\top$.
For inverse-polylogarithmic $\epsilon_L>0$, all numerical Lewis
vectors and all leverage/HeavyHitter structures in a fresh copy of
Algorithm 3 of~\cite{bllsssw21} can be initialized in
$\widetilde O(\nnz(A)+n+d^2)$ work, with prescribed inverse-polynomial
failure and coupling errors. This does not reinitialize $P$.
\end{lemma}

\begin{proof}
Use the contraction
\[
 v_{{\rm tra},i}:=(v_i^{2/p-1}\widehat\ell_i)^{p/2},\qquad
 \ell(v):=\sigma(\operatorname{Diag}(v^{1/2-1/p}g)A^\top)+z.
\]
If $v\approx_e\ell(v)$ and $\widehat\ell\approx_\xi\ell(v)$,
Lemma 5.3 of~\cite{bllsssw21} gives a new residual at most
$(1-p/2)e+(2-p/2)\xi$. Here $1-p/2\leq3/4$.
The initial residual is $O(\epsilon_0)$: coordinatewise closeness
of $v$ and $\tau(g)$ compares their column weights and Gram matrices,
and hence their regularized leverage scores. Take
$\xi=\epsilon_L/C$ and perform $O(\log(1/\epsilon_L))$ rounds.
The residual becomes $O(\epsilon_L)$, and the intermediate iterates
remain within a fixed constant factor of $\tau(g)$; this follows
also by summing the geometrically decreasing step bounds in
Lemma 5.3. Thus every round's full Gram is preconditioned by $P$.

Let $q\geq1$ bound the number of batches covered by the failure allocation,
and let $\beta\in(0,1/4)$ be its total failure budget.
For a round's factor $C\in\R^{n\times d}$, draw
$R\in\R^{n\times r}$ with independent entries in
$\{-1/\sqrt r,1/\sqrt r\}$, where
$r=O(\xi^{-2}\log(nq/\beta))$. Then $\|R\|_F^2=n$.
Form $B=C^\top R$ and use
Lemma~\ref{lem:tall_private_preconditioner_batch} with $L=C$.
Choose its relative error $\delta\leq\xi/(C\sqrt n)$.
The row bound in Lemma~\ref{lem:tall_pathwise_jl_perturbation}, followed
by the sign-JL estimate for the exact projection, yields all
$\widehat\ell_i$ to the required multiplicative accuracy. Add $z_i$
after squaring the row norms; zero leverage rows cause no problem.
Forming $B$ and $CY$ costs $O(\nnz(A)r)$, and the solve costs
$\widetilde O(\nnz(A)+n+d^2)$. Thus the complete contraction has that
soft cost. Its randomness is fresh between rounds.

Algorithm 4, line 16, of~\cite{bllsssw21} uses a fixed
constant for the inter-query change parameter in its Algorithm 3.
The number of recursive Lewis layers, which is logarithmic in that
parameter, is therefore an absolute constant. Initializing each layer uses the
same projected-sign computation and $O(\log(2q))$ HeavyHitter
instances. Their bulk initialization costs $\widetilde O(\nnz(A)+n)$,
not $n$ coordinate calls of cost $d$. All initial layer weights are
constant-close to $\tau(g)$, so $P$ also covers these solves.
Allocate failure and coupling errors by a union bound over the
polylogarithmic number of batches. Only responses, not $P$ or
intermediate iteration products, are released.
\end{proof}

\subsection{Recursive updates and query costs}
\label{subsec:tall_lewis_maintenance_recursive_updates_and_query_costs}

Once the layers are initialized, we bound the cost of their updates, queries, and archive copies over one epoch in terms of the weighted number of changes to the input scales.

\begin{lemma}[Lewis maintenance work]\label{lem:tall_lewis_epoch_cost}
Use the general-matrix HeavyHitter of~\cite[Lemma B.1]{bllsssw21},
with coordinate costs $c_i=\widetilde O(d)$,
$C:=\sum_i c_i=\widetilde O(nd)$, and query cost $Q=\widetilde O(d)$. Set $k_i:=\nnz(a_i)$.
Suppose the epoch starts with freshly initialized layer histories,
contains $T$ valid short steps, and has external weighted scale
charge $A_g=\widetilde O(T^2)$.
Use the endpoint preconditioners of
Lemma~\ref{lem:tall_layer_target_coverage} and batched solves with
sparse-target overhead $\Psi=\widetilde O(d^2+n)$.
Excluding initialization and actual persistent inverse updates,
the recursive layer updates, queries, and endpoint copies cost
$\widetilde O(T(d^2+n)+nT^2)$ in expectation, stopped at the first
failed invariant.
\end{lemma}
\begin{proof}
Here is the internal-layer charge, which must not be omitted.
Let $A_j$ denote the total leverage-weighted number of changes to
layer $j$'s scale vector, and $A_g$ the corresponding input-scale
charge. Lemma C.12 and the proof of Theorem 5.2 of~\cite{bllsssw21} give
\[
 A_1=\widetilde O(A_g+T^2),\qquad
 A_{j+1}\leq\poly \log(n)(A_j+A_g).
\]
Let $L_{\rm lay}=O(1)$ be the number of layers. Then
$\sum_{j=1}^{L_{\rm lay}}A_j=\widetilde O(A_g+T^2)=\widetilde O(T^2)$.
In the first bound, the $T^2$ term is the thresholded feedback of
the last layer into the first, not a new independent assumption.
The terms accounting for forced-in columns of $A$ in historical samples are
included in Lemmas C.12--C.13; they are charged to these same
weighted changes. Each layer's Scale cost is
$\widetilde O((C/d)\sigma_i+c_i)$, hence
$\widetilde O(n\tau_i)$. Consequently all recursive Scale calls
also cost $\widetilde O(nT^2)$.

Use Lemmas C.11 and C.13 of~\cite{bllsssw21} before the restart optimization in
Theorem C.2, and Eqs.~(31)--(34) in the proof of Theorem 5.2.
In our dimensions each of C.11 and C.13 charges a query
$\widetilde O(\Psi+d\max_i k_i+(C/d)T+Q)$, and charges a changed
column of $A$ $\widetilde O(c_i+(C/d)\sigma_i)$.
Substitute $C=\widetilde O(nd)$, $k_i\leq d$, and $Q=\widetilde O(d)$.
With sparse-target
solve overhead $\Psi=\widetilde O(d^2+n)$, their remaining query
cost over the block is
$\widetilde O(T(d^2+n)+nT^2)$, together with the already charged
forced-column products. Lemma~\ref{lem:tall_layer_target_coverage} supplies
the required endpoint preconditioner, and
Lemma~\ref{lem:tall_pathwise_jl_perturbation} justifies the use of
approximate, right-hand-side-dependent batch operators.
No cold arbitrary-weight inverse initializer is invoked inside
the epoch. Copies of all required inverse and scale histories
cost $\widetilde O(T(d^2+n))$.

\end{proof}

\section{Proportional sampling data structure}\label{sec:tall_proportional_sampling}

In Section~\ref{subsec:tall_proportional_sampling_main_data_structure}, we specify the sampling probabilities and the quantities maintained by \textsc{SparseSampler}.
In Section~\ref{subsec:tall_proportional_sampling_initialization_updates_and_tree_sampling}, we initialize and update a private sketch tree for adaptive sampling queries.
In Section~\ref{subsec:tall_proportional_sampling_sparsity_balanced_proportional_sampling}, we use the tree to implement proportional sampling and bound the work in terms of the column sparsities.

\subsection{Main data structure}
\label{subsec:tall_proportional_sampling_main_data_structure}

\textsc{SparseSampler} maintains column scales and Lewis masses so that it can sample for a new direction on each call. The sketches in its sparse JL tree remain private.

\begin{theorem}[SparseSampler]\label{thm:tall_sparse_sampler}
Use the dimensions, parameters, valid input conditions, horizon,
and update cap of Lemma~\ref{lem:tall_sparsity_balanced_sampler},
with $\eta\in[d^{-1/2},1]$.
There is a randomized data structure
(\textnormal{\textsc{SparseSampler}} in
Algorithms~\ref{alg:tall_sparse_sampler_update} and~\ref{alg:tall_sparse_sampler_sample})
that maintains fixed $A\in\R^{d\times n}$, a positive diagonal
matrix $C$, and a positive Lewis estimate $\bar\tau\in\R^n$.
Write $D:=\operatorname{Diag}(\bar\tau)^{-1}C$.
It supports the following procedures.
\begin{enumerate}[label=(\alph*)]
\item \textnormal{\textsc{Init}}$(A,C,\bar\tau)$ initializes the
sparse JL tree for $DA^\top$ and the scalar sum tree for $\bar\tau$
in $\widetilde O(n+\nnz(A))$ time.
\item \textnormal{\textsc{Update}}$(i\in[n],c\in\R_{>0},\theta\in\R_{>0})$
sets $C_{ii}=c$ and $\bar\tau_i=\theta$, updates $D_{ii}$ and both
trees, and takes $\widetilde O(1+\nnz(a_i))$ time.
\item \textnormal{\textsc{Sample}}$(v\in\R^d)$, for a valid supplied
direction, produces the sparse diagonal sample $R$, its actual
update list $h=-CRDA^\top v$, and the product $Ah$.
The complete released outputs have the proportional law and
adaptive coupling guarantee of Lemma~\ref{lem:tall_sparsity_balanced_sampler}.
\end{enumerate}
For at most $q$ sample calls and coordinate-update charge $W$, the
expected total work is
$\widetilde O(n+\nnz(A)+W+q(n/\eta+\eta(n+\nnz(A))+d^2))$.
The supplied directions and Lewis estimates are computed outside
this data structure. The choices of sketches, tree paths, and
intermediate estimates remain private.
\end{theorem}

\begin{proof}
Lemma~\ref{lem:tall_adaptive_jl_nonleakage} proves the sparse tree's
initialization, coordinate-update, and adaptive sampling guarantees.
A scalar sum tree initializes in $O(n)$ time and changes one mass
in $O(\log n)$ time. These statements give Parts (a) and (b).
Lemma~\ref{lem:tall_sparsity_balanced_sampler} proves Part (c),
including the three proposal branches, exact selected values,
actual update lists, residual products, and total expected work.
Its coupling applies to the complete released response, so it also
covers the outputs of the procedures in the displayed algorithms.
\end{proof}

The sampler uses one tree to sample according to squared directional entries and another to sample according to Lewis masses. Both trees must reflect the current scales.

\begin{definition}[Sparse-sampling invariant]\label{def:tall_sparse_sampler_state}
The stored diagonal scales satisfy $D_{ii}=C_{ii}/\bar\tau_i$.
For every node with interval $I_v$, its private cache is
$Q_v=J_vD_{I_v,I_v}(A_{:,I_v})^\top$,
with the sparse support representation of
Lemma~\ref{lem:tall_adaptive_jl_nonleakage}.
The scalar sum tree stores the masses $\bar\tau_i$ and their subtree
sums. Both trees represent the same current coordinate scales.
\end{definition}

A change at coordinate $i$ affects only the path from leaf $i$ to the root in each tree. Algorithm~\ref{alg:tall_sparse_sampler_update} updates the sketches and mass sums along these paths.

\begin{algorithm}[!ht]
\caption{\textsc{SparseSampler} data structure: initialization and updates}\label{alg:tall_sparse_sampler_update}
\begin{algorithmic}[1]
\DataStructure{SparseSampler} \Comment{Theorem~\ref{thm:tall_sparse_sampler}}
\Members
\State $A\in\R^{d\times n}$; positive diagonal $C,D$; $\bar\tau\in\R_{>0}^n$.
\State Fixed sampling parameters; private sparse JL tree and scalar sum tree.
\EndMembers
\State
\Procedure{Init}{$A,C,\bar\tau$} \Comment{Lemma~\ref{lem:tall_adaptive_jl_nonleakage}}
\State Store $A,C,\bar\tau$; $D\gets\operatorname{Diag}(\bar\tau)^{-1}C$.
\State Draw the node sketches and initialize all sparse caches $Q_v$.
\State Initialize the scalar sum tree with leaf masses $\bar\tau_i$.
\EndProcedure
\State
\Procedure{Update}{$i\in[n],c\in\R_{>0},\theta\in\R_{>0}$} \Comment{Lemma~\ref{lem:tall_adaptive_jl_nonleakage}}
\State $\Delta\gets c/\theta-D_{ii}$.
\For{every node $v$ on the path from leaf $i$ to the root}
\State $Q_v\gets Q_v+(J_ve_{\operatorname{loc}_v(i)})\Delta a_i^\top$.
\EndFor
\State Change leaf mass $i$ to $\theta$ and update its ancestor sums in the scalar tree.
\State $C_{ii}\gets c$; $\bar\tau_i\gets\theta$; $D_{ii}\gets c/\theta$.
\EndProcedure
\EndDataStructure
\end{algorithmic}
\end{algorithm}

Algorithm~\ref{alg:tall_sparse_sampler_sample} combines directional, uniform, and Lewis-mass samples and computes the product $Ah$ from the selected columns.

\begin{algorithm}[!ht]
\caption{\textsc{SparseSampler} data structure: sampling}\label{alg:tall_sparse_sampler_sample}
\begin{algorithmic}[1]
\DataStructure{SparseSampler} \Comment{Theorem~\ref{thm:tall_sparse_sampler}}
\Procedure{Sample}{$v\in\R^d$} \Comment{Lemma~\ref{lem:tall_sparsity_balanced_sampler}}
\State $U^{\rm dyn}\gets eC_1n/(\eta d)$; $U^{\rm unif}\gets C_2\eta n$.
\State $U^{\rm lev}\gets C_3\gamma^{-2}\log n\sum_i\bar\tau_i$.
\State Choose $S\ge U^{\rm dyn}+U^{\rm unif}+U^{\rm lev}$ with padding less than $1/C_0$ and $N:=C_0S$ integral.
\State Initialize empty sparse lists $R,h$ and residual accumulator $r={\bf 0}_d$.
\For{$j=1,\ldots,N$}
\State Choose the dynamic, uniform, leverage, or padding branch with probabilities $U^{\rm dyn}/S,U^{\rm unif}/S,U^{\rm lev}/S$, and the remaining mass.
\State Obtain an index $i$ or the empty outcome from that branch as in Lemma~\ref{lem:tall_sparsity_balanced_sampler}.
\If{an index $i$ is returned}
\State $u_i\gets D_{ii}a_i^\top v$; $p_i\gets C_1u_i^2/\eta+C_2\eta+C_3\bar\tau_i\gamma^{-2}\log n$.
\State Append $(i,1/(C_0p_i))$ to $R$ and $(i,-C_{ii}u_i/(C_0p_i))$ to $h$.
\State $r\gets r-a_iC_{ii}u_i/(C_0p_i)$.
\EndIf
\EndFor
\State \Return $(R,h,r)$.
\EndProcedure
\EndDataStructure
\end{algorithmic}
\end{algorithm}

\subsection{Initialization, updates, and tree sampling}
\label{subsec:tall_proportional_sampling_initialization_updates_and_tree_sampling}

The next direction may depend on earlier sampled indices. To justify reusing the sketches, we compare the entire sequence of replies with draws from the prescribed sampling law.

\begin{lemma}[Adaptive non-leaking sparse JL tree]
\label{lem:tall_adaptive_jl_nonleakage}
Let $A\in\mathbb R^{d\times n}$, $n\geq2$, be fixed. At each of at most $M$ calls, a public input supplies a column scale $g\in\R^n$, a direction $v\in\R^d$, and $\mathcal{B}>0$ with $\mathcal{B}\geq e\|\operatorname{Diag}(g)A^\top v\|_2^2$. Inputs and scale changes may depend on earlier released outputs but not on private preprocessing or traces. Let $N_{\rm sc}$ be the number of reported scale changes, and let $i_e$ be the coordinate changed by event $e$. Define
\[
 W:=\sum_{e=1}^{N_{\rm sc}}(1+\nnz(a_{i_e})).
\]
There is a data structure whose complete output is an index $i$, together with its exact value, or $\emptyset$. Its joint transcript can be coupled, with error at most $\delta$, to the exact adaptive kernel
\[
 K((g,v,\mathcal{B}),i):=\frac{(\operatorname{Diag}(g)A^\top v)_i^2}{\mathcal{B}},\qquad
 K((g,v,\mathcal{B}),\emptyset):=1-\frac{\|\operatorname{Diag}(g)A^\top v\|_2^2}{\mathcal{B}}.
\]
Initialization, all reported scale changes, and $N\leq M$ traversals cost $\widetilde O(n+\nnz(A)+W+dN)$, with logarithmic dependence on $M/\delta$. Node estimates, paths, and data-dependent timing remain private. Operations are capped to remain defined and bounded on bad preprocessing objects.
\end{lemma}
\begin{proof}
Pad $A$ with zero columns so that their number is a power of two,
and form a dyadic tree on these columns with height $h$. Number the root by $1$ and the children of node $v$ by $2v$ and $2v+1$. Put $\zeta:=1/(8h)$. For each node with column interval $I_v$, draw an independent sign-JL matrix $J_v$ with $k=O(\zeta^{-2}\log(nM/\delta))$ rows and store
\[
 Q_v:=J_v\operatorname{Diag}(g_{I_v})(A_{:,I_v})^\top.
\]
Store only the columns of $Q_v$ indexed by
$S_v:=\bigcup_{i\in I_v}\operatorname{supp}(a_i)\subseteq[d]$.
At each level $\ell\in\{0,\ldots,h\}$, $\sum_{v=2^\ell}^{2^{\ell+1}-1}|S_v|\leq \nnz(A)$. Sparse additions of
the transposed input columns initialize all $Q_v$ in $\widetilde O(n+\nnz(A))$ work and space, including the signs and tree. A change $\Delta g_i$ updates every ancestor by
\[
 Q_v\gets Q_v+(J_ve_{\operatorname{loc}_v(i)})\Delta g_i a_i^\top.
\]
This costs $\widetilde O(1+\nnz(a_i))$ and preserves each cache exactly.

Fix one input independently of the sketches and put $y:=\operatorname{Diag}(g)A^\top v$. A fixed-vector JL bound and a union bound over nodes show, except with probability $\delta/M$, that $\|Q_vv\|_2^2\approx_\zeta\|y_{I_v}\|_2^2$ at every node. Start at the root and choose each child with probability proportional to its squared sketched norm. If both norms vanish, return $\emptyset$. For a reached leaf $i$, let $p_i$ be the product of the branching probabilities. Each branching ratio is within $e^{2\zeta}$ of its exact energy ratio. Multiplying along the path telescopes and gives, for $y_i\ne0$,
\[
 e^{-1/4}\frac{y_i^2}{\|y\|_2^2}\leq p_i\leq e^{1/4}\frac{y_i^2}{\|y\|_2^2}.
\]
Compute $y_i$ exactly and accept with probability $y_i^2/(\mathcal{B}p_i)$, which is at most one. The probability of releasing $i$ is exactly $y_i^2/\mathcal{B}$ for every fixed good sketch. The remaining outcome is $\emptyset$. For $y=0$ the outcome is always empty. Each traversal costs $\widetilde O(d)$, including the exact value $g_i a_i^\top v$. On a bad sketch clip probabilities to $[0,1]$ and stop after one tree path.

To justify adaptive reuse, construct an ideal execution from the prescribed kernels using randomness independent of all sketches. Its next input, including all preceding scale changes, is independent of the sketches. Consequently the probability that its $j$th input violates a nodewise JL bound is at most $\delta/M$. A union bound over its at most $M$ inputs gives failure probability at most $\delta$. Couple the implemented and ideal responses identically whenever their preceding transcripts agree and the current input is good: their complete released kernels then coincide. Their first disagreement is contained in the first bad ideal-input event. This proves the adaptive statement without conditioning on future success. Exact postprocessing of released values preserves it. Private traces must not enter any later input. Summing the deterministic work charges gives the stated bound.
\end{proof}

\subsection{Sparsity-balanced proportional sampling}
\label{subsec:tall_proportional_sampling_sparsity_balanced_proportional_sampling}

Directional and uniform samples contribute differently to the cost of multiplying by sparse columns. We choose their relative weights to balance these costs.

\begin{lemma}[Sparsity-balanced proportional sampler]
\label{lem:tall_sparsity_balanced_sampler}
Let $A\in\mathbb R^{d\times n}$ have no zero columns and $n\geq d\geq2$. Work in exact real arithmetic.
Put $k_i:=\nnz(a_i)$. At every active step the public
pre-seed state supplies $C_t$, $D_t:=\bar T_t^{-1}C_t$,
$\bar\tau_t$, and $v_t$, with
\[
 u_t:=D_tA^\top v_t,\qquad \|u_t\|_2^2\leq n/d,
 \qquad \sum_{i=1}^n\bar\tau_{t,i}=O(d).
\]
Assume $C_t$ and $\bar T_t:=\operatorname{Diag}(\bar\tau_t)$ are positive diagonal and $\bar\tau_t$ dominates, up to an absolute constant, the leverage scores required in Definition 4.13 of~\cite{bllsssw21}. Use its source validity parameter and inverse-polylogarithmic $\gamma$.
Index the reported scale changes by $e=1,\ldots,N_{\rm sc}$, and let $i_e$ be the changed coordinate. Let $W$ be a deterministic cap on
$\sum_{e=1}^{N_{\rm sc}}(1+k_{i_e})$, and stop before
exceeding it. The directions may change arbitrarily and adaptively
between steps. They need not have a common span known in advance,
small increments, or small total energy.
\begin{itemize}[leftmargin=2.1em,labelwidth=1.5em,labelsep=0.6em]
\item[(a)]\phantomsection\label{lem:tall_sparsity_balanced_sampler_a}
For any fixed $d^{-1/2}\leq\eta\leq1$, replace the source masses by
\[
 p_{t,i}:=\frac{C_1}{\eta}(u_t)_i^2+C_2\eta
            +C_3\bar\tau_{t,i}\gamma^{-2}\log n.
\]
The fresh proportional law for these masses is $C_{\rm valid}$-valid.
Over at most $q$ steps, it has an exact-output adaptive implementation
with whole-execution coupling error at most $\delta_{\rm JL}$ and
expected work
\[
 \widetilde O(n+\nnz(A)+W
       +q(n/\eta+\eta(n+\nnz(A))+d^2)).
\]
This includes all proposals, exact selected values, the actual lists
$h_t:=-C_tR_tu_t$, their explicit products $Ah_t$, and scale
updates. The total list length satisfies, pathwise,
\[
 L_h\leq\widetilde O(q(n/(\eta d)+\eta n+d)).
\]
The computation of the supplied directions, Lewis weights, and
inverse actions is excluded.

\item[(b)]\phantomsection\label{lem:tall_sparsity_balanced_sampler_b}
Choose $\eta:=\sqrt{n/\nnz(A)}$. For
$q=\widetilde O(\sqrt d)$ the complete sampler costs
\[
 \widetilde O(n+W+\sqrt{\nnz(A) nd}+d^{5/2}),
 \qquad L_h=\widetilde O(n+\sqrt{\nnz(A) nd}+d^{3/2}).
\]
The dynamic and uniform branches, including their exact column dot products and residual contributions, cost $\widetilde O(\sqrt{\nnz(A) nd})$; the $d^{5/2}$ term pays for direct actions on the leverage branch. The coordinate-update charge $W$ is retained explicitly.
\end{itemize}
This is a supplied-interface result for a changed sampling law.
It neither constructs the startup, solve, or Lewis interfaces nor
transfers a runtime guarantee for those interfaces from the old
trajectory to the new one.
\end{lemma}

\begin{proof}
{\bf Proof of (a).}
For every real $z$ and every $\eta>0$, the arithmetic--geometric
mean inequality gives
\[
 C_1z^2/\eta+C_2\eta\geq2\sqrt{C_1C_2}|z|.
\]
The constants in Corollary 4.43 of~\cite{bllsssw21} therefore give
the same domination of $|z|$ as for its choice $\eta=d^{-1/2}$.
The leverage term is unchanged. In particular
$p_{t,i}\geq |(u_t)_i|+\sigma_{t,i}$, where $\sigma_{t,i}$ is the
leverage score required by Definition 4.13 of~\cite{bllsssw21}.
Lemma 4.42 there proves all five validity conditions for the new
proportional law. This is not an assertion that the old law is
unchanged.

Maintain the sparse JL tree for $D_tA^\top$ from
Lemma~\ref{lem:tall_adaptive_jl_nonleakage}, and a scalar sum tree for
$\bar\tau_t$. At a fixed step define
\[
 U^{\rm dyn}:=\frac{eC_1n}{\eta d},\qquad
 U^{\rm unif}:=C_2\eta n,\qquad
 U_t^{\rm lev}:=C_3\gamma^{-2}\log n\sum_{i=1}^n\bar\tau_{t,i}.
\]
Let $S_t$ be their sum, enlarged by less than $1/C_0$ so that
$N_t:=C_0S_t$ is an integer. The enlargement is empty-outcome mass.
Generate $N_t$ independent proposals. Choose each branch with
probability its upper mass divided by $S_t$; any padding returns
empty. On the dynamic branch call the exact-output tree with
denominator $en/d\geq e\|u_t\|_2^2$. On the uniform branch choose
an index uniformly from $[n]$. On the leverage branch use the scalar
tree to choose $i$ with probability
$\bar\tau_{t,i}/\sum_{j=1}^n\bar\tau_{t,j}$. Thus one proposal returns
index $i$ with probability exactly $p_{t,i}/S_t$.

For a returned index compute $(u_t)_i$ exactly, using the value
already obtained by a dynamic traversal or by computing $D_{t,ii}a_i^\top v_t$
on either other branch. Its contribution to $R_{t,ii}$ is
$1/(C_0p_{t,i})$. Summing the independent proposals is precisely
the law of Lemma 4.42 of~\cite{bllsssw21}. Accumulate the scalar
contribution $-C_{t,ii}(u_t)_i/(C_0p_{t,i})$ in the actual sparse
list and multiply it by $a_i$ into a length-$d$ residual accumulator.
Combining repetitions is optional. There is no uncharged
selected-value or matrix-product oracle.

The sparse tree and scalar tree cost
$\widetilde O(n+\nnz(A)+W)$ to initialize and maintain. A dynamic
traversal, including its exact value and residual contribution,
costs $\widetilde O(d)$. Conditional on the public history before
this step, the expected number of these traversals is
\[
 N_t U^{\rm dyn}/S_t=C_0U^{\rm dyn}
 =\widetilde O(n/(\eta d)).
\]
Thus their expected work is $\widetilde O(n/\eta)$ per step.
A uniform-branch occurrence of column $i$ costs
$\widetilde O(1+k_i)$, including its exact action, scalar output,
and residual contribution. Its expected multiplicity is
$C_0C_2\eta$, so the total uniform work per step is
\[
 \widetilde O(\eta\sum_{i=1}^n(1+k_i))
 =\widetilde O(\eta(n+\nnz(A))).
\]
Likewise the leverage branch costs
\[
 \widetilde O(\sum_{i=1}^n(1+k_i)\bar\tau_{t,i})
 \leq\widetilde O(d^2)
\]
per step, using $k_i\leq d$ and $\sum_{i=1}^n\bar\tau_{t,i}=O(d)$.
All scalar proposal processing is covered by these bounds; reading
$v_t$ and initializing the residual accumulator costs $O(d)$ and
is also covered. Summing conditional expectations proves the work
bound without conditioning the sampling coins on future success.
Index the executed steps by $t=1,\ldots,q_{\rm run}$, where $q_{\rm run}\leq q$. The list length is at most $\sum_{t=1}^{q_{\rm run}}N_t$, which proves its pathwise
bound.

All approximate node norms, paths, and timings remain private.
For each good fixed preprocessing object the complete released
response has the prescribed exact proportional kernel, independent
of that object. The adaptive non-leakage coupling in
Lemma~\ref{lem:tall_adaptive_jl_nonleakage} therefore applies even when
every next direction depends on previous outputs. The declared
horizon, scale-update cap, and scalar mass bounds give a polynomial
worst-case work bound, since each proposal costs at most
$\widetilde O(d)$. Choose the JL failure probability below both
$\delta_{\rm JL}$ and the reciprocal of that bound. Failed coupling
then adds at most one to expected work, and changes only logarithmic
factors. Use bounded, capped probabilities to define tree operations
also on a bad preprocessing object.

\medskip
{\bf Proof of (b).}
If $n\leq \nnz(A)\leq nd$, then $d^{-1/2}\leq\eta\leq1$ and
\[
 n/\eta=\eta \nnz(A)=\sqrt{n\nnz(A)},\qquad
 \eta n\leq\sqrt{n\nnz(A)},\qquad
 \nnz(A)\leq\sqrt{\nnz(A) nd}.
\]
No square-root primitive is necessary: start at $\eta=1$ and halve
it while $\nnz(A)(\eta/2)^2\geq n$. The resulting dyadic value lies
between $\sqrt{n/\nnz(A)}$ and twice that value, still lies in the
permitted interval, and gives the same bounds within absolute factors.
Part~\hyperref[lem:tall_sparsity_balanced_sampler_a]{(a)} with
$q=\widetilde O(\sqrt d)$ proves the cost. Its list bound follows
from
\[
 qn/(\eta d)=\widetilde O(\sqrt{n\nnz(A)/d}),\qquad
 q\eta n=\widetilde O(n\sqrt{nd/\nnz(A)})
          \leq\widetilde O(\sqrt{\nnz(A) nd}).
\]
In particular the additional uniform samples are actually processed
and do not retain the old, smaller list bound by assumption.

\end{proof}
\section{Primal and gradient data structure}\label{sec:tall_primal_gradient}

In Sections~\ref{subsec:tall_primal_gradient_main_data_structure} and~\ref{subsec:tall_primal_gradient_initialization}, we specify and initialize the state used for primal and gradient queries.
In Sections~\ref{subsec:tall_primal_gradient_entry_update} and~\ref{subsec:tall_primal_gradient_accuracy_update}, we update entries and their accuracy scales.
In Sections~\ref{subsec:tall_primal_gradient_gradient_product_query} and~\ref{subsec:tall_primal_gradient_primal_sum_query}, we compute the gradient product and report changes to the accumulated primal vector.
In Section~\ref{subsec:tall_primal_gradient_exact_sum_and_potential_queries}, we recover the exact primal vector and query the potential.

\subsection{Main data structure}
\label{subsec:tall_primal_gradient_main_data_structure}
The data structure \textnormal{\textsc{PrimalGradient}} maintains the gradient product and
accumulated primal increments used in Section~\ref{sec:tall_sparse_lp}.
It uses the sum observer of~\cite[Lemma D.3]{bllsssw21}.

\begin{theorem}[PrimalGradient]\label{thm:tall_primal_gradient}
Fix $A\in\R^{d\times n}$, $n\geq d\geq2$, $\rho=d/n$, and
$0<\alpha<1/4$. Fix the potential parameter $\lambda>0$ and the norm
constant $C_{\rm norm}>0$, and write
$\Psi(r):=\sum_i\cosh(\lambda r_i)$ and
$\|h\|_{\theta+\infty}:=\|h\|_\infty+C_{\rm norm}\|h\|_\theta$.
There is a deterministic data structure
(\textnormal{\textsc{PrimalGradient}} in Algorithms~\ref{alg:tall_primal_gradient_updates}
and~\ref{alg:tall_primal_gradient_queries}) that maintains signed scales
$g\in\R^n$, residuals $r\in[-1,1]^n$, weights $\rho\leq\theta_i\leq4$,
positive accuracy scales $e\in\R_{>0}^n$, and the invariant in
Definition~\ref{def:tall_primal_gradient_invariant}.
It supports the following procedures.
\begin{enumerate}[label=(\alph*)]
\item \textnormal{\textsc{Init}}$(A,x^{\rm init},g,\theta,r,e,\alpha)$
sets the accumulated primal vector to $x^{\rm init}\in\R^n$ and takes
$\widetilde O(n+\nnz(A))$ time.
\item \textnormal{\textsc{Update}}$(i\in[n],g^{\rm new}\in\R,
\theta^{\rm new}\in[\rho,4],r^{\rm new}\in[-1,1])$
replaces $g_i,\theta_i,r_i$ by the supplied values in
$\widetilde O(1+\nnz(a_i))$ amortized time.
\item \textnormal{\textsc{SetAccuracy}}$(i\in[n],e^{\rm new}>0)$
replaces $e_i$ by $e^{\rm new}$ in $\widetilde O(1)$ amortized time.
\item \textnormal{\textsc{QueryProduct}}$()$ returns
$A\operatorname{Diag}(g)h$, where $h$ maximizes
$\langle\nabla\Psi(\widehat r),h\rangle$ over
$\|h\|_{\widehat\theta+\infty}\leq1$, with
$\|r-\widehat r\|_\infty\leq\alpha$ and
$\widehat\theta\approx_\alpha\theta$.
\item \textnormal{\textsc{QuerySum}}$(u\in\R^n)$ takes a sparse list
for $u$, adds $\operatorname{Diag}(g)h+u$ to the accumulated primal
vector, and returns a pointer to its coordinatewise approximation
with error at most $\alpha e_i$, together with the changed-coordinate list.
\item \textnormal{\textsc{ComputeExactSum}}$()$ returns the accumulated
primal vector exactly in $\widetilde O(n)$ time.
\item \textnormal{\textsc{Potential}}$()$ returns $\Psi(\widehat r)$ in
$O(1)$ time, for the same rounded residual used by \textnormal{\textsc{QueryProduct}}.
\end{enumerate}
Calls to \textnormal{\textsc{QueryProduct}} and \textnormal{\textsc{QuerySum}} form consecutive
pairs, with no Update or SetAccuracy between the two calls of a pair.
Let $v^{(t)}:=\operatorname{Diag}(g^{(t)})h^{(t)}$ be the dense
increment in pair $t$, and let $L_u:=\sum_{t=1}^T\|u^{(t)}\|_0$.
The total cost of $T$ query pairs is
\begin{equation}\label{eq:tall_primal_query_cost}
 \widetilde O(Td+L_u+T\alpha^{-2}
 \sum_{t=1}^T\|v^{(t)}/e^{(t-1)}\|_2^2).
\end{equation}
The accuracy-vector indexing in this bound is the indexing of
Lemma D.3 of~\cite{bllsssw21}. Initialization and coordinate changes
are charged separately. The hidden factors are polynomial in
$\alpha^{-1}$ and logarithmic in the dimension and declared horizon.
\end{theorem}
\begin{proof}
Lemma~\ref{lem:tall_primal_init} establishes the invariant and initialization
bound. Lemmas~\ref{lem:tall_primal_update} and~\ref{lem:tall_primal_accuracy}
prove the update guarantees. Lemma~\ref{lem:tall_primal_product} proves the
gradient-product guarantee. Lemma~\ref{lem:tall_primal_sum} proves the
sum guarantee and Eq.~\eqref{eq:tall_primal_query_cost}.
Lemmas~\ref{lem:tall_primal_exact} and~\ref{lem:tall_primal_potential}
prove the remaining queries. Induction on the valid procedure calls
proves preservation of the invariant.
\end{proof}

Coordinates with the same rounded residual and weight share a gradient coefficient. The product query and the sum observer must use the same coefficients.

\begin{definition}[Primal-gradient invariant]\label{def:tall_primal_gradient_invariant}
Round $r_i$ to an additive grid of mesh at most $\alpha/4$ in $[-1,1]$
and $\theta_i$ to a geometric grid of ratio $1+\alpha/4$ covering
$[\rho,4]$. Write the rounded values as $\widehat r_i,\widehat\theta_i$.
The invariant consists of the following conditions.
\begin{enumerate}[label=(\alph*)]
\item The buckets $I_1,\ldots,I_K$ group equal rounded pairs, with
$K=O(\alpha^{-2}\log(4/\rho))$. Their stored sizes satisfy $m_k=|I_k|$, and their sums satisfy
$b_k=\sum_{i\in I_k}a_i g_i$.
\item The stored potential is $\Psi_0=\sum_k|I_k|\cosh(\lambda\widehat r_k)$.
\item The stored sum observer uses the same partition and scales $g$,
with accuracy vector $\alpha e$. After $t$ completed query pairs its
exact target is
$x^{(t)}=x^{\rm init}+\sum_{j=1}^t(v^{(j)}+u^{(j)})$.
\item Between \textnormal{\textsc{QueryProduct}} and its following \textnormal{\textsc{QuerySum}},
the stored bucket coefficients $h_k$ are those used in the returned
product. The following sum query uses these same coefficients.
\end{enumerate}
The exact target is an analytic description of the sum observer;
its historical dense increments are not materialized.
\end{definition}

Algorithm~\ref{alg:tall_primal_gradient_updates} builds the buckets and the sum observer, then maintains them as entries and accuracy scales change.

\begin{algorithm}[!ht]
\caption{\textnormal{\textsc{PrimalGradient}}: initialization and updates}\label{alg:tall_primal_gradient_updates}
\begin{algorithmic}[1]
\DataStructure{PrimalGradient}\Comment{Theorem~\ref{thm:tall_primal_gradient}}
\Members
\State Fixed $A\in\R^{d\times n}$; parameters $\alpha,\lambda,C_{\rm norm}$; vectors $g,\theta,r,e$.
\State Rounded-pair buckets $I_k$, membership pointers, sizes $m_k$, sums $b_k$; potential $\Psi_0$; coefficients $h_k$.
\State Sum observer $\mathrm{sum}$ of~\cite[Lemma D.3]{bllsssw21}.
\EndMembers
\State
\Procedure{Init}{$A,x^{\rm init},g,\theta,r,e,\alpha$}\Comment{Lemma~\ref{lem:tall_primal_init}}
\State Store the inputs and form the buckets, sums, and $\Psi_0$ in Definition~\ref{def:tall_primal_gradient_invariant}.
\State \Call{\ensuremath{\mathrm{sum}}.Init}{$x^{\rm init},g,(I_k)_{k=1}^K,\alpha e$}.\Comment{\cite[Lemma D.3]{bllsssw21}}
\EndProcedure
\State
\Procedure{Update}{$i\in[n],g^{\rm new}\in\R,\theta^{\rm new}\in[\rho,4],r^{\rm new}\in[-1,1]$}\Comment{Lemma~\ref{lem:tall_primal_update}}
\State Let $k$ be the current bucket of $i$ and $k'$ the bucket of the new rounded pair.
\State $b_k\gets b_k-a_i g_i$; decrease $m_k$ by one.
\State $\Psi_0\gets\Psi_0-\cosh(\lambda\widehat r_i)+\cosh(\lambda\widehat r_i^{\rm new})$.
\State \Call{\ensuremath{\mathrm{sum}}.Scale}{$i,g^{\rm new}$}.
\State \Call{\ensuremath{\mathrm{sum}}.Move}{$i,k'$}.
\State $g_i\gets g^{\rm new}$; $\theta_i\gets\theta^{\rm new}$; $r_i\gets r^{\rm new}$; record the new rounded pair and bucket.
\State $b_{k'}\gets b_{k'}+a_i g_i$; increase $m_{k'}$ by one.
\EndProcedure
\State
\Procedure{SetAccuracy}{$i\in[n],e^{\rm new}>0$}\Comment{Lemma~\ref{lem:tall_primal_accuracy}}
\State \Call{\ensuremath{\mathrm{sum}}.SetAccuracy}{$i,\alpha e^{\rm new}$}.
\State $e_i\gets e^{\rm new}$.
\EndProcedure
\EndDataStructure
\end{algorithmic}
\end{algorithm}

Algorithm~\ref{alg:tall_primal_gradient_queries} saves the coefficients from each gradient product for the next primal-sum query. It also provides access to the exact sum and the stored potential.

\begin{algorithm}[!ht]
\caption{\textnormal{\textsc{PrimalGradient}}: queries}\label{alg:tall_primal_gradient_queries}
\begin{algorithmic}[1]
\DataStructure{PrimalGradient}\Comment{Members in Algorithm~\ref{alg:tall_primal_gradient_updates}}
\Procedure{QueryProduct}{$ $}\Comment{Lemma~\ref{lem:tall_primal_product}}
\State Set $h_k\gets0$ for every empty bucket.
\State Compute the coefficients $h_k$ for nonempty buckets by the clipped-ratio construction in Lemma~\ref{lem:tall_primal_product}; store them.
\State \Return $\sum_{k=1}^K b_kh_k$.
\EndProcedure
\State
\Procedure{QuerySum}{$u\in\R^n\text{ given as a sparse list}$}\Comment{Lemma~\ref{lem:tall_primal_sum}}
\State \Return \Call{\ensuremath{\mathrm{sum}}.Query}{$(h_k)_{k=1}^K,u$}.
\EndProcedure
\State
\Procedure{ComputeExactSum}{$ $}\Comment{Lemma~\ref{lem:tall_primal_exact}}
\State \Return \Call{\ensuremath{\mathrm{sum}}.ComputeExactSum}{$ $}.
\EndProcedure
\State
\Procedure{Potential}{$ $}\Comment{Lemma~\ref{lem:tall_primal_potential}}
\State \Return $\Psi_0$.
\EndProcedure
\EndDataStructure
\end{algorithmic}
\end{algorithm}

The sum-observer calls are those of Lemma D.3 of~\cite{bllsssw21};
we write Init for its Initialize procedure. The arguments of
Scale, Move, SetAccuracy, Query, and ComputeExactSum are unchanged.

\subsection{Initialization}
\label{subsec:tall_primal_gradient_initialization}

A single pass over the columns forms the initial bucket sums, and the sum observer starts from the given primal vector.

\begin{lemma}[Initialization]\label{lem:tall_primal_init}
The procedure \textnormal{\textsc{Init}} in Algorithm~\ref{alg:tall_primal_gradient_updates}
establishes Definition~\ref{def:tall_primal_gradient_invariant} in
$\widetilde O(n+\nnz(A))$ time, for arbitrary positive accuracy scales.
\end{lemma}
\begin{proof}
Form the rounded-pair grids and buckets, and add $a_i g_i$ to the sum
of the bucket containing $i$. The total sparse-addition cost is
$O(\nnz(A))$, and the remaining grid, bucket, and potential work is
$\widetilde O(n)$. Initialize the sum observer with additive base
$x^{\rm init}$ and zero increment histories.

The restriction that accuracy scales be at most one is unnecessary
for the sum observer. In Algorithm 8 of~\cite{bllsssw21},
the comparison endpoints are $f_k\pm|\epsilon_i/(10g_i)|$.
Multiplying the accumulated vector, $g$, sparse increments, and
$\epsilon_i$ by the same positive scalar leaves all comparisons and
operation counts unchanged. For $g_i=0$, use infinite endpoints
until $g_i$ next changes. For any finite input sequence,
a single rescaling makes every positive accuracy scale at most one.
The algorithm itself uses the original scales; this rescaling
is only a proof of its correctness and operation counts.
Thus Lemma D.3 applies to the positive vector $\alpha e$.
The initialization assignments establish all parts of the invariant.
\end{proof}

\subsection{Entry update}
\label{subsec:tall_primal_gradient_entry_update}

An update can move entry $i$ from one bucket to another. Updating the two bucket sums requires only column $i$.

\begin{lemma}[Coordinate update]\label{lem:tall_primal_update}
The procedure \textnormal{\textsc{Update}} preserves
Definition~\ref{def:tall_primal_gradient_invariant} and takes
$\widetilde O(1+\nnz(a_i))$ amortized time.
\end{lemma}
\begin{proof}
Subtracting the old contribution and adding the new contribution
preserve both bucket sums. The size and potential assignments
preserve their defining sums. The calls to Scale and Move in
Lemma D.3 materialize the old accumulated contribution before
changing its scale or bucket, and hence preserve the exact target.
Finding the rounded bucket costs $O(\log K)$ time, the sparse vector
additions cost $O(\nnz(a_i))$ time, and the two observer calls cost
$\widetilde O(1)$ amortized time. This proves the claim.
\end{proof}

\subsection{Accuracy update}
\label{subsec:tall_primal_gradient_accuracy_update}

The tolerance for one coordinate can change between queries. The sum observer records the new tolerance without changing the bucket sums.

\begin{lemma}[Accuracy update]\label{lem:tall_primal_accuracy}
The procedure \textnormal{\textsc{SetAccuracy}} preserves the exact
accumulated primal vector and changes its coordinate accuracy to
$\alpha e^{\rm new}$ in $\widetilde O(1)$ amortized time.
\end{lemma}
\begin{proof}
This is SetAccuracy in Lemma D.3 of~\cite{bllsssw21}, using the
positive-scale extension proved in Lemma~\ref{lem:tall_primal_init}.
It changes neither the rounded pairs nor the bucket sums.
\end{proof}

\subsection{Gradient-product query}
\label{subsec:tall_primal_gradient_gradient_product_query}

Because the gradient coefficient is constant within each bucket, we can compute the product from the bucket sums without scanning all columns.

\begin{lemma}[Gradient-product query]\label{lem:tall_primal_product}
The procedure \textnormal{\textsc{QueryProduct}} returns the product
in Theorem~\ref{thm:tall_primal_gradient} in
$O(K\log K+Kd)=\widetilde O(d)$ time.
\end{lemma}
\begin{proof}
We give the gradient construction explicitly. This avoids imposing
the printed condition $1/2\leq r_i\leq2$ from Theorem D.1
of~\cite{bllsssw21} on a centered residual, which may be zero or negative.
Fix an internal accuracy $\alpha>0$. Round $r_i$ to an additive
grid of mesh at most $\alpha/4$ in $[-1,1]$, and round $\theta_i$
to a geometric grid of ratio $1+\alpha/4$ covering $[\rho,4]$.
Denote the rounded quantities by $\widehat r_i,\widehat\theta_i$.
They satisfy $\|r-\widehat r\|_\infty\leq\alpha$ and
$\widehat\theta\approx_\alpha\theta$. Group equal pairs in buckets
$I_k$. There are $K=O(\alpha^{-2}\log(4/\rho))$ buckets.
For each nonempty bucket store its size and
$b_k:=\sum_{i\in I_k}a_i g_i$, where $g$ is the current signed
primal scale. A coordinate change updates these sums and its bucket
in $O(\nnz(a_i)+\log K)$ time.

Write $\Psi(r):=\sum_i\cosh(\lambda r_i)$ and
$\|h\|_{\theta+\infty}:=\|h\|_\infty+C_{\rm norm}\|h\|_\theta$,
as in Definition 4.9 of~\cite{bllsssw21}. We can compute a maximizer
of $\langle\nabla\Psi(\widehat r),h\rangle$ over
$\|h\|_{\widehat\theta+\infty}\leq1$ using only the $K$ bucket records.
Indeed, averaging coordinates within a bucket preserves the objective
and does not increase either term of the norm. Thus a maximizer is
constant on each bucket. Put $q_k:=|\lambda\sinh(\lambda\widehat r_k)|$
and $w_k:=\widehat\theta_k$. If all $q_k=0$, use $h=0$.
Otherwise the multiplier conditions for the weighted Euclidean norm
and the infinity bound give a maximizer of the form
\[
 h_i=\frac{\operatorname{sign}(\widehat r_k)\min\{q_k/w_k,t\}}
 {t+C_{\rm norm}(\sum_j|I_j|w_j\min\{q_j/w_j,t\}^2)^{1/2}}
 \quad(i\in I_k)
\]
for some $0<t\leq\max_k q_k/w_k$. To find it, sort the $K$ ratios.
On an interval between consecutive ratios, set
\[
 A_0:=\sum_{q_k/w_k\geq t}|I_k|q_k,\qquad
 B_0:=\sum_{q_k/w_k\geq t}|I_k|w_k,\qquad
 D_0:=\sum_{q_k/w_k<t}|I_k|q_k^2/w_k.
\]
The objective is
$(A_0t+D_0)/(t+C_{\rm norm}\sqrt{B_0t^2+D_0})$.
Its interior stationary points satisfy
$C_{\rm norm}(A_0-B_0t)=\sqrt{B_0t^2+D_0}$ when $D_0>0$.
Squaring gives a quadratic; retain only roots satisfying the
unsquared equation and the interval constraints. Check these roots
and all interval endpoints with $t>0$. If $D_0=0$, the objective is constant
on that interval, so choose any positive interior point. This also
attains its limit at zero without dividing by zero. Hence
arithmetic operations, comparisons, and square roots find a maximizer
in $O(K\log K)$ time. The multiplier conditions follow by fixing
$\|h\|_\infty$, aligning signs with the objective, and applying the
optimality conditions for a weighted Euclidean ball with coordinate
upper bounds. They also cover intervals on which all positive
coordinates reach the upper bound.

The product $A\operatorname{Diag}(g)h=\sum_k b_kh_k$ costs $O(Kd)$.
The same bucket coefficients are supplied to the sum observer of
Lemma D.3 of~\cite{bllsssw21}; thus its product and primal increment
use the same perturbed residual and weight vector. Taking $\alpha$
below the source error margin gives the required gradient interface.
The stored potential is $\sum_k|I_k|\cosh(\lambda\widehat r_k)$.
Bulk formation of the buckets and sums costs
$\widetilde O(n+\nnz(A))$.
\end{proof}

\subsection{Primal-sum query}
\label{subsec:tall_primal_gradient_primal_sum_query}

The product and sum queries describe the same primal step. The sum observer therefore uses the coefficients saved by the preceding product query.

\begin{lemma}[Primal-sum query]\label{lem:tall_primal_sum}
The procedure \textnormal{\textsc{QuerySum}} uses the same dense
increment as the paired product query and returns a pointer to an
approximation of the accumulated primal vector to coordinate accuracy $\alpha e_i$,
together with the changed-coordinate list.
The total time for $T$ query pairs satisfies
Eq.~\eqref{eq:tall_primal_query_cost}.
\end{lemma}
\begin{proof}
The stored coefficients, scales, partition, and accuracy vector are
unchanged between the paired calls.
Thus the product and primal increment use the same perturbed residual
and weight vector. Lemma D.3 of~\cite{bllsssw21}, with accuracy
vector $\alpha e$, proves the approximation guarantee and charges
\[
 O(TK+L_u\log n+T\alpha^{-2}\log n
       \sum_{t=1}^T\|v^{(t)}/e^{(t-1)}\|_2^2).
\]
Lemma~\ref{lem:tall_primal_product} adds $O(T(K\log K+Kd))$ for
the product queries. These bounds give
Eq.~\eqref{eq:tall_primal_query_cost}. All sums use the source
finite-horizon indexing; update and initialization costs are separate.
\end{proof}

\subsection{Exact sum and potential queries}
\label{subsec:tall_primal_gradient_exact_sum_and_potential_queries}

A bulk reset needs the exact accumulated primal vector as the starting point for the next epoch.

\begin{lemma}[Exact sum]\label{lem:tall_primal_exact}
The procedure \textnormal{\textsc{ComputeExactSum}} returns
$x^{(t)}$ exactly in $\widetilde O(n)$ time.
\end{lemma}
\begin{proof}
This is ComputeExactSum in Lemma D.3 of~\cite{bllsssw21}.
At a bulk reset its returned vector becomes the new additive base
and all increment histories start at zero.
\end{proof}

All coordinates in a bucket contribute equally to the potential. Maintaining the sum of these contributions gives a constant-time potential query.

\begin{lemma}[Potential query]\label{lem:tall_primal_potential}
The procedure \textnormal{\textsc{Potential}} returns
$\Psi(\widehat r)$ in $O(1)$ time and leaves the state unchanged.
\end{lemma}
\begin{proof}
Definition~\ref{def:tall_primal_gradient_invariant} gives
$\Psi_0=\sum_k|I_k|\cosh(\lambda\widehat r_k)=\Psi(\widehat r)$.
The procedure reads this stored scalar.
\end{proof}
\section{Dual slack data structure}\label{sec:tall_dual_slack}

In Sections~\ref{subsec:tall_dual_slack_main_data_structure} and~\ref{subsec:tall_dual_slack_initialization}, we describe and initialize the data structure for dual slacks.
In Sections~\ref{subsec:tall_dual_slack_accuracy_update} and~\ref{subsec:tall_dual_slack_dual_update}, we change the accuracy of a coordinate and report changes to the dual slack after an update.
In Section~\ref{subsec:tall_dual_slack_exact_query_and_bulk_reset}, we recover the exact dual vector and use it to reset the data structure.

\subsection{Main data structure}
\label{subsec:tall_dual_slack_main_data_structure}

\textnormal{\textsc{DualSlack}} tracks the dual increments of Section~\ref{sec:tall_sparse_lp} with a positive error scale for each signed slack. It uses the finite-horizon observer of~\cite[Lemma E.5]{bllsssw21} with the general-matrix HeavyHitter of~\cite[Lemma B.1]{bllsssw21}.

\begin{theorem}[DualSlack]\label{thm:tall_dual_slack}
Fix $A\in\R^{d\times n}$, $n\geq d\geq2$, an internal accuracy
$0<\alpha<1/4$, and a declared finite horizon $1\leq T\leq n$.
There is a randomized data structure
(\textnormal{\textsc{DualSlack}} in Algorithm~\ref{alg:tall_dual_slack})
that maintains a dual vector $s_{\rm dual}\in\R^n$, positive accuracy
scales $e\in\R_{>0}^n$, and the invariant in
Definition~\ref{def:tall_dual_slack_invariant}.
It supports the following procedures, with the source HeavyHitter
failure budgets prescribed in Section~\ref{sec:tall_sparse_lp}.
\begin{enumerate}[label=(\alph*)]
\item \textnormal{\textsc{Init}}$(A,s^{\rm init},e,\alpha,T)$ sets
$s_{\rm dual}=s^{\rm init}\in\R^n$ and all increment histories to
zero in $\widetilde O(n+\nnz(A))$ time.
\item \textnormal{\textsc{SetAccuracy}}$(i\in[n],e^{\rm new}>0)$
replaces $e_i$ by $e^{\rm new}$ in $\widetilde O(d)$ amortized time.
\item \textnormal{\textsc{Add}}$(h\in\R^d)$ replaces
$s_{\rm dual}$ by $s_{\rm dual}+A^\top h$. It returns a pointer to
an approximation $\bar s$ with $|\bar s_i-s_{{\rm dual},i}|\leq\alpha e_i$
for every $i$, and the changed-coordinate list.
\item \textnormal{\textsc{ComputeExact}}$()$ returns the current
$s_{\rm dual}$ exactly in $O(n+\nnz(A))$ time.
\end{enumerate}
The approximation guarantee holds whenever the underlying HeavyHitter
calls satisfy their guarantees. Over $t\leq T$ calls to \textnormal{\textsc{Add}},
with inputs $h^{(1)},\ldots,h^{(t)}$, their total work is
\begin{equation}\label{eq:tall_dual_query_cost}
 \widetilde O(td+t\alpha^{-2}
       \sum_{j=1}^t\|(A^\top h^{(j)})/e^{(j)}\|_{c}^2),
 \qquad c_i=d.
\end{equation}
Initialization and accuracy changes are charged separately.
A bulk reset supplies the exact current vector as $s^{\rm init}$;
it preserves the dual vector and starts fresh increment histories.
\end{theorem}
\begin{proof}
Lemma~\ref{lem:tall_dual_init} establishes the invariant and the
initialization bound. Lemma~\ref{lem:tall_dual_accuracy} proves the
accuracy-update bound. Lemma~\ref{lem:tall_dual_add} proves the Add
guarantee and Eq.~\eqref{eq:tall_dual_query_cost}.
Lemma~\ref{lem:tall_dual_exact} proves exact materialization and the
bulk-reset assertion. These results preserve the invariant by
induction on the valid calls.
\end{proof}

The initial slack and the sum of the subsequent directions determine the exact dual vector. The observer measures its approximation against this same vector.

\begin{definition}[Dual-slack invariant]\label{def:tall_dual_slack_invariant}
Let $h^{(1)},\ldots,h^{(t)}$ be the Add inputs since the latest Init.
The data structure maintains the following conditions.
\begin{enumerate}[label=(\alph*)]
\item The additive base $s^{\rm init}$ and the length-$d$ sum
$f=\sum_{j=1}^t h^{(j)}$ represent the exact dual vector
$s_{\rm dual}=s^{\rm init}+A^\top f$.
\item The observer has the same additive base, accuracy scales $e$,
and accumulated increments. Its dyadic histories, masks, and
HeavyHitter instances satisfy the invariant of the finite-horizon
implementation in~\cite[Appendix E]{bllsssw21}.
\item After every Add, its returned approximation and changed-index
list have the guarantees in Theorem~\ref{thm:tall_dual_slack}.
\end{enumerate}
The histories cover at most $T$ Add calls. The next epoch begins
with a new Init using the current exact vector as its additive base.
\end{definition}

Algorithm~\ref{alg:tall_dual_slack} adds each direction to a running sum and passes it to the observer, which reports the coordinates whose approximations change.

\begin{algorithm}[!ht]
\caption{\textnormal{\textsc{DualSlack}} data structure}\label{alg:tall_dual_slack}
\begin{algorithmic}[1]
\DataStructure{DualSlack}\Comment{Theorem~\ref{thm:tall_dual_slack}}
\Members
\State Fixed $A\in\R^{d\times n}$; additive base $s^{\rm init}\in\R^n$; sum $f\in\R^d$; accuracy scales $e>0$; parameters $\alpha,T$.
\State Finite-horizon dual observer $\mathrm{observer}$ of~\cite[Lemma E.5]{bllsssw21}.
\EndMembers
\State
\Procedure{Init}{$A,s^{\rm init},e,\alpha,T$}\Comment{Lemma~\ref{lem:tall_dual_init}}
\State Store the inputs; $f\gets{\bf 0}_d$.
\State \Call{\ensuremath{\mathrm{observer}}.Init}{$A^\top,s^{\rm init},e,\alpha,T$}.\Comment{\cite[Appendix E]{bllsssw21}}
\EndProcedure
\State
\Procedure{SetAccuracy}{$i\in[n],e^{\rm new}>0$}\Comment{Lemma~\ref{lem:tall_dual_accuracy}}
\State \Call{\ensuremath{\mathrm{observer}}.SetAccuracy}{$i,e^{\rm new}$}.
\State $e_i\gets e^{\rm new}$.
\EndProcedure
\State
\Procedure{Add}{$h\in\R^d$}\Comment{Lemma~\ref{lem:tall_dual_add}}
\State $f\gets f+h$.
\State \Return \Call{\ensuremath{\mathrm{observer}}.Add}{$h$}.
\EndProcedure
\State
\Procedure{ComputeExact}{$ $}\Comment{Lemma~\ref{lem:tall_dual_exact}}
\State \Return $s^{\rm init}+A^\top f$.
\EndProcedure
\EndDataStructure
\end{algorithmic}
\end{algorithm}

\subsection{Initialization}
\label{subsec:tall_dual_slack_initialization}

Each epoch starts from the current dual slack, with no increments yet in the observer's history.

\begin{lemma}[Initialization]\label{lem:tall_dual_init}
The procedure \textnormal{\textsc{Init}} establishes
Definition~\ref{def:tall_dual_slack_invariant} in
$\widetilde O(n+\nnz(A))$ time for arbitrary positive accuracy scales.
\end{lemma}
\begin{proof}
The imported observer uses the dyadic histories of Algorithm 9
of~\cite{bllsssw21}, with the declared horizon $T$ and fresh zero
histories at Init. Its automatic epoch restart is handled by the
explicit bulk reset here, whose additive base is retained.

Store the additive base, set the accumulated sum to zero, and bulk
initialize the dyadic HeavyHitter instances with matrix
$\operatorname{Diag}(e)^{-1}A^\top$. There are logarithmically many
instances, each with bulk cost $\widetilde O(n+\nnz(A))$.
All increment histories start at zero, so the initial target is
exactly $s^{\rm init}$.

The restriction that accuracy scales be at most one is unnecessary
for the dual observer. Simultaneously rescale its initial vector,
matrix, and accuracy vector; its HeavyHitter matrix
$\operatorname{Diag}(e)^{-1}A^\top$ and all normalized changes are
unchanged. For any finite input sequence, a single rescaling makes
every positive accuracy scale at most one. The algorithm itself
uses the original scales; this rescaling is only a proof of its
correctness and operation counts. Lemma E.5 of~\cite{bllsssw21}
therefore applies to arbitrary positive scales.
\end{proof}

\subsection{Accuracy update}
\label{subsec:tall_dual_slack_accuracy_update}

Changing the tolerance of one coordinate changes its reporting threshold while leaving the exact dual vector unchanged.

\begin{lemma}[Accuracy update]\label{lem:tall_dual_accuracy}
The procedure \textnormal{\textsc{SetAccuracy}} changes the requested
accuracy scale, preserves the exact dual vector, and takes
$\widetilde O(d)$ amortized time.
\end{lemma}
\begin{proof}
For the dual observer, SetAccuracy updates the polylogarithmically
many HeavyHitter instances; their coordinate cost is
$\widetilde O(d)$ by Lemma B.1 of~\cite{bllsssw21}.
The procedure records the new scale and flags the affected coordinate
in the dyadic histories, as in Algorithm 9 of~\cite{bllsssw21}.
Its subsequent coordinate verification is charged to this update,
as in the proof of Lemma E.5. Neither the additive base nor $f$ changes.
\end{proof}

\subsection{Dual update}
\label{subsec:tall_dual_slack_dual_update}

Adding a direction $h$ changes the dual slack by $A^\top h$. The observer reports which entries of its approximation change after this increment.

\begin{lemma}[Dual update]\label{lem:tall_dual_add}
The procedure \textnormal{\textsc{Add}} preserves
Definition~\ref{def:tall_dual_slack_invariant}, returns the stated
approximation and changed-coordinate list, and has the total cost
in Eq.~\eqref{eq:tall_dual_query_cost}.
\end{lemma}
\begin{proof}
The assignment $f\gets f+h$ changes its represented target by
exactly $A^\top h$. The finite-horizon observer receives the same
increment. Its correctness proof in Appendix E of~\cite{bllsssw21}
gives the approximation and changed-coordinate guarantees.
Lemma E.5, with query cost $Q=\widetilde O(d)$ and coordinate
costs $c_i=d$, gives Eq.~\eqref{eq:tall_dual_query_cost}.
The additional length-$d$ sum update costs $O(d)$ per call and is
included in the $td$ term. This bound uses the positive accuracy
scales $e_i$ and does not divide by the signed dual slack.
\end{proof}

\subsection{Exact query and bulk reset}
\label{subsec:tall_dual_slack_exact_query_and_bulk_reset}

At the end of an epoch, one sparse matrix multiplication and a vector addition recover the exact dual slack for the next epoch.

\begin{lemma}[Exact query]\label{lem:tall_dual_exact}
The procedure \textnormal{\textsc{ComputeExact}} returns the current
dual vector in $O(n+\nnz(A))$ time. Reinitializing with this vector
as the additive base preserves the target.
\end{lemma}
\begin{proof}
Definition~\ref{def:tall_dual_slack_invariant} gives
$s_{\rm dual}=s^{\rm init}+A^\top f$. A sparse product and vector
addition evaluate this expression in the stated time.
On a bulk reset the additive base is the exact current vector,
and all increment histories start at zero. In particular, a dual
reset retains its additive base instead of replacing it by only
the accumulated matrix product.
\end{proof}

\bibliographystyle{alpha}
\bibliography{ref}

\section*{Acknowledgements and AI Disclosure}

The author would like to thank several people who taught the author many techniques while the author was a Ph.D. student. 
The author is grateful to David Woodruff and Eric Price for teaching many foundational ideas in sketching, streaming, and sampling. The author also thanks Rajesh Jayaram for his guidance on $\ell_2$ sampling, Josh Alman for sharing his expertise in matrix multiplication, Alex Andoni for his guidance on inner product search and nearest neighbor search, and Yintat Lee for teaching techniques in convex optimization. Part of this work was carried out during a visit to Microsoft Research, hosted by Victor Reis.

The author began using AI tools extensively on this problem in late July 2026, initially working with GPT Codex 5.6 and Claude Fable 5, and switching to GPT 6 Astra and Fable 5.1 following their release in September. After several days of intensive interaction, the author identified efficient algorithms for discrepancy problems as a promising direction toward the result sought here. The questions considered involved Spencer's theorem, the matrix Spencer conjecture, Kadison--Singer, the Beck--Fiala conjecture, and the Koml\'os conjecture. After a subsequent visit to Microsoft Research, discussions with Victor Reis clarified which algorithmic and existential questions remained open in these areas. The author decided to try to solve those discrepancy problems first. As progress was made on discrepancy and related problems, the author then have a decent interaction with the Codex session devoted to LP so that the techniques could be explored in the context of the present problem.

A clearer direction emerged from finding an efficient polynomial-time algorithm for Kadison--Singer \cite{sy26}, seeing a solution of the Koml\'os conjecture \cite{gfl26a,gfl26b}, and discussing with Victor Reis how to simplify Spencer's proof \cite{rs26}. These developments led the author to recognize the choice of potential function as a central ingredient in these approaches. The author therefore asked Codex to explore a broad range of potential functions within iterative algorithmic frameworks with $O(1)$, $\widetilde{O}(1)$,  $O(\sqrt{n})$, $O(n)$, $O(n^2)$, and more general polynomial iteration counts. This exploration included the Bregman-acceleration framework from the author's work on Lewis weights \cite{s26c}, which is also used to establish the result in this paper. The suggested starting point was to investigate potential functions in the SDP formulation of Kadison--Singer, motivated by the fact that LP is a special case of SDP. The author guided Codex’s exploration of these directions for more than a month. By October 5, 2026, the author had obtained the result presented here. The author would have preferred to release this result once the matrix multiplication exponent  had been reduced to 2, but decided to release it at this point due to the rapid advances in the field and in AI tools.

Beyond the exploratory work described above, AI tools assisted with developing proof details, checking numerical calculations and illustrative examples, preparing figures, and improving the exposition. Every proof was carefully reviewed and verified by the author, with further scrutiny provided through cross-checks using multiple AI systems. The author also used AI tools to verify the proof correctness with Lean. The author takes full responsibility for all mathematical claims and written content, including material developed with AI assistance, and for the accuracy, originality, and integrity of the paper as a whole.

\end{document}